\documentclass[a4paper,oneside,12pt]{report}
\usepackage[T1]{fontenc}
\usepackage[utf8]{inputenc}
\usepackage[english]{babel}
\usepackage{amsmath}
\usepackage{amsthm}
\usepackage{amssymb}
\usepackage{amsfonts}
\usepackage{mathrsfs}
\usepackage[bf,footnotesize]{caption}
\usepackage{booktabs}
\usepackage{graphicx}
\usepackage{subfig}
\usepackage[a4paper]{geometry}
\usepackage{indentfirst}
\usepackage{microtype}
\usepackage{sidecap}
\usepackage{bm}
\usepackage{braket}
\usepackage{slashed}
\usepackage{siunitx}
\usepackage{empheq}
\usepackage{color,colortbl}
\usepackage[usenames,dvipsnames]{xcolor}
\usepackage{float}

\usepackage{setspace} 

\usepackage[backref=page]{hyperref}
\usepackage[version=3]{mhchem}
\usepackage{makeidx}
\usepackage{listings}
\usepackage{chngpage}
\usepackage{enumitem}
\setlist{parsep=0pt,listparindent=\parindent}
\usepackage{slashed}
\usepackage{pgfplots}
\pgfplotsset{compat=newest}
\usetikzlibrary{pgfplots.groupplots}

\usepackage{xcolor}

\usepgfplotslibrary{fillbetween}

\usepackage{bbm}

\usepackage{relsize}

\usepackage{hologo}

\usepackage{feynmp}
\usepackage{ifpdf}
\ifpdf
\fi

\makeatletter
\def\endfmffile{%
  \fmfcmd{\p@rcent\space the end.^^J%
          end.^^J%
          endinput;}%
  \if@fmfio
    \immediate\closeout\@outfmf
  \fi
  \IfFileExists{\thefmffile.mp}{\immediate\write18{mpost \thefmffile}}{}
  \let\thefmffile\relax
}
\makeatother

\usepackage{cite}

\usepackage{tabularx}
\usepackage{multirow}

\usepackage{algorithm}
\usepackage[noend]{algpseudocode}

\usepackage{fancyhdr}
\renewcommand{\chaptermark}[1]{\markboth{#1}{}}

 \fancypagestyle{plain}{
\fancyhead{}
}
\newcommand{\tn}{\textnormal}

\usepackage{xspace}
\newcommand{\snr}[1][]{signal-to-noise ratio#1 (SNR#1)\renewcommand{\snr}[1][]{SNR##1\xspace}\xspace}

\newcommand{\bns}[1][]{binary neutron star#1 (BNS#1)\renewcommand{\bns}[1][]{BNS##1\xspace}\xspace}

\renewcommand*{\backref}[1]{}
\renewcommand*{\backrefalt}[4]{%
    \ifcase #1 (Not cited.)%
    \or        (Cited on page~#2.)%
    \else      (Cited on pages~#2.)%
    \fi}

\usepackage{doi}

\usepackage{nameref}

\begin{document}

\interfootnotelinepenalty=10000

\unitlength = 1mm

\newgeometry{top=2cm,left=2cm,right=2cm,bottom=2cm}

\begin{titlepage}

      \begin{minipage}{0.45\textwidth}
  \includegraphics[width=\textwidth]{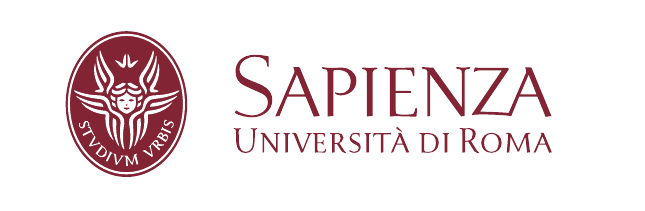} 
\end{minipage}
      \begin{minipage}{0.5\textwidth}
    \begin{center}
 \large
 Dipartimento di Fisica   \\ 
 Scuola di Dottorato ``Vito Volterra''\\
Sapienza -- Università di Roma 
    \end{center}
\end{minipage}   

\hrule

     \begin{center}
    \vspace{4cm} 
      \LARGE
     \textbf{\color{Maroon}{Tidal deformations of compact objects \\ and gravitational wave emission}}

        \vspace{3cm}
        \Large
        \textbf{Tiziano Abdelsalhin}
        
         \vspace{2cm}
        
       \large
        Thesis submitted for the degree of\\
        Doctor of Philosophy in Physics
        
        \vspace{2cm}
        
        Advisor:\\
        Prof. Leonardo Gualtieri
           \vspace{5cm}
\hrule
   \vspace{0.5cm}
        Defended on 12th February 2019
        
    \end{center}
\end{titlepage}

\restoregeometry

\tableofcontents

\chapter*{Introduction}
\chaptermark{Introduction}
\addcontentsline{toc}{chapter}{Introduction}
More than a century ago, just one year after the formulation of his revolutionary theory of gravitation --\emph{General Relativity}--, Einstein himself predicted the existence of \emph{gravitational waves}, tiny spacetime oscillations which propagate through time and space at the speed of light. Gravitational waves interact so weakly with matter that Einstein wondered if their detection would ever have been possible. Indeed, due to their weakness, catastrophic events and ultra-sensitive instruments are required to produce and observe gravitational waves.

One hundred years later, after decades of experimental and theoretical efforts this goal has been achieved. In 2015, the gravitational waves emitted from the coalescence of a binary black hole was detected by the LIGO interferometers~\cite{Abbott:2016blz}, two ground-based gravitational wave detectors. This first gravitational wave event, dubbed GW150914, has started the era of gravitational wave astronomy, opening a new window to look at the high-energy phenomena of our Universe.

During the first LIGO observation run (O1), other two gravitational wave signals produced by binary black hole coalescences were detected~\cite{Abbott:2016nmj,TheLIGOScientific:2016pea}. In 2017, during the second observation run (O2), two binary black hole mergers were observed by the LIGO sites~\cite{Abbott:2017vtc,Abbott:2017gyy}. Moreover, the Virgo interferometer finally joined the quest, greatly improving the capability of the network of detectors to localize the sky-position of the gravitational sources. This led to the first three-detector observation of a double black hole system~\cite{Abbott:2017oio}. Recently, the LIGO/Virgo collaboration has reported the observation of gravitational waves from other four binary black hole mergers detected in O2~\cite{LIGOScientific:2018mvr}.

Lastly, in August 2017, a gravitational wave signal from the coalescence of a binary neutron star system was detected for the first time by the LIGO and Virgo interferometers\cite{TheLIGOScientific:2017qsa}. This event, dubbed GW170817, represents a milestone in gravitational wave astronomy. Indeed, GW170817 is the first event observed simultaneously in both the gravitational and electromagnetic bands. Several telescopes reported the observation of a short gamma ray burst in coincidence with the gravitational wave signal~\cite{GBM:2017lvd,Monitor:2017mdv}, marking the birth of the gravitational wave multi-messenger astronomy. Furthermore, few hours after the merger, various teams (see, e.g.,~\cite{Coulter:2017wya}) detected a bright optical afterglow --the so-called \emph{kilonova}-- powered by the radioactive decay of heavy $r$-process nuclei synthesized in the ejecta of the neutron star merger.

During the next few years (starting with the third observation run (O3) in 2019), many other gravitational wave signals from the coalescence of compact binaries are expected by the LIGO/Virgo collaboration~\cite{Abbott:2017vtc,TheLIGOScientific:2017qsa}. This incoming flood of data will be extremely precious to test gravity in the highly-relativistic/strong-curvature regime, and to investigate the behavior of matter in extreme conditions. In particular, with more binary neutron star coalescence signals, it will be possible to explore the structure and composition of matter at supranuclear densities~\cite{Hinderer:2009ca,Maselli:2013rza,DelPozzo:2013ala,Lackey:2014fwa,Wade:2014vqa,Harry:2018hke}.

Indeed, a precise description of matter in this regime is currently unavailable, due to the lack of experimental data and the complexity of modeling strong interactions among hadrons above the nuclear saturation point. Therefore, the so-called \emph{equation of state} of matter (i.e., the thermodynamical relation between pressure and density) inside neutron star cores is currently uncertain, and represents an open problem in nuclear astrophysics. Various theoretical models have been developed so far, which predict different scenarios for the nuclear matter~\cite{Lattimer:2000nx}.

As GW170817 has already shown~\cite{TheLIGOScientific:2017qsa,Abbott:2018wiz,Abbott:2018exr,De:2018uhw,Annala:2017llu,Most:2018hfd,Bauswein:2017vtn,Raithel:2018ncd}, gravitational wave detections from coalescing binary neutron star systems will provide information on neutron star matter complementary to that coming from electromagnetic observations~\cite{Steiner:2010fz,Guillot:2014lla,Ozel:2015fia} (see~\cite{Ozel:2016oaf} for a review), shedding new light on the neutron star interior and constraining the equation of state. On this purpose, it is crucial to develop more accurate models of binary systems in order to best exploit the potentiality of current second-generation detectors and future third-generation interferometers (such as the Einstein Telescope~\cite{Hild:2010id} and the Cosmic Explorer~\cite{Evans:2016mbw}).

Gravitational wave searches and parameter estimation pipelines rely on gravitational waveform approximants that describe the inspiral, merger, and post-merger phases of the coalescence. While the early-inspiral (low-frequency) phase is accurately described by the post-Newtonian theory~\cite{Arun:2008kb,Buonanno:2009zt,Mishra:2016whh} (i.e., a weak-field/slow-velocity expansion of Einstein field equations, see~\cite{Blanchet:2013haa} for a review), this description breaks down when the two compact objects get closer, and finite-size and strong-gravity effects start to gain importance. 

Thus, a major challenge in the parameter estimation of neutron star binaries is the modeling of the gravitational signal during the late-inspiral, merger, and post-merger phases. This is typically achieved by using gravitational wave templates obtained either phenomenologically (the so-called Phenom approximants~\cite{Santamaria:2010yb,Hannam:2013oca,Husa:2015iqa,Khan:2015jqa}) or using the effective-one-body (EOB) approach~\cite{Buonanno:1998gg,Bernuzzi:2014owa,Bernuzzi:2015rla,Hinderer:2016eia} (the so-called SEOBNR approximants~\cite{Pan:2013rra,Purrer:2014fza,Bohe:2016gbl}), recalibrated by fitting to numerical relativity solutions~\cite{Dietrich:2017feu,Dietrich:2017aum,Dietrich:2018uni}. However, though these templates correct the deviations in the high-frequency regime, they are constrained to recover the analytical post-Newtonian solutions at low frequencies. Therefore, an accurate description of the early-inspiral phase as described by the post-Newtonian formalism is an essential ingredient of these waveform approximants. Any new post-Newtonian term included in the expansion would also propagate to the full waveform templates.

Within the post-Newtonian formalism, the inspiralling dynamics of the binary is driven by the loss of energy through gravitational wave emission, and the two bodies are modeled as two point-particles~\cite{Blanchet:2013haa} endowed with a series of multipole moments~\cite{Damour:1990pi,Damour:1991yw} and with finite-size tidal corrections~\cite{Vines:2010ca}. The latter are encoded in the so-called \emph{tidal Love numbers}/\emph{tidal deformabilities}, a set of coupling constants which characterize the multipolar deformation of the star induced by the external tidal field generated by its companion~\cite{Hinderer:2007mb} (see~\cite{Poisson:2014} for a review of the theory of tidal Love numbers in Newtonian gravity).

The relativistic theory of tidal Love numbers for non-spinning compact objects is well established in the literature~\cite{Hinderer:2007mb,Damour:2009vw,Binnington:2009bb,Landry:2015cva}. It was shown that the tidal Love numbers of a Schwarzschild black hole vanish exactly, whereas those of a neutron star depend on its equations of state. In recent years, the theory of tidal Love numbers has been extended to spinning objects~\cite{Pani:2015hfa,Pani:2015nua,Landry:2015zfa,Landry:2017piv,Gagnon-Bischoff:2017tnz}. The coupling between the tidal fields and the angular momentum introduces a new family of tidal Love numbers, dubbed \emph{rotational tidal Love numbers}. Also in the spinning case, it was shown that the tidal Love numbers of a slowly rotating Kerr black are precisely zero, while those a of slowly spinning neutron star depend on the equation of state.

Tidal deformations of neutron stars introduce a correction (starting at the fifth post-Newtonian order~\cite{Damour:1982wm,Damour:1984rbx}) to the waveform phase of the gravitational radiation emitted from binary systems~\cite{Flanagan:2007ix,Vines:2010ca,Vines:2011ud,Damour:2012yf,Yagi:2013sva,Banihashemi:2018xfb}. Up to now, this correction was computed only for non-spinning objects, i.e., neglecting the coupling between the angular momentum of one body and the tidal field produced by its companion. The tidal correction is proportional to the star tidal deformabilities, which are the only parameter that encodes the dependence of the gravitational waveform on the neutron star internal structure during the inspiral phase. Therefore, the measurement of the tidal deformability from gravitational wave detections by ground-based interferometers allows us to discriminate among equations of state proposed in the literature.

The first binary neutron star coalescence detected, GW170817, has already allowed us to constrain the neutron star equation of state (ruling out some of the proposed models), by extracting the leading-order, quadrupolar tidal deformability term from the gravitational waveform~\cite{TheLIGOScientific:2017qsa,Abbott:2018wiz,Abbott:2018exr}. With other gravitational wave signals from coalescing neutron star binaries expected in the next future~\cite{TheLIGOScientific:2017qsa}, it will be possible, through the measurement of tidal deformabilities, to put accurate bounds on the equation of state~\cite{DelPozzo:2013ala}.

To the present day, systematic studies to infer the features of the neutron star internal composition from astrophysical observations have been possible only in the electromagnetic band, in the so-called \emph{relativistic inverse stellar structure problem}~\cite{1992ApJ...398..569L,Ozel:2009da,Lindblom:2012zi,Lindblom:2013kra}: namely, reconstruct the microscopical properties of the equation of state from the measurement of macroscopical neutron star observables. Various groups exploited electromagnetic observations of neutron star masses and radii to constrain the high-density region of the equation of state~\cite{Steiner:2010fz,Guillot:2014lla,Ozel:2015fia}. The same approach is possible also in the gravitational band~\cite{Lackey:2014fwa}, where the radii, whose measurements are in general affected by large uncertainties, are replaced by tidal deformabilities, which can in principle provide tighter bounds as more events are observed by the advance generation of detectors~\cite{DelPozzo:2013ala}.

\ 

In this thesis, I study the tidal deformations of compact objects in binary systems, and the corresponding gravitational radiation emitted, within two lines of research. In the first one, I improve the post-Newtonian modeling of inspiralling compact binary systems, by computing the leading-order tidal corrections to the dynamics of spinning binaries, and to the corresponding waveform phase of the gravitational radiation emitted, to linear order in the spin (see section~\ref{sec:truncation}). So far, these corrections have been computed only for non-rotating objects.

The corrections arising from the spin-tidal couplings that affect the dynamics of two orbiting bodies belong to two classes: i) terms coming from the interaction between the ordinary tidal terms and the point-particle terms (namely, the spins), which depend on the standard tidal Love numbers; ii) terms depending on the rotational tidal Love numbers of spinning bodies recently introduced in the literature. 

The spin-tidal terms could be included in phenomenological/EOB models to obtain more accurate gravitational waveform templates. Although neutron stars in coalescing binaries are expected to rotate rather slowly~\cite{Brown:2012qf,Kastaun:2013mv,TheLIGOScientific:2017qsa}, neglecting the spin-tidal coupling might introduce systematic errors in the parameter estimation of the gravitational sources~\cite{Favata:2013rwa,Harry:2018hke}. This is specially important for the estimate of the tidal deformability, which affects the gravitational signal at relative high frequencies, where these spin-tidal higher-order corrections are larger. In this regard, I also estimate the impact of the new spin-tidal terms by analizing the parameter bias induced on GW170817-like events, assuming second and third-generation detectors (see section~\ref{sec:res}). 

Lastly, spin-tidal corrections might be important to improve current tests of the real nature of black holes (against other models of exotic compact objects (ECO) proposed in the literature~\cite{Cardoso:2017cqb,Cardoso:2017njb}) using the tidal effects in the inspiral~\cite{Cardoso:2017cfl,Maselli:2017cmm,Sennett:2017etc,Johnson-McDaniel:2018uvs}.

In the second line of research, I am instead interested in solving the inverse stellar problem (i.e., constrain the neutron star equation of state) using detections of gravitational wave signals emitted by coalescing binary neutron stars. I show the feasibility of reconstructing the parameters of a phenomenological representation of the equation of state from measurements of the stellar masses and tidal deformabilities. 

Phenomenological parametrizations~\cite{Read:2008iy,Lindblom:2010bb,Steiner:2010fz} of the neutron star equation of state provide an effective approach to solve the inverse stellar problem~\cite{Ozel:2009da,Lindblom:2012zi,Lackey:2014fwa,Ozel:2016oaf}, since they allow us to describe a large class of equation of state models through a relatively small set of coefficients, to be constrained by astrophysical data. These representations can be used to combine measurements of different neutron star observables, resulting specially suited to obtain multi-band constraints on the equation of state. Furthermore, it might be possible that the \emph{true} neutron star equation of state differs from the models proposed in the literature so far. Then, a phenomenological approach would be extremely useful to constrain the main properties of the correct equation of state.

I perform a Bayesian analysis of simulated masses and tidal deformabilities, modeling the neutron star equation of state through a piecewise polytropic parametrization~\cite{Read:2008iy}. I assume to detect gravitational wave signals emitted from coalescing neutron star binaries by a network of advanced interferometers at design sensitivity. My results suggest that a small number of gravitational wave detections would allow us to constrain the equation of state parameters, and to perform a
model selection among various equations of state proposed in the literature (see section~\ref{sec:setup}).

\ 

The results obtained in this thesis on the spin-tidal interactions are published in~\cite{Abdelsalhin:2018reg} and~\cite{Jimenez-Forteza:2018buh}, whereas those on the inverse stellar problem in~\cite{Abdelsalhin:2017cih}. Furthermore, though not explicitly discussed in this thesis, during my PhD I have contributed to other two publications related to tidal effects, namely Refs.~\cite{Maselli:2017cmm} and~\cite{Pani:2018inf}.

\ 

The structure of the thesis is the following. In Chapter~\ref{sec:star} I review the main features of neutron stars and their tidal deformations. In Chapter~\ref{sec:binary} I describe the tidal deformations in compact binary systems within the post-Newtonian formalism. I present my results on the spin-tidal interactions, and the following parameter estimation analysis of spinning neutron star binaries. In Chapter~\ref{sec:inverse} I introduce the inverse stellar problem, and show the results on the inference of the phenomenological parameters of the equation of state from gravitational wave detections. Finally, in~``\nameref{sec:conclu}'' I draw my conclusions and present possible extensions of my work.

Furthermore, there are three appendices. In Appendix~\ref{sec:appA} I provide some useful equations to computing numerically mass, radius and tidal deformability of a neutron star. In Appendix~\ref{sec:appB} I summarize the main properties of the spherical harmonics. Lastly, in Appendix~\ref{sec:appC} the reader can find some additional material on the study carried on in Chapter~\ref{sec:inverse}.

\section*{Notation}
\addcontentsline{toc}{section}{Notation}
\label{sec:notation}
In this section we summarize the main conventions adopted in this thesis. However, we explicitly describe any new symbol in the text, when it is introduced for the first time, or some variation occurs. Also, reminders and links to this section are often present in the text, when needed.

We use the spacetime metric with signature (-,+,+,+). Greek indices run over all four-dimensional coordinates, whereas Latin indices run only over three-dimensional spatial coordinates. We adopt the Einstein convention, i.e., repeated indices are implicitly summed over. We denote ordinary derivatives by $\partial_{\mu} \equiv \frac{\partial}{\partial x^{\mu}}$ and covariant derivatives by $\nabla_{\mu}$. Derivatives with respect to radial coordinates are also expressed by primes, and derivatives with respect to time coordinates by overdots. 

We denote the metric tensor by $g_{\mu \nu}$; the Riemann curvature tensor by $R^{\mu}_{\ \nu \alpha \beta}$,
\begin{equation*}
R^{\mu}_{\ \nu \alpha \beta} = \partial_{\alpha} \Gamma^{\mu}_{\ \nu \beta} -  \partial_{\beta} \Gamma^{\mu}_{\ \nu \alpha} + \Gamma^{\mu}_{\ \alpha \gamma} \Gamma^{\gamma}_{\ \nu \beta} - \Gamma^{\mu}_{\ \beta \gamma} \Gamma^{\gamma}_{\ \nu \alpha } \,,
\end{equation*}
where $\Gamma^{\alpha}_{\ \beta \gamma}$ are the Christoffel symbols
\begin{equation*}
\Gamma^{\alpha}_{\ \beta \gamma} = \frac{1}{2} g^{\alpha \delta} \left(\partial_{\beta} g_{\gamma \delta}+ \partial_{\gamma} g_{\beta \delta}-\partial_{\delta} g_{\beta \gamma}  \right) \,;
\end{equation*}
the Ricci tensor by $R_{\mu \nu} = R^{\alpha}_{\ \mu \alpha \nu}$; the scalar curvature by $R=R^{\alpha}_{\ \alpha}$; the Einstein tensor by $G_{\mu \nu}= R_{\mu \nu}- \frac{1}{2} g_{\mu \nu} R $. 

The (three-dimensional) Kronecker delta is denoted by $\delta^{ij}$ and the (three-dimensional) complete antisymmetric Levi-Civita symbol by $\epsilon^{ijk}$. Following~\cite{Thorne:1980ru}, we adopt the multi-index notation: we use capital letters as shorthand for multi-indices, $T^L \equiv T^{a_1 \dots a_l}$. Round $(\ )$, square $[\ ]$, and angular $\langle \ \rangle$ brackets in the indices indicate symmetrization, antisymmetrization and trace-free symmetrization, respectively. For instance,
\begin{equation*}
\begin{aligned}
T^{(ab)} = &\frac{1}{2} \left( T^{ab} + T^{ba} \right) \\
T^{[ab]} =& \frac{1}{2} \left( T^{ab} - T^{ba} \right) \\
T^{\langle ab \rangle} =&  T^{(ab)} - \frac{1}{3} \delta^{ab} T^{cc} \,.
\end{aligned}
\end{equation*}
We also define $u^{ij\dots k} \equiv u^i u^j \dots u^k$, where $u^i$ is a generic vector.

We denote the speed of light in vacuum by $c$ and the gravitational constant by $G$. In Chapter~\ref{sec:star} and~\ref{sec:inverse}, we mainly use geometric units $c= G=1$ (any (rare) exception is explicitly reported in the text), whereas in Chapter~\ref{sec:binary} (see below) we set only $G=1$. The reduced Planck constant is expressed by $\hbar$ and the solar mass unit by $M_{\odot}$. In this thesis we use ``tidal Love number'' and ``tidal deformability'' as synonyms.

\subsection*{Notation of Chapter 2}
\addcontentsline{toc}{subsection}{Notation of Chapter 2}
\label{sec:notation_2}
Since in Chapter~\ref{sec:binary} there are many computations, and the notation is slightly different from that of the other parts of the thesis, for the sake of clarity we report in this section a notation especially dedicated to it, and valid \emph{only} there. The reader can refer to this section while going through Chapter~\ref{sec:binary} (therefore, (s)he can safely avoid reading the following before that time).

We denote the speed of light in vacuum by $c$ and set the gravitational constant $G=1$. Latin indices $i,j,k$, etc. run over three-dimensional spatial coordinates and are contracted with the Euclidean flat
metric $\delta^{ij}$. Since there is not distinction between upper and lower spatial indices, we use only the upper ones. The Levi-Civita symbol is denoted by $\epsilon^{ijk}$. We use capital letters in the middle of the alphabet $L,K$, etc. as shorthand for multi-indices $a_1 \dots a_l$, $b_1\dots b_k$, etc. Round $(\ )$, square $[\ ]$, and angular $\langle \ \rangle$ brackets in the indices indicate symmetrization, antisymmetrization and trace-free symmetrization, respectively (see above). We call \emph{symmetric trace-free} (STF) those tensors $T^{i_1\dots i_{l}}$ which are symmetric on all indices and whose contraction of any pair of indices vanishes
\begin{equation*}
\begin{gathered}
T^{\langle i_1\dots i_{l} \rangle }  = T^{(i_1\dots i_{l})} = T^{i_1\dots i_{l}}  \\
T^{i_1\dots i_k i_k \dots i_{l}} = 0 \quad  \forall k  \,.
\end{gathered}
\end{equation*}
The contraction of a STF tensor $T^L$ with a generic tensor $U^L$ is $T^L U^L= T^L U^{\langle L \rangle}$. For a generic vector $u^i$ we define $u^{ij\dots k} \equiv u^i u^j \dots u^k$ and $u^2 \equiv u^{ii}$. Derivatives with respect to the coordinate time $t$ are expressed by overdots.

For a generic body $A$, the mass and current multipole moments are denoted by $M^L_A$ and $J^L_A$, respectively. We indicate the electric and magnetic tidal moments, which affect the body $A$, respectively by $G^L_A$ and $H^L_A$. All of them are STF tensors on all indices.

Restricted to a two-body system, $A =1,2$, we define the mass ratios $\eta_A = {}^n M_A/M$, where $M = {}^n M_1 + {}^n M_2 $ is the total mass and ${}^n M_A$ is the mass monopole $M_A$ in the Newtonian limit. The symmetric mass ratio is $\nu = \eta_1 \eta_2$ and the reduced mass is $\mu = \nu M$. We define the dimensionless spin parameters $\chi_A = c J_A/(\eta_A M)^2$, where $J_A = \sqrt{J^i_A J^i_A}$ is the absolute value of the current dipole moment. The body position, velocity and acceleration vectors are denoted by $z_A^i$, $v_A^i = \dot{z}_A^i$ and $a_A^i = \ddot{z}_A^i$, respectively. We define the two-body relative position, velocity and acceleration vectors by $z^i = z_2^i - z_1^i$, $v^i= v_2^i - v_1^i$ and $a^i = a_2^i - a_1^i$, respectively. We also define the relative unit radial vector $n^i = z^i/r$, where $r = \sqrt{z^i z^i}$ is the orbital separation. We define the derivatives with respect to the spatial coordinates $z^i$ as $\partial_L = \partial_{i_1}\dots\partial_{i_l}$. In particular, we denote the derivatives with respect to $z_A^i$ by $\partial_L^{(A)}$.

We shall denote $\lambda_l$ ($\sigma_l$) the electric (magnetic) tidal Love numbers of multipolar order $l$, whereas $\lambda_{ll'}$ and $\sigma_{ll'}$ are the rotational tidal Love numbers. For our computation, it is sufficient to consider that the multipole moments higher than the dipole are induced only on the second body by the tidal field produced by its companion. For this reason, to avoid burdening the notation, we define the quadrupolar and octupolar moments as $Q^{ab} \equiv M_2^{ab}$, $Q^{abc} \equiv M_2^{abc}$, $S^{ab} \equiv J_2^{ab}$ and $S^{abc} \equiv J_2^{abc}$.

Finally, for a binary system in circular orbit we define the post-Newtonian (PN) expansion parameter $x = (\omega M)^{2/3}/c^2$, where $\omega$ is the orbital angular velocity.

\subsection*{Abbreviations}
\addcontentsline{toc}{subsection}{Abbreviations}
\begin{description}
\item[APR4] model of equation of state~\cite{Akmal:1998cf}
\item[CGS] Centimetre–Gram–Second
\item[COM] Center-Of-Mass
\item[ECO] Exotic Compact Object
\item[EOS] Equation Of State
\item[FIM] Fisher Information Matrix
\item[H4] model of equation of state~\cite{Lackey:2005tk}
\item[LHS] Left-Hand-Side
\item[MCMC] Markov Chain Monte Carlo
\item[NBMT] Nonrelativistic Many-Body Theory
\item[ODE] Ordinary Differential Equation
\item[PDF] Probability Density Function
\item[PM] Post-Minkowskian
\item[PN] Post-Newtonian
\item[QCD] Quantum ChromoDynamics
\item[RHS] Right-Hand-Side
\item[RMFT] Relativistic Mean Field Theory
\item[SNR] Signal-to-Noise Ratio
\item[STF] Symmetric Trace-Free
\item[TOV] Tolman–Oppenheimer–Volkoff
\end{description}

\chapter{Neutron stars and their tidal deformations}
\label{sec:star}
Neutron stars are one of the final products of the stellar evolution. At the end of the thermonuclear evolution of a star, the matter pressure can not support any longer the gravitational force, leading the star to collapse. If the progenitor mass is in the range $M \sim (8 \div 30) M_{\odot}$, then the internal temperature is high enough to ignite the burning of heavier and heavier elements through exothermic nuclear reactions, up to the formation of $^{\text{56}}\text{Fe}$ inside the star core. Together with the heavy core formation, several mechanisms arise which contribute to destabilize the star and produce a large number of neutrons.

Neutrinos are produced through the silicon burning
\begin{equation*}
2 {\,}^{\text{28}}\text{Si}  \to  {\,}^{\text{56}}\text{Fe} + 2 \text{e}^+ + 2 \nu_{\text{e}} + \gamma \,,
\end{equation*}
and by electron capture in the inverse $\beta$-decay
\begin{equation*}
\text{p} + \text{e}^-   \to  \text{n} + \nu_{\text{e}} \,.
\end{equation*}
Since neutrinos interact weakly with matter, they leave the star undisturbed, carrying away energy from the core. Furthermore, the replacement of relativistic electrons by neutrons decreases the pressure of the star. Formation of neutron-rich elements, heavier than iron, by neutron capture, subtracts more energy to the core. Lastly, the iron photodisintegration
\begin{equation*}
\gamma + {\,}^{\text{56}}\text{Fe} \to 13 {\,}^4 \text{He} + 4 \text{n} \,,
\end{equation*}
that is an endothermic nuclear reaction, removes further energy from the star.

All these processes damage the equilibrium of the star. When the mass of the core exceeds the Chandrasekhar limit $M \sim 1.4 M_{\odot}$, the electron degeneracy pressure can not balance the gravitational attraction anymore, and the core collapses, reaching densities comparable to those of the atomic nuclei, $\rho \sim 10^{14} \, \text{g/cm}^{3}$. At this point the core is composed mostly of neutrons, and reacts to further compressions due to infalling matter with a strong shock wave that ejects the outer layers of the star in the so-called supernova explosion. The remnant of the core is the newly born neutron star.

Neutron stars are very hot at birth. However, they cool fastly via neutrino emission, reaching temperatures $T < 10^{9} \, \text{K}$ after just few years. The most efficient cooling mechanism is provided by direct Urca processes, i.e., direct and inverse neutron $\beta$-decays
\begin{equation*}
\text{n}  \to   \text{p} + l^- + \bar{\nu}_{l} \qquad \text{p}  +  l^-  \to  \text{n} + \nu_{l} \,,
\end{equation*}
where $l$ represents either an electron or a muon. Direct Urca processes can occur only in the inner regions of the core, where the densities of protons and leptons are high enough to satisfy momentum conservation. Others cooling mechanisms are the modified Urca process
\begin{equation*}
\text{n} + \text{N} \to   \text{p} + \text{N} + l^- + \bar{\nu}_{l} \qquad \text{p} + \text{N} +  l^-  \to  \text{n} + \text{N} + \nu_{l} \,,
\end{equation*}
which differs from the direct Urca process due to the presence of an additional nucleon-spectator $\text{N}$ (either a proton or a neutron), and the neutrino bremsstrahlung due to nucleon-nucleon collisions
\begin{equation*}
\text{N} + \text{N}' \to  \text{N} + \text{N}' + \nu + \bar{\nu} \,,
\end{equation*}
which can produce neutrinos of any flavour.

\begin{figure}[]
\captionsetup[subfigure]{labelformat=empty}
\centering
{\includegraphics[width=0.4\textwidth]{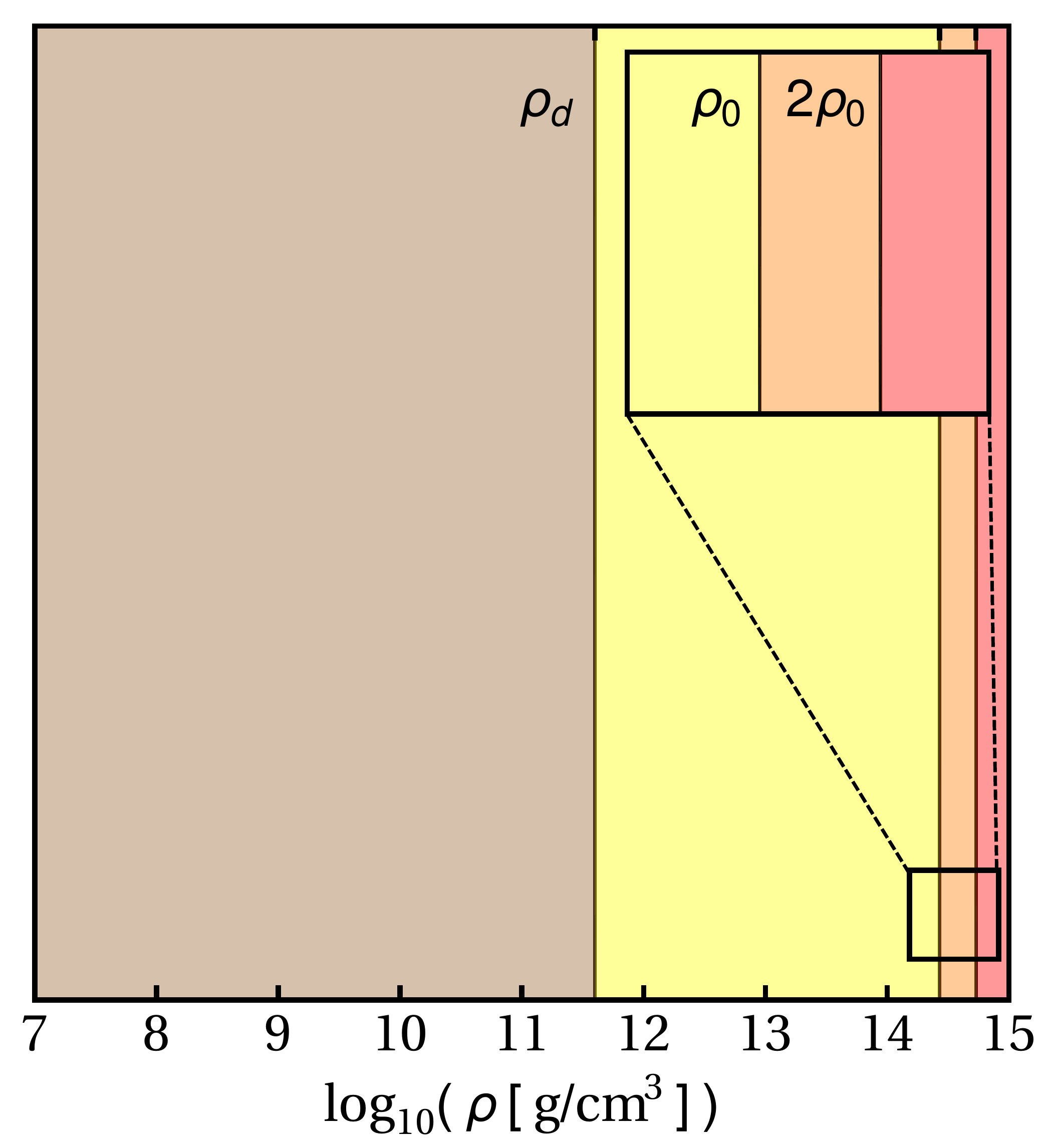}} \ \ \ 
{\includegraphics[width=0.4\textwidth]{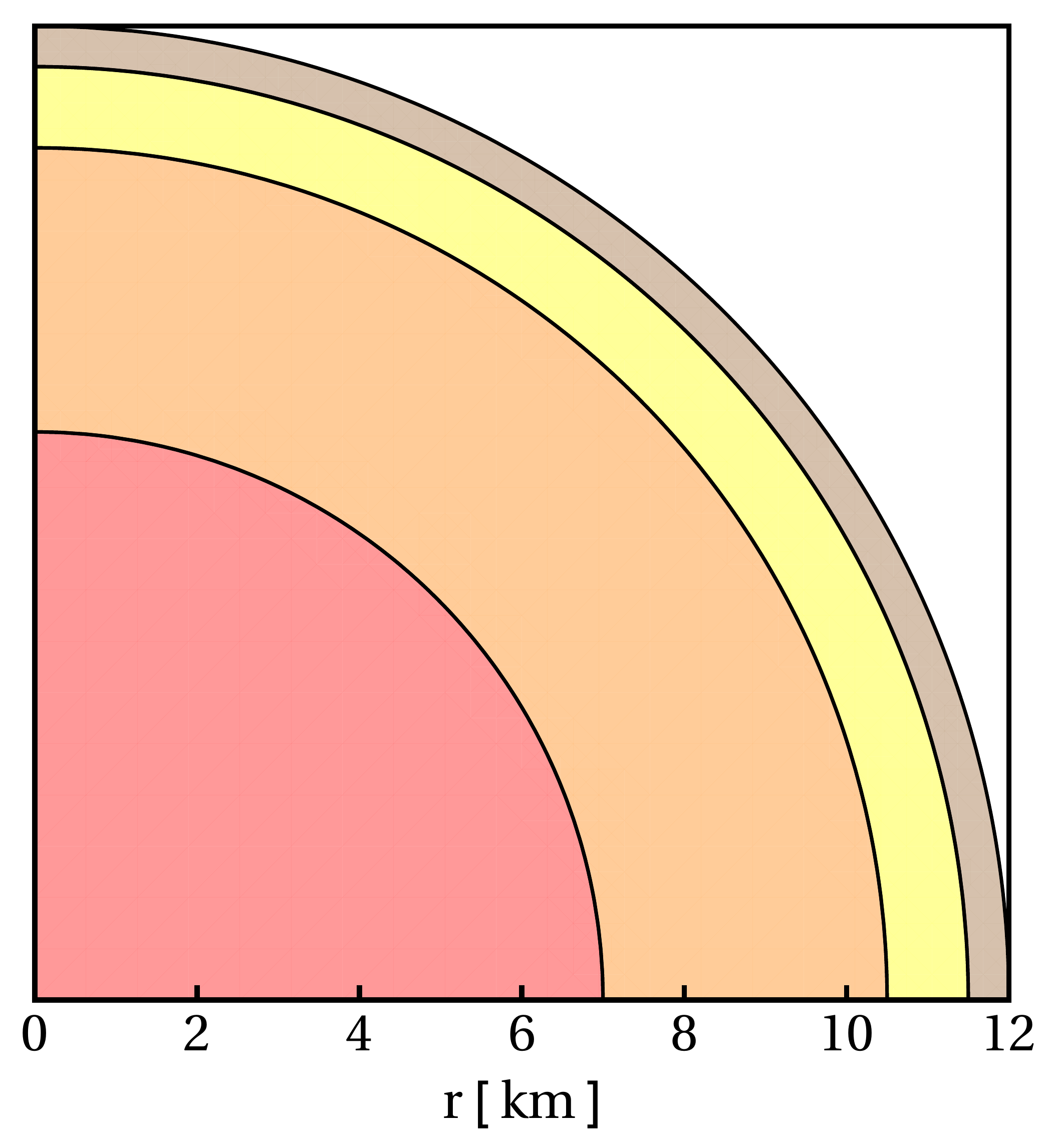}}
\caption{\textsl{Schematic plot of the interior of a neutron star. Density range (left) and thickness (right) of each layer are compared, see the text for details. We remark that the thickness of the outer and inner core can differ sensibly from those shown here, depending on the neutron star mass and on how matter is modeled (see section~\ref{sec:tabeos} and Chapter~\ref{sec:inverse}).}}
\captionsetup{format=hang,labelfont={sf,bf}}
\label{fig:neutron_star}
\end{figure}

The fast cooling justifies the assumption that the matter inside an ``old'' neutron star is cold, i.e., it behaves like matter at a temperature $T = 0\, \text{K}$. Indeed, the Fermi temperature of neutrons at densities typical of a neutron star is of order $T_F \sim (10^{11} \div 10^{12})\, \text{K}$, which is much larger than the temperature of a neutron star after just one year from birth. Therefore, the matter is strongly degenerate, and can be effectively considered as it were at the absolute zero.

In this chapter we review the structure and composition of cold, isolated neutron stars. In section~\ref{sec:compo} we describe the equation of state of cold nuclear matter and recall the equations of stellar structure of relativistic stars, taking into account also the rotation of the compact objects. Then, in section~\ref{sec:TLN} we review the theory of tidally deformed compact objects in General Relativity. We define the tidal Love numbers and describe the linear perturbations of a spherical background. Finally, we introduce the tidal deformations of a spinning object and describe some of the universal relations among the Love numbers.

\section{Structure and composition of neutron stars}
\label{sec:compo}
In this section, we mainly use the books of Haensel et al.~\cite{Haensel:2007} and Glendenning~\cite{Glendenning:2000} as references for the description of the internal composition of a neutron star and its the equation of state.

Neutron stars are astrophysical compact objects with a typical mass $M \sim 1.4 M_{\odot}$, radius of order $R \sim 10\, \text{km}$ and average density $\rho \sim 10^{14} \, \text{g/cm}^{3}$. The internal structure of a neutron star is modeled as a sequence of spherical shells with different density and composition. It can be divided in four main internal regions: outer crust, inner crust, outer core and inner core.
\begin{description}
\item[Outer Crust] is thick $\sim 0.5\, \text{km}$, from the surface of the star (where the density is around $\rho \sim 10^{7} \, \text{g/cm}^{3}$), up to a layer of density $\rho_d \sim 4\times  10^{11}\, \text{g/cm}^3$, the so-called neutron drip density. The matter in this region is mostly composed of a heavy nuclei lattice immersed in a degenerate electron gas. The latter provides the main contribution to the pressure in this region. Going towards the center of the star, as density increases, more and more neutrons are produced by inverse $\beta$-decay. At the neutron drip density $\rho_d$, all bound states in the nuclei for neutrons are filled and neutrons start leaking out. The properties of matter in this region are obtained directly by experimental data coming from nuclear physics experiments made in laboratory on Earth (for instance, heavy-ion collisions~\cite{Chen:2010qx}).
\item[Inner Crust] is thick $\sim 1\, \text{km} $ and the density ranges from $\rho_d$ to $\rho \sim \rho_0$, where $\rho_0 \sim 2.7\times 10^{14}\, \text{g/cm}^3$ is the equilibrium density of nuclear matter. The matter of the inner crust is composed of a mixture of two phases: neutron-rich nuclei and a degenerate neutron gas, besides the electron gas required to ensure charge neutrality. As the density increases, these two phases combine in different geometric structures, called pasta phases. At density $\rho \sim \rho_0$, the two phases are not separated any longer and form a homogeneous fluid. The properties of matter in the inner crust are based on extrapolations of the available empirical information, since such extreme densities can not be reproduced in a stable way on Earth.
\item[Outer Core] extends for several km and the density range is $\rho_0 \lesssim \rho \lesssim 2\rho_0$. All hadronic nuclear physics models generally agree that the matter in this region is composed of a homogeneous fluid of neutrons (for the most), protons, electrons and possibly muons in $\beta$-equilibrium, the so-called $\text{npe}\mu$ composition. Neutrons strongly interact with protons and can no longer be described as a perfect gas. It is indeed this strong interaction among nucleons the main source of pressure that prevents the star to collapse under gravitational attraction.
\item[Inner Core] extends for several km in the very central region of neutron stars. The density is larger than $2 \rho_0$, up to $\rho \sim \mathcal{O}\left(10^{15}\right)\, \text{g/cm}^3 $ in the center of the heaviest stars. The inner core can be absent in the lightest neutron stars, where instead the outer core extends up to the very center. The composition of matter in this region is very uncertain, and depends strongly on the underlying microscopic model assumed. Besides plain $\text{npe}\mu$ matter as in the outer core, the other main hypotheses are:
\begin{itemize}
\item formation of hyperons. Strange particles like $\Lambda^{\text{0}}$, $\Sigma^{\text{0}}$ and $\Sigma^-$ baryons can be produced through weak interactions like
\begin{align*}
\text{p} + \text{e}^- & \to \Lambda^{\text{0}} + \nu_{\text{e}} \\
\text{p} + \text{e}^- & \to \Sigma^{\text{0}} + \nu_{\text{e}} \\
\text{n} + \text{e}^- & \to \Sigma^- + \nu_{\text{e}} \,.
\end{align*} 
\item Bose-Einstein meson condensates, both without strangeness ($\pi$ mesons) or with it ($\text{K}$ mesons).
\item transition to deconfined quark matter, i.e., a new phase where $\text{u}$, $\text{d}$ and $\text{s}$ quarks are not confined any longer into nucleons. This may occur only if the matter density exceeds the nucleon density, $\rho \sim 10^{15}\ \text{g/cm}^3$. A neutron star which is modeled with a phase transition to quark matter in the very central region is also called hybrid star.
\end{itemize}
All these exotic models generally predict lower pressures than the plain $\text{npe}\mu$ matter ones. It is also possible that different phases are mixed together.
\end{description}
A schematic plot of the interior of a neutron star is shown in Fig.~\ref{fig:neutron_star}. The density range and the thickness of each region described above are compared. 

Finally, according to the Bodmer-Witten hypothesis~\cite{PhysRevD.4.1601,PhysRevD.30.272}, there is the possibility that strange stars, i.e., compact stars composed (almost) entirely of deconfined quarks, do exist. This can happen only if free quarks are the absolute ground state of hadronic matter. Some topics introduced here are discussed in more detail in section~\ref{sec:tabeos}. 

\subsection{The equation of state}
\label{sec:eos}
The \emph{equation of state} is a relation among the thermodynamical variables of a system, for instance density and pressure. It plays a fundamental role in determining the configuration of the hydrostatic equilibrium of a neutron star (cf. section~\ref{sec:tov}), encoding the information on the underlying microscopic nuclear interactions.

In General Relativity, the matter inside a neutron star is modeled as a fluid. We consider a system composed of different species. In a locally inertial frame comoving with the fluid element, the first law of thermodynamics states that the energy $dE$ contained in a fluid element of volume $dV$~\footnote{We assume that the fluid element is small with respect to the stellar length scale, but contains an amount of particles large enough to allow a statistical description of the system.}, evolves according to
\begin{equation}
\label{eq:thermo}
dE = -p dV + T dS + \sum_i \mu_i dN_i \,,
\end{equation}
where $p$ is the pressure, $T$ the temperature, $S$ the entropy and $\mu_i$ and $N_i$ are, respectively, the chemical potential and the number of particles of each specie $i$.

Assuming that the total number of particles $N$ is conserved,
\begin{equation}
N = \sum_i N_i = \mathrm{const} \,, 
\end{equation}
Eq.~\eqref{eq:thermo} can be recasted in the form
\begin{equation}
\label{eq:thermo2}
d \epsilon = \frac{\epsilon+p}{n} dn + n T ds + n \sum_i \mu_i dY_i\,,
\end{equation}
where $\epsilon=E/V$ is the energy density, $n=N/V$ the total particle number density, $s=S/N$ the entropy per particle and $Y_i=N_i/N$ the particle abundance of the specie $i$. This is a reasonable assumption if the star contains a negligible fraction of mesons and antimatter. Indeed, in this case, since also the contribution to the total energy coming from the leptons can be neglected, the total number of particles is the baryon number $N_B$, which is conserved by all physical interactions. In this way one can express the energy density as a function of the total particle number density, the entropy per particle and the particle abundances, $\epsilon=\epsilon(n,s,\{Y_i\})$~\footnote{We note that the particle abundances are not all independent, since $\sum_i Y_i =1$ by definition.}, the so-called equation of state. Once the latter is specified, all other thermodynamical quantities can be derived from this relation through Eq.~\eqref{eq:thermo2}.

Furthermore, we assume that the fluid is in chemical equilibrium with respect to some microscopic reactions which involve all the species of the system. Over the reaction timescales, the others thermodynamic variables are constant, which implies
\begin{equation}
\sum_i \mu_i dY_i =0 \,,
\end{equation}
and then Eq.~\eqref{eq:thermo2} reduces to
\begin{equation}
\label{eq:thermo3}
d \epsilon = \frac{\epsilon+p}{n} dn + n T ds \,.
\end{equation}
This assumption holds for cold neutron stars, where diffusive processes due to the neutrino flux can be neglected, and the very high pressure speeds up all the reactions (indeed, for instance, the neutron star matter is in $\beta$-equilibrium). This means that the energy density depends only on the particle number density and the entropy per baryon, $\epsilon=\epsilon(n,s)$, and the fluid composition is uniquely fixed by them~\cite{gravitation:1973}.

As we have said before, the matter inside a cold neutron star can be considered at $T = 0\, \mathrm{K}$. In the latter case, the second term in the right-hand-side of Eq.~\eqref{eq:thermo3} vanishes. Then Eq.~\eqref{eq:thermo3} further reduces to
\begin{equation}
\label{eq:thermo4}
d \epsilon = \frac{\epsilon+p}{n} dn \,.
\end{equation}
In the end, the energy density can be expressed as a function of a single variable only, $\epsilon=\epsilon(n)$. The latter is called a \emph{barotropic} equation of state. The pressure is determined from Eq.~\eqref{eq:thermo4} by
\begin{equation}
\label{eq:eospres}
p = n\frac{d\epsilon}{dn}-\epsilon \,.
\end{equation}
Replacing the particle number density in Eq.~\eqref{eq:eospres}, we can directly relate pressure and energy density, $p=p(\epsilon)$.

Finding the equation of state of cold, ultra-dense matter, i.e., determining the relation $p=p(\epsilon)$ or $\epsilon=\epsilon(n)$, is an open problem in nuclear physics. In the next section we discuss some of the models proposed in the literature.

\subsection{Models of equation of state}
\label{sec:tabeos}
The equation of state in the outer crust of a neutron star is based on the theory of strongly coupled Coulomb systems and it is constrained by data from atomic nuclei and nucleon scattering experiments. As the density increases, neutronization sets in, i.e., the nuclei become more and more massive and rich of neutrons through electron capture. At densities above the neutron drip density $\sim 4\times  10^{11}\, \text{g/cm}^3$, the experimental data are not available anymore, since a system with such densities cannot exist (stable) on Earth. The models used to describe the inner crust are based on extrapolations of the empirical information obtained in laboratory. The two matter phases in this regime, neutron-rich nuclei (phase I) and the neutron and electron gas (phase II) arrange themselves in different geometric structures as density increases, which are called pasta phases due to a resemblance to different types of pasta (gnocchi, spaghetti, lasagna, etc.). From lower to higher densities, recent models suggest that the configurations which minimize the energy are, respectively:
\begin{itemize}
\item[1)] spherical droplets of neutron-rich nuclei surrounded by the electron and neutron gas,
\item[2)] rods of matter in phase I immersed in matter in phase II,
\item[3)] alternating layers of matter in phase I and phase II.
\end{itemize}

At densities above the nuclear saturation density $\sim 2.7\times  10^{14}\, \text{g/cm}^3$, all hadronic models predict that in the outer core the main composition of matter is the $\mathrm{npe}\mu$ composition, a homogeneous fluid of neutrons, protons, electrons and muons in $\beta$-equilibrium. In this phase the nucleons strongly interacts between each other and can not be modeled like a perfect gas~\footnote{Treating protons and neutrons as a perfect gas we would find a maximum mass of about $ 0.7 M_{\odot}$, smaller than the Chandrasekahr limit, and incompatible with all astrophysical observations of neutron stars (see Chapter~\ref{sec:inverse}).}. As the density increases, the validity of these models is more and more uncertain. When the density is larger than $2 \rho_0$ the composition and the interactions are very model dependent, thus the structure of the inner core is essentially unknown.

In the following we summarize briefly the main theoretical models and approaches to determine the equation of state of the core of a neutron star ($\rho \gtrsim \rho_0$), assuming that the equation of state of the crust is well-known. Describing the properties of matter in this regime is a difficult problem, involving the complexity of both the strong interaction and a many-body system, whereas instead the Coulomb repulsion between protons can be neglected as a first approximation. In principle, one should start from quantum chromodynamics (QCD). However, even at the densities reached inside a neutron star core, the energies involved are not high enough to make a perturbative expansion in the strong coupling constant feasible (cf. section~\ref{sec:exotic}). Thus, it is necessary to use an effective theory, where quark degrees of freedom are not treated explicitly but are replaced by hadrons. Each model provides the energy density and the pressure of nuclear matter as a function of the baryon number density $n_B$ only (we recall that the total baryon number $N_B$ is conserved). For a given $n_B$, the particle fraction of each specie is uniquely fixed by the requirements of chemical (beta) equilibrium and charge neutrality. In neutron star cores, the neutron fraction is above the 90\%.

These models are mainly divided in two groups: \emph{nonrelativistic many-body theory} (NMBT) and \emph{relativistic mean field theory} (RMFT). For the sake of simplicity we start from nuclear matter composed of nucleons only. The interactions among nucleons are described through phenomenological effective nucleon-nucleon interactions. After that we briefly discuss the inclusion of hyperons and other exotic models.

\begin{description}
\item[Nonrelativistic many-body theory] \ 

\begin{figure}
\centering
\begin{fmffile}{yukawa}
\fmfframe (3,3)(3,3){
\begin{fmfgraph*}(80,50)
\fmfleft{i1,i2}
\fmfright{o1,o2}
\fmflabel{$N$}{i1}
\fmflabel{$N$}{i2}
\fmflabel{$N$}{o1}
\fmflabel{$N$}{o2}
\fmf{fermion}{i1,v1}
\fmf{fermion}{i2,v2}
\fmf{fermion}{v1,o1}
\fmf{fermion}{v2,o2}			
\fmf{dashes,label=$\pi$,label.side=right}{v1,v2}
\end{fmfgraph*}
}
\end{fmffile}
\caption{\textsl{Feynman diagram representing the Yukawa one-pion-exchange process between two nucleons.}}
\captionsetup{format=hang,labelfont={sf,bf}}
\label{fig:yukawa}
\end{figure}
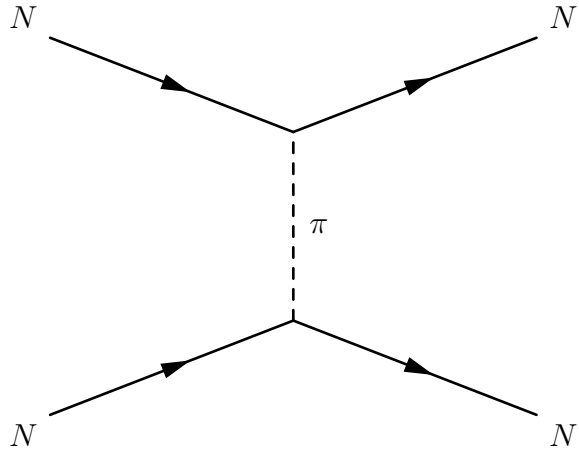

In NMBT nucleons are treated like an ensemble of pointlike particles. The dynamics is described by the non-relativistic Hamiltonian
\begin{equation}
\label{eq:ham}
H=\sum_i^A \frac{\mathbf{p}_i^2}{2m_N} + \sum_{i>j}^A v_{ij}+ \sum_{i>j>k}^A V_{ijk} \,,
\end{equation}
where $A$ is the number of nucleons of the system, $\mathbf{p}_i$ the three-momentum of the $i$-th nucleon and $m_N$ the nucleon mass (neglecting the difference of mass between neutrons and protons due mostly  to electromagnetic interactions). $v_{ij}$ and $V_{ijk}$ are two-body (NN) and three-body (NNN) potentials, respectively, describing the interaction among nucleons.

The Argonne v18 potential~\cite{Wiringa:1994wb}
\begin{equation}
\label{eq:argonne}
v^{ij}_{\mathrm{A18}} = \sum_{n=1}^{18} v_n(r_{ij}) \, \mathbb{O}_n^{ij} \,,
\end{equation}
is the form of the $v_{ij}$ NN potential general enough to reproduce, by constraining its phenomenological parameters, the experimental data from nucleon-nucleon scattering and the properties of $^2 \mathrm{H}$. It depends on the angular momentum, on the spin and (weakly) on the isospin of the nucleons through a set of eighteen operators $ \mathbb{O}_n^{ij} $, and it is local in coordinate space (i.e., it depends only on the relative distance between two nucleons $r_{ij}$). 

\begin{figure}
\centering
\begin{tikzpicture}
\begin{axis}[xmin=0,xmax=2.7,ymin=-230,ymax=700,
axis x line=middle,
axis y line=left,
width=0.8\textwidth,
height=0.6\textwidth,
xtick={0.5,1,1.5,2,2.5},
xlabel=$r_{ij} \, \text{[fm]}$,ylabel=$v^{ij}\, \text{[MeV]}$]
\addplot
[domain=0.4:2.7,samples=100,smooth,
thick,red]
{-10.463*exp(-0.75*x)/x/0.75 - 1650.6*exp(-4*0.75*x)/x/0.75 + 6484.2*exp(-7*0.75*x)/x/0.75};
\end{axis}
\end{tikzpicture}
\caption{\textsl{Schematic plot of the NN potential as a function of the radial distance between nucleons. At very short distance the nuclear force is characterized by an intense repulsive barrier. In the intermediate region, where the bottom of the well is around $0.8 \, \mathrm{fm}$, bound states can form. Lastly, at large distance the nuclear force is attractive, but it decays exponentially fast becoming insignificant beyond $2 \, \mathrm{fm}$.}}
\captionsetup{format=hang,labelfont={sf,bf}}
\label{fig:potential}
\end{figure}
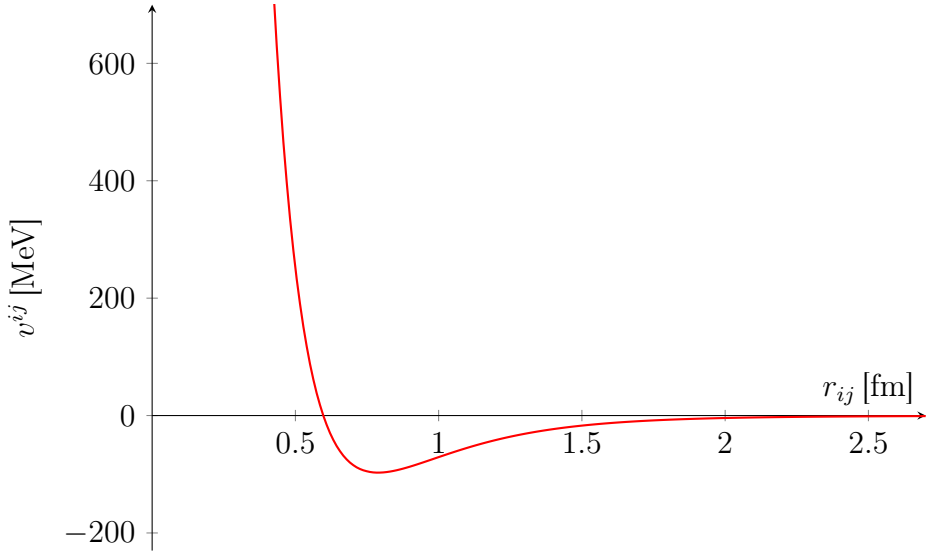

The NN potential $v^{ij}$ can be separated in
\begin{equation}
\label{eq:twobody}
v^{ij} = v^{ij}_{\pi} + v^{ij}_{\mathrm{IS}} \,.
\end{equation}
The $v^{ij}_{\pi}$ term represents the Yukawa potential which describes the long range part of the NN interaction due to one-pion-exchange processes (see Fig.~\ref{fig:yukawa}) 
\begin{equation}
\label{eq:yukawa}
v^{ij}_{\pi} \sim -g^2 \frac{\mathrm{e}^{-m_{\pi} r_{ij}}}{r_{ij}} \,,
\end{equation}
where $g$ is the coupling constant of the strong interaction and $m_{\pi}$ the mass of the $\pi$ meson. The functional form of the Yukawa potential in Eq.~\eqref{eq:yukawa} encodes the short-range nature ($\sim 1.4 \, \mathrm{fm}$) of the NN interaction. The phenomenological potential $v^{ij}_{IS}$ in Eq.~\eqref{eq:twobody} describes instead the intermediate and short range components of the nuclear force. A schematic plot of the NN potential is shown in Fig.~\ref{fig:potential}.

The three-body potential $V_{ijk}$ has to be introduced to reproduce the binding energies of $^3 \mathrm{H}$ and $^4\mathrm{He}$. By analogy with the decomposition of the two-body potential in Eq.~\eqref{eq:twobody}, also NNN potentials, such as the Urbana IX model~\cite{Pudliner:1995wk}, can be written as
\begin{equation}
\label{eq:threebody}
V^{ijk} = V^{ijk}_{2\pi} + V^{ijk}_{\mathrm{IS}} \,.
\end{equation}
The first term describes two-pion-exchange processes at large internucleon distance (see Fig.~\ref{fig:yukawa2}), while the latter one is purely phenomenological and accounts for intermediate and short range interactions.

\begin{figure}
\centering
\begin{fmffile}{yukawa2}
\fmfframe (3,3)(3,3){
\begin{fmfgraph*}(80,50)
\fmfleft{i1,i2,i3}
\fmfright{o1,o2,o3}
\fmflabel{$N$}{i1}
\fmflabel{$N$}{i2}
\fmflabel{$N$}{i3}
\fmflabel{$N$}{o1}
\fmflabel{$N$}{o2}
\fmflabel{$N$}{o3}
\fmf{fermion}{i1,v1}
\fmf{fermion}{i2,v2}
\fmf{fermion}{i3,v3}
\fmf{fermion}{v1,o1}
\fmf{fermion}{v4,o2}	
\fmf{fermion}{v3,o3}		
\fmf{dashes,label=$\pi$,label.side=right}{v1,v2}
\fmf{dashes,label=$\pi$,label.side=right}{v3,v4}
\fmf{double,label=$\Delta$,label.side=right}{v2,v4}
\end{fmfgraph*}
}
\end{fmffile}
\caption{\textsl{Feynman diagram of the two-pion-exchange process in three-body nuclear interactions. The intermediate double line represents an excited state of the nucleon, the $\Delta$ resonance.}}
\captionsetup{format=hang,labelfont={sf,bf}}
\label{fig:yukawa2}
\end{figure}
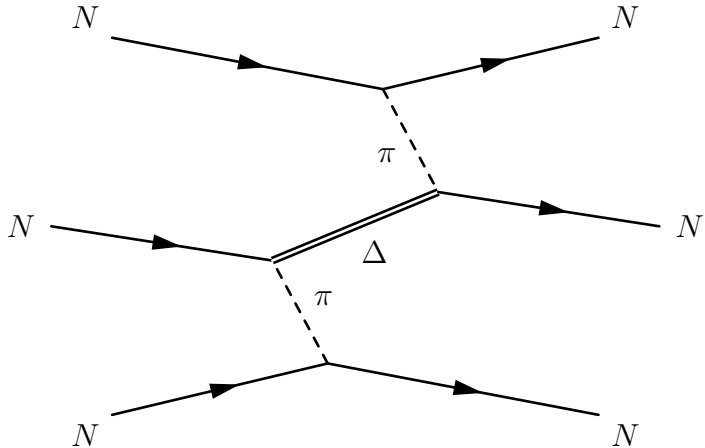

Since inside a neutron star the number of nucleons is $A \sim 10^{57} $, finding the ground state of the Hamiltonian~\eqref{eq:ham} for nuclear matter is a many-body problem which requires some approximations. Furthermore, many-body perturbation theory can not be directly applied, because the repulsive part of the potential at short internucleon distance is very strong. The solution is redefining either the interaction potential (Brueckner-Bethe-Goldstone G-matrix perturbation theory) or the basis states of the system (correlated basis function perturbation theory) in such a way that the resulting matrix elements are small. By applying these methods one can calculate the energy per baryon of the ground state as a function of the baryon number density, and then find the equation of state of nuclear matter. Non-relativistic many-body models perform fairy well around the equilibrium density, but, because of their non-relativistic nature, their validity starts to break down at higher densities, where relativistic effects are not negligible.

\item[Relativistic mean field theory] \ 

RMFT makes use of the Lagrangian formulation of quantum field theory. Nucleons are described as Dirac particles, which interact through meson exchange. These mesons do not need to be real existing particles as the Yukawa pion, but may be virtual states formed by other mesons. A modern version of the Lagrangian employed in RMFT calculations is the $\sigma$-$\omega$-$\rho$ model~\cite{Walecka:1974qa,Boguta:1977xi}, where the dynamics is described using a scalar field $\sigma$, a vector field $\omega_{\mu}$ and a isospin-triplet vector field $\rho_{\mu}^a$, with $a=\{-1,0,1\}$. In $c=\hbar=1$ units, the Lagrangian is written as
\begin{equation}
\label{eq:lagran}
\begin{aligned}
\mathcal{L}  \ = & \ \mathcal{L}_{\Psi} +  \mathcal{L}_{\sigma} +  \mathcal{L}_{\omega} +  \mathcal{L}_{\rho} +  \mathcal{L}_{\mathrm{int}} \\
\mathcal{L}_{\Psi} = &  \ \bar{\Psi}(\mathrm{i} \slashed{\partial}-m_N)\Psi   \quad   \qquad   \qquad \qquad \qquad \Psi = \binom{\psi_p}{\psi_n} \qquad \slashed{\partial} = \partial_{\mu} \gamma^{\mu} \\ 
\mathcal{L}_{\sigma} = &   \ \frac{1}{2}\partial_{\mu}\sigma\partial^{\mu}\sigma-\frac{1}{2}m_{\sigma}^2\sigma^2 - U(\sigma) \qquad U(\sigma) = \frac{1}{3} b \, m_N \,  (g_{\sigma} \sigma )^3 + \frac{1}{4} c \, (g_{\sigma} \sigma)^4 \\
\mathcal{L}_{\omega} = &  \ -\frac{1}{4}\omega_{\mu \nu}\omega^{\mu \nu} +\frac{1}{2}m_{\omega}^2\omega_{\mu}\omega^{\mu} \, \ \ \quad \qquad  \omega_{\mu \nu} =\partial_{\nu} \omega_{\mu}-\partial_{\mu} \omega_{\nu}\\
\mathcal{L}_{\rho} = &  \ -\frac{1}{4}\rho_{\mu \nu}^a \rho^{\mu \nu}_a +\frac{1}{2}m_{\rho}^2\rho_{\mu}^a\rho^{\mu}_a   \, \ \ \quad \qquad \ \  \rho_{\mu \nu}^a =\partial_{\nu} \rho_{\mu}^a-\partial_{\mu} \rho_{\nu}^a\\
\mathcal{L}_{\mathrm{int}} = & \  g_{\sigma} \bar{\Psi} \Psi \sigma - g_{\omega} \bar{\Psi} \gamma_{\mu} \Psi \omega^{\mu} - g_{\rho} \bar{\Psi} \gamma_{\mu} \tau^a \Psi \rho^{\mu}_a \,,
\end{aligned} 
\end{equation}
where $\bar{\Psi}=\Psi^{\dagger} \gamma^0$, with $\Psi^{\dagger}$ denoting the conjugate transpose of $\Psi$, $\psi_p$ ($\psi_n$) is the spinor field of the proton (neutron), $m_{\sigma}$, $m_{\omega}$ and $m_{\rho}$ the masses of the meson fields and $g_{\sigma}$, $g_{\omega}$ and $g_{\rho}$ their coupling constants with the nucleon field. $b$ and $c$ are the coupling constants of cubic and quartic self-interactions of the $\sigma$ field, respectively. Also, $\partial_{\mu} \equiv \partial / \partial x^{\mu}$, $\gamma^{\mu}$ are the Dirac Gamma matrices and $\tau^a$ the Pauli matrices.

Unfortunately, the equations of motion derived from the Lagrangian~(\ref{eq:lagran}) can be solved only within the \emph{mean-field approximation}, that consists in replacing the microscopical nucleon density distribution, within a volume element, by a mean, constant density (for instance, the ground state expectation value $\langle \bar\Psi \Psi \rangle = const$, independent of the coordinates $x^{\mu}$). As a consequence, the meson fields are replaced by their expectation values in the ground state of nuclear matter, i.e., they are treated as classical fields ($\sigma \to \langle \sigma \rangle$, etc.). The equation of state is then derived from the expectation value of the stress-energy tensor (cf. section~\ref{sec:tov})
\begin{equation}
T^{\mu \nu} = \frac{\partial \mathcal{L}}{\partial \left(\partial_{\mu} \Psi \right)} \partial^{\nu} \Psi - \mathcal{L} \,.
\end{equation}
The free parameters of the model can be estimated fitting the experimental properties of nuclear matter such as saturation density, symmetry energy, etc.

The physical meaning of the mean-field approximation is that the dynamics develops in a dense baryon medium and not in vacuum as in the NMBT. This approximation holds only in the limit of the baryon number density $n_B \to \infty$, which means that the average distance between nucleons has to be much smaller than the spatial range of mesons. This condition would require $n_B \gg 100 n_0$, where $n_0 \sim 0.16\, \mathrm{fm}^{-3}$ is the equilibrium density of nuclear matter. Clearly, this is not satisfied inside a neutron star. Furthermore, the quark degrees of freedom would appear anyway at much lower densities. In this sense, the RMFT approach has to be regarded as an effective way to parametrize the equation of state.
\end{description}

\subsubsection{Hyperons, Bose-Einstein meson condensates and quark matter}
\label{sec:exotic}
The process that may lead to the appearance of hyperons in neutron star matter is analogous to neutronization. As density increases, the production of heavy baryons through reactions like those in section~\ref{sec:compo} can become energetically favored. Since hyperons have larger masses and then are produced with lower kinetic energies than nucleons, their effect is to lower the pressure of the equation of state (with respect to models where matter is made of nucleons only). Both NMBT and RMFT can be extended to include the appearance of hyperons, even the full baryon octet, in neutron star cores. This is particularly straightforward within the RMFT. On the other hand, nucleon-hyperon and hyperon-hyperon interactions are poorly constrained by experimental data, making difficult to estimate their respective two-body potentials. Recent results suggest that also three-body interactions should be taken into account.

The hypothesis of the formation of a Bose-Einstein meson condensate is treated within the relativistic formulation of quantum field theory. In standard conditions the expectation value of (for example) $\pi$ or $K$ mesons in the ground state of nuclear matter vanishes, since the fermionic currents that source them vanish as well. In the first case this occurs because the pion has negative parity, while in the latter one because the kaon carries strangeness. However, at higher densities, it may be possible that different conditions do exist, changing the structure of the ground state in such a way that the expectation value of the current is finite, yielding as a consequence a non-vanishing expectation value for the mesons too. Negative $\pi$ or $K$ mesons may be energetically favored to ensure charge neutrality in nuclear matter, replacing the electrons, since the latter are fermions and their Fermi energetic level increases as density increases, whereas the former are bosons and can condensate in the lowest energetic level.

Free quarks in deconfined quark matter can not be described within perturbative QCD, because even the very large densities reached inside a neutron star are still too low to apply the QCD in the weak-coupling regime. The first models consisted in perturbative calculations at very high energy ($ \gg 1\, \text{GeV}$)~\footnote{For comparison, the Fermi energy of neutrons at densities typical of a neutron star is $E_F \sim 30 \, \mathrm{MeV}$.}, followed by an extrapolation at neutron star densities~\cite{Collins:1974ky}. A popular phenomenological approach is instead the \emph{MIT bag model}~\cite{Chodos:1974je}. In this model, quarks are assumed to be confined into a region of space, the bag, from where they can not escape. Inside the bag the interactions among quarks are weak and can be treated using perturbative techniques (common values of the strong coupling constant used are $\alpha_S \sim 0.2 \div 0.5 $). The volume of the bag is determined by the bag constant $B$, which represents the inward pressure of the QCD vacuum that balances the outward pressure generated by the quarks. Neutron stars composed entirely by deconfined quarks, i.e., strange stars, are self-bound, which means that they can exist also in absence of gravity. On the other hand, a star composed only by neutrons can not survive without gravity, since two neutrons do not form a bound system. 

All models of equation of state of neutron stars do no take into account the gravitational interaction, i.e., flat space is assumed, though neutron stars are relativistic objects. This is actually reasonable, since the radius of curvature of spacetime, even around and inside strong gravitating sources, is infinitely larger than both the size of hadrons and the spatial scale of strong interactions.

Some of the microscopic models of equation of state proposed in the literature are presented in Chapter~\ref{sec:inverse}, where we also discuss their impact on the macroscopic properties of neutron stars.

\subsection{The Tolman-Oppenheimer-Volkoff equations}
\label{sec:tov}
In this section we describe the relativistic equations of the hydrostatic equilibrium of a spherical star. We consider a static and spherically symmetric spacetime. In geometric units $G=c=1$, the spacetime metric is given in spherical coordinates $\{t,r,\theta,\varphi\}$ by the line element
\begin{equation}
\label{eq:line}
ds^2=-\mathrm{e}^{\nu (r)}dt^2+\mathrm{e}^{\lambda (r)}dr^2+r^2(d\theta^2 +\sin^2\theta d\varphi^2) \,,
\end{equation}
with metric tensor
\begin{equation}
g_{\mu \nu} = \left( \begin{array}{cccc}
 -\mathrm{e}^{\nu (r)} & 0 & 0 & 0   \\
0 & \mathrm{e}^{\lambda (r)} & 0 & 0   \\
0 & 0 & r^2 & 0   \\
0 & 0 & 0 & r^2 \sin^2\theta
\end{array} \right) \,,
\end{equation}
where $\nu(r)$ and $\lambda(r)$ are unknown functions of the radial coordinate only. According to the Birkhoff uniqueness theorem, in vacuum the solution must reduce to the Schwarzschild metric 
\begin{equation}
\mathrm{e}^{\nu (r)} = \mathrm{e}^{-\lambda (r)} = 1-\frac{2M}{r} \,,
\end{equation}
where $M$ is the total gravitational mass of the star. Inside the source, the functions $\nu(r)$ and $\lambda(r)$ are determined solving the Einstein equations (which imply the conservation law for the stress-energy tensor $T_{\mu \nu}$) together with the equation of state of matter (see below),
\begin{equation}
\label{eq:einstein}
\left \{
\begin{aligned}
& G_{\mu \nu}=8\pi T_{\mu \nu} \\
& \nabla_{\nu} T^{\mu \nu}=  0
\end{aligned}
\right.\,,~\footnote{We stress that the two equations in~\eqref{eq:einstein} are not independent. The conservation law for the stress-energy tensor follows from the Einstein field equations and it is considered only to simplify the calculations.}
\end{equation}
where $G_{\mu \nu}= R_{\mu \nu}-\frac{1}{2}g_{\mu \nu}R$ is the Einstein tensor, whereas $R_{\mu \nu}$ and $R$ are the Ricci tensor and the scalar curvature, respectively.

To go on with the computation, we need to specify how the matter of the star is modeled, which means that we have to provide explicitly the expression for the stress-energy tensor. In the neutron star case, the matter is modeled as a \emph{perfect fluid}, i.e. a fluid with zero viscosity and heat flow. Furthermore, the fluid is \emph{isotropic}, which means that the strength of the pressure is the same in every direction. The expression of the stress-energy tensor compatible with these assumptions is
\begin{equation}
\label{eq:fluid}
T^{\mu \nu}=(\epsilon + p)u^{\mu}u^{\nu}+p g^{\mu \nu} \,,
\end{equation}
where $u^{\mu}=(\mathrm{e}^{-\nu(r)/2},0,0,0)$, $\epsilon$ and $p$ are the four-velocity, the energy density and the pressure of the fluid, respectively. Indeed, in a locally inertial frame comoving with the fluid element, the stress-energy tensor in Eq.~\eqref{eq:fluid} reduces to
\begin{equation}
T^{\mu\nu} = 
\left( \begin{array}{cccc}
 \epsilon & 0 & 0 & 0   \\
0 & p & 0 & 0   \\
0 & 0 & p & 0   \\
0 & 0 & 0 & p
\end{array} \right) \,.
\end{equation}
For comparison, the full structure of $T^{\mu \nu}$ in the case of a non-perfect fluid is shown in Fig.~\ref{fig:stress}~\cite{Rezzolla:2013}.
\begin{figure}
\centering
\begingroup
\Large
\begin{equation*}
 T^{\mu \nu} = \left(
  \arraycolsep=5pt\def\arraystretch{1.5}
  \begin{array}{cccc}
     \cellcolor{green!30}{T^{00}}  &  \cellcolor{red!30}{T^{01}}  &  \cellcolor{red!30}{T^{02}} &  \cellcolor{red!30}{T^{03}} \\ 
    *   &  \cellcolor{blue!30}{T^{11}} & \cellcolor{yellow!30}{T^{12}}  & \cellcolor{yellow!30}{T^{13}}  \\
    *  & *   & \cellcolor{blue!30}{T^{22}} &  \cellcolor{yellow!30}{T^{23}}  \\
    *   & *  & *  & \cellcolor{blue!30}{T^{33}} \\
  \end{array}\right)
\end{equation*}
\endgroup
\caption{\textsl{Physical meaning of each component of the stress-energy tensor describing a generic (non-perfect) fluid. The $00$-component (green) is the energy density. The $0i$-components (red) represent the energy (for instance, heat) flow in the $i$-th direction. The off-diagonal $ij$-elements (yellow) are the shear stresses (flux of $i$-th component of the momentum in the $j$-th direction), which vanish for non-viscous fluids. Finally, the diagonal $ii$-elements (blue) represent the pressures in each direction. In anisotropic fluids at least one component of the pressure differs from the others. In perfect fluids all the off-diagonal terms vanishes. If the fluid is also isotropic the blue terms are all equal. The star denotes the elements obtained by symmetry.}}
\captionsetup{format=hang,labelfont={sf,bf}}
\label{fig:stress}
\end{figure}
We note that the perfect-fluid approximation is consistent with the assumption that the matter of cold neutron star is described by a barotropic equation of state $p=p(\epsilon)$ (cf. section~\ref{sec:eos}), in which, indeed, there is no heat exchange~\footnote{In many cases the perfect-fluid approximation holds even if the equation of state is non-barotropic, i.e., it depends also on the entropy or temperature, like for hot neutron stars. This occurs when the heat flows through the star on a timescale much larger than the hydrodynamical scale, i.e., the timescale in which the stars rearranges its own structure reaching the equilibrium. In these situations the flux of heat is negligible and the evolution of the star is a sequence of states in thermodynamic equilibrium.}.

Using the expression of the stress-energy tensor in Eq.~\eqref{eq:fluid} and defining the enclosed mass function
\begin{equation}
\label{eq:massfunc}
m(r)=\frac{r}{2}\left( 1-\mathrm{e}^{-\lambda(r)} \right) \,,
\end{equation}
it can be shown that the system in Eq.~\eqref{eq:einstein} reduces to the Tolman-Oppenheimer-Volkoff (TOV) equations~\cite{Tolman:1934,Oppenheimer:1939ne}
\begin{equation}
\label{eq:tov}
\left \{
\begin{aligned}
\frac{dm}{dr} & =4\pi r^2 \epsilon \\
\frac{dp}{dr}& = -\frac{(\epsilon+p)(m+4\pi r^3 p)}{r(r-2m)} \\
\frac{d\nu}{dr} & = \frac{2(m+4\pi r^3 p)}{r(r-2m)} \\
\end{aligned}
\right.\,.
\end{equation}
The first two of the Eqs.~\eqref{eq:tov} are the relativistic generalization of the Newtonian equations of stellar hydrostatic equilibrium, while the third one refers to the gravitational potential. The system~\eqref{eq:tov} is not closed. We need to specify a supplementary condition to integrate it, that is the relation between pressure and energy density, i.e., the so-called \emph{equation of state}
\begin{equation}
\label{eq:eos}
p=p(\epsilon) \,.
\end{equation}
Once the latter is specified, we can integrate numerically the TOV equations~\eqref{eq:tov} along with the equation of state~\eqref{eq:eos}, from the center of the star $r=0$ to its surface $r=R$, with $R$ the radius of the star, where we match the internal solution to the external Schwarzschild metric. The boundary conditions to impose are: \ \\ [-0.3cm]

\textbf{At the center of the star} $\mathbf{r=0}$
\begin{itemize}
\item the enclosed mass must vanish, $m(0)=0$
\item the central pressure (or equivalently the central energy density) can be freely specified, $p(0)=p_0$
\end{itemize}

\textbf{At the surface of the star} $\mathbf{r=R}$
\begin{itemize}
\item the pressure (and the energy density) must vanish, $p(R)=0$
\item the $\nu(r)$ function must reduce to $\mathrm{e}^{\nu (r)}  = 1-2M/r$, with $M$ the gravitational mass of the star given by
\begin{equation}
\label{eq:grmass}
M = m(R)=4\pi \int_{0}^{R} r^2\epsilon \, dr \,.
\end{equation}
\end{itemize}
Additional details on the numerical integration of the TOV equations are given in the Appendix~\ref{sec:appA}.

\begin{figure}
\centering
\begin{tikzpicture}
\begin{axis}[
axis x line=bottom,
axis y line=left,
enlargelimits,
width=0.6\textwidth,
height=0.45\textwidth,
xlabel=$R \, \text{[km]}$,ylabel=$M \, \text{[}M_{\odot}\text{]}$]
\addplot[smooth,thick,blue] table {chapter_1/figures/apr4_stable.dat};
\addplot[smooth,thick,blue,dashed] table {chapter_1/figures/apr4_unstable.dat};
\end{axis}
\end{tikzpicture}
\caption{\textsl{An example of a neutron star mass-radius diagram. The $M=M(R)$ curve is obtained integrating the TOV equations for different values $p_0$ of the central pressure with a given equation of state. The profile of the curve depends in general on the underlying equation of state, but its main features and the order of magnitude of masses and radii are the same. The solid line denotes the stable configurations, while the dashed line the unstable ones. The typical range of values of the central pressure is $p_0 \sim (10^{34} \div 10^{36})\, \mathrm{dyn/cm}^2$, or equivalently in geometric units, $p_0 \sim (10^{-5} \div 10^{-3})\, \mathrm{km}^{-2}$.}}
\captionsetup{format=hang,labelfont={sf,bf}}
\label{fig:mr}
\end{figure}
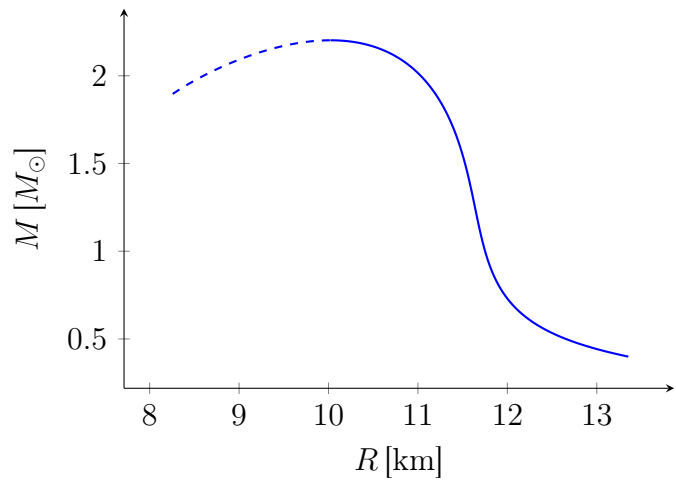

Since the initial condition on the central pressure, i.e., the value of $p_0$, can be chosen arbitrarily, the TOV equations admit a one-parameter family of solutions that depends on the equation of state used to close the system~\footnote{We deeply discuss the macroscopic effects of different models of equation of state in Chapter~\ref{sec:inverse}.}. For a given equation of state, varying the value of the central pressure gives rise to different equilibrium configurations for the star, which means different values of mass and radius as a function of $p_0$, $\{M(p_0),R(p_0) \}$. We show this in Fig.~\ref{fig:mr}, where we plot a typical example of different configurations of stellar equilibrium, specified by the values of mass and radius of the neutron star and parametrized by its central pressure. The result is a curve $M=M(R)$, the so-called \emph{mass-radius diagram} of neutron stars. The main features of the neutron star mass-radius diagram explained in the following are common to all models of equation of state (cf. Chapter~\ref{sec:inverse}). Larger values of $p_0$ correspond to smaller radii in the plot. We note that, like white dwarfs, neutron stars admit a maximum mass. Furthermore, we stress that not all the equilibrium configurations are \emph{stable}. Roughly speaking, all the configurations lying on the branch at the left of the maximum mass (dashed line) are unstable, which means that small density perturbations will grow exponentially in time, leading the star to collapse or expand. On the other hand, the configurations to the right of the maximum mass (solid line) are stable, i.e., small deviations from the state of equilibrium are restored by pressure or gravity~\footnote{A rigorous study of the stellar stability is more complicated. The criterion explained in the main text involves only the static equilibrium configurations of a star, and thus it is a necessary but not sufficient condition. The analysis of the modes of dynamical oscillations of a star is required to establish its stability.}. We discuss in more detail the mass-radius diagram of neutron stars in Chapter~\ref{sec:inverse}.

\subsection{Rotation}
\label{sec:rotation}
In this section we extend the previous discussion to rotating neutron stars, describing the equations governing the structure of a slowly-rotating relativistic star. We consider an object rotating with uniform angular velocity $\Omega$, as seen by an observer at rest at some fixed point in the spacetime located by the coordinates $(t,r,\theta,\varphi)$, and solve the resulting Einstein equations perturbatively. For our purposes (see section~\ref{sec:RTLN} and Chapter~\ref{sec:binary}), it is sufficient to consider only effects linear in the angular velocity $\Omega$. The spacetime metric is given by~\cite{Hartle:1967he,Hartle:1968si,Hartle:1973zza,Benhar:2005gi}
\begin{equation}
\label{eq:rotmetric}
ds^2=-\mathrm{e}^{\nu (r)}dt^2+\mathrm{e}^{\lambda (r)}dr^2+r^2(d\theta^2 +\sin^2\theta d\varphi^2) -2 \omega(r) r^2 \sin^2\theta dt d\varphi \,,
\end{equation}
with
\begin{equation}
g_{\mu \nu} = \left( \begin{array}{cccc}
 -\mathrm{e}^{\nu (r)} & 0 & 0 & -\omega(r) r^2 \sin^2\theta  \\
0 & \mathrm{e}^{\lambda (r)} & 0 & 0   \\
0 & 0 & r^2 & 0   \\
-\omega(r) r^2 \sin^2\theta & 0 & 0 & r^2 \sin^2\theta
\end{array} \right) \,,
\end{equation}
where the function $\omega(r)$ is of order $\mathcal{O}(\Omega)$ and the other terms are those of the non-rotating configuration in Eq.~\eqref{eq:line}. Outside the star, in vacuum, the exterior solution is given by
\begin{equation}
\mathrm{e}^{\nu (r)} = \mathrm{e}^{-\lambda (r)} = 1-\frac{2M}{r}  \qquad \omega(r) = \frac{2 J}{r^3} \,,
\end{equation}
where $M$ and $J$ are the mass and the total angular momentum of the source, respectively. The stress-energy tensor of the fluid is still given by Eq.~\eqref{eq:fluid}, but now the four-velocity is $u^{\mu}=(\mathrm{e}^{-\nu(r)/2},0,0,\Omega \mathrm{e}^{-\nu(r)/2})$. Note that there is no variation in the pressure or the energy density of the fluid with respect to the non-rotating case, since we are neglecting terms of order $\mathcal{O}(\Omega^2)$ and higher. As a consequence, to linear order in the angular velocity the shape of the star remains spherical and its mass and radius are the same of the unperturbed configuration. In particular, we stress that there is no \emph{spin-induced} quadrupole moment (not to be confused with the tidally induced quadrupole moment, cf. section~\ref{sec:TLN}).

To first order in perturbation theory the system of equations~\eqref{eq:einstein} reduces to the TOV equations~\eqref{eq:tov} plus the equation for the perturbative function $\bar{\omega}(r) = \Omega- \omega(r)$,
\begin{equation}
\label{eq:drag}
\frac{d^2 \bar{\omega}}{dr^2} -\left[\frac{4 \pi r^2 \left( \epsilon + p\right)}{r-2m} -\frac{4}{r}  \right] \frac{d \bar{\omega}}{dr} -\left[ \frac{16 \pi r \left( \epsilon + p\right)}{r-2m} \right] \bar\omega =0 \,.
\end{equation}
Here, the quantity $\omega$ identifies the angular velocity acquired by an observer freely falling from infinity towards the star (the angular velocity of the locally inertial frames), and then $\bar\omega$ denotes the angular velocity of the fluid as seen by a freely falling observer. Thus, Eq.~\eqref{eq:drag} describes the dragging of the locally inertial frames. The boundary conditions to impose are
\begin{itemize}
\item[1)] the solution must be regular at the center of the star $r=0$. This is always true in general: a star does not have any singularity, therefore the physical quantities can not diverge anywhere (cf. section~\ref{sec:polar}). This request gives
\begin{equation}
\label{eq:rotreg}
\bar\omega (0) = \bar\omega_0 \qquad \frac{d\bar \omega }{dr} \bigg{|}_{r=0} = 0 \,,
\end{equation}
where $\bar\omega_0$ can be arbitrarily chosen and its value determines the angular velocity of the star $\Omega$ and its total angular momentum $J$.
\item[2)] at the surface of the star $r=R$, the interior solution must be matched to the exterior one
\begin{equation}
\bar\omega(r) = \Omega- \frac{2 J}{r^3} \quad r \geq R \,.
\end{equation}
\end{itemize}
The angular velocity and momentum of the star are then determined by integrating the TOV equations~\eqref{eq:tov} together with Eq.~\eqref{eq:drag} and evaluating $\bar \omega$ and its first derivative at the star surface,
\begin{equation}
J = \frac{R^4}{6} \frac{d\bar \omega }{dr} \bigg{|}_{r=R} \qquad \Omega = \bar \omega(R) + \frac{2 J}{R^3} \,.
\end{equation}
Finally, the moment of inertia of the object is given to first order in $\Omega$ by
\begin{equation}
\label{eq:inertia}
I = \frac{J}{\Omega} \,.
\end{equation}

\section{Tidal deformations of neutron stars}
\label{sec:TLN}
The relativistic theory of tidal Love numbers has been developed by Hinderer~\cite{Hinderer:2007mb}, Damour and Nagar~\cite{Damour:2009vw}, Binnington and Poisson~\cite{Binnington:2009bb}, and Landry and Poisson~\cite{Landry:2015cva}, and extended to slowly spinning compact objects by Pani, Gualtieri, Maselli and Ferrari~\cite{Pani:2015hfa,Pani:2015nua} and Landry and Poisson~\cite{Landry:2015zfa,Landry:2015snx,Landry:2017piv}. In this section, and in the next one, we refer to their works.

\begin{figure}[]
\centering
\includegraphics[width=0.9\textwidth]{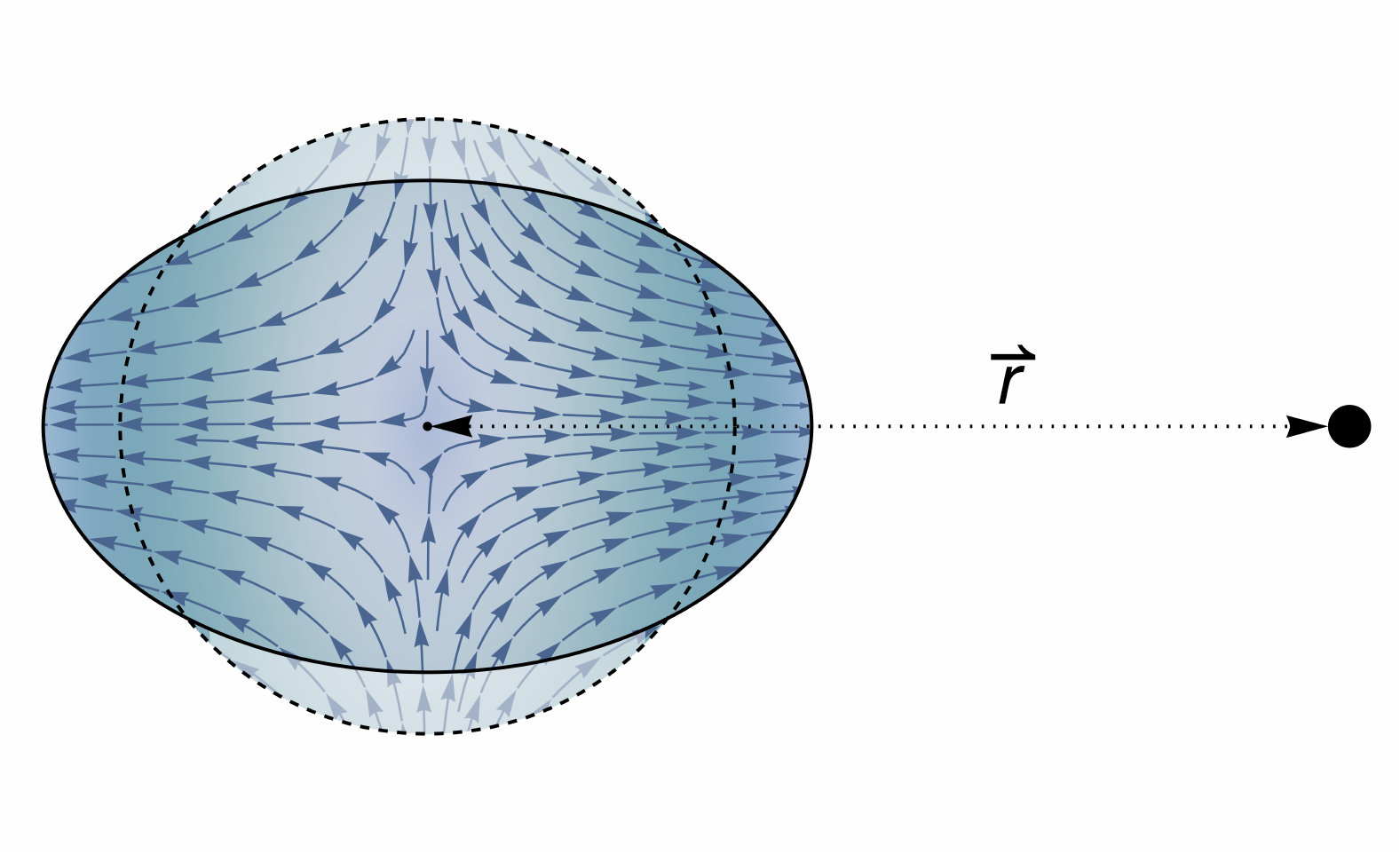}
\caption{\textsl{Schematic representation of the quadrupolar tidal deformation of an extended-body induced by a pointlike gravitational source at distance $r$. The original spherical body is stretched in the source direction, assuming the shape of a prolate ellipsoid. The vector field lines represent the tidal field in Eq.~\eqref{eq:tides}.}}
\captionsetup{format=hang,labelfont={sf,bf}}
\label{fig:tides}
\end{figure}

Tidal effects are finite-size effects arising on extended bodies when they are immersed in an external gravitational field. To fix the ideas let us consider a spherical body exposed to the gravitational field generated by a pointlike source in Newtonian gravity, as shown in Fig.~\ref{fig:tides}. The rigid translational motion of the body is completely determined by the acceleration of its center of mass. However, the side of the object which is closer to (respectively, farthest from) the external source experiences a gravitational attraction larger (respectively, smaller) than that felt by the center of mass. The overall result is that the structure of the body is deformed from the spherical shape and stretched in the direction of the gravitational source. Tidal forces are then due to the gradient of the gravitational acceleration on the volume of the body, and therefore they vanish for pointlike objects.

We can formally describe the picture depicted above within Newtonian gravity, explaining the pattern of the field lines plotted in Fig.~\ref{fig:tides}. We define the tidal field acting on a extended body as follows
\begin{equation}
\label{eq:tidalforce}
T_i(\mathbf{x}) = g_i(\mathbf{x}) - g_i(\mathbf{0}) \,,
\end{equation}
where $g_i(\mathbf{x}) = -  \partial_i \phi(\mathbf{x}) $ is the gravitational field, $\phi(\mathbf{x})$ the gravitational potential generated by the external source and $\mathbf{x}$ the position vector denoting a generic point inside the volume of the body, in a coordinate system with origin in the center of mass of the object $\mathbf{x} =\mathbf{0} $. The tidal field is then defined as the difference of the gravitational field in a given point and in the body center of mass. Series expanding Eq.~\eqref{eq:tidalforce} around the center of mass $\mathbf{x} =\mathbf{0} $ we get
\begin{equation}
\label{eq:tidalforce2}
T_i(\mathbf{x}) = - x^j \left[ \partial_j \partial_i \phi(\mathbf{x}) \right] |_{\mathbf{x}= \mathbf{0}} + \mathcal{O}(\mathbf{x}^2) \,.
\end{equation}
The potential associated to the vector tidal field is given by
\begin{equation}
\label{eq:tidalpot}
W(\mathbf{x}) = \frac{1}{2} x^i x^j \left[ \partial_i \partial_j \phi(\mathbf{x}) \right] |_{\mathbf{x}= \mathbf{0}} + \mathcal{O}(|\mathbf{x}|^3) \,.
\end{equation}

In the case of a pointlike external source, the gravitational potential is given by
\begin{equation}
\phi(\mathbf{x}) = -\frac{G M}{|\mathbf{x}-\mathbf{r}|} \,,
\end{equation}
where $M$ and $\mathbf{r}$ are the mass and the position of the source, respectively, and $G$ is the gravitational constant. The equations~\eqref{eq:tidalforce2}-\eqref{eq:tidalpot} read
\begin{equation}
\begin{aligned}
\mathbf{T}(\mathbf{x}) &= - \frac{GM}{|\mathbf{r}|^3} \left(\mathbf{x} - \frac{3 (\mathbf{x} \cdot \mathbf{r})  \mathbf{r} }{\mathbf{r}^2} \right) \\
W(\mathbf{x})& = - \frac{GM}{2 |\mathbf{r}|^3} \left( \frac{3 (\mathbf{x} \cdot \mathbf{r})^2 }{\mathbf{r}^2}  - \mathbf{x}^2\right) \,.
\end{aligned}
\end{equation}
Without loss of generality we can assume that the external source lies on the $z$-axis, $\mathbf{r} = \{0,0,r\}$. Thus the above equations further reduces to
\begin{equation}
\label{eq:tides}
\begin{aligned}
\mathbf{T}(\mathbf{x}) &= - \frac{GM}{r^3} \{x,y,-2z \} \\
W(\mathbf{x})& = - \frac{GM}{2 r^3} \left( 2z^2-x^2-y^2 \right) \,,
\end{aligned}
\end{equation}
which correspond to the vector field lines in Fig.~\ref{fig:tides}.

The naive description just presented has been deeply extended and studied in the literature within the theory of the \emph{gravitational tidal Love numbers}~\cite{1911spge.book.....L}. The latter are introduced to describe in a rigorous way how the multipolar structure of an extended body is modified by the presence of an external tidal field. In the following sections we discuss the full relativistic formulation of the theory.

\subsection{Tidal Love numbers in General Relativity}
\label{sec:TLNgr}
Let us consider a static (i.e., non-spinning), spherically symmetric star immersed in an external stationary tidal field. The object will be deformed by the tidal forces, developing a multipolar structure in response to the tidal field. As we largely discuss in Chapter~\ref{sec:binary}, this kind of situation occurs in coalescing binary systems, where each component is tidally deformed by the gravitational field of its companion. In this case the tidal field is not stationary. However, if the two objects are well separated, as it is in the inspiral phase, then the source of the tidal field affecting each of the bodies is very far away and it is slowly varying in time. This means that the timescale of the variation of the external tidal field $T_{\mathrm{ext}}$, which in this case is associated to the orbital dynamics, is much larger than the timescale on which the deformed object rearranges its own structure $T_{\mathrm{int}}$, that is related to the internal fluid dynamics of the body~\cite{Poisson:2014}. Under the assumption $T_{\mathrm{int}} \ll T_{\mathrm{ext}}$, the evolution of the tidal field along the orbital motion is adiabatically slow and we can effectively consider it as stationary. This approximation breaks down close to the merger of the binary system, when the orbital dynamics is much faster, and dynamical tides must be taken into account~\cite{Flanagan:2007ix,Maselli:2012zq,Steinhoff:2016rfi}.

Furthermore, consistently with the assumption that the tidal source is far away from the object, as we said it is in inspiralling binary systems, we make use of the approximation that the induced multipolar deformation of the star is linear in the strength of the external tidal field~\footnote{This statement can be derived (and not assumed a priori) in Newtonian gravity. Applying first-order perturbation theory to the fluid of a self-gravitating body affected by an external gravitational field, one finds that the multipole moments developed by the object are linearly proportional to the spatial derivatives of the external gravitational potential. A derivation, under appropriate assumptions, is possible also in General Relativity within a Post-Newtonian framework, through a Lagrangian approach. We show this in Chapter~\ref{sec:binary}.}. Thus, the tidal Love numbers are defined in General Relativity as the constants of proportionality between the tidally induced multipole moments of the object and the tidal moments of external gravitational field (henceforth we use geometric units $G=c=1$):
\begin{equation}
\label{eq:adiabaticrel}
\begin{aligned}
Q_L & = \lambda_l G_L \\
S_L & = \sigma_l H_L 
\end{aligned} \qquad l \geq 2 \,,
\end{equation}
where $Q_L$ ($S_L$) are the mass (current) multipole moments of order $l$ of the object, $G_L$ ($H_L$) the electric (magnetic) tidal multipole moments of order $l$, and we use the Latin capital letters as shorthand for multi-indices, $L \equiv a_1 \dots a_l $, see the~\nameref{sec:notation}. $\lambda_l$ ($\sigma_l$) are the electric (magnetic) tidal deformabilities related to the dimensionless tidal Love numbers $k^E_l$ ($k^M_l$) through the relations
\begin{equation}
\label{eq:lovetidal}
\begin{aligned}
\lambda_l & = \frac{2}{(2l-1)!!}    R^{2l+1} k^E_l \\
\sigma_l & = \frac{l-1}{4(l+2)(2l-1)!!}   R^{2l+1}  k^M_l
\end{aligned} \qquad l \geq 2 \,,
\end{equation}
where $R$ is the radius of the star. The electric and magnetic tidal deformabilities are the gravitational analog of the electric polarizability and the magnetic susceptibility, respectively. The quadrupolar ones ($l=2$) are the leading-order terms, and give the main contribution to the stellar deformations (cf. Chapter~\ref{sec:binary}).

The Eqs.~\eqref{eq:adiabaticrel} are called \emph{adiabatic relations} for the reasons explained above. The above multipole moments are symmetric and trace-free and are discussed in more detail in Chapter~\ref{sec:binary}. They can be extracted through an asymptotic expansion at spatial infinity of the spacetime metric of the stationary object perturbed by the external tidal source. We use the definition of multipole moments given by Thorne~\cite{Thorne:1980ru}, which has been shown to be equivalent~\cite{1983GReGr..15..737G} to the definition given by Geroch and Hansen~\cite{Geroch:1970cc,Geroch:1970cd,Hansen:1974zz}. In asymptotically Cartesian mass centered (ACMC) coordinates and using geometric units, the time-time and time-space components of the metric read~\cite{1986PhRvD..34.3617S}
\begin{equation}
\begin{aligned}
\label{eq:metrictidal}
g_{00} = & -1 + \frac{2M}{r} + \sum_{l \geq 2}  \left[  \frac{1}{r^{l+1}} \left( \frac{2(2l-1)!!}{l!} Q_L n_L + A_{l'<l} \right)  \right. \\
& \left. + r^l \left(\frac{2}{l!} G_L n_L+ A_{l'<l} \right) \right] \\
g_{0i} = & -\frac{2 \epsilon_{ijk} J_j n_k}{r^2}  + \sum_{l \geq 2}  \left[  \frac{1}{r^{l+1}} \left( -\frac{4l(2l-1)!!}{(l+1)!} \epsilon_{ija_l} S_{jL-1} n_L + A_{l'<l}  \right) \right. \\
& + \left. r^{l+1} \left( \frac{l}{(l+1)!}\epsilon_{ija_l} H_{jL-1} n_L + A_{l'<l} \right) \right] \,,
\end{aligned}
\end{equation}
where repeated spatial indices are summed using the flat Euclidean metric, $r = \delta_{ij} x_i x_j $ is the radial coordinate, $n_i = x_i/r$ is the unit radial vector and we have defined $n^L = n^{a_1}\dots n^{a_l}$, see the~\nameref{sec:notation}. $M$ and $J_i$ are, respectively, the mass and angular momentum of the central object. The symbol $A_{l'<l}$ denotes terms independent of $r$, with angular dependence proportional to spherical harmonics of order $l'<l$. We note that the mass dipole of the object identically vanishes, being the coordinates mass centered. Also, in our particular case of interest there is no angular momentum, $J_i=0$, since we are working with a non-spinning star (cf. section~\ref{sec:RTLN}). Removing the central object from the problem (i.e., setting $M=0$), it is also possible to express the tidal multipole moments in Eq.~\eqref{eq:metrictidal} in terms of the Weyl curvature tensor and its derivatives~\cite{Thorne:1984mz,1986PhRvD..34..991Z}
\begin{equation}
\begin{aligned}
G_L & = -\frac{l(l-1)}{2}\langle [ \nabla_{ L-2} \, C_{0 a_1 0 a_2} ]|_{r=0} \rangle \\
H_L & = \frac{3(l-1)}{2}\langle \epsilon_{ a_1 ij} [ \nabla_{L-2} \, C^{ij}_{\ \ a_2 0} ]|_{r=0} \rangle
\end{aligned} \qquad l \geq 2 \,,
\end{equation}
where 
\begin{equation}
C_{\mu \nu \alpha \beta} = R_{\mu \nu \alpha \beta} + \frac{1}{2} \left(R_{\mu \beta} g_{\nu \alpha} -R_{\mu \alpha} g_{\nu \beta}+R_{\nu \alpha} g_{\mu \beta}-R_{\nu \beta} g_{\mu \alpha}  \right) + \frac{1}{6} R \left(g_{\mu \alpha} g_{\nu \beta} -g_{\mu \beta} g_{\nu \alpha} \right) \,,
\end{equation}
$R_{\mu \nu \alpha \beta}$ is the Riemann tensor, $R_{\mu \nu}$ the Ricci tensor, $R$ the scalar curvature and the angular brackets denote trace-free symmetrization on the $L$ indices (see the~\nameref{sec:notation}).

We stress that the separation between the multipolar response of the central object (decaying solution for $r \to \infty$ in the metric in Eq.~\eqref{eq:metrictidal} and the external tidal field (growing solution) is not trivial when we relax some of the assumptions made, for instance in the case of a spinning object (see section~\ref{sec:RTLN}), but also when the tidal field is not weak or when the time dependence of the environment can not be neglected. In these cases there is no clear separation of the two solutions, and the definition of the Love numbers is ambiguous~\cite{Gralla:2017djj}. However, all these ambiguities disappear in the case that we are discussing, i.e., a static (non-spinning), spherically symmetric object perturbed by a weak, slowly varying in time tidal field. Furthermore, it was shown that the relativistic Love numbers are gauge-invariant~\cite{Binnington:2009bb}.

Looking at the adiabatic relations~\eqref{eq:adiabaticrel} we see that tidal Love numbers in General Relativity are divided in two distinct sectors with different parity, \emph{electric} (even) and \emph{magnetic} (odd), related to tidal deformations of mass and current distributions, respectively. The electric sector is the relativistic generalization of the Newtonian Love numbers. Indeed, in the Newtonian limit we have
\begin{equation}
\label{eq:newtlovetidal}
Q_L = \int_V \rho(\mathbf{x}) x_{\langle L \rangle} \, d^3x  \qquad G_L = -[ \partial_L \phi(\mathbf{x}) ] |_{\mathbf{x}=\mathbf{0}} \,,
\end{equation}
where $\rho(\mathbf{x})$ is the matter distribution of the object, $V$ its volume and $\phi(\mathbf{x})$ the external gravitational potential. The nature of the magnetic sector is instead fully relativistic, since current distributions do not gravitate in Newtonian theory, and therefore they can not excite any deformation in the star. This neat separation between the electric and magnetic sector breaks down when we take into account the spin of the central object. We describe this case in section~\ref{sec:RTLN}. In the next section we discuss instead how to compute the tidal Love numbers of a non-spinning object.

\subsection{Linear perturbations of a non-spinning object}
\label{sec:linear}
The tidal deformabilities $\lambda_l$ and $\sigma_l$ depend on the internal structure and composition of the object perturbed by the external tidal field. In the neutron star case, for a fixed compactness $C=M/R$, they depend only on the equation of state. Since, as we discuss in the next chapters, the tidal deformability can be directly measured through the detection of the gravitational signal emitted by coalescing binary neutron stars, it represents a powerful tool to discriminate among the models of equation of state proposed in the literature. This possibility is deeply discussed in Chapter~\ref{sec:inverse}. Also, it has been shown that the tidal deformabilities of any order $l$, both electric and magnetic, vanish in the black hole limit, $C=1/2$. In other words, the multipolar structure of a black hole is not affected by the tidal field. This can be view as a corollary of the no-hair theorem, and recently it has been proved beyond the perturbative level~\cite{Gurlebeck:2015xpa}~\footnote{Saying that a black hole does not develop a multipolar response to the tidal field does not mean that the metric around it is the same as in the unperturbed configuration. Indeed, for instance, the geometry of the event horizon changes in presence of a tidal field. This is encoded in the so-called \emph{surficial} Love numbers~\cite{Damour:2009vw,Damour:2009va,Landry:2014jka}, which express how the surface of an object is affected by a tidal field, and do not vanish even for a black hole. Furthermore, it has be shown that black holes develop a multipolar response if the tidal field is time-dependent~\cite{Poisson:2004cw,Fang:2005qq}. This is related to the phenomenon of \emph{tidal heating}~\cite{Thorne:1997kt,Purdue:1999gk}, i.e., the absorption of energy and angular momentum by the black hole due to the interaction of the tidal field with the horizon, which acts like a fictitious viscous membrane~\cite{Poisson:2009di}.}.

In this section we describe how to compute the tidal deformabilities, or equivalently the tidal Love numbers, of a non-rotating neutron star. The definitions of the tidal deformabilities/Love numbers given in Eqs.~\eqref{eq:adiabaticrel} and~\eqref{eq:lovetidal} and of the multipole moments in Eqs.~\eqref{eq:metrictidal} coincide exactly with the convention used by Damour and Nagar~\cite{Damour:2009vw}, who worked in the Regge-Wheeler gauge. Binnington and Poisson adopted instead the light-cone gauge~\cite{Binnington:2009bb}. They results are in agreement, which is consistent with the gauge-invariant property of the Love numbers. However, one has to take care of the different conventions used to avoid spurious overall constant factors.

We start with the unperturbed equilibrium configuration of the star given by the background metric~\eqref{eq:line}, the stress-energy tensor~\eqref{eq:fluid} and the TOV equations~\eqref{eq:tov}, that we report below for convenience
\begin{gather}
\label{eq:unpert}
ds^2=-\mathrm{e}^{\nu (r)}dt^2+\mathrm{e}^{\lambda (r)}dr^2+r^2(d\theta^2 +\sin^2\theta d\varphi^2) \\[0.3cm]
\label{eq:unpertfluid}
T^0_{\mu \nu}=(\epsilon + p)u_{\mu}u_{\nu}+p g^0_{\mu \nu} \qquad u^{\mu}=(\mathrm{e}^{-\nu(r)/2},0,0,0) \\[0.3cm]
\left \{
\begin{aligned}
\frac{dm}{dr} & =4\pi r^2 \epsilon \\
\frac{dp}{dr}& = -\frac{(\epsilon+p)(m+4\pi r^3 p)}{r(r-2m)} \\
\frac{d\nu}{dr} & = \frac{2(m+4\pi r^3 p)}{r(r-2m)} \\
\end{aligned}
\right.\,.
\end{gather}
Note that we have added the $0$-superscript to the metric and stress-energy tensors to highlight that now they describe the unperturbed star only. Then, we apply the techniques of perturbation theory, solving the linearized Einstein equations to first-order in the perturbations.

The full spacetime metric is given by
\begin{equation}
g_{\mu \nu}= g_{\mu \nu}^0 + h_{\mu \nu} + \mathcal{O}(|h_{\mu \nu}|^2) \,,
\end{equation}
where $g_{\mu \nu}^0 $ is the background metric of the unperturbed object in Eq.~\eqref{eq:unpert}, whereas $h_{\mu \nu}$ is a small perturbation due to the tidal field, that in an appropriate frame takes the form
\begin{equation}
|h_{\mu \nu}| \ll |g_{\mu \nu}^0| \,.
\end{equation}
We decompose the stress-energy tensor of the fluid in the same way,
\begin{equation}
T_{\mu \nu}= T_{\mu \nu}^0 + \delta T_{\mu \nu} \,,
\end{equation}
with $T_{\mu \nu}^0$ given by Eq~\eqref{eq:unpertfluid} and $ \delta T_{\mu \nu}$ that depends linearly on the metric perturbation $h_{\mu \nu}$ and on the Eulerian~\footnote{The variation of any quantity of the fluid as seen by an observer lying at a fixed coordinate point is called \emph{Eulerian}, whereas the variation measured by an observer comoving with the fluid element is said to be \emph{Lagrangian}.} perturbations of energy $\delta \epsilon$, pressure $\delta p$ and four-velocity $\delta u_{\mu}$ of the fluid. The explicit expression of $\delta T_{\mu \nu}$ reads
\begin{equation}
\delta T_{\mu \nu} = ( \delta \epsilon + \delta p) u_{\mu} u_{\nu} + (\epsilon + p)( \delta u_{\mu}  u_{\nu} + u_{\mu} \delta  u_{\nu}) + \delta p \, g_{\mu \nu}^0 + p \, h_{\mu \nu} \,,
\end{equation}
with $\delta u^{\mu} = \{h_{00} \mathrm{e}^{-3 \nu/2}/2 , \delta u^r, \delta u^{\theta}, \delta u ^{\varphi}\}$ and $\delta u^i = d \xi^i /d \tau$, where $\xi^i $ is the spatial displacement of the fluid element due to the perturbations and $\tau$ the proper time. Note that all the perturbative functions of the metric/fluid are independent of time, since we have assumed that the tidal field is stationary. Therefore they depend only on the spatial coordinates $\{r, \theta ,\varphi \}$.

The Einstein equations to solve read (we recall that the stress-energy tensor conservation law is a consequence of the field equations)
\begin{equation}
\label{eq:perteinstein}
\left \{
\begin{aligned}
& \delta G_{\mu \nu}=8\pi \delta T_{\mu \nu} \\
& \delta \left( \nabla_{\nu} T^{\mu \nu} \right)=  0
\end{aligned}
\right.\,,
\end{equation}
where $\delta G_{\mu \nu}$ is the perturbed part of the Einstein tensor and the $\delta$ in the second line denotes that only the perturbative terms of the four-divergence should be consider. We can greatly simplify the above system of equations choosing the Regge-Wheeler gauge~\cite{Regge:1957td}~\footnote{We can use the gauge freedom of General Relativity to set to zero four components of the metric, corresponding to the four arbitrary coordinate transformations.} and expanding the perturbative functions in spherical harmonics (in the Appendix~\ref{sec:appB} we recall the main properties of scalar, vector and tensor spherical harmonics). In this gauge the metric perturbation takes the form
\begin{equation}
\label{eq:pertmetric}
\begin{aligned}
h_{\mu\nu} &=  
\left( \begin{array}{cccc}
 \mathrm{e}^{\nu} H_{0,lm}(r) & H_{1,lm}(r) & 0 & 0   \\
* & \mathrm{e}^{\lambda} H_{2,lm}(r) & 0 & 0   \\
* & * & r^2 K_{lm}(r) & 0   \\
* & * & * & r^2 \sin^2{\theta} K_{lm}(r)
\end{array} \right) Y_{lm}\left( \theta ,\varphi \right) \\[0.4cm]
& + \left( \begin{array}{cccc}
 0 & 0 & h_{0,lm}(r) S_{\theta,lm}\left( \theta, \varphi \right) &  h_{0,lm}(r) S_{\phi,lm}\left( \theta ,\varphi \right) \\
* & 0 & h_{1,lm}(r) S_{\theta,lm}\left( \theta ,\varphi \right) & h_{1,lm}(r)  S_{\phi,lm}\left( \theta, \varphi \right)  \\
* & * & 0 & 0   \\
* & * & * & 0
\end{array} \right) \,,
\end{aligned}
\end{equation}
where $Y_{lm}$ and $\{S_{\theta,lm},S_{\phi,lm}\} = \{ -\partial_{\varphi}Y_{lm}/\sin{\theta}, \sin{\theta} \partial_{\theta} Y_{lm}  \}$ are the scalar and odd vector harmonics, respectively, and the sum over the indices $l,m$ is implicit. The star denotes the components obtained by symmetry. In the same way, we can decompose the fluid perturbations as~\cite{1967ApJ...149..591T}
\begin{equation}
\label{eq:pertfluid}
\begin{aligned}
\delta \epsilon & = \delta \epsilon_{lm}(r) Y_{lm}\left( \theta ,\varphi \right)  \\
\delta p & = \delta p_{lm}(r) Y_{lm}\left( \theta ,\varphi \right)  \\
\delta u^r & = W_{lm}(r) Y_{lm}\left( \theta ,\varphi \right) \\
\delta u^{\theta} & =  V_{lm}(r) R_{\theta,lm}\left( \theta, \varphi \right) +  U_{lm}(r) S_{\theta,lm}\left( \theta, \varphi \right)\\
\delta u^{\varphi} & = \frac{1}{\sin^2{\theta}} \left[ V_{lm}(r) R_{\varphi,lm}\left( \theta, \varphi \right) +  U_{lm}(r) S_{\varphi,lm}\left( \theta, \varphi \right) \right] \,,
\end{aligned}
\end{equation}
with $\{ R_{\theta,lm},R_{\varphi,lm} \}= \{\partial_{\theta}Y_{lm}, \partial_{\varphi}Y_{lm} \}$ the even vector spherical harmonics. We remark that Eqs.~\eqref{eq:pertmetric} and~\eqref{eq:pertfluid} are independent of time, because we have assumed that the perturbations are stationary.

The structure of the above expansion reflects the fact that there are two sectors of perturbations, \emph{polar}, or \emph{electric}, and \emph{axial}, or \emph{magnetic}, which are, respectively, even ($Y_{lm}, R_{A,lm},\  A=\{\theta ,\varphi \}$) and odd ($S_{A,lm},\  A=\{\theta ,\varphi \}$) under parity transformations. The polar sector is related to the electric tidal Love numbers, whereas the axial sector to the magnetic ones. Since the background spacetime is spherically symmetric, these two sectors are completely decoupled and can be solved independently. The expansion of the perturbative quantities in spherical harmonics reduces the problem to a system of equations depending only on the radial coordinate $r$. Furthermore, the spherical symmetry of the unperturbed configuration ensures also that: (i) the radial equations are independent of the index $m$, (ii) perturbations with different values of the index $l$ do not couple to each other. Therefore, we can solve the equations for any given order $l$ of the spherical harmonics independently. 

In the end, within the polar sector we obtain a system of ODEs in the radial coordinate $r$ for the functions $H_0(r), H_1(r), H_2(r), K(r), \delta \epsilon(r), \delta p(r), W(r) \text{ and } V(r)$, whereas for the axial one we obtain a system for the perturbations $h_0(r),h_1(r)$ and $U(r)$. We have dropped the $lm$-subscript in the radial functions to not burden the notation. In the following, we show how to calculate the electric and magnetic tidal Love numbers solving the polar and axial equations, respectively. Henceforth we restrict our analysis to $l \geq 2$, since no lower-order tidal fields do exist.

\subsubsection{The polar sector: electric tidal Love numbers}
\label{sec:polar}
In the polar sector, we can eliminate the function $\delta \epsilon(r)$ from the problem using the relation
\begin{equation}
\delta \epsilon = \frac{d\epsilon}{dp} \delta p \,,
\end{equation}
which is valid for a barotropic equation of state $p = p(\epsilon)$. Here $dp/d\epsilon = c_s^2$, where $c_s$ is the speed of sound in the fluid. Also, the conservation law for the stress-energy tensor provides the relation
\begin{equation}
\delta p = \frac{1}{2} (\epsilon+p) H_0 \,,
\end{equation}
that we can use to remove $\delta p(r)$ from the remaining equations. The Einstein equations also imply that $H_2(r) = H_0(r)$ (that we use to eliminate $H_2(r)$), arriving finally to a single second-order ODE for the function $H_0(r)$
\begin{equation}
\label{eq:polars}
\begin{aligned}
\dfrac{d^2 H_0(r)}{dr^2} & + \left(\frac{2\left[r-m+2 \pi r^3(p-\epsilon) \right]}{r(r-2m)} \right) \frac{d H_0(r)}{dr} + \left( \frac{4 \pi r^2 \left[p (9-16 \pi r^2 p) + 5 \epsilon\right] }{(r-2m)^2 }\right.\\
&  + \frac{ 4 \pi r^3 \left[ (r-2m)(p+\epsilon)/c_s^2 -2m(13p + 5 \epsilon) \right]-4m^2 }{r^2(r-2m)^2}\\
& \left. -\frac{l(l+1)}{r (r-2m)}\right) H_0(r) =0 \,.
\end{aligned}
\end{equation}
Then, the function $K(r)$ is given by a linear combination of $H_0(r)$ and its first derivative
\begin{equation}
\begin{aligned}
\label{eq:polars2}
K(r) = &  \left(- \frac{2 \left\{2m^2+ 4 \pi r^4 (p-8 \pi r^2 p^2+ \epsilon )+ m r \left[l(l+1)-4-8 \pi r^2(3p + \epsilon) \right]    \right\}  }{[l(l+1)-2]r(r-2m)} \right. \\
& \left.+ \frac{r}{r-2m} \right) H_0(r)  + \left( \frac{2(m+4 \pi r^3 p)}{l(l+1)-2}\right) H_0'(r) \,,
\end{aligned}
\end{equation}
where the prime denotes a derivative with respect to $r$. From the other Einstein equations, one can see that the function $H_1(r)$ vanishes identically in vacuum, and since the solution is unique, it must vanish everywhere, $H_1(r)=0$. As a consequence the fluid velocity perturbations vanish as well, $W(r)=V(r)=0$.

The Eq.~\eqref{eq:polars} can be integrated numerically, together with the TOV Eqs.~\eqref{eq:tov}, from the center of the star towards its surface~\cite{Hinderer:2007mb}. Requiring that $H_0(r)$ is regular~\footnote{See the discussion above Eq.~\eqref{eq:rotreg}.} at $r=0$ gives the boundary condition
\begin{equation}
\label{eq:H0init}
H_0(r) = a_0 \, r^{l} \left[1+ \mathcal{O} \left(r^2 \right) \right] \qquad r \to 0 \,.
\end{equation}
The constant $a_0$ can be chosen freely, since the Love numbers are independent of its value. Indeed, $a_0$ affects in the same way both the strength of the tidal field and the size of the induced multipolar deformation, and therefore it cancels out in the Love numbers, which are defined as the ratio of the two quantities. More details on the numerical integration are given in the Appendix~\ref{sec:appA}. At the star surface $r=R$, the function $H_0(r)$ should reduce to the vacuum solution, which is given by a linear combination of associated Legendre polynomials with $m=2$~\footnote{Note that usually the Legendre polynomials are defined on the unit complex disk, while we work with $x=r/M-1 >1$. This requires that we change the sign in the argument of the logarithmic terms of $Q_{l2}(x)$, or equivalently, we take their real part.}
\begin{equation}
\label{eq:extpolar}
H_0(r) = c_P \,  P_{l2}\left( \frac{r}{M}-1 \right) + c_Q \, Q_{l2}\left( \frac{r}{M}-1 \right)  \qquad r \geq R\,,
\end{equation}
where $M=m(R)$ is the mass of the star, and the integration constants $c_P$ and $c_Q$ are determined in terms of $H_0(R)$ and $H'_0(R)$ by matching to the interior solution.

The asymptotic expansion of the solution~\eqref{eq:extpolar} for $r \to \infty$ takes the form
\begin{equation}
\label{eq:h0ext}
H_0(r)=  \left[  C_P \,r^{l} + C_Q \, \frac{1}{r^{l+1}} \right] \left[ 1+ \mathcal{O} \left( \frac{M}{r} \right) \right] \qquad r \to \infty \,,
\end{equation}
where we have used the uppercase to distinguish the coefficients $C_P$ and $C_Q$ from before, since now they include also the dependence on the mass $M$ and the numerical factors coming from the expansion. Plugging the above result into the $g_{00}$ component of the metric we obtain 
\begin{equation}
\label{eq:extlarger}
g_{00} \sim -1 + \frac{2M}{r} + \sum_{l \geq 2,m} \left( \frac{1}{r^{l+1}} \, C_{Q,lm} + r^{l} \,  C_{P,lm}     \right) Y_{lm}    \,,
\end{equation}
where we have restored the $lm$-subscript in the radial solution. The last step is comparing the above equation with the asymptotic expansion of the metric in Eqs.~\eqref{eq:metrictidal}, which we rewrite as
\begin{equation}
\label{eq:extrinf}
g_{00} \sim -1 + \frac{2M}{r} + \sum_{l \geq 2,m} \left( \frac{1}{r^{l+1}}  \frac{2(2l-1)!!}{l!} Q_{lm} + r^{l} \frac{2}{l!} G_{lm}    \right) Y_{lm}  \,,
\end{equation}
where we have decomposed the multipole moments using symmetric trace-free tensors defined by Thorne~\cite{Thorne:1980ru}
\begin{equation}
\begin{aligned}
Q_L =& \sum_m Q_{lm} \mathcal{Y}^{lm}_L \\
G_L = & \sum_m G_{lm} \mathcal{Y}^{lm}_L \,,
\end{aligned}
\end{equation}
which satisfy the property $\mathcal{Y}_L^{lm}n_L = Y_{lm}$. Matching Eqs.~\eqref{eq:extlarger} and~\eqref{eq:extrinf}, we identify the growing solution in $H_0(r)$ with the tidal field and the decreasing one with the multipolar deformation of the object, respectively. Then, we extract the multipole moments of the object and of the tidal field in terms of the coefficients $C_{Q,lm}$ and $C_{P,lm}$, respectively. Using the adiabatic relations~\eqref{eq:adiabaticrel}, we finally obtain the electric tidal Love numbers from the ratio
\begin{equation}
\label{eq:electricTLN}
\lambda_l = \frac{Q_{lm}}{G_{lm}} \,.
\end{equation}

Two important comments have to be remarked. First, the LHS of Eq.~\eqref{eq:electricTLN} is independent of the index $m$ by definition, while the RHS seems to depend on it. Actually, also the ratio $Q_{lm}/G_{lm} \propto C_{Q,lm}/C_{P,lm}$ does not depend on $m$. Indeed, since the radial equation for $H_0(r)$ is independent of $m$, the only way in which different values of $m$ can affect the solution is through the initial condition $a_0$, that can be chosen independently for each value of $m$. However, as we said, the choice of the boundary condition at the center of the star contributes in the same way to both the growing and the decreasing part of metric perturbations at large distance, cancelling out in the ratio of the coefficients. Therefore, we can restrict without loss of generality to the axisymmetric case $m=0$, which is enough to compute the Love numbers.

The second remark concerns the match of the asymptotic solution in Eq.~\eqref{eq:extlarger} to the metric~\eqref{eq:extrinf}. Doing so, we have neglected the relativistic corrections in $H_0(r)$, i.e. the terms of order $\mathcal{O} \left(M/r \right)$ in the square brackets of Eq.~\eqref{eq:h0ext}. One may wonder if the higher-order terms of the growing solution can mix with the leading-order term of the decreasing solution. If this is the case, the multipolar deformation would be contaminated by the tidal field, and the definition of the Love numbers would suffer of ambiguities. However, if the central object is non-rotating, it has been shown that it is always possible to distinguish the tidal field from the multipolar response, through an analytic continuation in the number of dimensions of the spacetime $d$~\cite{Kol:2011vg} or in the multipolar index $l$~\cite{Chakrabarti:2013lua}~\footnote{The separation of the two parts of the solution is achieved recognizing that the general solution of $h_{00}$ at large distance reads
\begin{equation}
h_{00} \sim A r^l(1+ \dots) + B \frac{1}{r^{l+d-3}} (1+ \dots) \,,
\end{equation}
where the ellipsis represents a series in $1/r$ which may also contain logarithmic terms and $A$ and $B$ are two integration constants. Comparing the above solution with the Newtonian potential allows us to uniquely identify the first term with the tidal part and latter one with the multipolar response. Treating the indices $d$ and $l$ as real numbers, the two solutions can never mix, even if they have common terms in the series expansion.}. The result of this procedure is in agreement with the separation of the two contributions in Eq.~\eqref{eq:extlarger}.

The proof that all the electric tidal Love numbers vanish in the black hole case is given by requiring that the curvature invariants are regular on the event horizon $r= 2M$~\footnote{In the Regge-Wheeler gauge, requiring that the metric perturbations $h_{\mu \nu}$ are regular at the horizon is not sufficient to prove that the spacetime is well-behaved, since the background metric itself is singular at $r=2M$ in this coordinate system. For instance, in this case $h_{00}\sim (1-2M/r) H_0(r) $ is regular at the horizon (though the solution of $H_0(r)$ given in Eq.~\eqref{eq:extpolar} is not, due to the presence of $Q_{l2}\left( \frac{r}{M}-1 \right)$).}. Imposing that the Kretschmann scalar $R_{\mu \nu \alpha \beta}R^{\mu \nu \alpha \beta}$ is regular at the horizon returns $c_Q=0$, which kills the decreasing solution corresponding to the multipolar response of the object. This implies the vanishing of the multipole moments $Q_{lm}$, and then of the Love numbers.

\paragraph{Quadrupolar electric tidal Love number}
\label{sec:polarsub}
Lastly, we give the explicit expression of the electric tidal Love number for $l=2$, which can be obtained through the procedure outlined above. The quadrupolar one is the main contribution to the tidally induced deformations of neutron stars, and can be exploited to shed light on their internal structure, as we discuss in Chapter~\ref{sec:inverse}. The result for the quadrupolar electric Love number is
\begin{equation}
\label{eq:lovenumber2el}
\begin{aligned}
k_2^E = &  \frac{8C^5}{5} (1-2C)^2 \left[2+2C(y-1)-y \right] \Big\{ 2C\left[6-3y+3C(5y-8) \right]    \\
&  + 4C^3 \left[13-11y+C(3y-2)+2C^2(1+y) \right] \\
&  + 3(1-2C)^2 \left[2-y+2C(y-1) \right] \log{(1-2C)} \Big\}^{-1} \,,
\end{aligned}
\end{equation}
where $C=M/R$ is the compactness of the star and
\begin{equation}
y = \frac{R H_0'(R)}{H_0(R)}
\end{equation}
is evaluated at the surface $r=R$ integrating Eq.~\eqref{eq:polars} in the neutron star interior.

\begin{figure}
\captionsetup[subfigure]{labelformat=empty}
\centering
{\begin{tikzpicture}[baseline]
\begin{axis}[
axis x line=bottom,
axis y line=left,
enlargelimits,
width=0.45\textwidth,
height=0.45\textwidth,
scaled y ticks = false,
ytick={2000,4000,6000,8000,10000,12000},
yticklabels={$2$,$4$,$6$,$8$,$10$,$12$},
xlabel=$M \, \text{[}M_{\odot}\text{]}  $,
ylabel=$\lambda_2 \, \text{[} 10^3 \times \text{km}^5\text{]} $]
\addplot[smooth,thick,blue] table {chapter_1/figures/lambda2.dat};
\end{axis}
\end{tikzpicture}
}
{
\begin{tikzpicture}[baseline]
\begin{axis}[ymin=0,xmin=0,ymax=0.27,xmax=0.33,
axis x line=bottom,
axis y line=left,
width=0.45\textwidth,
height=0.45\textwidth,
legend style={draw=none},
ytick={0,0.05,0.1,0.15,0.2,0.25},
yticklabels={$0$,$0.05$,$0.1$,$0.15$,$0.2$,$0.25$}, 
xlabel=$C $, ylabel=$k_2^E$]
\addplot[smooth,thick,blue] table {chapter_1/figures/k2.dat};
\addplot[smooth,thick,red,dashed] table {chapter_1/figures/k2_poly.dat};
\addlegendentry{real. EOS};
\addlegendentry{poly. EOS};
\end{axis}
\end{tikzpicture}
}
\caption{\textsl{(Left) Quadrupolar electric tidal deformability as a function of the mass for a realistic neutron star equation of state. (Right) Quadrupolar electric Love number as a function of the compactness for realistic and polytropic equations of state.}}
\captionsetup{format=hang,labelfont={sf,bf}}
\label{fig:electricquad}
\end{figure}
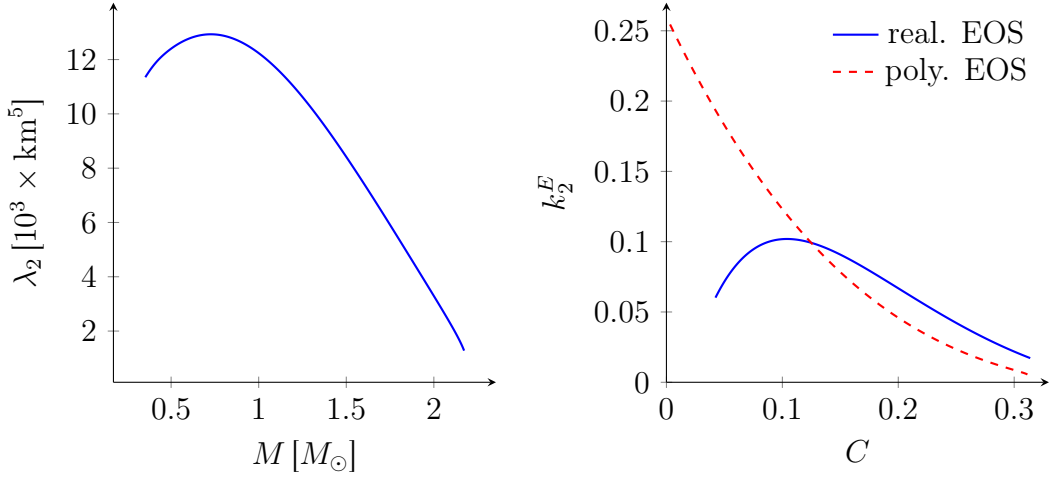

In Fig.~\ref{fig:electricquad}, we show, as examples, the quadrupolar deformabilities and Love numbers of different equilibrium configurations of a neutron star. In the left panel we plot $\lambda_2$ as a function of the mass of the neutron star, for a given equation of state. The impact of different equations of state is discussed in Chapter~\ref{sec:inverse}. In the right panel we compare $k_2^E$, as a function of the compactness, for ``realistic'' and polytropic equations of state (see section~\ref{sec:polytropicstate} for a review of polytropic equations of state). We can see that in the polytropic case the dimensionless Love number tends to a constant value in the limit $C \to 0$, which corresponds to the Newtonian limit~\footnote{The limit $C \to 0$ for ``realistic'' equations of state has no real physical meaning, since no stable neutron star can form with such low mass. In this case, the fact that the Love number tends to zero for small values of the compactness is a consequence of the small values assumed by the adiabatic index $\Gamma = (\epsilon+p)c_s^2/p$ (see section~\ref{sec:polytropicstate}) at small densities. Indeed, when $\Gamma \to 6/5$, the radius of a Newtonian polytropic star tends to infinity, and then the Love number, which scales with $R$, vanishes~\cite{Damour:2009vw}.}. In the next section instead, we show that the magnetic Love numbers vanish in the Newtonian limit, since they are a relativistic effect. Note that within the definition used, the electric deformabilities/Love numbers are always positive. This is true at any multipolar order (cf. section~\ref{sec:axialsub}).

\subsubsection{The axial sector: magnetic tidal Love numbers}
\label{sec:axial}
In the axial sector the Einstein equations~\eqref{eq:perteinstein} return $h_1(r) =0 $ and a single second-order ODE for the function $h_0(r)$,
\begin{equation}
\label{eq:axials}
\begin{aligned} 
\frac{d^2h_0(r)}{dr^2} & - \left( \frac{4 \pi r^2(p+\epsilon)}{r-2m} \right)  \frac{dh_0(r)}{dr} - \left(\frac{l(l+1)r-4m + 8 \pi r^3(p+ \epsilon)}{r^2(r-2m)} \right)  h_0(r) \\
&= \left(\frac{16 \pi r^3 (p + \epsilon) \mathrm{e}^{\nu/2}}{r-2m} \right) U(r) \,.
\end{aligned}
\end{equation}
Differently from the polar sector, the problem is not determined: we have to specify the internal dynamics of the fluid to get the metric perturbation. This passage is crucial, because $h_0(r)$ is the radial part of the time-angular component of the metric, which will be used to extract the multipole moments. In other words, the magnetic tidal Love numbers depend on the internal dynamics of the object. Two different assumptions can be made for the motion of the fluid:
\begin{description}
\item[Static fluid] \ \\[-0.3cm]

The first assumption is that the fluid remains strictly \emph{static} also in the perturbed configuration. A fluid is static if the spatial components of the four-velocity vanish, as it is in the unperturbed (non-rotating) configuration. If we assume the fluid to be static also in presence of the tidal field, all the spatial perturbations of the fluid four-velocity must vanish: $U(r)=0$ (we have seen in the previous section that $W(r)$ and $V(r)$ vanish in any case because of the Einstein equations). Thus, the tidal field does not excite any internal motion in the star. 

\item[Irrotational fluid] \ \\[-0.3cm]

In the second case the fluid is assumed to be in an \emph{irrotational} state. A fluid is said to be irrotational if the circulation of the spatial part of the four-velocity around any spatial closed circuit vanishes, i.e., it is vorticity-free~\cite{Favata:2005da,Landry:2015cva}. Indeed, the vorticity four-vector $\omega^{\alpha}=\frac{1}{2} \epsilon^{\alpha \beta \mu \nu} u_{\nu} \nabla_{\mu} u_{\beta}$~\cite{Rezzolla:2013} (where $\epsilon^{\alpha \beta \mu \nu}$ is the Levi-Civita pseudo-tensor) vanishes identically in irrotational configurations. This is surely true in the unperturbed configuration. Furthermore, since the circulation is conserved, if a fluid is in an irrotational state at a given time, it remains irrotational at all times, even if the external perturbation is time-dependent. For these reasons, the assumption of an irrotational fluid is more realistic. Indeed, static fluids would be compatible only with the idealistic assumption of a true stationary tidal field, while irrotational fluids can sustain also a slowly varying in time tidal field. Recent simulations of binary neutron star mergers from Numerical Relativity assume the fluids of the stars to be in an irrotational state~\cite{Bonazzola:1998yq,Bernuzzi:2011aq,Haas:2016cop,Tichy:2016vmv}.

In the irrotational case, it can be shown that $U(r)$ is given by the expression
\begin{equation}
\label{eq:uirr}
U(r) = -\frac{\mathrm{e}^{-\nu/2} }{r^2} h_0(r) \,.
\end{equation}
We can see that the difference between static/irrotational fluids does not affect the polar sector and then the electric Love numbers ($W(r)=V(r)$ always, whatever the state of the fluid is). However, this is not the case for the axial sector. The different expressions for $U(r)$ lead the equation~\eqref{eq:axials} for $h_0(r)$ to differ in the star interior for the static/irrotational cases. As we show below, this gives rise to different magnetic tidal Love numbers. In conclusion, magnetic deformations depend on the state of the fluid. On the other hand, electric deformation are insensible to the internal motion of the fluid, determining uniquely (for a given equation of state) the electric Love numbers.

A last, very important remark concerns the expression of $U(r)$ in the irrotational fluid case given in Eq.~\eqref{eq:uirr}. If we allow the metric/matter perturbations to vary in time, the problem is fully determined and the fluid perturbations are fixed by the metric. In this case, it can be shown that $U(t,r)$, which now is time-dependent, assumes \emph{exactly} the same expression as in the irrotational fluid case, Eq.~\eqref{eq:uirr}. This means that taking the stationary limit of a time-dependent perturbation put the fluid in an irrotational state~\cite{Pani:2018inf}. The latter result further supports the choice of an irrotational fluid over a static one.
\end{description}

Replacing the static/irrotational expressions for $U(r)$ in Eq.~\eqref{eq:axials} we get
\begin{equation}
\label{eq:axials2}
\frac{d^2h_0(r)}{dr^2} - \left( \frac{4 \pi r^2(p+\epsilon)}{r-2m} \right)  \frac{dh_0(r)}{dr} - \left(\frac{l(l+1)r-4m \pm 8 \pi r^3(p+ \epsilon)}{r^2(r-2m)} \right)  h_0(r) =0 \,,
\end{equation}
where the plus sign refers to static fluids and the minus sign to irrotational fluids. The procedure to compute the magnetic Love numbers is analog to that of the electric sector.

The Eq.~\eqref{eq:axials2} is integrated numerically, with boundary conditions at the center of the star given by
\begin{equation}
\label{eq:h0init}
h_0(r) = b_0 \, r^{l+1} \left[1+ \mathcal{O} \left(r^2 \right) \right] \qquad r \to 0 \,,
\end{equation}
for both irrotational and static fluids. As in the electric case, the constant $b_0$ cancels out in the definition of the Love numbers and therefore its value is irrelevant. More details are given in the Appendix~\ref{sec:appA}. At the star surface $r=R$, we match the interior solution with the exterior one, which is given by
\begin{equation}
\label{eq:extaxials}
\begin{aligned}
h_0(r) =& d_P \left(\frac{r}{2M} \right)^{l+1} \,_2F_1 \left(-l+1,-l-2,-2l;\frac{2M}{r} \right) \\
+& d_Q   \left(\frac{2M}{r} \right)^l \,_2F_1 \left(l-1,l+2,2l+2;\frac{2M}{r} \right)  \qquad r \geq R\,,
\end{aligned}
\end{equation}
where $_2F_1 \left(a,b,c; x \right)$ is the hypergeometric function. The integration constants $d_P$ and $d_Q$ are given by the matching procedure in terms of $h_0(R)$ and $h_0'(R)$.

For $r \to \infty$, the asymptotic expansion of Eq.~\eqref{eq:extaxials} reads
\begin{equation}
\label{eq:h0ext2}
h_0(r)=  \left[  D_P \,r^{l+1} + D_Q \, \frac{1}{r^{l}} \right] \left[ 1+ \mathcal{O} \left( \frac{M}{r} \right) \right] \qquad r \to \infty \,,
\end{equation}
where, alike the electric case, we have included the mass dependence and the numerical factors in the coefficients $D_P$ and $D_Q$. Restoring the $lm$-subscript in the radial solution, the $g_{0\varphi}$ component of the metric takes the form
\begin{equation}
\label{eq:extlarger2}
g_{0\varphi} \sim  \sum_{l \geq 2,m} \left( \frac{1}{r^{l}} \, D_{Q,lm} + r^{l+1} \,  D_{P,lm}     \right) S_{\varphi,lm}  \,.
\end{equation}
Finally, we match the above expansion to the time-angular component of metric in Eqs.~\eqref{eq:metrictidal}
\begin{equation}
\label{eq:extrinf2}
g_{0\varphi} \sim  \sum_{l \geq 2,m} \left( \frac{1}{r^{l}}  \frac{4(2l-1)!!}{(l+1)!} S_{lm} - r^{l+1} \frac{1}{(l+1)!} H_{lm}    \right) S_{\varphi,lm}  \,,
\end{equation}
which we have obtained from the time-space components $g_{0i}$ using the relations
\begin{equation}
\begin{aligned}
S_L =& \sum_m S_{lm} \mathcal{Y}^{lm}_L \\
H_L = & \sum_m H_{lm} \mathcal{Y}^{lm}_L \,,
\end{aligned}
\end{equation}
and the spherical harmonic properties described in the Appendix~\ref{sec:appB}. Comparing Eqs.~\eqref{eq:extlarger2} and~\eqref{eq:extrinf2}, one identifies the growing solution in $h_0(r)$ with the tidal field, and the decreasing solution with the multipolar response of the object, respectively. The multipole moments of the object and of the tidal field are extracted in terms of the coefficients $D_{Q,lm}$ and $D_{P,lm}$, respectively. By means of the adiabatic relations~\eqref{eq:adiabaticrel} the magnetic tidal Love numbers are defined as
\begin{equation}
\label{eq:magneticTLN}
\sigma_l = \frac{S_{lm}}{H_{lm}} \,.
\end{equation}

The whole discussion below Eq.~\eqref{eq:electricTLN} is equally valid in the magnetic case. Since the details have been already given in previous section, we only summarize here the main points. (i) The magnetic Love numbers do not depend on the spherical harmonics index $m$, and we can restrict to the axisymmetric case without loss of generality. (ii) The separation of the growing/decreasing solutions in Eq.~\eqref{eq:h0ext2} defines unambiguously the magnetic Love numbers, without any mixing of the higher-order terms in $M/r$. (iii) The magnetic Love numbers of a black hole vanish at any mulitpolar order $l$, because the coefficients $d_Q$ vanish as well from the requirement that the Chern-Pontryagin scalar $\frac{1}{2} \epsilon^{\mu \nu}_{\ \ \gamma \delta} R_{\mu \nu \alpha \beta} R^{\alpha \beta \gamma \delta}$ (where $\epsilon^{\mu \nu \gamma \delta}$ is the Levi-Civita pseudo-tensor) is regular at the horizon.

\paragraph{Quadrupolar magnetic tidal Love number}
\label{sec:axialsub}
\begin{figure}
\captionsetup[subfigure]{labelformat=empty}
\centering
{\begin{tikzpicture}[baseline]
\begin{axis}[
axis x line=bottom,
axis y line=left,
enlargelimits,
legend style={draw=none},
legend pos=south east,
width=0.45\textwidth,
height=0.45\textwidth,
xlabel=$M \, \text{[}M_{\odot}\text{]}  $,ylabel=$|\sigma_2| \, \text{[} \text{km}^5\text{]} $]
\addplot[smooth,thick,blue] table {chapter_1/figures/sigma2_irr.dat};
\addplot[smooth,thick,blue,dashed] table {chapter_1/figures/sigma2_stat.dat};
\addlegendentry{irro. fluid};
\addlegendentry{stat. fluid};
\end{axis}
\end{tikzpicture}
}
{
\begin{tikzpicture}[baseline]
\begin{axis}[
axis x line=center,
axis y line=left,
width=0.45\textwidth,
height=0.45\textwidth,
scaled y ticks = false,
legend style={draw=none},
legend style={at={(0.5,0)},anchor=north},
ytick={-0.02,0,0.02,0.04},
yticklabels={$-2$,$0$,$2$,$4$},
xlabel=$C $,ylabel=$k_2^M \times 10^{-2}$]
\addplot[smooth,thick,blue] table {chapter_1/figures/j2_irr.dat};
\addplot[smooth,thick,blue,dashed] table {chapter_1/figures/j2_stat.dat};
\addplot[smooth,thick,red,dotted] table {chapter_1/figures/j2_irr_poly.dat};
\addplot[smooth,thick,red,dashdotted] table {chapter_1/figures/j2_stat_poly.dat};
\addlegendentry{irro. fluid, real. EOS};
\addlegendentry{stat. fluid, real. EOS};
\addlegendentry{irro. fluid, poly. EOS};
\addlegendentry{stat. fluid, poly. EOS};
\end{axis}
\end{tikzpicture}
}
\caption{\textsl{(Left) Quadrupolar magnetic tidal deformability in the static/irrotational case plotted as a function of the mass, for a realistic neutron star equation of state. (Right) Quadrupolar magnetic Love number as a function of the compactness, for realistic and polytropic equations of state. Static and irrotational fluids are compared.}}
\captionsetup{format=hang,labelfont={sf,bf}}
\label{fig:magneticquad}
\end{figure}
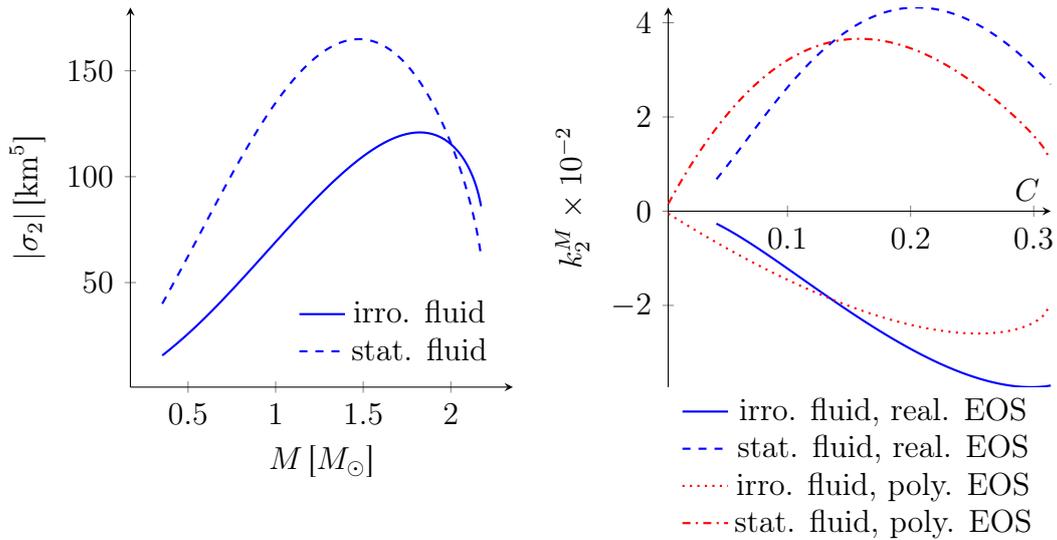

In the quadrupolar case $l=2$, the explicit expression for the magnetic tidal Love number is given by
\begin{equation}
\label{eq:lovenumber2mag}
\begin{aligned}
k_2^M = & \frac{96 C^5}{5} \left[ 3+ 2 C(y-2)-y \right] \Big\{ 2C\left\{9 -3y + C \left[ 3(y-1)+2C(C+y +C y) \right] \right\}  \\
& + 3 \left[ 3+2C(y-2)-y\right] \log{(1-2C)} \Big\}^{-1}\,,
\end{aligned}
\end{equation}
where $R$ and $C$ are, respectively, the radius and the compactness of the star and
\begin{equation}
y = \frac{R h_0'(R)}{h_0(R)} \,.
\end{equation}
The difference between static and irrotational fluids is encoded in the parameter $y$, which changes as a result of the integration of different differential equations in the star interior.

In Fig.~\ref{fig:magneticquad}, we compare the quadrupolar deformabilities and Love numbers for static and irrotational fluids. Remarkably, the magnetic Love numbers of a static fluid are positive, while those of an irrotational fluid are negative. This is a general feature, true at any multipolar order. In the left panel we show the absolute value of $\sigma_2$ in the two cases, for a realistic equation of state. Note that the magnitudes of the tidal deformabilities in the two cases are comparable (less than a factor 2), and are a factor $\sim 100$ smaller than the electric ones, see Fig.~\ref{fig:electricquad} (cf. also section~\ref{sec:universal}). In the right panel we show $k_2^M$ for the two fluid states and for realistic and polytropic equations of state. Note that the dimensionless Love number vanish for $C \to 0$, i.e., in the Newtonian limit (cf. section~\ref{sec:polarsub}). This result is consistent with the relativistic nature of the magnetic deformations.

\subsection{Rotational tidal Love numbers}
\label{sec:RTLN}
\begin{figure}[]
\centering
\includegraphics[width=0.7\textwidth]{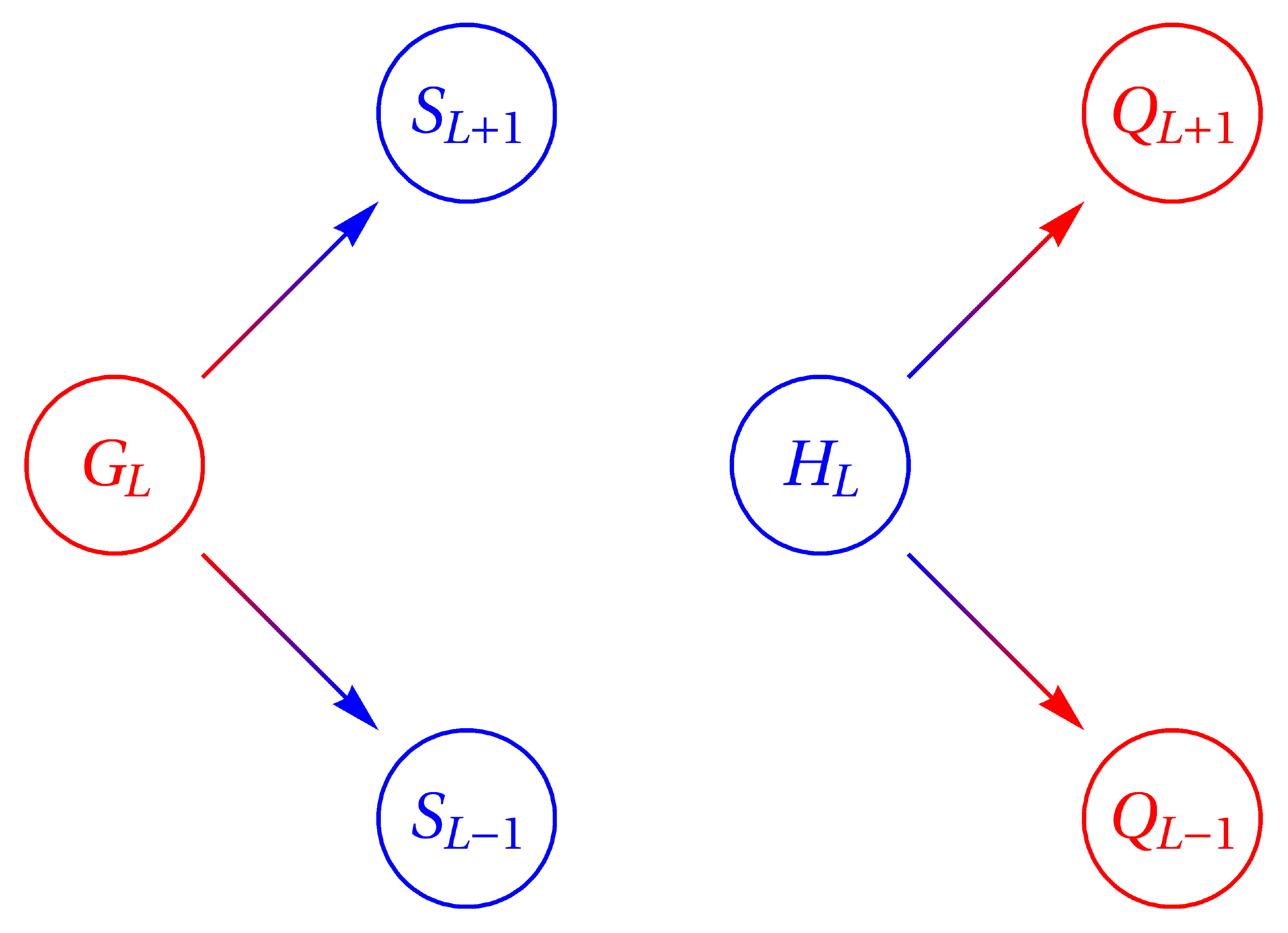}
\caption{\textsl{Scheme of the multipolar deformations induced by the interaction between the tidal field and the rotation of the central object.}}
\captionsetup{format=hang,labelfont={sf,bf}}
\label{fig:rotidal}
\end{figure}

The relativistic theory of tidal Love numbers has been extended to slowly spinning compact objects by Pani, Gualtieri, Maselli and Ferrari~\cite{Pani:2015hfa,Pani:2015nua} and Landry and Poisson~\cite{Landry:2015zfa,Landry:2015snx,Landry:2017piv}. The effect of the rotation is to couple the electric and magnetic sectors. To first order in the spin, the electric tidal fields (even parity) of multipolar order $l$ induce magnetic deformations (odd parity) with multipolar order $l \pm 1$, and viceversa. The scheme of these relations is shown in Fig.~\ref{fig:rotidal}. A new class of tidal Love numbers arises from the interaction between the spin and the tidal field, dubbed \emph{rotational tidal Love numbers}. Working to linear order in the tidal fields, these new Love numbers are defined as the constants of proportionality between the induced multipole moments of the object and the tidal multipole moments of different parity and multipolar order $l$. The adiabatic relations~\eqref{eq:adiabaticrel} are then modified as follows
\begin{equation}
\label{eq:adiabaticrelspin}
\begin{split}
& \begin{aligned}
Q_{ij} & = \lambda_2 \, G_{ij} + \lambda_{23} \, J_k \, H_{ijk} \\
S_{ij} & = \sigma_2 \,  H_{ij} + \sigma_{23} \, J_k \, G_{ijk}
\end{aligned}\\
& \begin{aligned}
Q_L & = \lambda_l \, G_L + \lambda_{l \, l+1} J_k \, H_{kL} +  \lambda_{l \, l-1} \, J_{\langle a_l} \, H_{L-1 \rangle} \\
S_L & = \sigma_l \, H_L + \sigma_{l \, l+1} \, J_k  \, G_{kL} +  \sigma_{l \, l-1} \, J_{\langle a_l} \, G_{L-1 \rangle}
\end{aligned} \qquad l \geq 3 \,,
\end{split}
\end{equation}
where $\lambda_{l \, l'}$ and $\sigma_{l \, l'}$ are the new rotational tidal Love numbers. The quadrupolar case $l=2$ is separated from the others, because the coupling can occur only with the octupolar tidal fields (no dipolar tidal field does exist).

It has been shown that also these new tidal Love numbers vanish in the black hole case. Furthermore, the extension of the theory to the rotating case has pointed out a new element to favor the irrotational fluid state over the static one. Indeed, Landry and Poisson~\cite{Landry:2015snx} showed that the fluid perturbation induced by a magnetic tidal field in a rotating star is time dependent, even if the magnetic tidal field is stationary. The time dependent perturbation is confined in the star interior, the external spacetime remains stationary. This is not possible if the fluid is imposed to be in a static state. Indeed, a static fluid would violate the Einstein equations, if the magnetic tidal field is not axisymmetric. This limitation of the static state further suggests that the correct fluid state (in the non-rotating limit) is the irrotational one.

Taking into account the rotation of the central object, computing the Love numbers gets more involved. Here we refer to the works of Pani et al.~\cite{Pani:2015hfa,Pani:2015nua}, outlining their procedure without giving all the details of the computation.

First, one derives the linearized Einstein equations arising from a small, stationary perturbation of the background metric. The unperturbed configuration is no longer spherically symmetric like the spacetime in Eq.~\eqref{eq:unpert}, but is given by the axisymmetric metric of a rotating object. Assuming rigid rotation, the background configuration to first order in the angular velocity $\Omega$~\footnote{Pani et al. considered also second-order corrections in the spin, which would give rise to other rotational tidal Love numbers.} (or equivalently in the spin of the body) is given in section~\ref{sec:rotation}:
\begin{gather}
ds^2=-\mathrm{e}^{\nu (r)}dt^2+\mathrm{e}^{\lambda (r)}dr^2+r^2(d\theta^2 +\sin^2\theta d\varphi^2) -2 \omega(r) r^2 \sin^2\theta dt d\varphi \\[0.3cm]
T^0_{\mu \nu}=(\epsilon + p)u_{\mu}u_{\nu}+p g^0_{\mu \nu} \qquad u^{\mu}=(\mathrm{e}^{-\nu(r)/2},0,0,\Omega \mathrm{e}^{-\nu(r)/2}) \\[0.3cm]
\left \{
\begin{aligned}
\frac{dm}{dr} & =4\pi r^2 \epsilon \\
\frac{dp}{dr}& = -\frac{(\epsilon+p)(m+4\pi r^3 p)}{r(r-2m)} \\
\frac{d\nu}{dr} & = \frac{2(m+4\pi r^3 p)}{r(r-2m)} \\
\frac{d^2 \bar{\omega}}{dr^2} & =\left[\frac{4 \pi r^2 \left( \epsilon + p\right)}{r-2m} -\frac{4}{r}  \right] \frac{d \bar{\omega}}{dr} +\left[ \frac{16 \pi r \left( \epsilon + p\right)}{r-2m} \right] \bar\omega  
\end{aligned} 
\right. \,.
\end{gather}
One perturbs the above spacetime exactly like in the previous section,
\begin{equation}
\begin{aligned}
g_{\mu \nu}&= g^0_{\mu \nu} + h_{\mu \nu} \\
T_{\mu \nu}& = T^0_{\mu \nu} + \delta T_{\mu \nu} \,,
\end{aligned}
\end{equation}
and expands the perturbative functions in spherical harmonics (Eqs.~\eqref{eq:pertmetric} and~\eqref{eq:pertfluid}). Plugging the above decomposition in the system
\begin{equation}
\left \{
\begin{aligned}
& \delta G_{\mu \nu}=8\pi \delta T_{\mu \nu} \\
& \delta \left( \nabla_{\nu} T^{\mu \nu} \right)=  0
\end{aligned}
\right.\,,
\end{equation}
one gets rid of the angular dependence, and finds the Einstein equations to first order in the spin for the radial part of the perturbative functions~\cite{Kojima:1992ie}. Several differences arise with respect to the non-rotating case:
\begin{itemize}
\item[i)] the radial equations depend on the azimuthal index $m$, so the degeneracy in $m$ is removed.
\item[ii)] the polar and axial sectors are not decoupled any longer. Polar perturbations with indices $\{l,m\}$ couple to axial perturbations with indices $\{l \pm 1 ,m\}$ and viceversa (cf. with the discussion above Eqs.~\eqref{eq:adiabaticrelspin}).
\end{itemize}
Next, we expand the generic radial function $f_{lm}(r)$ of the metric/fluid perturbations in powers of $\Omega$
\begin{equation}
f_{lm}(r) = f^0_{lm}(r) + f^1_{lm}(r) + \mathcal{O}\left(\Omega^2 \right) \,,
\end{equation}
where $f^0_{lm}(r)$ is zeroth-order in spin function, which solves the equations in the non-rotating case, and $f^1_{lm}(r)$ the first-order correction in $\Omega$. In this way, it can be shown that system reduces to a set of equations for the unknown functions $\{f^1_{lm}(r)\}$, with source terms depending on the non-spinning solutions $\{f^0_{lm}(r)\}$.

\begin{figure}
\centering
\begin{tikzpicture}
\begin{loglogaxis}[ymin=4,ymax=100,
axis x line=bottom,
axis y line=left,
width=0.8\textwidth,
height=0.55\textwidth,
ytick={1,10,100},
ylabel=$\bar{I}$,xlabel=$\Lambda_2$]
\addplot
[domain=1:20000,samples=100,smooth,
thick,red]
{exp(1.47+0.0817*ln(x)+0.0149*(ln(x))^2+2.87*(10^(-4))*(ln(x))^3-3.64*(10^(-5))*(ln(x))^4)};
\end{loglogaxis}
\end{tikzpicture}
\caption{\textsl{I-Love relation between the dimensionless moment of inertia $\bar{I}$ and the dimensionless electric quadrupolar tidal deformability $\Lambda_2$.}}
\captionsetup{format=hang,labelfont={sf,bf}}
\label{fig:ILQ}
\end{figure}
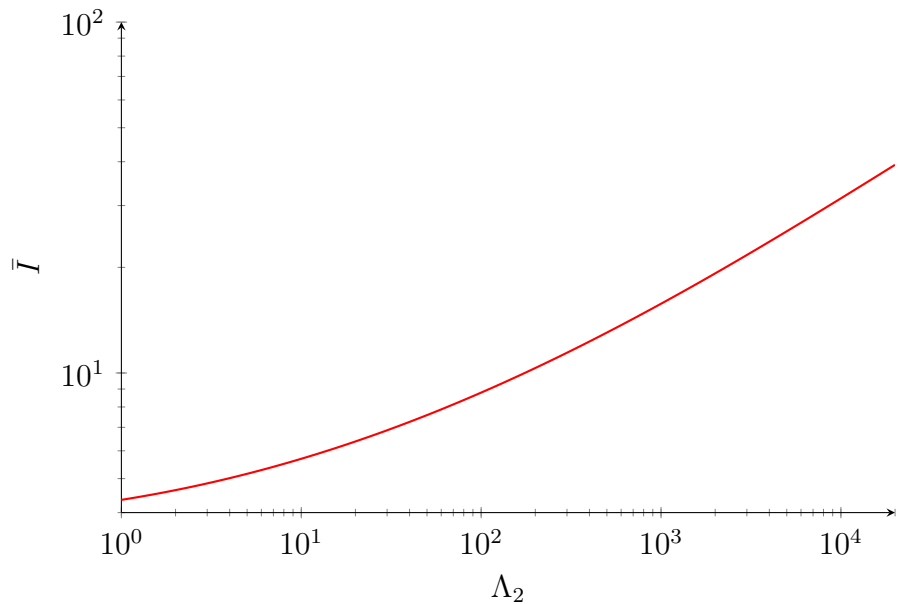

Then, to compute the Love numbers of a spinning object, one takes the following steps:
\begin{itemize}
\item[1)] integrate numerically the Einstein equations in the star interior.
\item[2)] match the interior solution to the exterior one at the star surface.
\item[3)] extract the spacetime multipole moments.
\end{itemize}
Note that the last step requires some prescription, because the tidal field and the multipolar response of the object are not clearly separated in the rotating case, as anticipated in section~\ref{sec:TLNgr}. Furthermore, instead of matching the exterior metric to the asymptotic expansion~\eqref{eq:metrictidal}, the multipole moments are defined in a gauge-invariant way following the Ryan approach~\cite{Ryan:1995wh,Ryan:1997hg,Pappas:2012nt}.

\subsection{Quasi-universal relations}
\label{sec:universal}

In the previous sections we have said that neutron star observables, such as mass, radius and tidal deformability, depend on the underlying equation of state of nuclear matter (cf. Chapter~\ref{sec:inverse}). Thus, for a given mass, different equations of state give rise to different radii and tidal deformabilities. In this sense, we say that the relation between the mass and the radius, or the mass and the tidal deformability, is EOS-dependent.

In 2013, Yagi and Yunes~\cite{Yagi:2013bca,Yagi:2013awa} found that the moment of inertia (defined in Eq.~\eqref{eq:inertia}), the spin-induced quadrupole moment and the electric quadrupolar tidal deformability of neutron stars can be linked to each other through three remarkable, almost EOS-independent relations, the so-called \emph{I-Love-Q relations}. This means that the knowledge of one of these quantities determines automatically the other two~\footnote{The I-Love-Q relations hold actually among the dimensionless versions of these observables, i.e., divided by appropriate powers of mass and spin of the star. Therefore, a measurement, for instance, of the moment of inertia alone, is not enough to determine the tidal deformability, if the mass of the star is unknown.}. The accuracy of these relations is at level of $\sim1\%$, and they hold even for quark stars. In Fig.~\ref{fig:ILQ} we show, as an example, the I-Love branch, i.e., the relation between the dimensionless moment of inertia $\bar{I}=I/M^3$ and the dimensionless tidal deformability $\Lambda_2 = \lambda_2/M^5$, where $M$ is the mass of the neutron star.

\begin{table}
\centering
\begin{tabular}{c|ccc}
\toprule
Fluid		 & $c_0$ & $c_1$ & $c_2$ \\
\midrule
Irrotational 	 & $-2.03$ & $4.87 \times 10^{-1}$ & $\ 9.69 \times 10^{-3}$ \\
Static		 & $-2.66$ & $7.86 \times 10^{-1}$ & $-1.00  \times 10^{-2}$  \\
\midrule
Fluid & $c_3$ & $c_4$ & $c_5$ \\
\midrule
Irrotational & $1.03 \times 10^{-3}$ &  $-9.37 \times 10^{-5}$ & $2.24 \times 10^{-6}$ \\
Static & $1.28 \times 10^{-3}$ & $-6.37 \times 10^{-5}$ & $1.18 \times 10^{-6}$ \\
\bottomrule
\end{tabular}
\caption{\textsl{Coefficients of the formula in Eq.~\eqref{eq:fit}, which fits the quasi-universal relation between quadrupolar eletric and magnetic dimensionless tidal deformabilities.}}
\captionsetup{format=hang,labelfont={sf,bf}}
\label{tab:fit}
\end{table}

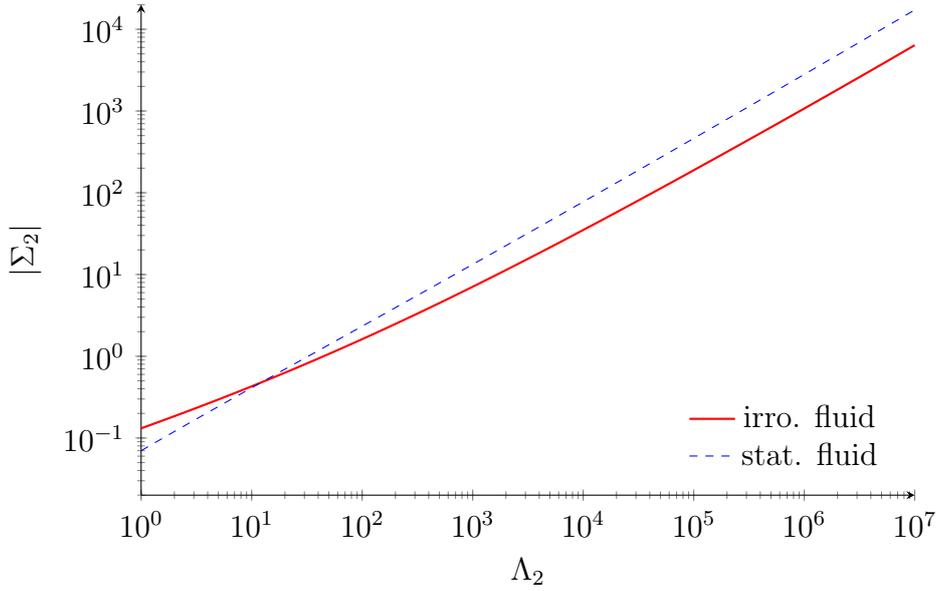
\begin{figure}
\centering
\begin{tikzpicture}
\begin{loglogaxis}[ymin=2*10^(-2),ymax=2*10^4,
axis x line=bottom,
axis y line=left,
legend style={draw=none},
legend pos=south east,
width=0.8\textwidth,
height=0.55\textwidth,
ylabel=$|\Sigma_2|$,xlabel=$\Lambda_2$]
\addplot
[domain=1:10^7,samples=100,smooth,
thick,red]
{exp(-2.03+4.87*(10^(-1))*ln(x)+9.69*(10^(-3))*(ln(x))^2+ 1.03*(10^(-3))*(ln(x))^3 -9.37*(10^(-5))*(ln(x))^4+2.24*(10^(-6))*(ln(x))^5)};
\addplot
[domain=1:10^7,samples=100,smooth,
dashed,blue]
{exp(-2.66+7.86*(10^(-1))*ln(x)-1*(10^(-2))*(ln(x))^2+ 1.28*(10^(-3))*(ln(x))^3 -6.37*(10^(-5))*(ln(x))^4+1.18*(10^(-6))*(ln(x))^5)};
\addlegendentry{irro. fluid};
\addlegendentry{stat. fluid};
\end{loglogaxis}
\end{tikzpicture}
\caption{\textsl{Quasi-universal relations between quadrupolar eletric and magnetic dimensionless tidal deformabilities.}}
\captionsetup{format=hang,labelfont={sf,bf}}
\label{fig:ILQem}
\end{figure}

In the past years, it was found that quasi-universal relations exist also in the magnetic sector and for higher-order tidal deformabilities~\cite{Yagi:2013sva,Delsate:2015wia}, and even among the rotational tidal Love numbers~\cite{Gagnon-Bischoff:2017tnz} (see~\cite{Yagi:2016bkt} for a review of neutron star quasi-universal relations). Here, we report the quasi-universal relations between the quadrupolar electric and magnetic tidal deformabilities (both in the static and the irrotational case)
\begin{equation}
\label{eq:fit}
\log{\left(\pm \Sigma_2\right)} = \sum_{n=0}^5 c_n \left[\log{\left(\Lambda_2 \right)}\right]^n \,,
\end{equation}
where $\Sigma_2 = \sigma_2/M^5$ and the plus (minus) sign refers to the static (irrotational) configuration. The coefficients $c_n$ are given for both cases in Table~\ref{tab:fit}~\cite{Jimenez-Forteza:2018buh}. We compare the two relations in Fig.~\ref{fig:ILQem}. We make use of this result in Chapter~\ref{sec:binary}.

\begin{figure}
\centering
\begin{tikzpicture}
\begin{semilogxaxis}[
axis x line=bottom,
axis y line=left,
enlargelimits,
legend style={draw=none},
legend pos=south east,
width=0.8\textwidth,
height=0.55\textwidth,
ylabel=$C$,xlabel=$\Lambda_2$]
\addplot
[domain=1:10^4,samples=100,smooth,
thick,red]
{3.71*10^(-1)-3.91*10^(-2)*ln(x)+1.056*10^(-3)*(ln(x))^2};
\end{semilogxaxis}
\end{tikzpicture}
\caption{\textsl{Quasi-universal relation between the compactness and the quadrupolar eletric tidal deformability.}}
\captionsetup{format=hang,labelfont={sf,bf}}
\label{fig:lambdaC}
\end{figure}
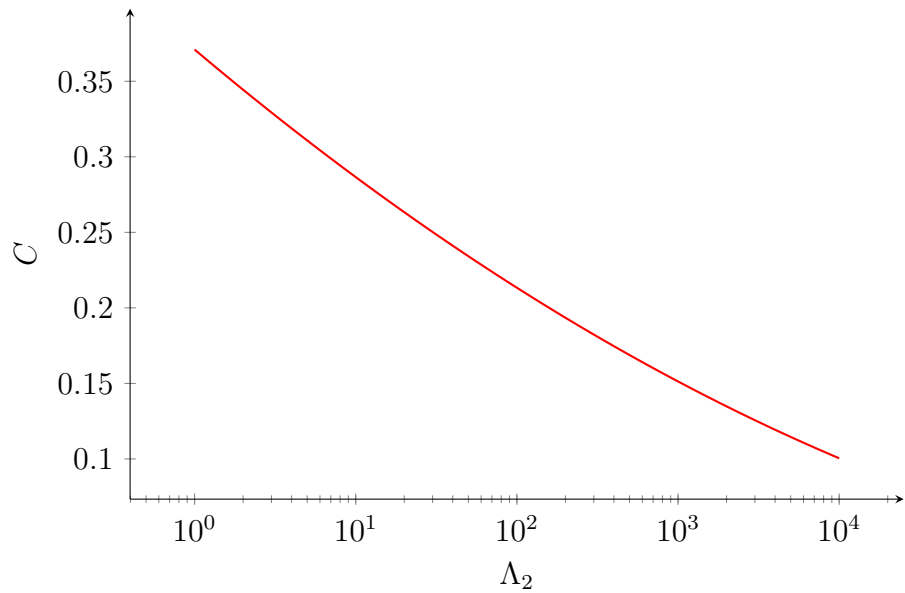

Lastly, another useful, even if less accurate, quasi-universal relation exists also between the electric, quadrupolar tidal deformability and the compactness of a neutron star~\cite{Maselli:2013mva}. Such relation is given by the fitting formula
\begin{equation}
C = 3.71 \times 10^{-1} -3.91 \times 10^{-2} \log{\Lambda_2} + 1.056 \times 10^{-3} \left( \log{\Lambda_2} \right) ^2 \,,
\end{equation}
that we show in Fig.~\ref{fig:lambdaC}. If the mass of the neutron star is known, through the latter relation, it is possible to obtain the radius of the star from a measurement of its tidal deformability. Indeed, the LIGO/Virgo collaboration used this relation to estimate the radii of the observed binary neutron star, from the bounds on the tidal deformabilities obtained through the detection of the gravitational wave event GW170817 (see section~\ref{sec:impact})~\cite{Abbott:2018exr}.

\chapter{Tidal deformations in binary systems}
\label{sec:binary}
Coalescences of compact binary systems are an outstanding source of gravitational waves. In particular, stellar-mass compact binary systems, either binary black holes (up to $50\mathrm{-}100 M_{\odot}$), binary neutron stars or mixed black hole-neutron star binaries, emit gravitational radiation detectable by ground-based interferometers. These phenomena occur inside galaxies. The emission of gravitational radiation subtracts energy to the binary system, shrinking its orbit. Since gravity is extremely weak, the emission of gravitational waves can effectively set in, driving the inspiral of stellar-mass binaries and leading them to merge within an Hubble time $\sim 14.4 \,\mathrm{Gyr}$, only when the orbital separation of the compact objects reaches the astronomical unit scale~\cite{Peters:1963ux,Peters:1964zz}. At larger scales other dissipative processes cause the shrinking of the orbit~\cite{Barack:2018yly}. Regarding this, the two major formation channels for stellar-mass compact binaries are isolated massive binary stars in galactic fields~\cite{Belczynski:2010tb,Dominik:2013tma} and dynamical environments in dense globular or nuclear star clusters~\footnote{A third channel for the formation of binary black holes comes from primordial black holes in the early universe~\cite{Garcia-Bellido:2017imq,Sasaki:2018dmp}.}~\cite{PortegiesZwart:1999nm,Benacquista:2011kv}. 

From the latest gravitational wave detections, the LIGO/Virgo collaboration has estimated the merger rates of binary black holes and neutron stars in the local universe (cosmological redshift $z\sim 0$) to be $9.7\div 101 \, \mathrm{Gpc}^{-3} \, \mathrm{yr}^{-1} $~\cite{LIGOScientific:2018mvr,LIGOScientific:2018jsj,Abbott:2017vtc,TheLIGOScientific:2016pea,Abbott:2016nhf,Abbott:2016drs} and $110\div 3840 \, \mathrm{Gpc}^{-3} \, \mathrm{yr}^{-1} $~\cite{LIGOScientific:2018mvr,TheLIGOScientific:2017qsa} at $90\%$ confidence level, respectively.

The coalescence of a compact binary system can be divided in three different phases: inspiral, merger and post-merger. During the inspiral phase, the orbital separation of the system $d$ is much larger than the size of the single objects $R$, i.e., the neutron star radius or the black hole horizon. At this stage the compact objects are effectively modeled as pointlike massive particles, and the orbital dynamics is described using the post-Newtonian theory (see the next section). As we discuss extensively in the following, the internal structure of the compact objects enters at this level only as a small perturbative correction. When the objects make closer, the approximation $d \gg R$ breaks down, and Numerical Relativity simulations, which integrate the fully non-linear Einstein equations, are needed to describe the merger phase~\cite{Pretorius:2005gq}. Here, the internal structure of the objects plays a fundamental role in determining the characteristics of the emitted gravitational radiation and the final product of the coalescence. In the neutron star case, this means that there is a strong dependence on the equation of state. However, due to the high energies involved in the process, it has to be stressed that in this phase the assumption of cold nuclear matter is not appropriate. Therefore, a non-barotropic equation of state, depending also on the temperature, is usually used in the simulations (see, e.g.,~\cite{Radice:2017lry}).

\begin{figure}[]
\centering
\includegraphics[width=0.6\textwidth]{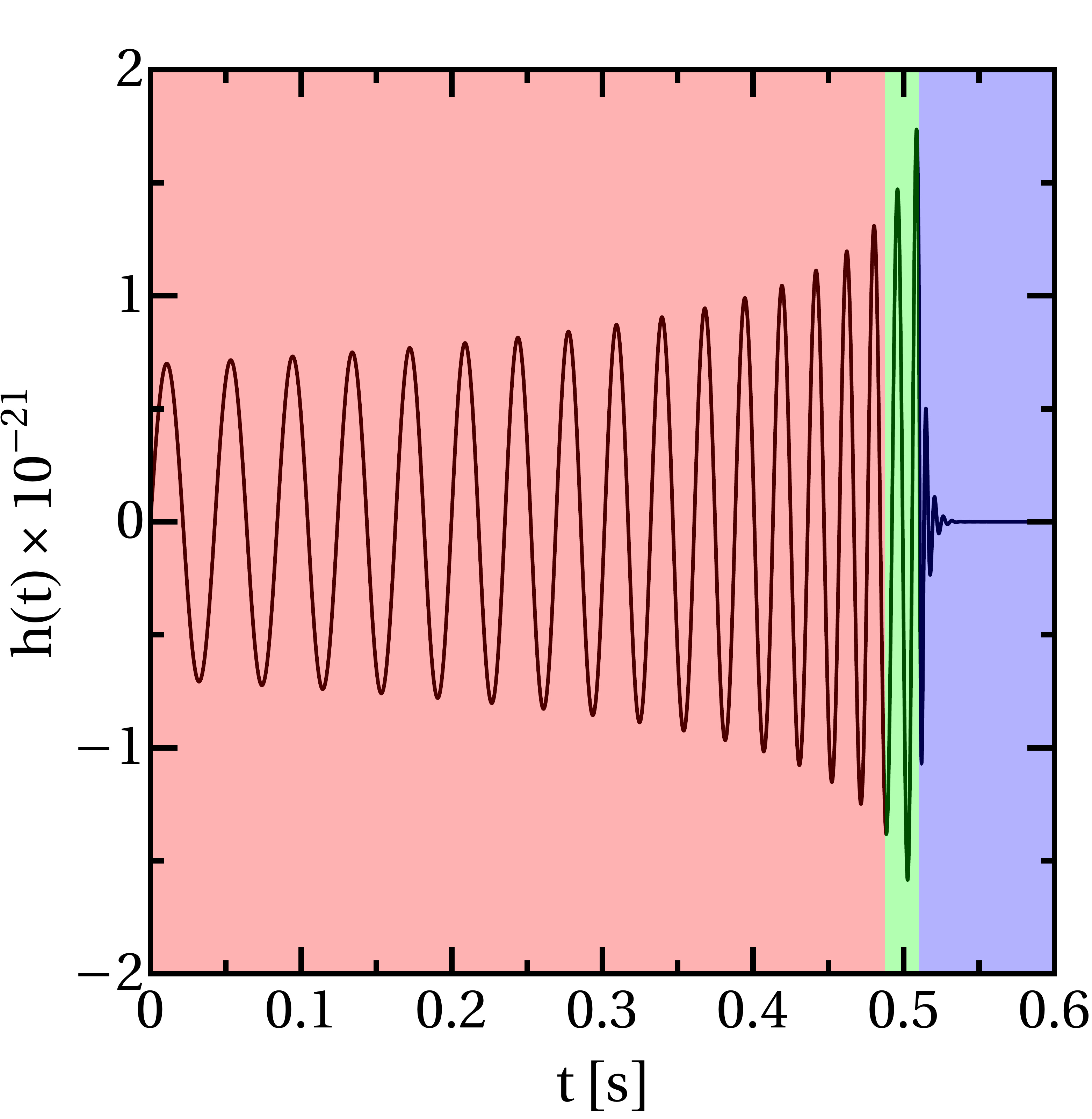}
\caption{\textsl{An example of the gravitational signal emitted by a GW140915-like black hole binary (gravitational strain as a function of time). The starting time $t=0\,$s corresponds to the typical frequency of the signal when it enters the sensitive band of second-generation interferometers, $\sim 20\,$Hz. The three different phases of coalescence are highlighted: inspiral (red), merger (green) and ringdown (blue).}}
\captionsetup{format=hang,labelfont={sf,bf}}
\label{fig:signal}
\end{figure}

The post-merger phase is different depending on the nature of the components of the binary system. For binary black holes, this stage is called ringdown, and can be treated using perturbation theory methods~\cite{Price:1994pm}. The final black hole resulting from the merger oscillates according to its quasi-normal mode frequencies, dissipating energy through gravitational wave emission and relaxing to a stationary equilibrium configuration. The outcome of mixed black hole-neutron star systems is slightly different. If the neutron star is compact enough to not be torn apart by tidal forces, it plunges into the horizon, being absorbed by the black hole~\cite{Foucart:2013psa}. On the other hand, if the star is more deformable, it is tidally disrupted during the merger and forms an accretion disk surrounding the black hole~\cite{Kyutoku:2013wxa,Foucart:2014nda,Deaton:2013sla}. Finally, in the case of binary neutron stars, many scenarios are possible, depending on the masses of the components and the equation of state of matter~\cite{Shibata:2006nm,Abbott:2018wiz,Radice:2018xqa}. Systems with larger masses and less deformable matter result to a prompt collapse ($\sim 1 \, \mathrm{ms}$) to a black hole right after the merger~\cite{Hotokezaka:2011dh}. Binaries with smaller masses and more deformable matter lead to the formation of an unstable, possibly long-lived, remnant. Hypermassive neutron stars\cite{Baumgarte:1999cq}, which have a mass larger than the maximum mass of uniformly rotating stars, are supported against gravitational collapse by fast, differential rotation, and survive for $\lesssim 1 \, \mathrm{s}$ before forming a black hole~\cite{Shapiro:2000zh}. Supramassive neutron stars\cite{1992ApJ...398..203C,Cook:1993qj,Cook:1993qr}, whose mass exceeds the maximum mass of the non-rotating configuration, are supported by uniform rotation, and can survive on the spin-down timescale (from seconds up to hours) before collapsing to a black hole~\cite{Giacomazzo:2013uua,Hotokezaka:2013iia,Ravi:2014gxa}. Lastly, systems with very low masses form a stable, massive neutron star. 

In Fig.~\ref{fig:signal} we show a typical gravitational signal emitted by binary black holes, highlighting the three phases of the coalescence. It has to be mentioned that the full gravitational waveforms can be correctly produced (to a few percent level) using the effective-one-body (EOB) approach~\cite{Buonanno:1998gg,Buonanno:2006ui}. The EOB method provides a reformulation of the post-Newtonian dynamics by mapping the original two-body problem in General Relativity into a one-body problem in an effective metric, which is also improved using resummation techniques. The dynamics obtained in the EOB framework is an extension of the perturbative post-Newtonian results. The EOB waveforms generated are then matched to the analytical waveforms of the ringdown phase, and calibrated by fitting to Numerical Relativity solutions~\cite{Damour:2009kr}. In this way, templates which describe the gravitational wave emission during the entire coalescence are produced with low computational cost. The EOB formalism has been applied also to binary neutron stars~\cite{Damour:2009wj,Bernuzzi:2014owa} (see section~\ref{sec:phase}).

We have said that tidal effects are dominant during the merger of binary neutron stars, when the orbital separation is comparable with the size of the compact objects. However, tidal deformations play a fundamental role also in the inspiral phase, even if they are subdominant with respect to the point-particle contributions (see sections~\ref{sec:pntidal} and~\ref{subsec:waveform}). Indeed, as we deeply discuss in Chapter~\ref{sec:inverse}, they can be used to shed light on the internal composition of neutron stars~\footnote{Besides the tidal effects, also the spin-induced quadrupole moment affects the dynamics of the inspiral phase, and, in principle, it can provide information on the internal composition of the star too (as we said in section~\ref{sec:universal}, the spin-induced quadrupole moment can be written as a function of the electric quadrupolar tidal Love number using the quasi-universal relations). However, it is hard to extract the information on the internal structure contained in the spin-induced quadrupole with current gravitational wave detectors, because it is degenerate with other point-particle contributions. On the other hand, writing the spin-induced quadrupole moment in terms of $\lambda_2$, increases instead the accuracy on the measurement of the spins~\cite{Yagi:2013awa}.}. Therefore, in this chapter we study only the inspiral phase of the coalescence. Although in the following we focus on neutron star binaries (see section~\ref{sec:impact}), the post-Newtonian formalism that we use is independent of the internal structure of the compact objects. Thus, our results can apply as well to other exotic compact objects (ECO)~\cite{Cardoso:2017cqb,Cardoso:2017njb}, i.e., models alternative to black holes, whose putative quantum corrections would prevent the formation of the event horizon. Indeed, it has been shown that tidal effects in the late inspiral phase can be used to distinguish ECOs from black holes in supermassive compact binary coalescences~\cite{Maselli:2017cmm}.

This chapter is organized as follows. In section~\ref{sec:pntidal} we introduce the post-Newtonian framework to describe the tidal interactions of a $N$-body system. In section~\ref{sec:truncation} we present the original results obtained in this thesis, applying the post-Newtonian formalism to a spinning binary system. We compute the contribution to the gravitational waveform phase due to the tidal deformations of rotating objects (Abdelsalhin \emph{et al.}~\cite{Abdelsalhin:2018reg}). In section~\ref{sec:impact} we estimate the impact of these new spin-tidal effects on the parameter estimation of binary neutron stars (Jimenez-Forteza, Abdelsalhin \emph{et al.}~\cite{Jimenez-Forteza:2018buh}).

\section{Post-Newtonian tidal interactions}
\label{sec:pntidal}
The orbital dynamics and the gravitational wave emission of a compact binary system in the inspiral phase can be described within the post-Newtonian (PN) framework. The PN approximation consists in an expansion of the Einstein equations around the Newtonian limit~\cite{Will:1981}, in terms of a small parameter $\epsilon$ of order
\begin{equation}
\epsilon \sim \frac{v^2}{c^2} \sim \frac{R_S}{d} \ll 1 \,,
\end{equation}
where $v$ is the scale of velocity of a system with typical size $d$, and $R_S=2 G M/c^2$ is the Schwarzschild radius associated to its total mass $M$. $c$ and $G$ are the speed of light in vacuum and gravitational constant, respectively. One may regard the two ratios in the above equation as independent, attempting to perform two separated expansions in $v/c$ and $R_S/d$. However, for self-gravitating bodies the relation $(v/c)^2 \sim R_S/d $ follows from the virial theorem. In other words, when we consider corrections to Newtonian gravity due to the strength of the gravitational field (measured by $R_S/d $), we must take into account, for consistency, relativistic deviations from the classic kinematics (measured by $v/c$)~\cite{Maggiore:2008}. The PN approximation is then suitable to study slowly moving, weakly self-gravitating systems. The more relativistic the source, the higher the order of the expansion that we need to correctly describe it. Henceforth, we design a quantity of order $\mathcal{O}\left( \epsilon^n \right)\sim \mathcal{O} \left( \left(   v/c \right)^{2n}  \right)$ as a term of $n$-PN order. Thus, the Newtonian limit, $\epsilon^0$, is the $0$PN order, $\epsilon^1$ the $1$PN correction, etc.~\footnote{Note that with such definition, also terms of half-integer order may show up in the expansion.}

\begin{figure}[]
\centering
\includegraphics[width=0.9\textwidth]{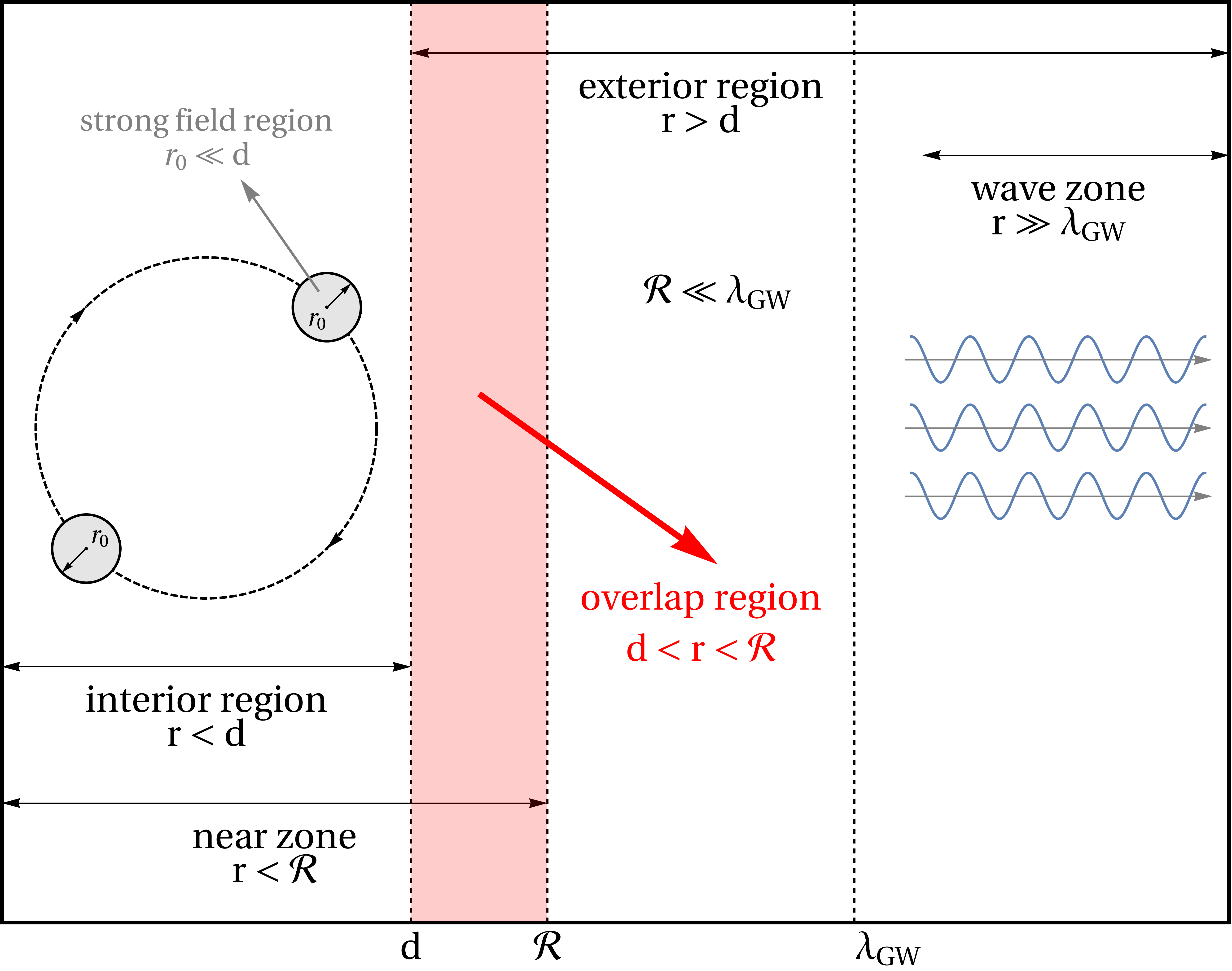}
\caption{\textsl{Partitioning of the space around a PN source (referring in particular to a binary system) in the Blanchet-Damour approach. The PN expansion holds in the near zone, whereas the PM expansion is convergent in the exterior region. The two solutions are matched in the overlap region (in red). The strong field region surrounding the compact objects is shown in gray. See the text for explanation.}}
\captionsetup{format=hang,labelfont={sf,bf}}
\label{fig:PN}
\end{figure}

In the past, the computation of higher-order terms in the PN approximation presented technical and conceptual issues which prevented its application to relativistic sources. These problems were solved by the groups of Blanchet and Damour~\cite{Damour:1986ny,Damour:1982wm,Blanchet:1985sp,Blanchet:1989ki,Blanchet:1995fr,Blanchet:1995fg,Blanchet:1996pi} and of Will, Wiseman and Pati~\cite{Will:1996zj,Pati:2000vt,Will:2005sn}. It was shown that the methods of the two groups are completely equivalent, thus nowadays PN techniques are successfully applied to study the gravitational emission from compact binaries. 

Here, we briefly outline the ideas of the Blanchet-Damour approach~\cite{Maggiore:2008,Blanchet:2013haa,Buonanno:2014aza}. Let us consider a source of size $d$ and typical velocity $v$. Such a system emits gravitational waves at frequency $\lambda_{GW} \sim (c/v) d$. We can then divide the space around the source in a \emph{near zone} $r \ll \lambda_{GW}$, and a \emph{far} or \emph{wave zone} $r \gg \lambda_{GW}$ (see Fig.~\ref{fig:PN}). Since we are assuming $v/c \ll 1$, it follows that $\lambda_{GW} \gg d$. We introduce a length $\mathcal{R} \ll \lambda_{GW}$ to denote the boundary which limits the near zone, $r< \mathcal{R}$. In this region the effects of time-retardation are negligible, and we can consider the gravitational potentials as instantaneous. On the other hand, the propagation of waves occurs in the far zone, where the time delay is significant and retarded potentials must be taken into account. Furthermore, we call \emph{exterior region} the space outside the source, $r>d$. The partitioning of the space just described has not to be confused with the spacetime partition that we use in section~\ref{sec:1pn_app}.

Since we are considering weakly gravitating sources, we can solve the Einstein equations inside the near zone using a PN expansion. This expansion breaks down at $r \sim \lambda_{GW}$~\footnote{One can understand why the PN expansion fails in the wave zone from the following argument. Let us consider a generic function of the retarded time, and expand it for a small retardation
\begin{equation}
\begin{aligned}
f \left( t-\frac{r}{c} \right) & \sim f(t) - \frac{r}{c} \dot{f}(t) + \frac{1}{2} \left( \frac{r}{c}\right)^2 \ddot{f}(t) + \dots \\
& \sim f(t) \left[ -\frac{r}{c\, T} + \frac{1}{2} \left( \frac{r}{c\, T}\right)^2 + \dots \right] \\
& \sim f(t) \left[ -\frac{r}{\lambda_{GW}} + \frac{1}{2} \left( \frac{r}{\lambda_{GW}}\right)^2 + \dots \right] \,,
\end{aligned}
\end{equation}
where we have used $\dot{f} \sim f(t)/T$ and $\lambda_{GW} \sim c \, T$, with $T$ denoting the typical timescale of the source. Therefore, the PN expansion is actually an expansion in $r/\lambda_{GW}$, and then it breaks down in the wave zone $r \gg \lambda_{GW}$.}. On the other hand, in the exterior region $r>d$, the matter sources vanish. In this region we can solve the vacuum Einstein equations performing a \emph{post-Minkowskian} (PM) expansion, i.e., an expansion around the flat spacetime in powers of the gravitational constant $G$. Within the PM expansion, the velocities can be arbitrary close to the speed of light, but the gravitational potential is weak. This expansion is well-behaved in the wave zone, but breaks down at $r \sim d$. Let us now assume that the near zone extends into the exterior region, i.e., $\mathcal{R}>d$ (see again Fig.~\ref{fig:PN}). This means that an overlap region $d<r<\mathcal{R}$ does exist. In this region both the PN and the PM asymptotic expansions are regular. Matching the two expansions in the overlap region, we obtain a solution well-behaved everywhere, which completely describes the system. In the optimal case, $\mathcal{R} \gg d$, and we have a large overlap region for matching the two solutions. However, in general the extension of the near zone depends on the PN order that we are considering. For higher PN orders, the near zone falls inside the source, $\mathcal{R} < d$, and the overlap region does not exist~\footnote{The $n$-th PN order is associated with the emission of gravitational radiation with wavelength $\lambda_n \sim \mathcal{O} (1/n) \lambda_{GW}$. Thus, for higher PN orders the condition $\lambda_n \gg d$ is not fulfilled any longer.}. In these cases we are not able to compute the full solution. On the other hand, the higher PN orders (which we can not evaluate) are not negligible when $v/c \to 1$, explaining why the PN approximation fails in this limit.

We have described the PN approximation referring to weakly gravitating systems. One may wonder if we are allowed to apply the above approach to compact binaries composed of black holes and/or neutron stars, which are very relativistic sources, characterized by strong gravitational fields. For a compact binary inspiralling with slow orbital velocity $v \ll c$ the answer is yes, and it can be understood looking at Fig.~\ref{fig:PN}. Indeed, even for such a system, the gravitational field is actually strong only inside a region of radius $r_0$ surrounding the compact objects, with $r_0 \ll d$ since we are in the inspiral phase. It can be shown that the PN expansion can be performed also when the near zone, $r < \mathcal{R}$, contains strong gravitational sources. This is obtained through the evaluation of surface integrals far away from the strong sources, at a distance $r_0 \ll r < d$, where the gravitational field is weak. Another way to see it comes from the \emph{strong equivalence principle}, which states that self-gravitating bodies fall in an external gravitational fields as test-particles. Thus, we can model strong gravitating sources as pointlike particles moving along the geodesics of a regularized metric~\footnote{The regularization takes care of the divergent gravitational field of a pointlike mass.}. 

In definitive, we can safely apply the PN approximation to inspiralling compact binaries. This has been done, computing the phase (amplitude) of the waveform of the gravitational radiation emitted by a non-spinning binary system in circular orbit up to $3.5$PN ($3$PN) order~\footnote{Recently, the conservative dynamics of spinless compact binaries has been extended up to $4$PN order~\cite{Jaranowski:2012eb,Damour:2016abl,Bernard:2015njp,Bernard:2017ktp}.}. Spin-orbit effects (linear terms in the spins) in the gravitational wave phase are included up to $3.5$PN order, whereas quadratic spin-spin contributions are known up to $2$PN order (see, e.g.,~\cite{Khan:2015jqa,Santamaria:2010yb}).

We have said that the PN approximation can be applied to binary systems whether the component objects are weak or strong gravitating sources. This is strictly related to the \emph{effacement principle}, i.e., the fact that the internal structure of the bodies contributes to the PN series expansion only at very high order. The difference between a extended-body and a pointlike massive particle is due to tidal interactions. Contrary to point-particles, extended-bodies are deformed by tidal forces. We can estimate at which PN order the tidal effects show up. In the previous chapter, we have shown that the leading contribution is the quadrupolar deformation. Let us consider two compact objects with typical masses $M$ and radii $R$, separated by a distance $r$. At Newtonian order, the gravitational force which acts on each body reads
\begin{equation}
\begin{aligned}
F & \sim  \frac{GM^2}{r^2} + \frac{G M Q}{r^4} + \mathcal{O} \left( Q^2\right) \\
& \sim \frac{GM^2}{r^2} + \frac{G M^2 R^5}{r^7} + \mathcal{O} \left( Q^2\right) \\
& \sim \frac{GM^2}{r^2} \left[ 1+ \left( \frac{R}{r}\right)^5 + \mathcal{O} \left( Q^2\right)\right] \,,
\end{aligned}
\end{equation}
where $Q$ is the tidally induced quadrupole moment, and we have used (cf. Eqs.~\eqref{eq:adiabaticrel}, \eqref{eq:lovetidal} and~\eqref{eq:newtlovetidal})
\begin{equation}
\label{eq:pnimpact}
Q \sim R^5 \frac{d^2 U}{dr^2} \qquad U \sim \frac{M}{r} \,.
\end{equation}
For a compact object, $R \sim G M/c^2 $, and since for the virial theorem $GM/r \sim v^2$, the correction to the Newtonian force due to tidal interactions is of order $(v^2/c^2)^5$. Therefore, the internal structure of the bodies affects the PN approximation at $5$PN order, and it is encoded in the tidal deformations of the compact objects. For a neutron star, this means that the tidal effects can provide information on the equation of state.

We stress that the fact that tidal effects enter at such high PN order does not necessary imply that they are negligible. Indeed, even if the tidal contribution is a $5$PN order term, the series coefficient in front of it is proportional, as we will see, to the inverse of the star compactness to the fifth power, $1/C^5$. For a typical neutron star this factor is of order $10^3 \div 10^4$, and then the tidal term is comparable to the $3.5$PN  order point-particle contributions~\cite{Vines:2011ud}. For a non-relativistic binary such as the Earth-Moon system, the above factor is huge, and indeed in this case the tidal interactions dominate all the PN series.

Although currently we lack a PN expansion up to the $5$PN order, in the next section we show how it is possible to include effectively the tidal interactions in the dynamics, and in the consequent gravitational emission, of compact binary system. We follow the works of Vines, Flanagan and Hinderer~\cite{Vines:2010ca,Vines:2011ud}.

\subsection{Post-Newtonian approximation for a system of structured bodies}
\label{sec:1pn_app}
In this section we describe the PN theory of a system of interacting, arbitrarily structured bodies, which has been developed in the works of Damour, Soffel and Xu~\cite{Damour:1990pi,Damour:1991yw} and Racine, Flanagan and Vines~\cite{Racine:2004hs,Vines:2010ca}. Throughout this chapter we set the gravitational constant $G=1$, and keep the speed of light $c$ as the formal parameter of the PN expansion. We work to the 1PN order, i.e., $\mathcal{O}(c^{-2})$. Also, throughout this chapter we contract the spatial indices using the Euclidean flat metric $\delta^{ij}$, therefore there is no distinction between upper and lower indices, and we use the upper ones only. Furthermore, we use the multi-index notation $T^L \equiv T^{a_1 \dots a_l}$, and for a generic (three-)vector $v^i$, we define $v^{a_1 \dots a_l} \equiv v^{a_1} \dots v^{a_l}$ and $v^2 = v^{ii}$, see the~\nameref{sec:notation_2}.

Choosing conformally Cartesian coordinates~\footnote{Conformally Cartesian coordinates are a special case of isotropic coordinates and require $g_{00} g_{ij} = -\delta_{ij} + \mathcal{O}(c^{-4})$~\cite{Damour:1990pi}.}, the spacetime metric of General Relativity to 1PN order reads
\begin{equation}
\label{eq:metric1pn}
ds^2 = - \left(1+ \frac{2 \Phi}{c^2} + \frac{2 \Phi^2}{c^4} \right) c^2 dt^2 + \frac{2 \zeta^i}{c^3} c dt dx^i + \left(1-\frac{2 \Phi}{c^2} \right) \delta^{ij} dx^i dx^j + \mathcal{O}(c^{-4}) \,.
\end{equation}
The scalar field $\Phi(t,\boldsymbol{x})$ can be decomposed into the Newtonian potential $\phi(t,\boldsymbol{x})$ and its 1PN correction $\psi(t,\boldsymbol{x})$, as $\Phi=\phi+c^{-2}\psi$. The three-vector field $\zeta^i(t,\boldsymbol{x})$ is the 1PN gravito-magnetic potential. Adopting the harmonic gauge condition~\footnote{The harmonic, or Lorenz, or de Donder gauge is defined by $\partial_{\mu}(\sqrt{-g} g^{\mu \nu})=0$, or equivalently $\Gamma_{\ \mu \nu}^{\alpha} g_{\mu \nu} =0$, where $\Gamma_{\ \mu \nu}^{\alpha}$ are the Christoffel symbols. For the metric~\eqref{eq:metric1pn}, this condition implies $4 \dot{\Phi}+ \partial_i \zeta^i = \mathcal{O}(c^{-2})$.}, the Einstein equations for the metric~\eqref{eq:metric1pn} reduce to
\begin{equation}
\begin{aligned}
\nabla^2 \Phi & = 4 \pi T^{tt} + c^{-2} \left( 4 \pi T^{ii} + \ddot{\Phi} \right)  + \mathcal{O}(c^{-4}) \\
\nabla^2 \zeta^i &= 16 \pi T^{ti} + \mathcal{O}(c^{-2}) \,,
\end{aligned}
\end{equation}
where overdots denote time derivatives and $\nabla^2 = \delta^{ij} \partial_i \partial_j$.

\begin{figure}[]
\centering
\includegraphics[width=0.8\textwidth]{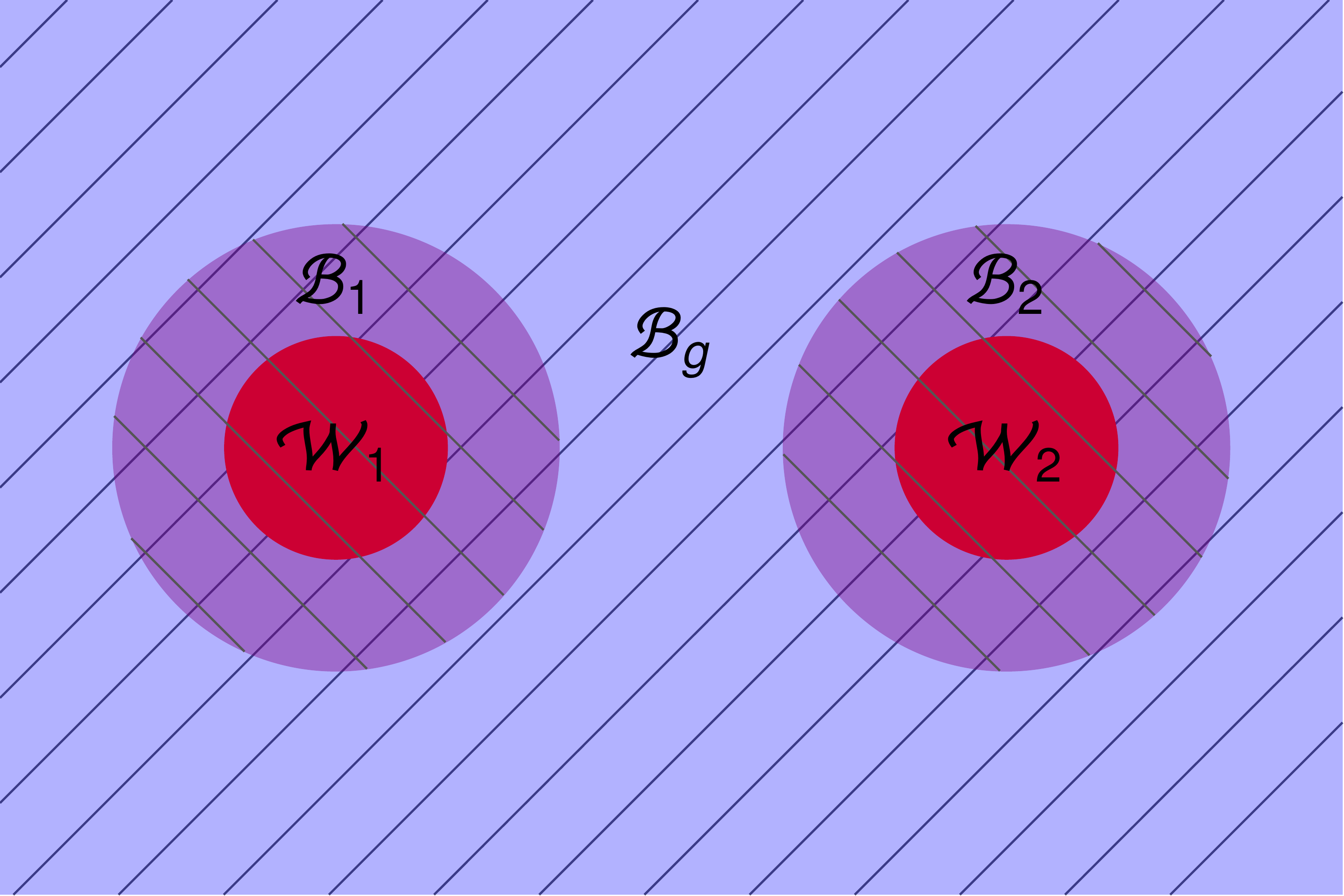}
\caption{\textsl{Illustration of the different charts covering the spacetime metric of the $N$-body system. The worldtube regions are shown in red, the buffer regions in purple and the space among them in blue. The global buffer region $\mathcal{B}_g$ is the sum of the blue and purple areas. The left-diagonal lines mark the regions covered by the local frame coordinates $(s_A,y_A^i)$, whereas the right-diagonal lines denote the areas where the global frame coordinates $(t,x^i)$ are defined. Note that the coordinate systems overlap in the buffer region of each body.}}
\captionsetup{format=hang,labelfont={sf,bf}}
\label{fig:buffer}
\end{figure}

Let us now consider a system of $N$ interacting, arbitrarily structured bodies. Each body $A$, with $A=1,\dots,N$, can have arbitrarily high velocity fields and/or strong gravity. For each body we assume the existence of a local coordinate system $(s_A,y^i_A)$, which covers the product of an open ball of radius $r_2$, $|\boldsymbol{y}_A| < r_2$, with an open interval of time $(s_1,s_2)$, $s_1 < s_A < s_2$. Moreover, we assume that: (i) all the body matter fields and/or strong gravity regions are contained in a ball $|\boldsymbol{y}_A| < r_1$, with $r_1 < r_2$, which we call the \emph{worldtube} $\mathcal{W}_A$. (ii) In the region $r_1 < |\boldsymbol{y}_A| < r_2$, which we call the \emph{buffer region} $\mathcal{B}_A$, the gravitational field is weak, the coordinates $(s_A,y^i_A)$ are harmonic and conformally Cartesian and the spacetime metric reduces to the form in Eq.~\eqref{eq:metric1pn}, with potentials $\Phi_A(s_A,\boldsymbol{y}_A)$ and $\zeta^i_A(s_A,\boldsymbol{y}_A)$. (iii) The buffer regions (and thus the worldtubes) of the bodies do not overlap. We call the coordinate system $(s_A,y^i_A)$ \emph{body frame} or \emph{local frame}. Furthermore, we assume the existence of a harmonic and conformally Cartesian coordinate system $(t,x^i)$ which covers the spatial region $\mathcal{B}_g$, composed of all the buffer regions of the bodies as well as the space among them, i.e., the entire spacetime except for the worldtubes of the bodies. In the region $\mathcal{B}_g$, the metric can be written in the form shown in Eq.~\eqref{eq:metric1pn} in terms of potentials $\Phi_g(t,\boldsymbol{x})$ and $\zeta^i_g(t,\boldsymbol{x})$. We call the coordinate system $(t,x^i)$ \emph{global frame}. A graphic visualization of the above system is shown in Fig.~\ref{fig:buffer}. Note that the spacetime partitioning is different from that of Fig.~\ref{fig:PN} (and has not to be confused with it).

We stress that each body buffer region $\mathcal{B}_A$ is covered by both the global coordinates $(t,x^i)$ and local coordinates $(s_A,y^i_A)$. In these regions the coordinate transformation between the two frames is given by~\footnote{The coordinate transformation in Eq.~\eqref{eq:coordtransf} is the most general transformation between two harmonic and conformally Cartesian coordinate systems with the metric given by Eq.~\eqref{eq:metric1pn}. The functions $\{z^i, R^i, \alpha,\beta \}$ can be freely specified, with the only condition $\nabla^2 \beta =0$.}
\begin{equation}
\begin{aligned}
x^i(s,\boldsymbol{y}_A) =& y_A^i + z_A^i(s_A) +\frac{1}{c^2} \left\{ \left[ \frac{1}{2} \dot{z}_A^{kk}(s_A) \delta^{ij}-\dot{\alpha}_A(s_A)\delta^{ij} +\epsilon^{ijk}R_A^k(s_A)\right. \right. \\
& \left. \left. + \frac{1}{2} \dot{z}_A^{ij}(s_A) \right] y_A^j + \left[\frac{1}{2} \ddot{z}^i_A(s_A) \delta^{jk}- \ddot{z}^k_A(s_A) \delta^{ij} \right] y_A^{jk}\right\} + \mathcal{O}(c^{-4}) \nonumber
\end{aligned}
\end{equation}
\begin{equation}
\label{eq:coordtransf}
\begin{aligned}
t(s,\boldsymbol{y}_A)  =& s_A + \frac{1}{c^2} \left[\alpha_A(s_A)+\dot{z}_A^j(s_A) y_A^j \right] + \frac{1}{c^4} \left[\beta_A(s_A,\boldsymbol{y}_A)+ \frac{1}{6}\ddot{\alpha}_A(s_A)y^{jj}_A \right. \\
& \left. + \frac{1}{10}\dddot{z}_A^j(s_A) y_A^{jkk} \right] + \mathcal{O}(c^{-6}) \,,
\end{aligned}
\end{equation}
where $\{z^i_A, R^i_A, \alpha_A,\beta_A \}$ are different functions for each body $A$, and $\epsilon^{ijk}$ is the Levi-Civita symbol. The vector $z^i_A$ ($R^i_A$) describes a time-dependent spatial translation (rotation) between the two frames.

Under the above assumptions, working in the body frame, we have to solve the vacuum Einstein equations in the buffer region $\mathcal{B}_A$, 
\begin{equation}
\label{eq:einvac}
\begin{aligned}
\nabla^2 \Phi_A & = c^{-2} \ddot{\Phi}_A   + \mathcal{O}(c^{-4}) \\
\nabla^2 \zeta^i_A &  = + \mathcal{O}(c^{-2}) \,.
\end{aligned}
\end{equation}
The general solution for the potentials $\Phi_A(s_A,\boldsymbol{y}_A)$ and $\zeta^i_A(s_A,\boldsymbol{y}_A)$ is given by
\begin{align}
\label{eq:pn_expansion}
  \Phi_A(s_A,\boldsymbol{y}_A)=& -\sum_{l=0}^\infty\frac{1}{l!}\left\{(-1)^lM_A^L(s_A)
  \partial_L\frac{1}{|\boldsymbol{y}_A|}\right.
  +G_A^L(s_A)y_A^L\nonumber\\
&+\frac{1}{c^2}\left[\frac{(-1)^l(2l+1)}{(l+1)(2l+3)}{\dot\mu}^L_A(s_A)
  \partial_L\frac{1}{|\boldsymbol{y}_A|}\right.
  +\frac{(-1)^l}{2}{\ddot M}^L_A(s_A)\partial_L|\boldsymbol{y}_A|\nonumber\\
  &\left.\left.-{\dot\nu}^L_A(s_A)y^L_A+\frac{1}{2(2l+3)}{\ddot G}^L_A(s_A)y_A^{jjL}\right]\right\}+\mathcal{O}(c^{-4}) \,, \nonumber\\
  \zeta^i_A(s_A,\boldsymbol{y}_A)=& -\sum_{l=0}^\infty\frac{1}{l!}\left\{(-1)^lZ_A^{iL}(s_A)
\partial_L\frac{1}{|\boldsymbol{y}_A|} +Y_A^{iL}(s_A)y_A^L\right\}+\mathcal{O}(c^{-2})\,,
\end{align}
where
\begin{align}
Z^{iL}_A(s_A)=&\frac{4}{l+1}{\dot M}^{iL}_A(s_A)-\frac{4l}{l+1}\epsilon^{ji \langle a_l}J_A^{L-1 \rangle j}(s_A)
+\frac{2l-1}{2l+1}\delta^{i\langle a_l}\mu_A^{L-1 \rangle}(s_A)+\mathcal{O}(c^{-2})\,,\\
Y_A^{iL}(s_A)=&\nu_A^{iL}(s_A)+\frac{l}{l+1}\epsilon^{ji \langle a_l}H_A^{L-1 \rangle j}(s_A)
-\frac{4(2l-1)}{2l+1}{\dot G}_A^{\langle L-1}\delta^{a_l \rangle i}(s_A)+\mathcal{O}(c^{-2})\,.
\end{align}
The above expansion defines the body and tidal multipole moments. The internal degrees of freedom of each body are described by its \emph{mass multipole moments} $M_A^L(s_A)$ and its \emph{current multipole moments} $J_A^L(s_A)$, with $l\ge0$ and $l\ge1$, respectively. Mass and current multipole moments take into account the distributions of energy and momentum inside the source. If the bodies are weakly gravitating objects, the 1PN approximation holds also inside the worldtube regions $\mathcal{W}_A$, and it is possible to write the body multipole moments as integrals over the source volume. On the other hand, if the bodies have a region of strong gravity, this is not possible. In the latter case, one can express the multipole moments in terms of surface integrals in the buffer region (cf. with the discussion in section~\ref{sec:pntidal}). The tidal field due to the other bodies $B\neq A$ is described by the electric tidal moments $G_A^L(s_A)$ and the magnetic tidal moments $H_A^L(s_A)$ (defined for $l\ge0$ and $l\ge1$, respectively). The tidal moments encode the gravitational fields generated by external sources and the inertial effects due to the motion of the local frame with respect to the global frame. Both the body and the tidal moments are symmetric trace-free (STF) tensors on all indices (see the~\nameref{sec:notation_2} for the definition of STF tensor). Note that the multipole moments are functions of the time coordinate alone, they are independent of the spatial variables. Mass and electric moments are defined up to 1PN order, while current and magnetic moments are defined just to Newtonian level~\footnote{We stress that there is no inconsistency with the definition of the multipole moments in Chapter~\ref{sec:star}, where we said that the current/magnetic multipole moments vanish in the Newtonian limit. Indeed, here we have already factored out the dependency $1/c^2$ in the definition of the multipole moments, thus even if the current/magnetic moments do not vanish in the Newtonian limit, they can affect the dynamics only at 1PN order.}. The quantities $\mu^L_A$, $\nu^L_A$ (defined for $l \geq 0$ and $l\geq 1$, respectively) are called internal and external gauge moments, respectively, because they do not contain gauge-invariant information (and indeed they can be set to zero through a coordinate transformation, see below). Within the 1PN approximation, the separation between the tidal field and the multipolar structure of the body is clear and unique (cf. with the discussion in section~\ref{sec:polar}). In Eq.~\eqref{eq:pn_expansion}, the terms with negative powers of $|\boldsymbol{y}_A|$ (which diverge for $|\boldsymbol{y}_A| \to 0$) depend on the body multipole moments, whereas the terms with positive powers of $|\boldsymbol{y}_A|$ (which diverge for $|\boldsymbol{y}_A| \to \infty$) depend on the tidal  moments. 

We can use the residual gauge freedom in the coordinate transformation~\eqref{eq:coordtransf} to choose the \emph{body-adapted gauge} for the local frame, which defines the body local asymptotic rest frame. This is achieved by imposing the following conditions:
\begin{equation}
\begin{aligned}
M^i_A(s_a) =& 0 \\
R^i_A(s_a) =& 0 \\
G_A(s_a) =& \mu_A(s_a)= 0 \\
\mu^L_A(s_a) =& \nu^L_A(s_a) = 0 \qquad l \geq 1 \,.
\end{aligned}
\end{equation}
The first condition, setting the body mass dipole $M^i_A$ to zero, ensures that the center of
mass of the body $A$ is located at $y^i_A=0$. The second condition, setting the rotation vector $R^i_A$ to zero, fixes the orientation of the local frame spatial axes to that of the global frame ones. The third condition ensures that, replacing the body by a freely falling observer at $y^i_A=0$, its proper time is measured by the coordinate $s_A$. Lastly, we can set to zero all the internal and external gauge moments, revealing their nature of pure gauge degrees of freedom. In the body-adapted gauge, setting $y^i_A=0$, the coordinate transformation~\eqref{eq:coordtransf} yields the equation $x^i=z^i_A(t)$, with $t = s_A + c^{-2} \alpha_A (s_A)  + \mathcal{O}(c^{-4})$. This relation describes the position of the body $A$ in the global frame, i.e., it parametrizes the location of the local frame of the body $A$ in the global coordinate system. The function $z^i_A(t)$ is called the center-of-mass (COM) worldline of the body $A$, even if in general it does not parametrize an actual worldline in the spacetime. Indeed, the global coordinate system $(t,x^i)$ is not defined in the worldtube region $\mathcal{W}_A$ of the body, and thus it is not defined in its center of mass. The vector $z^i_A(t)$ describes a real worldline only if the body is weakly gravitating and the 1PN approximations holds also inside the worldtube region.

In the region $\mathcal{B}_g$ the Einstein equations take the form in Eqs.~\eqref{eq:einvac}, but for the global potentials $\Phi_g(t,\boldsymbol{x})$ and $\zeta^i_g(t,\boldsymbol{x})$,
\begin{equation}
\begin{aligned}
\nabla^2 \Phi_g & = c^{-2} \ddot{\Phi}_g   + \mathcal{O}(c^{-4}) \\
\nabla^2 \zeta^i_g &  = + \mathcal{O}(c^{-2}) \,.
\end{aligned}
\end{equation}
Alike in the local frame, we can expand the global frame potentials in terms of the (mass and current) \emph{global body multipole moments} $M^L_{g,A}(t)$, $Z^{iL}_{g,A}(t)$ (defined for $l \geq 0$ and STF tensors on all (the last) $l$-indices)
\begin{align}
\label{eq:globmultexp}
  \Phi_g(t,\boldsymbol{x})=&-\sum_{A=1}^{N}\sum_{l=0}^\infty\frac{(-1)^l}{l!}\left\{
    M_{g,A}^L(t)\partial_L\frac{1}{|\boldsymbol{x}-\boldsymbol{z}_A(t)|}\right.\nonumber\\
  &\left.+\frac{1}{2c^2}\partial^2_t\left[M_{g,A}^L(t)\partial_L|\boldsymbol{x}-\boldsymbol{z}_A(t)|\right]\right\} + \mathcal{O}\left(c^{-4}\right)\,,\nonumber\\
  \zeta_g^i(t,\boldsymbol{x})=&-\sum_{A=1}^{N}\sum_{l=0}^\infty\frac{(-1)^l}{l!}Z^{iL}_{g,A}(t)
  \partial_L\frac{1}{|\boldsymbol{x}-\boldsymbol{z}_A(t)|} + \mathcal{O}\left(c^{-2}\right)\,.
\end{align}
Furthermore, the moments $Z^{iL}_{g,A}$ satisfy
\begin{equation}
\begin{aligned}
Z^{\langle iL \rangle }_{g,A} & = -\frac{4}{l+1} \dot{M}^L_{g,A}(t) \\
Z^{jjL}_{g,A} & = 0 \,.
\end{aligned}
\end{equation}
The first relation arises from the harmonic gauge condition, whereas the latter is equivalent to setting the global gauge moments $\mu_{g,A}^L$ (which we have not introduced) to zero. This expansion is the sum~\footnote{The sum over the bodies is justified by the linearity of the Einstein equations within the 1PN approximation.} of the contribution from each body, which is centered at the COM worldline $x^i=z^i_A(t)$. Note that there are no tidal terms, since the global frame extends up to spatial infinity, where the metric reduces to the flat metric, and no tidal force acts on the $N$-body system.

However, we can introduce the tidal moments for each body, as the result of the presence of the other bodies, also in the global frame. In the buffer region $\mathcal{B}_A$ of the body, the potentials $\Phi_g(t,\boldsymbol{x})$ and $\zeta^i_g(t,\boldsymbol{x})$ can be rewritten in a different way, including the contributions from the bodies $B\neq A$ in the (electric and magnetic) \emph{global tidal multipole moments} $G^L_{g,A}(t)$, $Y^{iL}_{g,A}(t)$ (defined for $l \geq 0$ and STF tensors on all (the
last) $l$-indices)
\begin{align}
\label{eq:globlocmultexp}
  \Phi_g(t,\boldsymbol{x})=&-\sum_{l=0}^\infty\frac{1}{l!}\left\{
   (-1)^l  M_{g,A}^L(t)\partial_L\frac{1}{|\boldsymbol{x}-\boldsymbol{z}_A(t)|}\right. \left.+G^L_{g,A}(t)[x-z_A(t)]^L  \right.\nonumber\\
    & \left.+\frac{1}{2c^2}\partial^2_t
    \bigg[M_{g,A}^L(t)\partial_L|\boldsymbol{x}-\boldsymbol{z}_A(t)|\right. \left.\left.+\frac{1}{2l+3}G_{g,A}^L(t)[x-z_A(t)]^{jjL}\right]\right\}  \nonumber\\
    & + \mathcal{O}\left(c^{-4}\right),\nonumber\\
  \zeta_g^i(t,\boldsymbol{x})=&-\sum_{l=0}^\infty\frac{1}{l!}\left\{(-1)^lZ^{iL}_{g,A}(t)
  \partial_L\frac{1}{|\boldsymbol{x}-\boldsymbol{z}_A(t)|}\right. +Y^{iL}_{g,A}(t)[x-z_A(t)]^L\bigg\}+ \mathcal{O}\left(c^{-2}\right).
\end{align}
Comparing the expansions~\eqref{eq:globmultexp} and~\eqref{eq:globlocmultexp}, the global tidal moments of the body $A$ can be expressed as functions of the global body multipole moments of the bodies $B \neq A$ and their COM worldlines.

Outside the $N$-body system, far away from all sources, it is possible to rewrite the potentials of the global frame through an expansion around the origin $x^i=0$, in terms of the multipole moments
of the entire system, i.e., the mass and current \emph{system multipole moments} $M^L_{sys}(t)$ and $J^L_{sys}(t)$ (STF tensors defined for $l \geq 0$ and $l \geq 1$, respectively)
\begin{align}
\label{eq:globmultexp2}
  \Phi_g(t,\boldsymbol{x})=&-\sum_{l=0}^\infty\frac{(-1)^l}{l!}\left\{
    M_{sys}^L(t)\partial_L\frac{1}{|\boldsymbol{x}|}\right. \left.+\frac{1}{c^2}\left[\frac{(2l+1)}{(l+1)(2l+3)}{\dot\mu}^L_{sys}
      \partial_L\frac{1}{|\boldsymbol{x}|}\right.\right.\nonumber\\
      &\left.\left.+\frac{1}{2}{\ddot M}^L_{sys}\partial_L|\boldsymbol{x}|\right]\right\} +  \mathcal{O}\left(c^{-4}\right)\,,\nonumber\\
    \zeta_g^i(t,\boldsymbol{x})=&-\sum_{l=0}^\infty\frac{(-1)^l}{l!}
    Z_{sys}^{iL}(t)\partial_L\frac{1}{|\boldsymbol{x}|}+  \mathcal{O}\left(c^{-2}\right)\,\,,
\end{align}
where
\begin{equation}
Z_{sys}^{iL} = \frac{4}{l+1} \dot{M}^{iL}_{sys}- \frac{4l}{l+1} \epsilon^{ji \langle a_l} J_{sys}^{L-1\rangle j}+\frac{2l-1}{2l+1} \delta^{i \langle a_l} \mu_{sys}^{L-1 \rangle} + \mathcal{O} (c^{-2}) 
\end{equation}
and $\mu_{sys}^L$, with $l \geq 0$, are the (non-vanishing) system gauge moments. The above expansion is alike that in Eq.~\eqref{eq:pn_expansion} for the local frame multipole moments, but without the tidal terms (see the discussion above). Comparing the expansions~\eqref{eq:globmultexp} and~\eqref{eq:globmultexp2}, we can express the system multipole moments in terms of the global multipole moments through the relations
\begin{align}
\label{sysM}
  M^L_{sys} =&  \sum_{A=1}^{N} \sum_{k = 0}^l  \binom{l}{k} \bigg[ M_{g,A}^{\langle L-K} z_A^{K \rangle}
    + \frac{1}{c^2} \frac{1}{2(2l+3)}\partial_t^2 \left(2 M_{g,A} ^{j \langle L- K } z_A^{K \rangle j}\right.\nonumber\\
  & \left.\left.+ M_{g,A}^{\langle L -K} z_A^{K \rangle jj} \right) \right. \bigg] - \frac{1}{c^2} \frac{2l +1}{(l+1)(2l+3)} \dot{\mu}^L_{sys} +  \mathcal{O}\left(c^{-4}\right)\,,\\
\label{sysS}
J^L_{sys}= & \frac{1}{4}Z_{sys}^{jk \langle L-1}\epsilon^{a_l \rangle kj}\,,
\end{align}
where
\begin{align}
\label{eq:sysmu}
\mu^L_{sys}= & Z^{jjL}_{sys} \,, \\
\label{sysZ}
  Z^{iL}_{sys} =&  \sum_{A=1}^{N} \sum_{k=0}^l  \binom{l}{k}
  Z_{g,A}^{i \langle L -K} z_A^{K \rangle}+  \mathcal{O}\left(c^{-2}\right) \,.
\end{align}

We have said that both the local and the global frame are defined in the buffer region of each body, and they are related by the coordinate transformation~\eqref{eq:coordtransf}. This means that the metrics in the two coordinate systems are related by
\begin{equation}
g^A_{\mu \nu} = \frac{\partial x^{\alpha}}{\partial y^{\mu}_A  } \frac{\partial x^{\beta}}{\partial y^{\nu}_A} g^g_{\alpha \beta} \,. 
\end{equation}
Through this equivalence, and using the expansions~\eqref{eq:pn_expansion} and~\eqref{eq:globlocmultexp}, one can determine (and eliminate from the problem) the functions $\alpha_A$ and $\beta_A$ in the transformation~\eqref{eq:coordtransf}, and obtain the relations between the local multipole moments and the global multipole moments, involving only the worldlines $z_A^i$. Henceforth, we define $v^i_A \equiv \dot{z}^i_A$, $a^i_A \equiv \ddot{z}^i_A$ and $\partial_{i}^{(A)} \equiv \partial / \partial z^i_A$. The global mass and current multipole moments $M^L_{g,A}$, $Z^{iL}_{g,A}$ and the electric and magnetic tidal moments $G^L_{g,A}$, $Y^{iL}_{g,A}$ can be expressed in terms of the local body multipole moments $M^L_{A}$, $J^{L}_{A}$ as
\begin{align}
\label{eq:Mg}
  M^L_{g,A} =&  M^L_A + \frac{1}{c^2} \bigg[\left(\frac{3}{2} v_A^2 -(l+1) G_{g,A} \right)M^L_A -\frac{2l^2+5l-5}{(l+1)(2l+3)} v^j_A \dot{M}^{jL}_A\nonumber\\
  &  -\frac{2l^3+7l^2+16l+7}{(l+1)(2l+3)} a^j_A M^{jL}_A  -\frac{2l^2+17l-8}{2(2l+1)} v^{j \langle a_l}_A M^{L-1 \rangle j}_A \nonumber\\
    &+\frac{4l}{l+1} v^j_A \epsilon^{jk \langle a_l} J^{L-1 \rangle k}_A \bigg]
  + \mathcal{O} \left(c^{-4} \right)\,,\\
  \label{eq:Zg}
  Z^{iL}_{g,A} =&  \frac{4}{l+1} \dot{M}^{iL}_A +4 v^i_A M^L_A -\frac{4(2l-1)}{2l+1} v^j_A M^{j \langle L-1}_A
  \delta^{a_l \rangle i} -\frac{4l}{l+1} \epsilon^{ji \langle a_l} J^{L-1 \rangle j}_A \nonumber\\
&  + \mathcal{O} \left(c^{-2} \right) \,, 
\end{align}
\begin{align}
\label{eq:Gg}
G^L_{g,A} =&  \sum_{B \neq A} \sum_{k=0}^{\infty} \frac{(-1)^k}{k!} M_B^K
\partial^{(A)}_{KL} \frac{1}{|\boldsymbol{z}_A-\boldsymbol{z}_B|} + \mathcal{O} \left( c^{-2}\right)\,,\\
\label{eq:Yg}
Y^{iL}_{g,A} =& \sum_{B \neq A} \sum_{k=0}^{\infty} \frac{(-1)^k}{k!} Z^{iK}_{g,B}\,
\partial^{(A)}_{KL} \frac{1}{|\boldsymbol{z}_A-\boldsymbol{z}_B| } + \mathcal{O} \left(c^{-2} \right)\,.
\end{align}
The local electric and magnetic tidal moments $G_A^L$, $H_A^L$ can be expressed in terms of the global tidal multipole moments as
\begin{align}
  \label{eq:G}
  G^{L}_{A} = & F_{g,A}^L- l! \Lambda^L_{\Phi,A} +\frac{1}{c^2}
  \bigg[ \dot{Y}^{\langle L \rangle}_{g,A}-v^j_A Y^{jL}_{g,A}
  +(2v^2_A-lG_{g,A})G^L_{g,A}-(l/2)v^{j\langle a_l}_A G^{L-1 \rangle j}_{g,A} \nonumber \\
  & +(l-4) v^{\langle a_l}_A \dot{G}^{L-1 \rangle }_{g,A}
  -(l^2-l+4) a^{\langle a_L}_A G^{L-1 \rangle}_{g,A}
  -(l-1)! \dot{\Lambda}_{\zeta,A}^{\langle L \rangle} \bigg] + \mathcal{O}
  \left(c^{-4} \right) \  l \geq 1 \,, \\
  \label{eq:H}
  H^L_A =& Y_{g,A}^{jk \langle L-1} \epsilon^{a_l \rangle jk}-4v_A^jG_{g,A}^{k\langle L-1}\epsilon^{a_l\rangle jk}-l!\Lambda_{\zeta,A}^{jk\langle L-1}\epsilon^{a_l\rangle jk}~\footnotemark+ \mathcal{O} \left(c^{-2} \right)
\qquad l  \geq 1\,,
\end{align}\footnotetext{Note that $\Lambda_{\zeta,A}^{jk\langle L-1}\epsilon^{a_l\rangle jk}= \mathcal{O}\left(c^{-2} \right) $ for $l \geq 2$, see Eqs.~\eqref{eq:Lambdaz}.}
where
\begin{align}
\label{eq:Fg}
  F^{L}_{g,A} = &\sum_{B \neq A} \sum_{k=0}^{\infty} \frac{(-1)^k}{k!}
  \bigg[ N^K_{g,B}\, \partial^{(A)}_{KL} \frac{1}{|\boldsymbol{z}_A-\boldsymbol{z}_B| }
    +\frac{1}{2c^2} P^K_{g,B} \, \partial^{(A)}_{K\langle L \rangle}
    |\boldsymbol{z}_A-\boldsymbol{z}_B|\bigg]+ \mathcal{O} \left(c^{-4} \right)\\
    \label{eq:Ng}
  N^{L}_{g,A} = &M^L_{g,A} + \frac{1}{(2l+3)c^2} [v^2_A M^L_A +2 v^j_A \dot{M}^{jL}_A
    +2lv^{j\langle a_l}_A M^{L-1 \rangle j}_A + a^j_A M^{jL}_A ] + \mathcal{O} \left(c^{-4} \right)\\
    \label{eq:Pg}
  P^{L}_{g,A} =&   \ddot{M}^L_{A} +2 l v_A^{\langle a_l} \dot{M}_A^{L-1 \rangle}
  +l a_A^{\langle a_l} M_A^{L-1 \rangle}
  + l(l-1)v_A^{\langle a_l a_{l-1}} M_A^{L-2\rangle} + \mathcal{O} \left(c^{-2} \right)
\end{align}
\begin{align}
  \label{eq:Lambdaph}
\Lambda^i_{\Phi,A} = & a^i_A + \frac{1}{c^2} \bigg[(v^2_A+G_{g,A})a^i_A+\frac{1}{2} v^{ij}_A a_A^j
  +2 \dot{G}_{g,A} v^i_A \bigg]  + \mathcal{O} \left(c^{-4} \right)\nonumber\\
\Lambda^{ij}_{\Phi,A} = &  \frac{1}{c^2} \left(-\frac{1}{2} a_A^{\langle ij \rangle} + v_A^{\langle i }
\dot{a}_A^{j \rangle} \right)+ \mathcal{O} \left(c^{-4} \right) \nonumber\\
\Lambda^{L}_{\Phi,A} = & \, 0 \qquad l \geq 3\\
\label{eq:Lambdaz}
\Lambda^i_{\zeta,A} = & -2 G_{g,A} v^i_A  + \mathcal{O} \left(c^{-2} \right)\nonumber\\
\Lambda^{ij}_{\zeta,A} = & -\frac{3}{2} v^{[i}_A a^{j]}_A -2 v^{\langle i}_A
a_A^{j \rangle}-\frac{4}{3} \dot{G}_{g,A} \delta^{ij} + \mathcal{O} \left(c^{-2} \right)\nonumber\\
\Lambda^{ijk}_{\zeta,A} = & -\frac{6}{5} \delta^{i \langle j}
\dot{a}_A^{k \rangle} + \mathcal{O} \left(c^{-2} \right)\nonumber \\
\Lambda^{L}_{\zeta,A} = & \, 0 \qquad l \geq 4\,.
\end{align}
Note that at Newtonian order
\begin{align}
\label{eq:Ma}
M_A^L = &  M_{g,A}^L + \mathcal{O}(c^{-2}) \,, \\
\label{eq:Ga}
  G^i_{A} = & G_{g,A}^i- a^i_A+\mathcal{O}(c^{-2})\,,\\
  \label{eq:G0}
  G^L_{A} = & G_{g,A}^L+\mathcal{O}(c^{-2}) \qquad l \geq2\,.
\end{align}
Through the above relations, it is possible to express any quantity in terms of the local body multipole moments $M_A^L$, $J_A^L$ and the worldlines $z^i_A$. Therefore, the dynamics of the $N$-body system is completely determined once the equations of motion for the latter are known.

The equations of motion of the single body $A$ are the laws which govern the rate of change of mass-energy, momentum and angular momentum of the body, due to the interaction with the gravitational fields generated by the other bodies. They were first derived by Damour, Soffel and Xu~\cite{Damour:1991yw} for weakly gravitating sources, by imposing the stress-energy conservation law in the body interior, and then extended to objects with strong gravity by Racine and Flanagan~\cite{Racine:2004hs}, through the 2PN vacuum Einstein equations in the body buffer region. In the body-adapted gauge, the equations of motion are written in terms of the local frame body and tidal multipole moments, and take the form
\begin{align}
\label{eq:monopole}
  \dot{M}_A  =&  -\frac{1}{c^2} \sum_{l=0}^{\infty} \frac{1}{l!} \left[(l+1) M^L_A \dot{G}^L_A
    + l \, \dot{M}^L_A G^L_A \right] + \mathcal{O}\left(c^{-4}\right) \,, \\
  \label{eq:dipole}
  \ddot{M}^i_A  = &\sum_{l=0}^{\infty} \frac{1}{l!} \bigg\{ M^L_A G^{iL}_A+ \frac{1}{c^2}
  \bigg[ \frac{1}{l+2} \epsilon^{ijk} M_A^{jL} \dot{H}^{kL}_A + \frac{1}{l+1} \epsilon^{ijk} \dot{M}_A^{jL} H^{kL}_A \nonumber\\
    &- \frac{2l^3+7l^2+15l+6}{(l+1)(2l+3)} M^{iL}_A \ddot{G}^L_A - \frac{2l^3+5l^2+12l+5}{(l+1)^2} \dot{M}^{iL}_A \dot{G}^L_A\nonumber\\
    &-\frac{l^2+l+4}{l+1} \ddot{M}^{iL}_A G^L_A +\frac{l}{l+1} J^L_A H^{iL}_A -\frac{4(l+1)}{(l+2)^2} \epsilon^{ijk} J^{jL}_A \dot{G}^{kL}_A \nonumber\\
    & - \frac{4}{l+2} \epsilon^{ijk} \dot{J}^{jL}_A G^{kL}_A \bigg] \bigg\} + \mathcal{O}\left(c^{-4}\right)\,,\\
    \label{eq:spin}
\dot{J}^i_A =&  \sum_{l=0}^{\infty} \frac{1}{l!} \epsilon^{ijk} M^{jL}_A G^{kL}_A + \mathcal{O}\left(c^{-2}\right)\,.
\end{align}
The orbital equation of motion for the body $A$, that is the translational equation of motion for the COM worldline $z^i_A$, can be obtained from the condition $\ddot{M}^i_A=0$, which follows from the gauge condition ${M}^i_A=0$. Imposing that the RHS of Eq.~(\ref{eq:dipole}) vanishes, and replacing the local tidal multipole moments of the body $A$ by the body multipole moments of other bodies $B \neq A$, through the relations~\eqref{eq:G}--\eqref{eq:Lambdaz}, yields a second-order ODE for the global frame COM worldline $z^i_A(t)$,
\begin{equation}
\label{eq:eomorbit}
  \ddot{z}^i_A(t)=\mathcal{F}^i_A \left( z^j_B,{\dot z}^j_B,M^L_B,{\dot M}^L_B,{\ddot M}^L_B,J^L_B,{\dot J}^L_B \right)\,.
\end{equation}
The above equation depends only on the worldlines and the local mass and current multipole moments of all bodies $B$. Making the same replacement in Eq.~(\ref{eq:monopole}) and (\ref{eq:spin}), yields, respectively, the equations of motion for mass $M_A(t)$ and spin $S_A(t)$, in terms of the body multipole moments and COM worldlines,
\begin{align}
\label{eq:massev}
\dot{M}_A(t) = & \mathcal{G}_A \left(z^j_B,\dot{z}^j_B,M^L_B,\dot{M}^L_B \right) \\
\label{eq:spinev}
\dot{S}^i_A(t) = & \mathcal{G}^i_A \left(z^j_B,M^L_B \right) \,.
\end{align}

The form of the equations of motion~\eqref{eq:monopole}--\eqref{eq:spin} is valid in any coordinate frame where the metric takes the form in Eq.~\eqref{eq:metric1pn}, and the potentials that in Eq.~\eqref{eq:pn_expansion}. This means that the above equations of motion can be applied also to the evolution of the entire $N$-body system (indeed, Eq.~\eqref{eq:pn_expansion} and Eq.~\eqref{eq:globmultexp2} have the same form). Since there are no tidal terms when considering the system as a whole, replacing the body multipole moments by the system multipole moments in the above equations, we get
\begin{align}
  \dot{M}_{sys}  =&   \mathcal{O}\left(c^{-4}\right) \,, \\
  \label{eq:sysdipolelaw}
  \ddot{M}^i_{sys}  = & \mathcal{O}\left(c^{-4}\right)\,,\\
\dot{J}^i_{sys} =&    \mathcal{O}\left(c^{-2}\right)\,.
\end{align}
The above equations express the conservation of mass-energy, momentum and angular momentum (at 1PN order) for an isolated system. This result is useful to check explicitly the correctness of the equations of motion for the single body. In the following we make use of Eq.~\eqref{eq:sysdipolelaw} to check our computations. 

Using Eqs.~\eqref{eq:massev} and~\eqref{eq:spinev}, we can eliminate the time derivatives on masses and spins from Eq.~\eqref{eq:eomorbit}, obtaining
\begin{equation}
\label{eq:eomorbit2}
  \ddot{z}^i_A(t)=\mathcal{F}^i_A \left( z^j_B,{\dot z}^j_B,M_B,J^i_B,Q^L_B,{\dot Q}^L_B,{\ddot Q}^L_B,S^L_B,{\dot S}^L_B \right)\,,
\end{equation}
where we have defined
\begin{equation}
\begin{aligned}
Q^L_A \equiv & M^L_A \\
S^L_A \equiv & J^L_A 
\end{aligned} \qquad l \geq 2 \,.
\end{equation}
Henceforth, we adopt this notation to denote the higher-order body multipole moments. However, this is not enough to fully determine the dynamics of all bodies. To close the system, the orbital equation of motion~\eqref{eq:eomorbit2} has to be supplemented also by the evolution equations for the multipole moments $Q_A^L, S_A^L$, which depend on the internal dynamics of the bodies. In other words, we need to provide a model for the interior degrees of freedom. Since we are interested in studying the tidal deformations of compact objects, we assume that the body multipole moments are tidally induced by the (local) tidal moments. Therefore, within the \emph{adiabatic approximation}, the equations for the higher-order multipole moments in the local frame are given by the adiabatic relations defined in Eqs.~\eqref{eq:adiabaticrelspin}, that we report here
\begin{equation}
\label{eq:adiabaticrelspin2}
\begin{aligned}
Q^L_A & = \lambda_l^{(A)} \, G^L_A + \frac{\lambda_{l \, l+1}^{(A)}}{c^2} J^k_A \, H^{kL}_A +  \frac{\lambda_{l \, l-1}^{(A)}}{c^2} \, J^{\langle a_l}_A \, H^{L-1 \rangle}_A \\
S^L_A & = \frac{\sigma_l^{(A)}}{c^2} \, H^L_A + \sigma_{l \, l+1}^{(A)} \, J^k  \, G^{kL}_A +  \sigma_{l \, l-1}^{(A)} \, J^{\langle a_l}_A \, G^{L-1 \rangle}_A
\end{aligned} \ .
\end{equation}
Note the peculiar presence of the factor $1/c^2$ together with the magnetic tidal moments $H^L_A$. This is due to the fact that we have defined the magnetic tidal moments to be non-zero also at Newtonian order, i.e., we have factored out the 1PN order dependence in the metric, by writing the gravito-magnetic potential as $\zeta^i/c^2$. In other words, a magnetic tidal field can source the current moments only at 1PN order. On the other hand, the electric tidal fields can induce (by means of the spin) current deformations also at Newtonian order, but the current moments would affect the dynamics only at 1PN order. This means that at Newtonian level, gravity can affect the internal motion of a body, but instead the momentum distributions do not gravitate, like it must be. These arguments are in perfect agreement with the discussion in~\cite{Gagnon-Bischoff:2017tnz}. In section~\ref{sec:truncation} we apply explicitly the adiabatic approximation to a compact binary. 

\subsubsection{Newtonian gravity}
\label{sec:newtgrav}
At Newtonian order, Eqs.~\eqref{eq:monopole}--\eqref{eq:spin} reduce to (we recall that $M_A^i =0$ and $G_A=0$)
\begin{align}
\label{eq:monopole_newt}
  \dot{M}_A  =&   \mathcal{O}\left(c^{-2}\right) \,, \\
  \label{eq:dipole_newt}
  \ddot{M}^i_A  = & M_A G^{i}_A+ \sum_{l=2}^{\infty} \frac{1}{l!}  Q^L_A G^{iL}_A  + \mathcal{O}\left(c^{-2}\right)\,,\\
    \label{eq:spin_newt}
\dot{J}^i_A =&  \sum_{l=1}^{\infty} \frac{1}{l!} \epsilon^{ijk} Q^{jL}_A G^{kL}_A + \mathcal{O}\left(c^{-2}\right)\,.
\end{align}
We can see that the mass is conserved at Newtonian order. Imposing the condition $\ddot{M}_A^i =0$, and using the relations~\eqref{eq:Ga} and~\eqref{eq:G0}, we obtain the translation equations of motion for the worldlines $z_A^i$,
\begin{equation}
  \label{eq:accel_newt}
  M_A a^i_A  =  M_A G^{i}_{g,A}+ \sum_{l=2}^{\infty} \frac{1}{l!}  Q^L_A G^{iL}_{g,A}  + \mathcal{O}\left(c^{-2}\right)\,,
\end{equation}
where the global frame tidal moments $G^L_{g,A}$ are given in Eqs.~\eqref{eq:Gg}. Replacing the local frame tidal moments through Eq.~\eqref{eq:G0}, the equation for the spin becomes
\begin{equation}
  \label{eq:spin_newt2}
\dot{J}^i_A = \sum_{l=1}^{\infty} \frac{1}{l!} \epsilon^{ijk} Q^{jL}_A G^{kL}_{g,A} + \mathcal{O}\left(c^{-2}\right)\,.
\end{equation}
To fully constrain the dynamics, we need to provide only the equations of motion for the $l \geq 2$ mass multipole moments $Q^L_A$. In the adiabatic approximation these are given by
\begin{equation}
\label{eq:adiabnewt}
Q^L_A = \lambda_l^{(A)} G^L_A = \lambda_l^{(A)} G^L_{g,A} + \mathcal{O}\left(c^{-2}\right) \,.
\end{equation}
From this, we see that, if the mass moments are tidally induced, the spin is conserved
\begin{equation}
  \label{eq:spin_newt3}
\dot{J}^i_A = \sum_{l=1}^{\infty} \frac{\lambda_l^{(A)}}{l!} \epsilon^{ijk}  G^{jL}_{g,A} G^{kL}_{g,A} + \mathcal{O}\left(c^{-2}\right) =  \mathcal{O}\left(c^{-2}\right) \,,
\end{equation}
where the last equality follows from the contraction of the Levi-Civita symbol (which is antisymmetric under the exchange of the $j,k$ indices) with $G^{jL}_{g,A} G^{kL}_{g,A}$ (that instead is symmetric under the exchange $j \leftrightarrow k$).

\subsection{Binary systems}
\label{sec:bns}
In this section, we summarize the steps needed to obtain the phase of the gravitational waveform emitted by an inspiralling compact binary system, where the objects are tidally deformed. We start from the equations of motion~\eqref{eq:eomorbit2} and~\eqref{eq:adiabaticrelspin2}. In the next section, this procedure is applied in details to a spinning binary where the single components have tidally induced, mass and current, quadrupolar and octupolar multipole moments. Note that we are describing the dynamics of the system to 1PN order, but the tidal effects start to affect the waveform at 5PN order. There is no inconsistency in that, because we have defined the 1PN approximation through the formal parameter $1/c^2$, which can always be set to $c=1$. The PN order of any term is given by the powers of the orbital velocity, which is the true small parameter (with respect to speed of light) of the expansion.

We can define the COM (global) frame as the coordinate system for which the mass dipole of the system vanishes, $M^i_{sys}=0$. For binary systems ($N=2$), the conservation of momentum allows us to describe the dynamics in the COM frame as a function of the orbital separation alone, $z^i(t)=z^i_2(t)-z^i_1(t)$. Thus, the equation of motion of the orbital separation takes the form
\begin{equation}
\label{eq:eom_com}
{\ddot z}^i={\ddot z}_2^i-{\ddot z}^i_1=\mathcal{F}^i_2-\mathcal{F}^i_1\,.
\end{equation}
After the equation for $z^i$ is solved, one can go back to the single worldlines $z^i_1, \ z^i_2$, using the condition $M^i_{sys}=0$.

Within the adiabatic approximation, the equation of motion~\eqref{eq:eom_com}, together with the equations for the multipole moments (the adiabatic relations~\eqref{eq:adiabaticrelspin2}), can be derived from a generalized action principle, in terms of a Lagrangian function $\mathcal{L}(z^i,{\dot z}^i,{\ddot z}^i,Q_A^L,{\dot Q}_A^L,S_A^L)$. Since this Lagrangian does not depend explicitly on time, $\partial \mathcal{L}/ \partial t = 0$, the total energy $E$ of the two-body system is a conserved quantity (neglecting the gravitational wave emission~\footnote{The emission of gravitational radiation produce back-reaction forces which act on the system, ensuring that the total energy is conserved. This \emph{radiation reaction} affects the conservative dynamics of a system only starting at 2.5PN order, which is well beyond our approximation~\cite{Maggiore:2008,Blanchet:2013haa}.}), which can be obtained using the standard techniques of Lagrangian mechanics. Indeed, let us consider a generalized Lagrangian which depends on a set of variables $\{q_i(t)\}$ as well as their time derivatives up to the $n$-th order~\cite{DeLeon:1985},
\begin{equation}
\mathcal{L} = \mathcal{L} \left(q_i, \dot{q}_i , \ddot{q}_i , \dots, q^{(n)}_i \right) \,.
\end{equation}
The Euler-Lagrange equations of motion read
\begin{equation}
\label{eq:euler}
\sum_{k=0}^n (-1)^k \frac{d^k}{dt^k} \frac{\partial \mathcal{L}}{\partial q_i^{(k)}} = 0 \,.
\end{equation}
Assuming that the Lagrangian does not depend explicitly on time, it can be shown that the conserved energy is given by
\begin{equation}
\label{eq:euleren}
E = \sum_i \sum_{j=1}^{n} p_{i,j} q_i^{(j)} - \mathcal{L}
\end{equation}
where $\{p_{i,j} \}$ are the $j$-th order momenta
\begin{equation}
\label{eq:euleren2}
p_{i,j} = \sum_{k=0}^{n-j} (-1)^k \frac{d^k}{dt^k} \frac{\partial \mathcal{L}}{\partial q^{(j+k)}_i} \qquad j=1,\dots,n \,.
\end{equation}

Within the PN approximation, the gravitational radiation emitted by a system is due to the presence of time-varying multipole moments. Because of the non-linearity of the gravitational interaction, the gravitational wave flux can be split in two contributions: an \emph{instantaneous} term due to the gravitational wavefront, and a hereditary \emph{tail} term which arrives later~\footnote{We are neglecting the so-called \emph{non-linear memory} hereditary term, which affects the instantaneous flux at 2.5PN order~\cite{Maggiore:2008,Blanchet:2013haa}.}. Let us consider the total radiated energy at infinity at the time $t_0$. The instantaneous term corresponds to the energy loss by the system at the retarded time $u_0=t_0-r/c$ (where $r$ is the distance between the source and the observer). The tail term encodes instead the part of the gravitational radiation emitted at all times $u < u_0$, which is scattered by the curved background of the system, and therefore accumulates delay. At $1.5$PN order~\footnote{Even if we are working within the 1PN approximation, we need the gravitational wave flux at 1.5PN order to derive the leading-order tail-tidal contribution to the waveform (the only 1.5PN order term in the flux is indeed the tail term). This is consistent, because other higher-order corrections to the metric, beyond the 1PN order, would affect the waveform beyond the 1.5PN order. A similar situation occurs for the spin, whose contribution enters to the waveform at 1.5PN order, but it is derived from the metric at 1PN order. We see this explicitly in the next section (see also the discussion above Eq.~\eqref{eq:eom_com}).}, the gravitational wave flux is given by~\cite{Blanchet:2013haa}
\begin{equation}
\label{eq:flux}
\begin{aligned}
F = &F_{\mathrm{inst}} + F_{\mathrm{tail}} + \mathcal{O}\left(c^{-9} \right) \\
  F_{\mathrm{inst}} =& \frac{1}{5c^5} \dddot{M}^{ij}_{sys}\dddot{M}^{ij}_{sys} + \frac{1}{189 c^7} \ddddot{M}^{ijk}_{sys} \ddddot{M}^{ijk}_{sys} + \frac{16}{45c^7} \dddot{J}^{ij}_{sys} \dddot{J}^{ij}_{sys} \\
  F_{tail}=& \frac{2}{5c^8}\dddot{M}^{ij}_{sys}\dot{U}^{ij}_{tail} \,,
\end{aligned}
\end{equation}
where
\begin{equation}
\label{eq:tail}
  U^{ij}_{tail}(U) = 2M_{sys} \int_0^{\infty} \ddddot{M}^{ij}_{sys}(U-\tau)
  \left[ \log{\left(\frac{c \tau}{2 r_0} \right)+ \frac{11}{12}} \right] d\tau \,.
\end{equation}
In the above equation, $U=t-r/c- \left( 2M_{sys}/c^3 \right) \log{\left( r/r_0\right)}$ is the retarded time in radiative coordinates, and $r_0$ a gauge-dependent arbitrary constant due to the freedom of choice of the radiative coordinates themselves.

We turn now to the evaluation of the gravitational waveform phase. At large distance from the source, the gravitational radiation emitted by a system is described as a small perturbation $h_{\mu \nu}$ of the flat metric. In the transverse-traceless (TT) gauge, the asymptotic waveform $h_{ij}^{TT}$ (with $h_{0 \mu}^{TT}=0$) has only two degrees of freedom: $h_{+}$ and $h_{\times}$, corresponding to the two polarizations of the gravitational waves. The two polarization states are obtained projecting the waveform onto the orthonormal triad $(\vec{\mathrm{e}}_r,\vec{\mathrm{e}}_{\theta},\vec{\mathrm{e}}_{\varphi})$, made of the spatial basis vectors of spherical radiative coordinates $(t,r,\theta,\varphi)$~\cite{Kidder:2007rt}. The (scalar) waveform $h(t,r,\theta,\varphi)$ is defined as the complex scalar $h= h_{+}-\mathrm{i} h_{\times} $. It can be shown that $h$ can be expanded in spin-weighted spherical harmonics with spin-weight $s=-2$,
\begin{equation}
h(t,r,\theta,\varphi) = \sum_{l \geq 2} \sum_{m} h_{lm}(t,r) \ _{-2}Y_{lm}(\theta,\varphi)
\end{equation}
(see the Appendix~\ref{sec:appB} for the definition of the spin-weighted spherical harmonics). 

Let us consider now a non-precessing binary system in circular orbit with orbital frequency $\omega/(2\pi)$. Since the orbital motion is planar, it can be shown that in this case $h_{l\,m} = (-1)^l h^*_{l\, -m}$~\cite{Blanchet:2013haa}. We can use the quadrupolar approximation, since for such a system the contribution of the higher-order modes is negligible. At the leading quadrupolar order $l=2$, only the modes with $|m|=2$ are different from zero, and the gravitational radiation is emitted at twice the orbital frequency, $\omega_{\mathrm{GW}}= 2 \omega$. Since $h_{2\, -2} = h^*_{2\, 2}$, besides the angular dependence, we can focus on the $l=m=2$ mode, $h \sim h_{2\,2}$~\footnote{The spin-weighted spherical harmonics $\ _{-2} Y_{2 \, \pm 2} \sim (1 \pm \cos{(\theta)})^2 \mathrm{e}^{\mathrm{i} 2 \varphi}$, where for a binary system the angle $\theta$ is the angle between the line of sight and the normal of the orbital plane, and we can put $\varphi=0$ without loss of generality (it is equivalent to redefine the overall phase constant of the gravitational waveform, see Eq.~\eqref{eq:phasedef}). Thus, $h \sim (1 + \cos{(\theta)})^2 h_{2\, 2}+(1 - \cos{(\theta)})^2 h^*_{2\, 2} $.}. Thus, the waveform can be written as $h(t) = A(t) \mathrm{e}^{-\mathrm{i} \phi(t)}$, where the gravitational wave phase $\phi(t)$ is given by
\begin{equation}
\label{eq:phasedef}
\phi(t) = \int^t \omega_{GW}(t') \, dt'  = \int_{t_0}^t 2\omega(t') \, dt' + \phi_{0} \,,
\end{equation}
where $\phi_0$ and $t_0$ are two constants.

The total energy of the binary system and the emitted gravitational wave flux are related through the energy balance relation
\begin{equation}
\label{eq:enbal}
\dot{E}=-F\,.
\end{equation}
If the binary system evolves adiabatically slow in time, the gravitational wave phase can be extracted from the above equation. Assuming that the energy and the gravitational flux can be written as functions of the orbital frequency (without any explicit time dependence), we write
\begin{equation}
\frac{dE}{dt} = \frac{dE}{d\omega} \frac{d\omega}{dt} =  \frac{1}{2} \frac{dE}{d\omega} \frac{d^2\phi}{dt^2} \,,
\end{equation}
that replaced in Eq.~\eqref{eq:enbal} gives
\begin{equation}
\frac{d^2\phi}{dt^2} = -\frac{2F}{dE/d\omega} \,,
\end{equation}
which can be recast into the system
\begin{equation}
\label{eq:phasesystem}
\left \{
\begin{aligned}
\frac{d \phi}{dt} = & 2 \omega \\
\frac{d\omega}{dt} = & -\frac{F}{dE/d\omega}
\end{aligned}
\right. \,.
\end{equation}
The above system can be solved through different methods, giving rise to the so-called TaylorT1--TaylorT4 gravitational waveform approximants~\cite{Creighton:2011}. We follow the TaylorT2 approach. Exploiting the equation
\begin{equation}
dt = - \frac{1}{F }\frac{dE}{d\omega} d\omega\,,
\end{equation}
the solution to the system~\eqref{eq:phasesystem} can be written as
\begin{equation}
\label{eq:phasesystemsol}
\left \{
\begin{aligned}
t(\omega) = & t_c + \int_{\omega}^{\omega_c} \frac{1}{F(\omega')}\frac{dE(\omega')}{ d\omega'} \, d\omega' \\[0.3cm]
\phi(\omega) = & \phi_c + \int_{\omega}^{\omega_c} \frac{2 \omega'}{F(\omega')} \frac{dE(\omega')}{ d\omega'} \, d\omega' 
\end{aligned}
\right. \,,
\end{equation}
where $t_c$, $\omega_c$ and $\phi_c$ are constants.

Next, we transform the solution for the waveform phase $\phi$ in the frequency domain. The Fourier transform of the gravitational signal is defined as
\begin{equation}
\label{eq:fourier}
\tilde{h}(f) = \int^t h(t) \mathrm{e}^{\mathrm{i} 2 \pi ft} \, dt=  \int_{-\infty}^{+\infty} A(t) \mathrm{e}^{\mathrm{i}(2\pi ft-\phi(t))} \, dt \,.
\end{equation}
During the inspiral phase, both the amplitude $A$ and the frequency $\omega_{\mathrm{GW}}=d\phi/dt $ vary slowly with respect to the phase $\phi$ ($d\log{A}/dt \ll d\phi/dt $ and $d\omega_{\mathrm{GW}}/dt \ll (d\phi/dt)^2 $). Under these assumptions, we can solve the integral~\eqref{eq:fourier} using the stationary phase approximation, which states that the only non-negligible contribution to the integral comes from the region around the point where the phase in the exponential is stationary. Everywhere else the integrand is highly oscillating, and averages to zero. In our case, the stationary point $t^*$ is given by $\dot{\phi}(t^*) = 2 \pi f$, which means that the largest contribution to the Fourier transform $\tilde{h}(f)$, at a given $f$, comes from the instant of time $t^*$ for which the gravitational frequency is $\omega_{\mathrm{GW}} =2 \pi f$ (as expected). Around the stationary point, we expand $\phi(t) = \phi(t^*) + 1/2 \ddot{\phi}(t^*) (t-t^*)^2 + \mathcal{O}(t^3)$, obtaining
\begin{equation}
\tilde{h}(f) \sim A(t^*) \mathrm{e}^{\mathrm{i}(2\pi ft^*-\phi(t^*))}  \int_{-\infty}^{+\infty}  \mathrm{e}^{-\mathrm{i}  \ddot{\phi}(t^*) (t-t^*)^2/2} \, dt \,.
\end{equation}
Using the Fresnel integral $\int_{-\infty}^{+\infty} \mathrm{e}^{-\mathrm{i} x^2}dx = \sqrt{\pi} \mathrm{e}^{-\mathrm{i} \pi/4}$, we obtain
\begin{equation}
\label{eq:amplitudefreq}
\begin{aligned}
\tilde{h}(f) \sim & \mathcal{A}(f) \,  \mathrm{e}^{\mathrm{i} \psi(f)} \\
\mathcal{A}(f)=& A(t^*) \sqrt{\frac{2 \pi }{\ddot{\phi}{(t^*)}}} \\
\psi(f) = & 2\pi ft^*-\phi(t^*)  -\frac{\pi}{4} \,.
\end{aligned}
\end{equation}
Note that $t^*$ depends implicitly on $f$, because $\omega_{\mathrm{GW}}(t^*) = 2\pi f$.

The gravitational wave phase in the time domain is given by the TaylorT2 approximant~\eqref{eq:phasesystemsol}. Recalling that $\omega_{\mathrm{GW}}= 2\omega$ we get
\begin{equation}
\label{eq:phasefreq}
\psi(f) =  2\pi f t_c-\phi_c  -\frac{\pi}{4} + \int^{\omega_c}_{\pi f}\frac{2(\pi f-\omega')}{F(\omega')} \frac{dE(\omega')}{d\omega'} \, d\omega' \,,
\end{equation}
which is the expression of the TaylorF2 approximant. Besides the two integration constants $t_c$ and $\phi_c$ (the time and the phase at the coalescence, respectively), the piece of the phase $\psi$, which contains the physical information from the source, is given by~\cite{Tichy:1999pv}
\begin{equation}
\label{eq:wav}
\frac{d^2\psi}{d\omega^2}=-\frac{2}{F}\frac{dE}{d\omega}\,, 
\end{equation}
where we have used the change of variable $\omega = \pi f$.

Here, we are not considering the gravitational wave amplitude (either in the time or frequency domain). The reason is that the estimation of the physical parameters of the source during the inspiral is more sensible to variations of the gravitational wave phase, than to those of the amplitude~\cite{Cutler:1994ys} (see section~\ref{sec:impact}). Therefore, we need to include the tidal effects only in the waveform phase, while we can use the amplitude at Newtonian order (i.e., that given by the Einstein quadrupole formula). We stress that this is no longer true during the merger, where the amplitude can differ significantly depending on the neutron star equation of state~\cite{Hotokezaka:2016bzh}.

\section{Tidal interactions of a spinning binary system}
\label{sec:truncation}
Working at Newtonian level, Flanagan and Hinderer computed for the first time the leading 5PN order contribution of the electric quadrupolar Love number $\lambda_2$ to the waveform phase~\cite{Flanagan:2007ix}. Then, Vines, Flanagan and Hinderer applied the approach described in the previous section to a non-spinning binary system, computing the next-to-leading, 6PN order, contribution of $\lambda_2$ to the waveform~\footnote{Vines and Flanagan~\cite{Vines:2010ca} derived the equations of motion, and the consequent Lagrangian, for a binary system where only one of the components is spinning. Then, since in the adiabatic approximation the spin is constant (see Eqs.~\eqref{eq:spin_newt3} and~\eqref{eq:spinconserv}), they set it to zero, neglecting its subleading contribution to the tidal part of the phase.}~\cite{Vines:2010ca,Vines:2011ud}. Damour, Nagar and Villain extended this result computing the next-to-next-to-leading, 6.5PN order, contribution of $\lambda_2$, due to the coupling to the tail part of the gravitational radiation~\cite{Damour:2012yf}. Yagi included the effect of the quadrupolar magnetic Love number $\sigma_2$, which enters at 6PN order~\cite{Yagi:2013sva} (see also~\cite{Banihashemi:2018xfb}).

In this thesis we extend their results including the effect of the spins of the compact objects. We consider a spinning binary system where the components have tidally induced mass and current, quadrupolar and octupolar multipole moments. We take into account the coupling between the spin and the standard tidal Love numbers, as well as the contribution of the rotational tidal Love numbers. We show how in this way we can obtain the complete tidal contribution to the gravitational waveform phase, up to 6.5PN order. As a by-product of our computation, we re-obtain the results previously derived by those authors, and we get also the leading-order contributions of the electric and magnetic octupolar Love numbers $\lambda_3$ and $\sigma_3$, which enter at higher PN orders.

We consider a binary system where the body $1$ is characterized by its mass $M_1$ and its spin $J_1^i$, whereas the the body $2$ is characterized by its mass $M_2$, its spin $J_2^i$, its mass quadrupole $Q_2^{ij}$, its current quadrupole $S_2^{ij}$, its mass octupole $Q_2^{ijk}$ and its current octupole $S_2^{ijk}$. All the other higher-order multipole moments vanish identically. Following Vines and Flanagan~\cite{Vines:2010ca}, we call this the $M_1$-$J_1$-$M_2$-$J_2$-$Q_2$-$S_2$-$Q_3$-$S_3$ truncation. We work to linear order in the spins and in the tidal fields (which means in the tidal Love numbers). We assume that the quadrupole and octupole moments are tidally induced, therefore we neglect the spin-induced quadrupole, etc. (which indeed are higher than linear order in the spin). Since the $l \geq 2$ multipole moments are tidally induced, working to linear order in the tidal fields means neglecting quadratic and higher-order terms in the $l \geq 2$ multipole moments. With all these assumptions, the contribution that would come from the multipole moments with $l=2,3$ of the body $1$ can be obtained at the end of the computation, simply exchanging the indices $A=1,2$ of the bodies. Because of this, we can drop the index $2$ in the multipole moments with $l=2,3$ of the body $2$, i.e., $Q^{ij} \equiv Q_2^{ij}$, $S^{ij} \equiv S_2^{ij}$, $Q^{ijk} \equiv Q_2^{ijk}$ and $S^{ijk} \equiv S_2^{ijk}$. We do the same with the Love numbers of the body $2$, i.e., $\lambda_2\equiv \lambda^{(2)}_2 $, etc. We restore the indices at the end of the computation.

\subsection{Equations of motion}
\label{subsec:eom}
Within our truncation, the equations of motion of the mass monopole (i.e., the mass), the mass dipole and the current dipole (i.e., the spin) of the two bodies, namely Eqs.~\eqref{eq:monopole}, \eqref{eq:dipole} and \eqref{eq:spin}, respectively, reduce to
\begin{align}
\label{eq:monopole_tr}
\dot{M}_1  = &  \, \mathcal{O}\left(c^{-4}\right) \,,\nonumber\\
  \dot{M}_2  = & -\frac{1}{c^2} \left(\frac{3}{2} Q^{ij} \dot{G}^{ij}_2 +  \dot{Q}^{ij} G^{ij}_2 \right.\left.+ \frac{2}{3} Q^{ijk} \dot{G}_{2}^{ijk} + \frac{1}{2}  \dot{Q}^{ijk} G_{2}^{ijk} \right) + \mathcal{O}\left(c^{-4}\right) \,,
  \end{align}
    \begin{align}
  \label{eq:dipole_tr}
\ddot{M}^i_1  = & \,  M_1 \, G^i_1 + \frac{1}{c^2} \left( \frac{1}{2} J^j_1 H^{ij}_1 - \epsilon^{ijk} J^{j}_1 \dot{G}^{k}_1
  - 2 \epsilon^{ijk} \dot{J}^{j}_1 G^{k}_1  \right)+ \mathcal{O}\left(c^{-4}\right)\,,\nonumber\\
  \ddot{M}^i_2  = & \,  M_2 G^{i}_2+ \frac{1}{2} Q^{jk} G^{ijk}_2 + \frac{1}{6} Q^{jka} G^{ijka}_2+ \frac{1}{c^2} \left( \frac{1}{3} \epsilon^{ijk} Q^{ja} \dot{H}^{ka}_2 + \frac{1}{2} \epsilon^{ijk} \dot{Q}^{ja} H^{ka}_2 \right.
    \nonumber\\
    &+ \frac{1}{4} \epsilon^{ijk} Q^{jab} \dot{H}^{kab}_2 + \frac{1}{3} \epsilon^{ijk} \dot{Q}^{jab} H^{kab}_2 - 3 Q^{ij} \ddot{G}^j_2 - 6 \dot{Q}^{ij} \dot{G}^j_2-3 \ddot{Q}^{ij} G^j_2 \nonumber \\
    &- \frac{80}{21} Q^{ijk} \ddot{G}^{jk}_2 - \frac{65}{9} \dot{Q}^{ijk} \dot{G}^{jk}_2- \frac{10}{3} \ddot{Q}^{ijk} G^{jk}_2+\frac{1}{2} J^j_2 H^{ij}_2 +\frac{1}{3} S^{jk} H^{ijk}_2 \nonumber\\
    &+\frac{1}{8} S^{jka} H^{ijka}_2 - \epsilon^{ijk} J^{j}_2 \dot{G}^{k}_2-\frac{8}{9} \epsilon^{ijk} S^{ja} \dot{G}^{ka}_2 - \frac{3}{8}\epsilon^{ijk} S^{jab} \dot{G}^{kab}_2  - 2 \epsilon^{ijk} \dot{J}^{j}_2 G^{k}_2 \nonumber\\
    & \left. - \frac{4}{3} \epsilon^{ijk} \dot{S}^{ja} G^{ka}_2- \frac{1}{2} \epsilon^{ijk} \dot{S}^{jab} G^{kab}_2 \right)  + \mathcal{O}\left(c^{-4}\right)\,, 
      \end{align}
    \begin{align}
    \label{eq:spin_tr}
\dot{J}^i_1 = & \,  \mathcal{O}\left(c^{-2}\right) \,, \nonumber\\  
\dot{J}^i_2 = & \,\epsilon^{ijk} Q^{ja} G^{ka}_2 + \epsilon^{ijk} Q^{jab} G^{kab}_2 + \mathcal{O}\left(c^{-2}\right)\,.
\end{align}

As discussed in section~\ref{sec:1pn_app}, the orbital equations of motion for the worldlines $z_1(t)$, $z_2(t)$ can be obtained by replacing Eqs.~\eqref{eq:dipole_tr} in the condition ${\ddot M}^i_A=0$, which is a consequence of the gauge condition $M^i_A=0$. In order to this, the local frame tidal moments in the RHS of Eqs.~\eqref{eq:dipole_tr} have to be expressed in terms of the body frame multipole moments. To this
aim, the expressions of $G^i_1$ and $G^L_2$ with $l=1,\dots,4$ are needed up to $1$PN order, while those of $H^{ij}_1$, $H^L_2$ with $l=2,\dots,4$ are needed up to $0$PN order. These are obtained through the relations~\eqref{eq:G}--\eqref{eq:Lambdaz}, which for our truncation reduce to~\footnote{We recall that $v^i_A \equiv \dot{z}^i_A$, $a^i_A \equiv \ddot{z}^i_A$ and $\partial_{i}^{(A)} \equiv \partial / \partial z^i_A$, see the~\nameref{sec:notation_2}.}
\begin{equation*}
\begin{aligned}
G^i_1 = & \, F^i_{g,1} - \Lambda^i_{\Phi,1} +\frac{1}{c^2} \Big[ \dot{Y}^{i}_{g,1} -v^j_1 Y^{ji}_{g,1} +(2v^2_1-G_{g,1})G^i_{g,1}-(1/2)v^{j i}_1 G^{j}_{g,1} -3 v^{ i}_1 \dot{G}_{g,1} \\
& -4 a^{ i}_1 G_{g,1}- \dot{\Lambda}_{\zeta,1}^{i} \Big] + \mathcal{O} \left(c^{-4} \right) \\
G^i_2 = & \, F^i_{g,2} - \Lambda^i_{\Phi,2} +\frac{1}{c^2} \Big[ \dot{Y}^{i}_{g,2} -v^j_2 Y^{ji}_{g,2} +(2v^2_2-G_{g,2})G^i_{g,2}-(1/2)v^{j i}_2 G^{j}_{g,2} -3 v^{ i}_2 \dot{G}_{g,2} \\
& -4 a^{ i}_2 G_{g,2}- \dot{\Lambda}_{\zeta,2}^{i} \Big] + \mathcal{O} \left(c^{-4} \right) \\
G^{ij}_{2} = & F_{g,2}^{ij}- 2 \Lambda^{ij}_{\Phi,2} +\frac{1}{c^2} \Big[ \dot{Y}^{\langle ij \rangle}_{g,2} -v^k_2 Y^{kij}_{g,2} +2(v^2_2-G_{g,2})G^{ij}_{g,2}-v^{k\langle j}_2 G^{i \rangle k}_{g,2} -2 v^{\langle j}_2 \dot{G}^{i \rangle }_{g,2} \\
& -6 a^{\langle j}_2 G^{i \rangle}_{g,2}- \dot{\Lambda}_{\zeta,2}^{\langle ij \rangle} \Big] + \mathcal{O} \left(c^{-4} \right) \\
G^{ijk}_{2} = & F_{g,2}^{ijk} +\frac{1}{c^2} \Big[ \dot{Y}^{\langle ijk \rangle}_{g,2} -v^a_2 Y^{aijk}_{g,2} +(2v^2_2-3G_{g,2})G^{ijk}_{g,2}-(3/2)v^{a\langle k}_2 G^{ij \rangle a}_{g,2} - v^{\langle k}_2 \dot{G}^{ij \rangle }_{g,2} \\
& -10 a^{\langle k}_2 G^{ij \rangle}_{g,2}-2 \dot{\Lambda}_{\zeta,2}^{\langle ijk \rangle} \Big] + \mathcal{O} \left(c^{-4} \right) 
\end{aligned}
\end{equation*}
\begin{equation}
\label{eq:gmom}
\begin{aligned}
G^{ijka}_{2} = & F_{g,2}^{ijka} +\frac{1}{c^2} \Big[ \dot{Y}^{\langle ijka \rangle}_{g,2} -v^b_2 Y^{bijka}_{g,2} +(2v^2_2-4G_{g,2})G^{ijka}_{g,2}-2v^{b\langle a}_2 G^{ijk \rangle b}_{g,2} \\
& -16 a^{\langle a}_2 G^{ijk \rangle}_{g,2} \Big] + \mathcal{O} \left(c^{-4} \right) \,,
\end{aligned}
\end{equation}
\begin{equation}
\label{eq:hmom}
\begin{aligned}
H^{ij}_{1} = & \, Y_{g,1}^{ ab \langle i} \epsilon^{j \rangle ab}-4 v_1^a G_{g,1}^{ b \langle i} \epsilon^{j \rangle ab} + \mathcal{O} \left(c^{-2} \right) \\
H^{ij}_{2} = & \, Y_{g,2}^{ ab \langle i} \epsilon^{j \rangle ab} -4 v_2^a G_{g,2}^{ b \langle i} \epsilon^{j \rangle ab}+ \mathcal{O} \left(c^{-2} \right) \\
H^{ijk}_{2} = & \, Y_{g,2}^{ ab \langle ij} \epsilon^{k \rangle ab} -4 v_2^a G_{g,2}^{ b \langle ij} \epsilon^{k \rangle ab}+ \mathcal{O} \left(c^{-2} \right) \\
H^{ijkc}_{2} = & \, Y_{g,2}^{ ab \langle ijk} \epsilon^{c \rangle ab} -4 v_2^a G_{g,2}^{ b \langle ijk} \epsilon^{c \rangle ab}+ \mathcal{O} \left(c^{-2} \right) \,,
\end{aligned}
\end{equation}
where
\begin{equation}
\label{eq:gnewt}
\begin{aligned}
G_{g,1} = & \,  \frac{M_2}{|\boldsymbol{z}_1-\boldsymbol{z}_2|} + \frac{1}{2} Q^{ij} \partial^{(1)}_{ij} \frac{1}{|\boldsymbol{z}_1-\boldsymbol{z}_2|} - \frac{1}{6} Q^{ijk} \partial^{(1)}_{ijk} \frac{1}{|\boldsymbol{z}_1-\boldsymbol{z}_2|} + \mathcal{O} \left( c^{-2}\right) \\
G^i_{g,1} = & \,  M_2 \partial^{(1)}_{i} \frac{1}{|\boldsymbol{z}_1-\boldsymbol{z}_2|} + \frac{1}{2} Q^{jk} \partial^{(1)}_{ijk} \frac{1}{|\boldsymbol{z}_1-\boldsymbol{z}_2|}  - \frac{1}{6} Q^{jka} \partial^{(1)}_{ijka} \frac{1}{|\boldsymbol{z}_1-\boldsymbol{z}_2|} + \mathcal{O} \left( c^{-2}\right) \\
G^{ij}_{g,1} = & \,  M_2 \partial^{(1)}_{ij} \frac{1}{|\boldsymbol{z}_1-\boldsymbol{z}_2|} + \frac{1}{2} Q^{ka} \partial^{(1)}_{ijka} \frac{1}{|\boldsymbol{z}_1-\boldsymbol{z}_2|}  - \frac{1}{6} Q^{kab} \partial^{(1)}_{ijkab} \frac{1}{|\boldsymbol{z}_1-\boldsymbol{z}_2|}  + \mathcal{O} \left( c^{-2}\right) \\
G_{g,2} = & \,  \frac{M_1}{|\boldsymbol{z}_1-\boldsymbol{z}_2|}+ \mathcal{O} \left( c^{-2}\right) \\
G^i_{g,2} = & \,  M_1 \partial^{(2)}_{i} \frac{1}{|\boldsymbol{z}_1-\boldsymbol{z}_2|} + \mathcal{O} \left( c^{-2}\right) \\
G^{ij}_{g,2} = & \,  M_1 \partial^{(2)}_{ij} \frac{1}{|\boldsymbol{z}_1-\boldsymbol{z}_2|} + \mathcal{O} \left( c^{-2}\right) \\
G^{ijk}_{g,2} = & \,  M_1 \partial^{(2)}_{ijk} \frac{1}{|\boldsymbol{z}_1-\boldsymbol{z}_2|} + \mathcal{O} \left( c^{-2}\right) \\
G^{ijka}_{g,2} = & \,  M_1 \partial^{(2)}_{ijka} \frac{1}{|\boldsymbol{z}_1-\boldsymbol{z}_2|} + \mathcal{O} \left( c^{-2}\right) \,,
\end{aligned}
\end{equation}
\begin{equation*}
\begin{aligned}
Y^{i}_{g,1} = & \,\sum_{k=0}^{3} \frac{(-1)^k}{k!} Z^{iK}_{g,2}\, \partial^{(1)}_{K} \frac{1}{|\boldsymbol{z}_1-\boldsymbol{z}_2| } + \mathcal{O} \left(c^{-2} \right) \\
Y^{ij}_{g,1} = & \,\sum_{k=0}^{3} \frac{(-1)^k}{k!} Z^{iK}_{g,2}\, \partial^{(1)}_{jK} \frac{1}{|\boldsymbol{z}_1-\boldsymbol{z}_2| } + \mathcal{O} \left(c^{-2} \right) \\
Y^{ija}_{g,1} = & \,\sum_{k=0}^{3} \frac{(-1)^k}{k!} Z^{iK}_{g,2}\, \partial^{(1)}_{jaK} \frac{1}{|\boldsymbol{z}_1-\boldsymbol{z}_2| } + \mathcal{O} \left(c^{-2} \right) \\
Y^{i}_{g,2} = & \,\sum_{k=0}^{1} \frac{(-1)^k}{k!} Z^{iK}_{g,1}\, \partial^{(2)}_{K} \frac{1}{|\boldsymbol{z}_1-\boldsymbol{z}_2| } + \mathcal{O} \left(c^{-2} \right) 
\end{aligned}
\end{equation*}
\begin{equation}
\begin{aligned}
Y^{ij}_{g,2} = & \,\sum_{k=0}^{1} \frac{(-1)^k}{k!} Z^{iK}_{g,1}\, \partial^{(2)}_{jK} \frac{1}{|\boldsymbol{z}_1-\boldsymbol{z}_2| } + \mathcal{O} \left(c^{-2} \right) \\
Y^{ija}_{g,2} = & \,\sum_{k=0}^{1} \frac{(-1)^k}{k!} Z^{iK}_{g,1}\, \partial^{(2)}_{jaK} \frac{1}{|\boldsymbol{z}_1-\boldsymbol{z}_2| } + \mathcal{O} \left(c^{-2} \right)  \\
Y^{ijab}_{g,2} = & \,\sum_{k=0}^{1} \frac{(-1)^k}{k!} Z^{iK}_{g,1}\, \partial^{(2)}_{jabK} \frac{1}{|\boldsymbol{z}_1-\boldsymbol{z}_2| } + \mathcal{O} \left(c^{-2} \right) \\
Y^{ijabc}_{g,2} = & \,\sum_{k=0}^{1} \frac{(-1)^k}{k!} Z^{iK}_{g,1}\, \partial^{(2)}_{jabcK} \frac{1}{|\boldsymbol{z}_1-\boldsymbol{z}_2| } + \mathcal{O} \left(c^{-2} \right) 
\end{aligned}
\end{equation}
and 
\begin{equation}
\begin{aligned}
F^{i}_{g,1} = & \sum_{k=0}^{3} \frac{(-1)^k}{k!}  N^K_{g,2}\, \partial^{(1)}_{iK} \frac{1}{|\boldsymbol{z}_1-\boldsymbol{z}_2| } +  \frac{1}{2c^2} \sum_{k=0}^{5} \frac{(-1)^k}{k!} P^K_{g,2} \, \partial^{(1)}_{iK} |\boldsymbol{z}_1-\boldsymbol{z}_2|  + \mathcal{O} \left(c^{-4} \right) \\
F^{i}_{g,2} = & \sum_{k=0}^{1} \frac{(-1)^k}{k!}  N^K_{g,1}\, \partial^{(2)}_{iK} \frac{1}{|\boldsymbol{z}_1-\boldsymbol{z}_2| } +  \frac{1}{2c^2} \sum_{k=0}^{2} \frac{(-1)^k}{k!} P^K_{g,1} \, \partial^{(2)}_{iK} |\boldsymbol{z}_1-\boldsymbol{z}_2|  + \mathcal{O} \left(c^{-4} \right) \\
F^{ij}_{g,2} = & \sum_{k=0}^{1} \frac{(-1)^k}{k!}  N^K_{g,1}\, \partial^{(2)}_{ijK} \frac{1}{|\boldsymbol{z}_1-\boldsymbol{z}_2| } +  \frac{1}{2c^2} \sum_{k=0}^{2} \frac{(-1)^k}{k!} P^K_{g,1} \, \partial^{(2)}_{\langle ij \rangle K} |\boldsymbol{z}_1-\boldsymbol{z}_2|  + \mathcal{O} \left(c^{-4} \right) \\
F^{ija}_{g,2} = & \sum_{k=0}^{1} \frac{(-1)^k}{k!}  N^K_{g,1}\, \partial^{(2)}_{ijaK} \frac{1}{|\boldsymbol{z}_1-\boldsymbol{z}_2| } +  \frac{1}{2c^2} \sum_{k=0}^{2} \frac{(-1)^k}{k!} P^K_{g,1} \, \partial^{(2)}_{\langle ija \rangle K} |\boldsymbol{z}_1-\boldsymbol{z}_2|  \\
&+ \mathcal{O} \left(c^{-4} \right) \\
F^{ijab}_{g,2} = & \sum_{k=0}^{1} \frac{(-1)^k}{k!}  N^K_{g,1}\, \partial^{(2)}_{ijabK} \frac{1}{|\boldsymbol{z}_1-\boldsymbol{z}_2| } +  \frac{1}{2c^2} \sum_{k=0}^{2} \frac{(-1)^k}{k!} P^K_{g,1} \, \partial^{(2)}_{\langle ijab \rangle K} |\boldsymbol{z}_1-\boldsymbol{z}_2|  \\
&+ \mathcal{O} \left(c^{-4} \right) \,,
\end{aligned}
\end{equation}
\begin{equation*}
\begin{aligned}
N_{g,1} = & \, M_{g,1} + \frac{M_1}{3c^2} v_1^2  + \mathcal{O} \left(c^{-4} \right) \\
N^{i}_{g,1} = & \, M^i_{g,1}   + \mathcal{O} \left(c^{-4} \right) \\
N^{L}_{g,1} = & \, \mathcal{O} \left(c^{-4} \right) \qquad l \geq 2 \\
N_{g,2} = & \, M_{g,2} + \frac{M_2}{3c^2} v_2^2  + \mathcal{O} \left(c^{-4} \right) \\
N^{i}_{g,2} = & \, M^i_{g,2} + \frac{1}{5c^2} \left(2 v_2^j \dot{Q}^{ji} +a_2^j Q^{ji} \right)  + \mathcal{O} \left(c^{-4} \right) \\
N^{ij}_{g,2} = & \, M^{ij}_{g,2} + \frac{1}{7c^2} \left( v_2^2 Q^{ij} +4 v_2^{a \langle j} Q^{i \rangle a} + 2 v_2^k \dot{Q}^{ijk} + a_2^k Q^{ijk}   \right)  + \mathcal{O} \left(c^{-4} \right) 
\end{aligned}
\end{equation*}
\begin{equation}
\begin{aligned}
N^{ijk}_{g,2} = & \, M^{ijk}_{g,2}  + \frac{1}{9c^2} \left(v^2_2 Q^{ijk}+ 6 v_2^{a \langle k} Q^{ij \rangle a} \right) + \mathcal{O} \left(c^{-4} \right) \\
N^{L}_{g,2} = & \, \mathcal{O} \left(c^{-4} \right) \qquad l \geq 4 \,,
\end{aligned}
\end{equation}
\begin{equation}
\begin{aligned}
P_{g,1} = & \,   \ddot{M}_1 + \mathcal{O} \left(c^{-2} \right) \\
P^{i}_{g,1} = & \,  a_1^i M_1 + \mathcal{O} \left(c^{-2} \right) \\
P^{ij}_{g,1} = & \,  2M_1 v_1^{\langle ij \rangle} + \mathcal{O} \left(c^{-2} \right) \\
P^{L}_{g,1} = & \,  \mathcal{O} \left(c^{-2} \right) \qquad l \geq 3 \\ 
P_{g,2} = & \,  \ddot{M}_2 + \mathcal{O} \left(c^{-2} \right) \\
P^{i}_{g,2} = & \,  2 v_2^i \dot{M}_2 + a_2^i M_2 + \mathcal{O} \left(c^{-2} \right) \\
P^{ij}_{g,2} = & \,  \ddot{Q}^{ij} + 2M_2 v_2^{\langle ij \rangle} + \mathcal{O} \left(c^{-2} \right) \\
P^{ijk}_{g,2} = & \,  \ddot{Q}^{ijk} + 6 v_2^{\langle k} \dot{Q}^{ij \rangle } + 3 a_2^{\langle k } Q^{ij \rangle}+ \mathcal{O} \left(c^{-2} \right) \\
P^{ijka}_{g,2} = & \,  8 v_2^{\langle a} \dot{Q}^{ijk \rangle} + 4 a_2^{\langle a} Q^{ijk \rangle}  + 12 v_2^{\langle ak } Q^{ij\rangle} + \mathcal{O} \left(c^{-2} \right) \\ 
P^{ijkab}_{g,2} = & \, 20 v_2^{\langle ab} Q^{ijk \rangle} + \mathcal{O} \left(c^{-2} \right) \\
P^{L}_{g,2} = & \,  \mathcal{O} \left(c^{-2} \right)  \qquad l \geq 6 \,,
\end{aligned}
\end{equation}
\begin{equation}
\begin{aligned}
\Lambda^i_{\Phi,1} = & a^i_1 + \frac{1}{c^2} \bigg[(v^2_1+G_{g,1})a^i_1+\frac{1}{2} v^{ij}_1 a_1^j
  +2 \dot{G}_{g,1} v^i_1 \bigg]  + \mathcal{O} \left(c^{-4} \right) \\
\Lambda^i_{\Phi,2} = & a^i_2 + \frac{1}{c^2} \bigg[(v^2_2+G_{g,2})a^i_2+\frac{1}{2} v^{ij}_2 a_2^j
  +2 \dot{G}_{g,2} v^i_2 \bigg]  + \mathcal{O} \left(c^{-4} \right) \\
\Lambda^{ij}_{\Phi,2} = &  \frac{1}{c^2} \left(-\frac{1}{2} a_2^{\langle ij \rangle} + v_2^{\langle i }
\dot{a}_2^{j \rangle} \right)+ \mathcal{O} \left(c^{-4} \right) \\
\Lambda^i_{\zeta,1} = & -2 G_{g,1} v^i_1  + \mathcal{O} \left(c^{-2} \right) \\
\Lambda^i_{\zeta,2} = & -2 G_{g,2} v^i_2  + \mathcal{O} \left(c^{-2} \right) \\
\Lambda^{ij}_{\zeta,2} = & -\frac{3}{2} v^{[i}_2 a^{j]}_2 -2 v^{\langle i}_2
a_2^{j \rangle}-\frac{4}{3} \dot{G}_{g,2} \delta^{ij} + \mathcal{O} \left(c^{-2} \right) \\
\Lambda^{ijk}_{\zeta,2} = & -\frac{6}{5} \delta^{i \langle j}
\dot{a}_2^{k \rangle} + \mathcal{O} \left(c^{-2} \right) \,,
\end{aligned}
\end{equation}
with
\begin{equation*}
\begin{aligned}
M_{g,1} = & M_1 +\frac{M_1}{c^2} \left(\frac{3}{2} v_1^2 -G_{g,1}   \right) + \mathcal{O} \left(c^{-4} \right) \\
M^i_{g,1} = & \frac{1}{c^2} \left(2 \epsilon^{ijk} v_1^j J_1^k \right) + \mathcal{O} \left(c^{-4} \right) \\
M^L_{g,1} = & \, \mathcal{O} \left(c^{-4} \right) \qquad l \geq 2  \\
M_{g,2} = & M_2 +\frac{M_2}{c^2} \left(\frac{3}{2} v_2^2 -G_{g,2}   \right) + \mathcal{O} \left(c^{-4} \right) 
\end{aligned}
\end{equation*}
\begin{equation}
\begin{aligned}
M^i_{g,2} = & \frac{1}{c^2} \left(-\frac{1}{5} v_2^j \dot{Q}^{ij} - \frac{16}{5} a_2^j Q^{ij} +2 \epsilon^{ijk} v_2^j J_2^k \right)+ \mathcal{O} \left(c^{-4} \right) \\
M^{ij}_{g,2} = & Q^{ij} + \frac{1}{c^2} \left[ 3\left( \frac{1}{2} v_2^2 -G_{g,2} \right) Q^{ij} -\frac{17}{5}v_2^{k \langle i}Q^{j \rangle k} +\frac{8}{3} v_2^k \epsilon^{ka\langle j} S^{i \rangle a} -\frac{13}{21} v_2^k \dot{Q}^{ijk} \right. \\
&\left. - \frac{83}{21} a_2^k Q^{ijk} \right] + \mathcal{O} \left(c^{-4} \right) \\
M^{ijk}_{g,2} = & Q^{ijk} + \frac{1}{c^2} \left[ \left(\frac{3}{2} v^2_2 - 4 G_{g,2} \right) Q^{ijk} - \frac{61}{14} v_2^{a \langle k} Q^{ij \rangle a} + 3 v_2^a \epsilon^{ab \langle k} S^{ij \rangle b} \right]  + \mathcal{O} \left(c^{-4} \right)  \\
M^L_{g,2} = & \, \mathcal{O} \left(c^{-4} \right) \qquad l \geq 4 \,,
\end{aligned}
\end{equation}
\begin{equation}
\begin{aligned}
Z^i_{g,1} = & \, 4M_1v_1^i + \mathcal{O} \left(c^{-2} \right) \\
Z^{ij}_{g,1} = & \, -2 \epsilon^{ijk} J_1^k + \mathcal{O} \left(c^{-2} \right) \\
Z^{iL}_{g,1} = & \, \mathcal{O} \left(c^{-2} \right) \qquad l \geq 2 \\
Z^i_{g,2} =  & \, 4M_2v_2^i + \mathcal{O} \left(c^{-2} \right) \\
Z^{ij}_{g,2} = & \, 2 \dot{Q}^{ij}-2 \epsilon^{ijk} J_2^k + \mathcal{O} \left(c^{-2} \right) \\
Z^{ijk}_{g,2} = & \, 4v_2^i Q^{jk} -\frac{12}{5} v_2^a Q^{a \langle j} \delta^{k \rangle i} -\frac{8}{3} \epsilon^{a i \langle k} S^{j \rangle a} + \frac{4}{3} \dot{Q}^{ijk} + \mathcal{O} \left(c^{-2} \right) \\
Z^{ijka}_{g,2} = & \, 4 v_2^i Q^{jka} -\frac{20}{7} v_2^b Q^{b \langle jk } \delta^{a \rangle i } -3 \epsilon^{bi \langle a} S^{jk \rangle b} + \mathcal{O} \left(c^{-2} \right) \\
Z^{iL}_{g,2} = & \, \mathcal{O} \left(c^{-2} \right) \qquad l \geq 4 \,.
\end{aligned}
\end{equation}

Using the above relations, and replacing the time derivatives on masses and spins by the evolution equations~\eqref{eq:monopole_tr} and~\eqref{eq:spin_tr}, we find the orbital equations of motion in the form
\begin{equation}
\label{eq:orbeom12}
\begin{aligned}
  M_1  a^i_1 = & F^i_{1,M} +  F^i_{1,J} + F^i_{1,Q2}+ F^i_{1,Q3} + F^i_{1,S2}  +  F^i_{1,S3}  \,,\\
  M_2  a^i_2 = & F^i_{2,M} +  F^i_{2,J} + F^i_{2,Q2}+ F^i_{2,Q3} +  F^i_{2,S2}  +  F^i_{2,S3}  \,,
\end{aligned}
\end{equation}
where we have separated different contributions coming from mass, spin, mass quadrupole, mass octupole, current quadrupole and current octupole, respectively. Defining
\begin{equation}
\begin{gathered}
z^i=z_2^i-z_1^i \qquad v^i=v_2^i-v_1^i \\
r =  |\boldsymbol z| \qquad n^i = \frac{z^i}{r}
\end{gathered}
\end{equation}
and using the relations
\begin{equation}
\begin{aligned}
n^{ii} =& 1 \\
\partial_L^{(2)} \frac{1}{r} = & (-1)^l \partial_L^{(1)} \frac{1}{r} = (-1)^l (2l-1)!! \frac{n^{ \langle L \rangle}}{r^{l+1}} \\
\partial_L^{(A)} r = & -\frac{r^2}{2l-1} \partial_{L}^{(A)} \frac{1}{r} + \frac{l(l-1)}{2l-1} \delta_{( a_l a_{l-1}} \partial_{L-2 )}^{(A)} \frac{1}{r} \,,
\end{aligned}
\end{equation}
the mass monopole contributions read
\begin{equation}
\begin{aligned}
  F^i_{1,M} = & \frac{M_1 M_2}{r^2} n^i + \frac{1}{c^2}
  \frac{M_1 M_2}{r^2} \left\{n^i \left[ 2 v^2 -v_1^2
    -\frac{3}{2} \left(n^a v^a_2\right)^2-\frac{5 M_1}{r}
    -\frac{4 M_2}{r} \right] \right.\\
    & + v^i n^a \left(4 v^a_1-3 v^a_2\right) \bigg\} +\mathcal{O}\left(c^{-4}\right)\,,\\
  F^i_{2,M} = & - \frac{M_1 M_2}{r^2} n^i - \frac{1}{c^2}
  \frac{M_1 M_2}{r^2} \left\{n^i \left[ 2 v^2 -v_2^2
    -\frac{3}{2} \left(n^a v^a_1\right)^2-\frac{4 M_1}{r} -\frac{5 M_2}{r} \right] \right. \\
&   - v^i n^a \left(4 v^a_2-3 v^a_1\right) \bigg\}+\mathcal{O}\left(c^{-4}\right)\,,
\end{aligned}
\end{equation}
while the spin contributions are
\begin{equation}
\begin{aligned}
  F^i_{1,J} = & \frac{1}{c^2} \frac{M_1}{r^3} \epsilon^{abc} J_2^c \left[\delta^{ai}
    \left(4 v^b-6n^{bd} v^d \right) -6 n^{ai} v^{b}\right]
  -\frac{1}{c^2} \frac{M_2}{r^3} \epsilon^{abc} J_1^c \left[3\delta^{ai} \left(n^{bd} v^d -v^b\right) \right. \\
  & \left. +6 n^{ai} v^{b}\right] +\mathcal{O}\left(c^{-4}\right) \,, \\
  F^i_{2,J} = & \frac{1}{c^2}\frac{M_1}{r^3} \epsilon^{abc} J_2^c
  \left[3\delta^{ai} \left(n^{bd} v^d -v^b\right) +6 n^{ai} v^{b}\right]
  -\frac{1}{c^2} \frac{M_2}{r^3} \epsilon^{abc} J_1^c \left[\delta^{ai} \left(4 v^b-6n^{bd} v^d \right)\right. \\
  & \left.  -6 n^{ai} v^{b}\right]+\mathcal{O}\left(c^{-4}\right)\,.
\end{aligned}
\end{equation}
Recalling that the local frame body multipole moments are STF tensors~\footnote{We remind that, if $T^L$ is a STF tensor, then $T^{ii L-2} = 0$, etc., and the contraction with a generic tensor $U^L$ satisfies $T^L U^L = T^L U^ {\langle L \rangle}$. See the~\nameref{sec:notation_2} for the properties of STF tensors.}, the mass quadrupole contributions are
\begin{equation}
\begin{aligned}
  F^i_{1,Q2} &=  \frac{3 M_1}{2 r^4} Q^{ab} \left(5n^{abi}-2n^a \delta^{bi} \right)
  + \frac{1}{c^2} \bigg( \frac{3 M_1}{2 r^4} Q^{ab} \bigg\{ 5 n^{abi} \left[2 v^2-v_1^2 -
    \frac{7}{2} \left( n^c v_2^c\right)^2 \right. \\
    &\left.-\frac{47 M_1}{5 r} -\frac{24 M_2}{5 r} \right]
  -2 n^a \delta^{bi} \left[2 v^2-v_1^2 -\frac{5}{2} \left(n^c v_2^c \right)^2 -\frac{19 M_1}{2 r}
    -\frac{4 M_2}{r} \right]   \\
  &+ n^a v_2^{bi} + \left(5 n^{ai}- \delta^{ai} \right) v_2^{bc} n^c 
   + v^i \left(5 n^{abc}-2n^a \delta^{bc} \right) \left(4 v_1^c -3 v_2^c \right) \bigg\} \\
   & + \frac{3 M_1}{2 r^3} \dot{Q}^{ab} \left[ n^{ab} \left(5 v_2^c\, n^{ci} + 3v^i \right) -4 v^a n^{bi}  - 2 \delta^{ai} n^{bc} \left(2v_1^c-v_2^c \right) \right] \\
   & -\frac{3 M_1}{4 r^2} \ddot{Q}^{ab}\left( n^{abi}+2 n^a \delta^{bi} \right) \bigg)  - \frac{3}{c^2} \epsilon^{icd} J_1^c \bigg\{ \frac{Q^{ab}}{r^5} \bigg[ \frac{5}{2} n^{ab} \left(7 n^{de} v^e- v^d\right) \\
   &  +\left(\delta^{ad} -5 n^{ad}\right) v^b -5 \delta^{ad} n^{be} v^e \bigg] + \frac{\dot{Q}^{ab}}{r^4}
  \left(\delta^{ad}-\frac{5}{2}n^{ad}\right) n^b \bigg\} \\
  & - \frac{3 \epsilon^{cde} J_1^a}{c^2} \bigg\{ \frac{Q^{bd}}{r^5} \bigg[ 5 n^{ab}
    \left( \delta^{ic} v^e - \delta^{ie} v^c \right) +
    5 n^{ac} \left( \delta^{ib} v^e - \delta^{ie} v^b\right) \\
    & + 5 n^{bc} \left( \delta^{ia} v^e - \delta^{ie} v^a \right)  + 35 n^{abc} \left( \delta^{ie} n^f v^f -n^i \right) + \delta^{ab} \left(\delta^{ie} v^c -\delta^{ic}  \right) \\
   &  + \delta^{ac} \left(\delta^{ie} v^b -\delta^{ib}  \right)  + 5 \left(\delta^{ab} n^c + \delta^{ac} n^b \right) \left(n^i -\delta^{ie} n^f v^f \right) \bigg] \\
   &- \frac{ \dot{Q}^{bd}}{r^4} \delta^{ie}
  \left(5 n^{abc}-\delta^{ab} n^c -\delta^{ac} n^b \right) \bigg\} +\mathcal{O}\left(c^{-4}\right)\,, \nonumber
\end{aligned}
\end{equation}
\begin{equation}
\begin{aligned}
  F^i_{2,Q2} = &  - \frac{3 M_1}{2 r^4} Q^{ab} \left(5n^{abi}-2n^a \delta^{bi} \right)
  + c^{-2} \bigg( \frac{3 M_1}{2 r^4} Q^{ab} \bigg\{ -5 n^{abi} \left[2 v^2-v_2^2 \right. \\
    & \left. - \frac{7}{2} \left( n^c v_1^c\right)^2 -\frac{8 M_1}{r} -\frac{6 M_2}{r} \right] +2 n^a \delta^{bi} \left[3 v^2-v_2^2 -5 \left(n^c v^c \right)^2  -\frac{5}{2} \left(n^c v_1^c \right)^2 \right. \\
    & \left. -\frac{8 M_1}{r} -\frac{11 M_2}{2r} \right] + n^i v^{ab} + 5 n^{aci}\left(2v^b v_1^c-  v_2^{bc}\right) \\
    & + v^i \left(5 n^{abc}-2n^a \delta^{bc} \right) \left(4 v_2^c -3 v_1^c \right) + n^a v_2^b \left(v_2^i-2 v_1^i\right)  \\
    & + \delta^{bi} n^c\left[\left(5 v_2^a-4v_1^a\right)v_2^c -6 v^a v_1^c\right]\bigg\} + \frac{3 M_1}{r^3} \dot{Q}^{ab} \left[ v^b \left(2 n^{ai}-\delta^{ai}\right) \right. \\
    &\left. +\delta^{ai} n^{bc} v^c  - 2 n^{ab} v^i \right] \bigg)  + \frac{3 M_1}{c^2M_2} \epsilon^{icd} J_2^c \bigg\{ \frac{Q^{ab}}{r^5} \bigg[ \frac{5}{2} n^{ab} \left(7 n^{de} v^e - v^d\right) \\
  & +\left(\delta^{ad} -5 n^{ad}\right) v^b -5 \delta^{ad} n^{be} v^e \bigg] + \frac{\dot{Q}^{ab}}{r^4} \left(\delta^{ad}-
  \frac{5}{2}n^{ad}\right) n^b \bigg\} \\
  & + \frac{3 \epsilon^{cde} J_1^a}{c^2} \bigg\{ \frac{Q^{bd}}{r^5}
  \bigg[ 5 n^{ab} \left( \delta^{ic} v^e - \delta^{ie} v^c \right) + 5 n^{ac} \left( \delta^{ib} v^e - \delta^{ie} v^b\right)\\
  &  + 5 n^{bc} \left( \delta^{ia} v^e- \delta^{ie} v^a \right) + 35 n^{abc} \left( \delta^{ie} n^f v^f -n^i \right) + \delta^{ab} \left(\delta^{ie} v^c -\delta^{ic}  \right) \\
    & + \delta^{ac} \left(\delta^{ie} v^b -\delta^{ib}  \right) + 5 \left(\delta^{ab} n^c + \delta^{ac} n^b \right) \left(n^i -\delta^{ie} n^f v^f \right) \bigg] \\
 &  - \frac{ \dot{Q}^{bd}}{r^4} \delta^{ie} \left(5 n^{abc}-\delta^{ab} n^c -\delta^{ac} n^b \right) \bigg\}+\mathcal{O}\left(c^{-4}\right)\,,
\end{aligned}
\end{equation}
the mass octupole contributions read
\begin{equation}
\label{eq:octucontr}
\begin{aligned}
F^i_{1,Q3} = & - \frac{5 M_1}{2r^5} Q^{abc} \left( 7 n^{iabc} - 3 \delta^{ic} n^{ab} \right) +\mathcal{O}(c^{-2})\,,\\
F^i_{2,Q3} = &  \frac{5 M_1}{2r^5} Q^{abc} \left( 7 n^{iabc} - 3 \delta^{ic} n^{ab} \right)+\mathcal{O}(c^{-2})\,,
\end{aligned}
\end{equation}
the current quadrupole contributions are
\begin{equation*}
\begin{aligned}
  F^i_{1,S2} = & - \frac{4 M_1 \epsilon^{bcd}}{c^2 }
  \bigg\{ \frac{S^{ad}}{r^4} \left[ n^a \left(\delta^{ib} v^c -\delta^{ic} v^b \right) +
n^b \left(\delta^{ia} v^c -\delta^{ic} v^a \right) \right. \\
& \left.  + 5 n^{ab} \left(\delta^{ic} n^e v^e -n^i v^c \right) \right]  - \frac{ \dot{S}^{ad}}{r^3} \delta^{ic} n^{ab}  \bigg\}  - \frac{ J_1^c}{c^2} \frac{S^{ab} }{r^5} \left[ 4 \delta^{bc} \left( 5 n^{ia} - \delta^{ia} \right) \right. \\
    & \left. + 10 \left( \delta^{ia} n^{bc} + \delta^{ib} n^{ac} + 
\delta^{ic} n^{ab}   \right)- 70 n^{iabc} \right]+\mathcal{O}\left(c^{-4}\right)\,,
\end{aligned}
\end{equation*}
\begin{equation}
\begin{aligned}
  F^i_{2,S2} = & \frac{4 M_1 \epsilon^{bcd}}{c^2 } \bigg\{ \frac{S^{ad}}{r^4}
  \left[ n^a \left(\delta^{ib} v^c -\delta^{ic} v^b \right) +
n^b \left(\delta^{ia} v^c -\delta^{ic} v^a \right) \right. \\
& \left. + 5 n^{ab} \left(\delta^{ic} n^e v^e -n^i v^c \right) \right]- \frac{ \dot{S}^{ad}}{r^3} \delta^{ic} n^{ab}  \bigg\}  + \frac{ J_1^c}{c^2} \frac{S^{ab}}{r^5} \left[ 4 \delta^{bc}
    \left( 5 n^{ia} - \delta^{ia} \right) \right. \\
    & \left. + 10 \left( \delta^{ia} n^{bc} + \delta^{ib} n^{ac} + \delta^{ic} n^{ab}   \right)- 70 n^{iabc} \right]+\mathcal{O}\left(c^{-4}\right)\,,
\end{aligned}
\end{equation}
and the current octupole contributions read
\begin{equation}
\begin{aligned}
  F^i_{1,S3} = & - \frac{15 M_1}{c^2} \bigg\{ \frac{S^{bde}}{r^5}  n^{ab}  v^c
  \bigg[\frac{7}{2}\left(\epsilon^{iae} n^{c}-\epsilon^{cae} n^{i}\right)n^d
    -\left( \epsilon^{iad}\delta^{ce} + \epsilon^{icd}\delta^{ae} +
    \epsilon^{acd}\delta^{ie} \right)\bigg] \\
    &- \frac{\dot{S}^{bde}}{2r^4} \epsilon^{iad} n^{abe} \bigg\}  - \frac{45 J_1^c}{4 c^2} \frac{S^{abc}}{r^6} \bigg[ \delta^{cd} \left(\delta^{ia} n^b + \delta^{ib} n^a -7 n^{iab} \right) \\
   & - \frac{7}{3} \left( \delta^{ia} n^{bcd} +\delta^{ib} n^{acd} +
    \delta^{ic} n^{abd} + \delta^{id} n^{abc} -9 n^{iabcd} \right) \bigg] +\mathcal{O}\left(c^{-4}\right)\,,\\
  F^i_{2,S3} = &  \frac{15 M_1}{c^2} \bigg\{ \frac{S^{bde}}{r^5}  n^{ab}  v^c
  \bigg[\frac{7}{2}\left(\epsilon^{iae} n^{c}-\epsilon^{cae} n^{i}\right)n^d
    -\left( \epsilon^{iad}\delta^{ce} + \epsilon^{icd}\delta^{ae} +
    \epsilon^{acd}\delta^{ie} \right)\bigg] \\
    &- \frac{\dot{S}^{bde}}{2r^4} \epsilon^{iad} n^{abe} \bigg\}  + \frac{45 J_1^c}{4 c^2} \frac{S^{abc}}{r^6} \bigg[ \delta^{cd}\left(\delta^{ia} n^b + \delta^{ib} n^a -7 n^{iab} \right)\\
   & - \frac{7}{3} \left( \delta^{ia} n^{bcd} +\delta^{ib} n^{acd} +
    \delta^{ic} n^{abd} + \delta^{id} n^{abc} -9 n^{iabcd} \right) \bigg]+\mathcal{O}\left(c^{-4}\right)\,.
\end{aligned}
\end{equation}
Deriving the above equations, we have used the Newtonian expression of any quantity explicitly multiplied by a factor $1/c^2$. In particular, we have made use of the Newtonian orbital equations of motion
\begin{equation}
\label{eq:accel}
\begin{aligned}
a_1^i = & \frac{M_2}{r^2} n^i+  \frac{3 }{2 r^4} Q^{ab} \left(5n^{abi}-2n^a \delta^{bi} \right)- \frac{5 }{2r^5} Q^{abc} \left( 7 n^{iabc} - 3 \delta^{ic} n^{ab} \right) + \mathcal{O}\left( c^{-2} \right)\\
a_2^i = & -\frac{M_1}{r^2} n^i-  \frac{3M_1 }{2 M_2r^4} Q^{ab} \left(5n^{abi}-2n^a \delta^{bi} \right)+ \frac{5 M_1}{2M_2r^5} Q^{abc} \left( 7 n^{iabc} - 3 \delta^{ic} n^{ab} \right) \\
&+ \mathcal{O}\left( c^{-2} \right) \,,
\end{aligned}
\end{equation}
which follow from Eqs.~\eqref{eq:accel_newt} in our approximation. We remind that we are working to linear order in the spins and in the $l \geq 2$ multipole moments. Therefore, we have neglected terms of order $\mathcal{O} \left(S^{i}_1 S^{j}_2 \right)$, $\ \mathcal{O} \left(Q^{ab} Q^{ij} \right)$, $ \ \mathcal{O} \left(Q^{abc} Q^{ijk} \right)$ and $\ \mathcal{O} \left(Q^{ab} Q^{ijk} \right)$.

We stress also that we have included the contribution of the mass octupole moment only at Newtonian order, neglecting the 1PN order terms $\sim \mathcal{O} \left(Q^{ijk}/c^2 \right)$ in Eqs.~\eqref{eq:octucontr}. This may seem inconsistent within the 1PN approximation. However, in the following we show that this is enough to derive the leading-order contributions to the gravitational waveform phase due to the mass octupole, which enter at 6.5PN order through the rotational tidal Love number $\lambda_{32}$. Next-to-leading order corrections to Eqs.~\eqref{eq:octucontr} would affect only the very subleading terms depending on the standard Love number $\lambda_3$ (8PN order and beyond).

As a consistency check of Eqs.~\eqref{eq:orbeom12}, we have computed the mass dipole of the
system $M_{sys}^i$, by applying Eq.~\eqref{sysM} to our truncation. Taking the second time derivative of the system mass dipole, and replacing the orbital equations of motion~\eqref{eq:orbeom12}, we found that ${\ddot  M}^i_{sys}= \mathcal{O}\left(c^{-4} \right)$, as expected from Eq.~\eqref{eq:sysdipolelaw}.

Next, we can obtain the equations of motion in the system COM frame by subtracting those for the individual accelerations, $\ddot{z}^i \equiv a^i = a^i_2-a^i_1$. Before going on with the computation, following Vines and Flanagan~\cite{Vines:2010ca}, we adopt a useful partitioning of the mass $M_2$ of the body $2$, which is the one with $l \geq 2$ multipole moments, that is tidally deformed. Indeed, while the mass $M_1$ of the body $1$ is conserved to 1PN order, the mass $M_2$ changes in time due to tidal interactions (see Eqs.~\eqref{eq:monopole_tr}). Therefore, we can divide the latter in two contributions: a conserved Newtonian mass and a time dependent part.

In order to do this, we notice that in Newtonian gravity the equations of motion for a generic $N$-body system can be derived from an action principle, through the Lagrangian function
\begin{equation}
\label{eq:lagrnewt}
  \mathcal{L}=\sum_{A=1}^N\left(\frac{1}{2}M_A{\dot z}_A^2 +\frac{1}{2} M_A G_{g,A} +\frac{1}{2}\sum_{l=2}^\infty\frac{1}{l!}Q_A^LG_{g,A}^L+\mathcal{L}_A^{\mathrm{int}}
  \right) + \mathcal{O} \left(c^{-2} \right) \,,
\end{equation}
where $\mathcal{L}_A^{\mathrm{int}}$ is the piece of the Lagrangian which describes the internal dynamics of the bodies. $\mathcal{L}_A^{\mathrm{int}}$ is independent of the worldlines $z^i_A$ by construction. It depends on a set of unspecified internal variables $q_A^{\alpha}$, as well as their time derivatives, $\mathcal{L}_A^{\mathrm{int}}(q_A^{\alpha},\dot{q}_A^{\alpha})$. These internal variables represent, for instance, the mass density, the matter velocity fields, etc. The mass multipole moments with $l \geq 2$ depend on the internal variables too, $Q^L_A(q^{\alpha}_A)$. Applying the Euler-Lagrange equations~\eqref{eq:euler} for the variables $z_A^i$ to the above Lagrangian, we get the Newtonian equations of motion~\eqref{eq:accel_newt}. Note that the tidal multipole moments depend on the worldlines $z_A^i$ (see Eq.~\eqref{eq:Gg}), while the mass multipole moments do not by construction. On the other hand, applying the action principle with respect to the internal variables, we get
\begin{equation}
\label{eq:eqint}
\frac{\partial \mathcal{L}_A^{\mathrm{int}}}{\partial q^{\alpha}_A}-\frac{d}{dt} \frac{\partial \mathcal{L}_A^{\mathrm{int}}}{\partial \dot{q}^{\alpha}_A} +  \sum_{l=2}^\infty\frac{1}{l!} G^L_{g,A}\frac{\partial Q^L_A}{\partial q^{\alpha}_A} = \mathcal{O}\left(c^{-2} \right) \,.
\end{equation}
We stress that the tidal moments $G^L_{g,A}$ of the body $A$ depend on the mass multipole moments $M_B, \ Q^L_B$ of the other bodies $B \neq A$, therefore they depend implicitly on the internal variables $q^{\alpha}_B$.

The conserved total energy of the system is given by
\begin{equation}
\begin{aligned}
E =&  \sum_{A=1}^{N} \left( \frac{\partial \mathcal{L}}{\partial \dot{z}^i_A} \dot{z}^i_A + \frac{\partial \mathcal{L}}{\partial \dot{q}^{\alpha}_A} \dot{q}^{\alpha}_A  \right)- \mathcal{L} +\mathcal{O}\left(c^{-2} \right) \\
  =&\sum_{A=1}^N\left(\frac{1}{2}M_A{\dot z}_A^2 -\frac{1}{2} M_A G_{g,A} -\frac{1}{2}\sum_{l=2}^\infty\frac{1}{l!}Q_A^LG_{g,A}^L+E_A^{\mathrm{int}}
  \right) + \mathcal{O} \left(c^{-2} \right) \,,
\end{aligned}
\end{equation}
where we have defined the internal energy of the $A$-th body as~\footnote{We notice that such a definition is not unique, because a residual dependence on the internal variables resides in the mass multipole moments $Q^L_A(q^{\alpha}_A)$. In other words, the mass multipole moments are the link between the internal dynamics of the bodies and the external orbital motion. One has the freedom to include the gravitational energy due the tidal interactions (i.e., the terms proportional to the $Q^L_A$) either to the gravitational energy of the orbit (as we do), or to the internal energy of the body. The latter convention is used, e.g., in~\cite{Alvi:2001mx,Poisson:2004cw}.}
\begin{equation}
E_A^{\mathrm{int}} = \frac{\partial \mathcal{L}_A^{\mathrm{int}}}{\partial \dot{q}^{\alpha}_A} \dot{q}^{\alpha}_A - \mathcal{L}_A^{\mathrm{int}} \,.
\end{equation}
Remarkably, without any assumption on the dependence of $\mathcal{L}_A^{\mathrm{int}}$ on the variables $q_A^\alpha$, it is possible to derive an equation for the internal energy. Taking its time derivative, and replacing the equations of motion~\eqref{eq:eqint}, we obtain
\begin{equation}
\label{eq:tidalheating}
\begin{aligned}
\dot{E}_A^{\mathrm{int}} = & -\dot{q}^{\alpha}_A \left(\frac{\partial \mathcal{L}_A^{\mathrm{int}}}{\partial q^{\alpha}_A}-\frac{d}{dt} \frac{\partial \mathcal{L}_A^{\mathrm{int}}}{\partial \dot{q}^{\alpha}_A} \right) \\
& =  \sum_{l=2}^\infty\frac{1}{l!} G^L_{g,A}\frac{\partial Q^L_A}{\partial q^{\alpha}_A} \dot{q}^{\alpha}_A \\
& =\sum_{l=2}^\infty\frac{1}{l!} G^L_{g,A} \dot{Q}^L_A + \mathcal{O} \left(c^{-2} \right) \,.
\end{aligned}
\end{equation}
The above equation expresses the work done on the bodies by the tidal forces, the so-called tidal heating~\footnote{The result in Eq.~\eqref{eq:tidalheating} holds even if the internal Lagrangian depends on higher-order time derivatives, $\mathcal{L}_A^{\mathrm{int}}(q_A^{\alpha},\dot{q}_A^{\alpha},\ddot{q}_A^{\alpha},\dots)$. The only requirement is that the mass multipole moments do not depend on the time derivatives of the internal variables ($Q^L_A=Q^L_A(q^{\alpha}_A)$ only).}.

With our truncation, the Lagrangian which gives the Newtonian equations~\eqref{eq:accel} reads
\begin{equation}
\label{lagrnewt}
  \mathcal{L}=\frac{1}{2}M_1{\dot z}_1^2+\frac{1}{2}M_2{\dot z}_2^2+\frac{M_1M_2}{r}-U_{Q2}- U_{Q3}+\mathcal{L}_2^{\mathrm{int}}+\mathcal{O}\left( c^{-2} \right)\,,
\end{equation}
where we have defined (see Eqs.~\eqref{eq:gnewt})
\begin{align}
  \label{newtonU}
  U_{Q2}=&-\frac{1}{2}Q^{ij}G^{ij}_{g,2}=-\frac{3M_1}{2r^3}n^{  ij  }Q^{ij} \,, \nonumber \\
  U_{Q3}=&-\frac{1}{6}Q^{ijk}G^{ijk}_{g,2}=\frac{5M_1}{2r^4}n^{  ijk }Q^{ijk}  \,.
\end{align}
$ U_{Q2}$ and $ U_{Q3}$ are the quadrupolar and octupolar Newtonian gravitational potential energy, respectively. Since the body $1$ is a pure mass monopole by construction, we can ignore its internal structure without loss of generality. For this reason, there is no $\mathcal{L}_1^{\mathrm{int}}$ in Eq.~\eqref{lagrnewt}. On the other hand, $\mathcal{L}_2^{\mathrm{int}}$ depends on the internal variables $q_2^{\alpha}$ (and their time derivatives). The internal energy of the body $2$ is
\begin{equation}
\label{defEint0}
E_2^{\mathrm{int}}={\dot q}_2^\alpha\frac{\partial\mathcal{L}_2^{\mathrm{int}}}{\partial{\dot q}_2^\alpha}-\mathcal{L}_2^{\mathrm{int}}\,.
\end{equation}
Replacing the Euler-Lagrange equations for the Lagrangian~\eqref{lagrnewt},
\begin{equation}
  \frac{\partial\mathcal{L}_2^{\mathrm{int}}}{\partial{q}_2^\alpha}-\frac{d}{dt}\frac{\partial\mathcal{L}_2^{\mathrm{int}}}{\partial{\dot q}_2^\alpha}+\frac{1}{2}G^{ij}_{g,2}\frac{\partial Q^{ij}}{\partial q_2^\alpha}+\frac{1}{6}G^{ijk}_{g,2}\frac{\partial Q^{ijk}}{\partial q_2^\alpha} =\mathcal{O}\left( c^{-2} \right)\,,
 \end{equation}
into the time derivative of Eq.~\eqref{defEint0}, yields
\begin{equation}
\label{eq:tidheat}
  {\dot E}_2^{\mathrm{int}}=\frac{1}{2}G^{ij}_{g,2}{\dot Q}^{ij}+\frac{1}{6}G^{ijk}_{g,2}{\dot Q}^{ijk}+\mathcal{O}\left( c^{-2} \right)\,.
\end{equation}

The evolution equation for the mass of the body $2$ is (Eq.~\eqref{eq:monopole_tr})
\begin{equation}
\dot{M}_2  =  -\frac{1}{c^2} \left(\frac{3}{2} Q^{ij} \dot{G}^{ij}_{g,2} +  \dot{Q}^{ij} G^{ij}_{g,2}  + \frac{2}{3} Q^{ijk} \dot{G}_{g,2}^{ijk} + \frac{1}{2}  \dot{Q}^{ijk} G_{g,2}^{ijk} \right)  + \mathcal{O}\left(c^{-4}\right) \,,
\end{equation}
and can be rewritten in terms of $U_{Q2}$, $U_{Q3}$ and $\dot{E}_2^{\mathrm{int}}$ as
\begin{equation}
\label{dotpartitioning}
  {\dot M}_2=\frac{1}{c^2}\left({\dot E}_2^{\mathrm{int}}+3{\dot U}_{Q2}+ 4{\dot U}_{Q3}\right)+ \mathcal{O}\left( c^{-4} \right)\,.
\end{equation}
The above equation provides a way to partition the mass $M_2$. Integrating it over time, one gets
\begin{equation}
\label{partitioning}
M_2=\null^nM_2+\frac{1}{c^2}\left({ E}_2^{\mathrm{int}}+3{ U}_{Q2}+4U_{Q3}\right)+\mathcal{O}\left( c^{-4} \right) \,, 
\end{equation}
where $\null^nM_2$ is the conserved Newtonian mass of body $2$. As we discuss below, this partitioning of $M_2$ is also useful to find an action principle for the system at 1PN order.

Now we can write the equations of motion in the system COM frame. We define the (Newtonian) total mass $M$, the mass ratios $\eta_1$, $\eta_2$, the symmetric mass ratio $\nu$ and the reduced mass $\mu$ as
\begin{equation}
\begin{gathered}
M = M_1 + \null^nM_2 \\
\eta_1 = \frac{M_1}{M} \qquad \eta_2 = \frac{\null^nM_2}{M} \\
\nu = \frac{M_1 \, \null^nM_2}{M^2} = \eta_1 \eta_2 \qquad \mu =  \frac{M_1\,  \null^nM_2}{M}= \eta_1 \eta_2 M = \nu M \,.
\end{gathered}
\end{equation}
With our truncation, the mass dipole of the system reads (see Eq.~\eqref{sysM})
\begin{equation}
\begin{aligned}
M_{sys}^i = & M \left(\eta_1 z_1^i + \eta_2 z_2^i \right)+ \frac{1}{c^2} \left[ \left( \frac{\eta_1 M v_1^2}{2} - \frac{\mu M}{2r} + \frac{U_{Q2}}{2}  \right)z_1^i \right. \\
&+ \left( \frac{\eta_2 M v_2^2}{2} - \frac{\mu M}{2r} + \frac{U_{Q2}}{2}  + E_2^{\mathrm{int}}\right) z_2^i  + \frac{3\eta_1 M}{2r^2} Q^{ij} n^{j} \\
&+ \epsilon^{ijk} \left(v_1^j J_1^k +v_2^j J_2^k \right) \bigg] + \mathcal{O} \left(c^{-2} Q^{ijk} \right) + \mathcal{O} \left( c^{-4}\right) \,,
\end{aligned}
\end{equation}
where we have used the partitioning of $M_2$ in Eq.~\eqref{partitioning}, and neglected 1PN order terms in the mass octupole. In the system COM frame, replacing $z^i=z_2^i-z_1^i$ in the condition $M^i_{sys}=0$, and solving for the single worldlines, we get~\footnote{We stress that including the mass octupole (or other higher-order mass moments) would not change the form of the Eqs.~\eqref{eq:newtconv}. Indeed, the form of the system mass dipole at 1PN order is
\begin{equation}
M^i_{sys} = M \left(\eta_1 z_1^i + \eta_2 z_2^i \right)+ c^{-2} P^i\left(z_1^j, z_2^j \right) + \mathcal{O} \left( c^{-4}\right) \,.
\end{equation}
Replacing, for instance, $z_2^i =z^i + z_1^i $, and solving for $z_1^i$ perturbatevely in $1/c^2$, we obtain
\begin{equation}
z_1^i = -\eta_2 z^i +  \frac{P^i \left( z^j \right)}{c^2M} + \mathcal{O}\left( c^{-4} \right) \,,
\end{equation}
from which
\begin{equation}
z_2^i = \eta_1 z^i +  \frac{P^i \left( z^j \right)}{c^2M} + \mathcal{O}\left( c^{-4} \right) \,.
\end{equation}
}
\begin{equation}
\label{eq:newtconv}
\begin{gathered}
z_1^i = -\eta_2 z^i + c^{-2} D^i+ \mathcal{O}\left( c^{-4} \right)  \qquad z_2^i = \eta_1 z^i + c^{-2} D^i + \mathcal{O}\left( c^{-4} \right) \\
v_1^i = -\eta_2 v^i + \mathcal{O}\left( c^{-2} \right)  \qquad v_2^i = \eta_1 v^i  + \mathcal{O}\left( c^{-2} \right) \,,
\end{gathered}
\end{equation}
where
\begin{equation}
\begin{aligned}
D^i = & \nu  \left[\left( \eta_2-\eta_1 \right) \left(\frac{v^2}{2} - \frac{M}{2r} - \frac{3}{4 \eta_2 r^3} Q^{jk}n^{jk}\right) - \frac{\eta_1}{M} E_2^{\mathrm{int}} \right] z^i \\
& -\frac{3\eta_1}{2r^2} Q^{ij} n^j + \frac{1}{M}\epsilon^{ijk} v^j \left(\eta_2 J_1^k - \eta_1 J_2^k \right) + \mathcal{O} \left( Q^{ijk} \right) \,.
\end{aligned}
\end{equation}
We make also use of the relations
\begin{equation}
\dot{r}= n^i v^i \qquad \eta_1 + \eta_2 = 1  \,. 
\end{equation}
Finally, taking the difference of the Eqs.~\eqref{eq:eomorbit}, and replacing Eq.~\eqref{partitioning}, we obtain the equation for the relative acceleration $a^i=a_2^i-a_1^i$,
\begin{equation}
\label{eq:orbitaleom}
  a^i=a^i_M+a^i_J+a^i_{Q2}+a^i_{Q3}+a^i_{S2}+a^i_{S3} \,.
\end{equation}
The mass contribution is
\begin{align}
\label{eq:orbitaleomm}
  a^i_M=&
  -\frac{M}{r^2} n^i
  -\frac{1}{c^2} \frac{M}{r^2} \left\{n^i \left[ (1+3 \nu) v^2-
    \frac{3 \nu}{2} \dot{r}^2-2(2+\nu) \frac{M}{r} \right]
  -2(2-\nu)\dot{r} v^i \right\} \nonumber \\
  &+\mathcal{O}\left(c^{-4}\right)\,.
  \end{align}
  The spin contribution is
  \begin{align}
  \label{eq:orbitaleomj}
  a^i_J=&
  \frac{\epsilon^{abc} J_2^c}{c^2 \eta_2 r^3} 
  \left[(3+\eta_2) v^a \delta^{bi}
    -3 (1+\eta_2) \dot{r} n^a \delta^{bi} + 6 n^{ai} v^b\right] \nonumber \\
   & +\frac{\epsilon^{abc} J_1^c}{c^2 \eta_1 r^3} \left[(3+\eta_1) v^a \delta^{bi} -3
    (1+\eta_1) \dot{r} n^a \delta^{bi} + 6 n^{ai} v^b \right]+\mathcal{O}\left(c^{-4}\right)\,.
\end{align}
The mass quadrupole contribution is
\begin{align}
\label{eq:orbitaleomq}
    a^i_{Q2}=&
     -\frac{3 Q^{ab}}{2 \eta_2 r^4} \left[5 n^{abi}-2n^a \delta^{bi}\right]+ 
    \frac{1}{c^2} \bigg\{ \frac{Q^{ab}}{r^4} \bigg[ n^{abi}
      \left(-\frac{15}{2 \eta_2} \left(1+3\nu\right) v^2+ \frac{105 \eta_1}{4} \dot{r}^2 \right. \nonumber \\
      & \left. + \frac{12}{\eta_2} \left(5-2\eta_2^2\right) \frac{M}{r} \right)   +n^a \delta^{bi} \left(\frac{3}{\eta_2}\left(2+2\eta_2-3\eta_2^2\right) v^2 -\frac{15}{2\eta_2}\left(2-\eta_2-\eta_2^2\right) \dot{r}^2 \right. \nonumber \\
      &\left. -\frac{3}{\eta_2} \left(8-\eta_2-3\eta_2^2\right) \frac{M}{r} \right)
      + \frac{15}{\eta_2} \left(2-\nu\right) \dot{r} n^{ab} v^i   -\frac{3}{2\eta_2}  \left(7-2\eta_2 + 3 \eta_2^2\right) n^a v^{bi} \nonumber\\ 
      &  -\frac{15 \eta_1}{2 \eta_2} \left(1+\eta_2\right) \dot{r} n^{ai} v^b 
      + \frac{3 \eta_1}{2 \eta_2} v^{ab} n^i + \frac{3 }{2 \eta_2} \left(5-4\eta_2-\eta_2^2\right) \dot{r} v^a \delta^{bi} \bigg] \nonumber\\
   &  + \frac{\dot{Q}^{ab}}{r^3} \left[-\frac{3}{2\eta_2} \left(4-\eta_2\right) n^{ab} v^i  -\frac{15 \eta_1}{2} \dot{r} n^{abi}
      +\frac{6}{\eta_2} n^{ai} v^b 
      - \frac{3 \eta_1}{\eta_2} v^a \delta^{bi} \right. \nonumber\\
   & \left.+ \frac{3}{\eta_2} \left(1-2\eta_2-\eta_2^2\right) \dot{r} n^a \delta^{bi}\right] + \frac{\ddot{Q}^{ab}}{r^2} \left[\frac{3}{4}n^{abi}+
      \frac{3}{2} n^a \delta^{bi}\right]- \frac{E_2^{\mathrm{int}}}{r^2}n^i \bigg \}\nonumber\\
    & + \frac{3 \eta_1}{c^2 M \eta_2^2} \epsilon^{icd} J_2^c \bigg\{ \frac{Q^{ab}}{r^5}
    \bigg[ \frac{5}{2} n^{ab} \left(7 \dot{r} n^{d} - v^d\right)
      +\left(\delta^{ad} -5 n^{ad}\right) v^b -5 \dot{r} \delta^{ad} n^{b} \bigg] 
    \nonumber\\
    & + \frac{\dot{Q}^{ab}}{r^4} \left(\delta^{ad}-\frac{5}{2}n^{ad}\right) n^b \bigg\} +\frac{3 }{c^2 M \eta_1} \epsilon^{icd} J_1^c \bigg\{ \frac{Q^{ab}}{r^5}
    \bigg[ \frac{5}{2} n^{ab} \left(7 \dot{r} n^{d} - v^d\right)
       \nonumber\\
  & +\left(\delta^{ad} -5 n^{ad}\right) v^b -5 \dot{r} \delta^{ad} n^{b} \bigg]   +
    \frac{\dot{Q}^{ab}}{r^4} \left(\delta^{ad}-\frac{5}{2}n^{ad}\right) n^b \bigg\} \nonumber \\
    &+ \frac{3 \epsilon^{cde} J_1^a}{c^2 M \nu}
    \bigg\{ \frac{Q^{bd}}{r^5} \bigg[ 5 n^{ab} \left( \delta^{ic} v^e - \delta^{ie} v^c \right) 
      + 5 n^{ac} \left( \delta^{ib} v^e - \delta^{ie} v^b\right) \nonumber\\
      & + 5 n^{bc} \left( \delta^{ia} v^e - \delta^{ie} v^a \right)
      + 35 n^{abc} \left( \dot{r}\delta^{ie}  -n^i \right) +\delta^{ab} \left(\delta^{ie} v^c -\delta^{ic}  \right) \nonumber \\
      &  + \delta^{ac} \left(\delta^{ie} v^b -\delta^{ib}  \right)
      + 5 \left(\delta^{ab} n^c + \delta^{ac} n^b \right) \left(n^i -\dot{r}\delta^{ie}  \right) \bigg] \nonumber\\
   & - \frac{ \dot{Q}^{bd}}{r^4} \delta^{ie} \left(5 n^{abc}-\delta^{ab} n^c -\delta^{ac} n^b \right) \bigg\} +\mathcal{O}\left(c^{-4}\right)    \,. 
\end{align}
The mass octupole contribution is
\begin{equation}
\label{eq:massoctu}
a^i_{Q3} = \frac{5Q^{abc}}{2 \eta_2 r^5} \left(7 n^{iabc}- 3 \delta^{ic} n^{ab} \right)+\mathcal{O}\left(c^{-2}\right) \,. 
\end{equation}
The current quadrupole contribution is
\begin{align}
\label{eq:quadcurr}
a^i_{S2} =  &  \frac{4  \epsilon^{bcd}}{c^2 \eta_2 }
\bigg\{ \frac{S^{ad}}{r^4} \bigg[ n^a \left(\delta^{ib} v^c -\delta^{ic} v^b \right) +
  n^b \left(\delta^{ia} v^c -\delta^{ic} v^a \right) + 5 n^{ab}
  \left(\dot{r}\delta^{ic} -n^i v^c \right) \bigg]\nonumber \\
& - \frac{ \dot{S}^{ad}}{r^3} \delta^{ic} n^{ab}  \bigg\} + \frac{ J_1^c}{c^2 M \nu} \frac{S^{ab}}{r^5}
\left[ 4 \delta^{bc} \left( 5 n^{ia} -\delta^{ia} \right) + 10 \left( \delta^{ia} n^{bc} +
  \delta^{ib} n^{ac} + \delta^{ic} n^{ab}   \right) \right. \nonumber \\
  &\left. - 70 n^{iabc} \right]+\mathcal{O}\left(c^{-4}\right) \,.
\end{align}
The current octupole contribution is
\begin{align}
\label{eq:finaleqf}
a^i_{S3} = & \frac{15}{c^2 \eta_2} \bigg\{ \frac{S^{bde}}{r^5}  n^{ab}  v^c
\bigg[\frac{7}{2}(\epsilon^{iae} n^{c}-\epsilon^{cae} n^{i})n^d -( \epsilon^{iad}
  \delta^{ce} + \epsilon^{icd}\delta^{ae} + \epsilon^{acd}\delta^{ie} )\bigg]  \nonumber\\
& -\frac{\dot{S}^{bde}}{2r^4} \epsilon^{iad} n^{abe} \bigg\} + \frac{45 J_1^c}{4 c^2 M \nu} \frac{S^{abc}}{r^6} \bigg[ \delta^{cd} \left(\delta^{ia} n^b + \delta^{ib} n^a -7 n^{iab} \right) \nonumber \\
 & - \frac{7}{3} \left( \delta^{ia} n^{bcd} +\delta^{ib} n^{acd} +
  \delta^{ic} n^{abd} + \delta^{id} n^{abc} -9 n^{iabcd} \right) \bigg] +\mathcal{O}\left(c^{-4}\right) \,.
\end{align}
Deriving these equations, we have made use of the relations~\eqref{eq:newtconv} to replace the body single velocities by the relative velocity, at 1PN order.

Note that we have included the contribution of the (Newtonian) internal energy $E_2^{\mathrm{int}}$ in the mass quadrupole term. We clarify this point in the following, when we derive an action principle in the adiabatic approximation. Indeed, the internal energy of body $2$ will result proportional to its mass quadrupole and octupole. For this reason, we have neglected terms of order $\mathcal{O} \left(Q^{ab}E_2^{\mathrm{int}} \right)$ and $\mathcal{O} \left(Q^{abc}E_2^{\mathrm{int}} \right)$. Also, like before, we have neglected quadratic terms in $ Q^{ij}$ and $Q^{ijk}$, and 1PN order terms proportional to the mass octupole. 

We stress that up to now we have not made (yet) any assumption on the internal dynamics of the bodies. The above equations are valid for a generic binary. We have never assumed that the $l \geq 2$ multipole moments are tidally induced. On the other hand, this prevents us to solve the problem: we need to provide the evolution equations for the higher-order moments. In the adiabatic approximation, the $l\ge2$ multipole moments are given by the algebraic relations (cf. Eqs.~\eqref{eq:adiabaticrelspin} and~\eqref{eq:adiabaticrelspin2})
\begin{equation}
\label{eq:adiabatic2}
\begin{split}
Q^{ab}  &=   \lambda_{2} G_2^{ab}  + \frac{\lambda_{23}}{c^2}  J_2^c H_2^{abc} \\
Q^{abc} &=   \lambda_{3} G_2^{abc} + \frac{\lambda_{32}}{c^2}  J_2^{\langle c} H_2^{ab \rangle} \\
S^{ab}  &=  \frac{\sigma_{2}}{c^2} H_2^{ab}   + {\sigma_{23}}   J_2^c G_2^{abc}  \\
S^{abc} &=   \frac{\sigma_{3}}{c^2} H_2^{abc}  + {\sigma_{32}}   J_2^{\langle c} G_2^{ab \rangle} \,.
\end{split}
\end{equation}
We note that, replacing the adiabatic relations~\eqref{eq:adiabatic2} in the evolution equation of the spin, Eq.~\eqref{eq:spin_tr}, it follows that
\begin{equation}
\label{eq:spinconserv}
\dot{J}_2 = \mathcal{O} \left( c^{-2} \right) \,,
\end{equation}
i.e., to Newtonian order the spin is conserved in the adiabatic approximation (cf. section~\ref{sec:newtgrav}, Eq.~\eqref{eq:spin_newt3}).

\subsection{Lagrangian}
In the this section we show that the orbital equation of motion~\eqref{eq:orbitaleom}, together with the adiabatic relations~\eqref{eq:adiabatic2}, can be derived from an action principle.

\subsubsection{Orbital dynamics}
The orbital equation of motion in the COM frame, $a^i=a^i_M+a^i_J+a^i_{Q2}+a^i_{Q3}+a^i_{S2}+a^i_{S3}$
(Eq.~\eqref{eq:orbitaleom}), can be derived from an action principle. One first writes the most
general Lagrangian consistent with the truncation and at most linear in the spin, which depends on a set of free coefficients. Then, applying the Euler-Lagrange equations to the Lagrangian, replacing
the evolution equations for the spins $J_1^i$, $J_2^i$ and the internal energy $E_2^{\mathrm{int}}$, Eqs.~\eqref{eq:spin_tr} and~\eqref{eq:tidheat}, and comparing with the orbital equations of motion, it is possible to find the values of the coefficients, which will depend only on the (Newtonian) masses of the two bodies. Following this approach, we find that the Lagrangian is
\begin{equation}
\label{eq:lagr_orb_trunc}
  \mathcal{L}_{\mathrm{orb}}(\boldsymbol{z},\boldsymbol{v},\boldsymbol{a})=
  \mathcal{L}_M+\mathcal{L}_J+\mathcal{L}_{Q2}+\mathcal{L}_{Q3}+{\cal
  L}_{S2}+\mathcal{L}_{S3}\,.
\end{equation}
The contribution of the mass is
\begin{equation}
\label{eq:LM}
\mathcal{L}_M = \frac{\mu v^2}{2} + \frac{\mu M}{r} + \frac{\mu}{c^2}
\left\{\frac{1-3 \nu}{8}v^4 + \frac{M}{2r} \left[\left(3+ \nu\right)v^2+ \nu
  \dot{r}^2-\frac{M}{r} \right] \right\} +\mathcal{O} \left( c^{-4} \right)\,.
  \end{equation}
The contribution of the spin is
  \begin{equation}
  \label{eq:LJ}
\mathcal{L}_J =\frac{\epsilon^{abc}}{c^2}   v^b \left[ \left( \eta_2 J_1^a
  + \eta_1 J_2^a \right)  \frac{2M}{r^2} n^c  + \left( \eta_2^2 J_1^a + \eta_1^2 J_2^a  \right) \frac{a^c}{2}  \right] +\mathcal{O} \left( c^{-4} \right)
  \,.
\end{equation}
The mass quadrupole term reads
\begin{align}
 \label{eq:LMQ}
  \mathcal{L}_{Q2} =& \frac{3 \eta_1 M}{2r^3} Q^{ab} n^{ab}+ \frac{1}{c^2} \bigg\{ \frac{M}{r^3} Q^{ab} \bigg[ n^{ab}
    \bigg( \frac{3 \eta_1}{4} (3+\nu) v^2 + \frac{15 \nu \eta_1}{4} \dot{r}^2 \nonumber\\
    &-\frac{3 \eta_1}{2} (1+3\eta_1)\frac{M}{r} \bigg)  +\frac{3 \eta_1^2}{2} v^{ab}  -\frac{3 \eta_1^2}{2}(3+\eta_2) \dot{r} n^a v^b \bigg]\nonumber\\
    &   - \frac{M}{r^2} \dot{Q}^{ab} \left[\frac{3 \nu }{2} n^a v^b+\frac{3 \nu }{4}\dot{r} n^{ab}\right]
  +E_2^{\mathrm{int}} \left[\frac{\eta_1^2}{2} v^2 + \eta_1 \frac{M}{r} \right] \bigg\} \nonumber \\
  & + \frac{3 }{c^2 r^4} \epsilon^{icd} J_1^a Q^{bd} \left(5 n^{abc}
  -\delta^{ab} n^{c} -\delta^{ac} n^b \right) v^i +\mathcal{O} \left( c^{-4} \right)\,.
\end{align}
The mass octupole term reads
\begin{equation}
\label{eq:LQ3}
\mathcal{L}_{Q3}  = -\frac{5 \eta_1 M}{2r^4} Q^{abc} n^{abc} +\mathcal{O} \left( c^{-2} \right)\,.
\end{equation}
The current quadrupole contribution is
\begin{equation}
\label{eq:LS2}
\mathcal{L}_{S2}  = \frac{4 \eta_1 M}{c^2 r^3} \epsilon^{bcd} n^{ab} S^{ad} v^c + \frac{2}{c^2 r^4} J_1^c S^{ab} \left( 5 n^{abc} -2 \delta^{bc} n^a  \right)+\mathcal{O} \left( c^{-4} \right) \,.
\end{equation}
The current octupole contribution is
\begin{equation}
\label{eq:LS3}
\mathcal{L}_{S3}  =  \frac{15 \eta_1 M}{2c^2r^4} \epsilon^{ade} S^{bcd}  n^{abc}v^e 
 +\frac{45}{4 c^2 r^5} J_1^d S^{abc} \left(\delta^{cd} n^{ab} -
\frac{7}{3} n^{abcd} \right) +\mathcal{O} \left( c^{-4} \right)\,.
\end{equation}
Note that Eq.~\eqref{eq:lagr_orb_trunc} is a \emph{generalized Lagrangian}, since it depends on the (relative) acceleration $a^i$, together with the (relative) position $z^i$ and velocity $v^i$. The action is stationary if the \emph{generalized Euler-Lagrange equations} are satisfied,
\begin{equation}
\label{eq:EL}
\left(\frac{\partial}{\partial z^i}-\frac{d}{dt}\frac{\partial}{\partial v^i}+\frac{d^2}{dt^2}\frac{\partial}{\partial a^i}
\right)\mathcal{L}_{\mathrm{orb}}=0\,.
\end{equation}
A generalized Lagrangian is needed in order to obtain the spin contribution of the orbital equation of motion, $a^i_J$, from an action principle~\footnote{The equations of motion (and then the Lagrangian) for a spinning two-body system depend on the spin supplementary condition assumed~\cite{Mikoczi:2016fiy}. Choosing a different spin supplementary condition (i.e., a different gauge), it is possible to make the Lagrangian independent of the acceleration.}~\cite{Vines:2010ca}. Applying the Eq.~\eqref{eq:EL} to the Lagrangian function~\eqref{eq:lagr_orb_trunc} reproduces the orbital equation of motion~\eqref{eq:orbitaleom}.

We stress that at Newtonian order the mass quadrupole and octupole contributions are actually given by
\begin{equation}
\label{eq:massqo}
\begin{aligned}
\mathcal{L}_{Q2} = & \frac{1 }{2} G_2^{ij} Q^{ij} +\mathcal{O} \left( c^{-2} \right)\\
\mathcal{L}_{Q3} = & \frac{1}{6} G_2^{ijk} Q^{ijk}  +\mathcal{O} \left( c^{-2} \right)\,,
\end{aligned}
\end{equation}
consistently with the Newtonian Lagrangian~\eqref{lagrnewt}. Remarkably, we found that the same occurs at 1PN order with the current quadrupole and octupole~\footnote{For a generic mass multipole moment of order $l$, the contribution to the Newtonian Lagrangian is $\mathcal{L}_{Ql}=\frac{1}{l!} G_2^L Q^L + \mathcal{O}(c^{-2})$. In the case of current multipole moments, the structure is akin to the Newtonian one, but at 1PN order, $\mathcal{L}_{Sl}=\frac{1}{c^2} \frac{1}{l!} \frac{l}{l+1} H_2^L S^L + \mathcal{O}(c^{-4})$.}
\begin{equation}
\label{eq:currentqo}
\begin{aligned}
\mathcal{L}_{S2} = & \frac{1 }{3 c^2} H_2^{ij} S^{ij}   +\mathcal{O} \left( c^{-4} \right)\\
\mathcal{L}_{S3} = & \frac{1}{8 c^2} H_2^{ijk} S^{ijk}  +\mathcal{O} \left( c^{-4} \right) \,.
\end{aligned}
\end{equation}

We remark that: (i) the mass monopole contribution to the acceleration $a^i_M$, Eq.~\eqref{eq:orbitaleomm}, is due only to the monopole term of the Lagrangian, $\mathcal{L}_M$. (ii) The spin contribution to the acceleration $a^i_J$, Eq.~\eqref{eq:orbitaleomj}, is due only to the spin term of the Lagrangian, $\mathcal{L}_J$. (iii) The mass quadrupole contribution to the acceleration $a^i_{Q2}$, Eq.~\eqref{eq:orbitaleomq}, arises from terms in $\mathcal{L}_M$, $\mathcal{L}_J$ and $\mathcal{L}_{Q2}$. We recall that both the Newtonian acceleration~\eqref{eq:accel} and the time derivative of the spin~\eqref{eq:spin_tr} are proportional to the mass quadrupole. For this reason, the contributions to $a^i_{Q2}$ can not be entirely encoded in $\mathcal{L}_{Q2}$, but must come, without a chance, also from $\mathcal{L}_M$ and $\mathcal{L}_J$. (iv) the mass octupole (which we need only at leading order) and the current quadrupole and octupole contributions to the acceleration $a^i_{Q3}$, $ a^i_{S2}$, $a^i_{S3}$, Eqs.~\eqref{eq:massoctu}--\eqref{eq:finaleqf}, arise from $\mathcal{L}_{Q3}$, $\mathcal{L}_{S2}$ and $\mathcal{L}_{S3}$, respectively.

\subsubsection{Internal dynamics}
It is possible to extend the Lagrangian $\mathcal{L}_{\mathrm{int}}$ in order to describe also the adiabatic evolution of the mass and current, quadrupole and octupole moments ($Q^{ij}$, $S^{ij}$, $Q^{ijk}$, $S^{ijk}$), i.e., to enforce the adiabatic relations~\eqref{eq:adiabatic2} from an action principle.

In this derivation we use the explicit expressions of the $l=2,3$ tidal moments of body $2$, which can be derived through Eqs.~\eqref{eq:gmom} and~\eqref{eq:hmom}, and read
\begin{align}
\label{eq:G2ab}
  G_2^{ab}=&\frac{3\eta_1M}{r^3}n^{\langle ab\rangle}
  +\frac{1}{c^2}\frac{3\eta_1M}{r^3}\left[\left(2v^2-\frac{5\eta^2_2}{2}
    {\dot r}^2-\frac{5+\eta_1}{2}\frac{M}{r}\right)n^{\langle ab\rangle}+v^{\langle ab\rangle} \right. \nonumber \\
    & \left. -\left(3-\eta_2^2 \right){\dot r}n^{\langle a}v^{b\rangle}\right]  + \frac{6  }{c^2 r^4}  J_1^d v^e \epsilon^{ec \langle a} \left( 5 n^{b \rangle cd}- \delta^{b \rangle d} n^c- n^{b \rangle }\delta^{cd} \right)  +\mathcal{O}(c^{-4})\,, \\
\label{eq:G2abc}
G_2^{abc}=&-\frac{15\eta_1M}{r^4}n^{\langle abc\rangle}+\mathcal{O}(c^{-2})\,,\\
\label{eq:H2ab}
H_2^{ab}=&\frac{12\eta_1M}{r^3}v^d n^{c \langle a} \epsilon^{b \rangle cd}+\frac{30J_1^c}{r^4}n^{\langle abc\rangle}+\mathcal{O}(c^{-2})\,,\\
\label{eq:H2abc}
H_2^{abc}=&-\frac{60\eta_1M}{r^4}v^e n^{d \langle ab} \epsilon^{c \rangle d e}-\frac{210 J_1^d}{r^5}n^{\langle abcd\rangle}+\mathcal{O}(c^{-2})\,.
\end{align}

We note that up to 1PN order, the mass quadrupole contribution $\mathcal{L}_{Q2}$~\eqref{eq:LMQ} can be written as~\cite{Vines:2010ca}
\begin{equation}
\label{eq:LQ2}
\mathcal{L}_{Q2}=U^{ab}Q^{ab}+V^{ab}{\dot Q}^{ab}+WE_2^{\mathrm{int}}+\mathcal{O}(c^{-4})\,,
\end{equation}
where $U^{ab}(\boldsymbol{z},\boldsymbol{v})$, $V^{ab}(\boldsymbol{z},\boldsymbol{v})$,
$W(\boldsymbol{z},\boldsymbol{v})$ are the coefficients appearing in Eq.~\eqref{eq:LMQ}, i.e.,
\begin{align}
\label{def:U}
  U^{ab} = & \, \frac{3\eta_1 M}{2r^3} n^{ab}  +  \frac{1}{c^2} \frac{M}{r^3}
  \bigg[ n^{ab} \bigg( \frac{3 \eta_1}{4} (3+\nu)  v^2+ \frac{15 \nu \eta_1}{4} \dot{r}^2  -\frac{3 \eta_1}{2} (1+3\eta_1)\frac{M}{r} \bigg)\nonumber\\
    &+\frac{3 \eta_1^2}{2} v^{ab}  -\frac{3 \eta_1^2}{2}(3+\eta_2) \dot{r} n^a v^b \bigg]  +  \frac{1}{c^2 } \frac{3}{r^4}\epsilon^{eca} J_1^d \left( 5 n^{bcd} -\delta^{bd} n^c -\delta^{cd} n^b\right) v^e\,,\\
  \label{def:V} 
V^{ab} = & \, \frac{1}{c^2} \frac{M}{r^2} \left[ -\frac{3 \nu }{2}n^a v^b -\frac{3 \nu }{4} \dot{r} n^{ab}\right]\,,  \\
\label{def:W}
W = & \, \frac{1}{c^2}  \left[\frac{\eta_1^2}{2}v^2 + \eta_1 \frac{M}{r} \right]\,.
\end{align}
Together with the Eqs.~\eqref{eq:massqo} and~\eqref{eq:currentqo}, this allows us to write the orbital Lagrangian in the form
\begin{equation}
\begin{aligned}
  \mathcal{L}_{\mathrm{orb}}=& \mathcal{L}_M+\mathcal{L}_J+ U^{ab}Q^{ab}+V^{ab}{\dot Q}^{ab}+WE_2^{\mathrm{int}} + \left( \frac{1}{6} G_2^{abc} + \left(c^{-2}  \right) \right) Q^{abc}  \\
  & + \frac{1 }{3 c^2} H_2^{ab} S^{ab} + \frac{1}{8 c^2} H_2^{abc} S^{abc} + \mathcal{O} \left(c^{-4} \right)\,.
  \end{aligned}
\end{equation}
As we said, we do not explicitly compute the 1PN corrections in the mass octupole contribution, because they do not affect the leading 6.5PN order tidal contribution to the GW phase, coming from the mass octupole. We recall that the mass and current moments, as well as the internal energy $E_2^{\mathrm{int}}$, are independent of the relative position vector $z^i$ (i.e., the system COM worldline) by definition: they only evolve in time. The same is not true for the tidal moments, which instead depend on orbital degrees of freedom (see Eqs.~\eqref{eq:G2ab}--\eqref{eq:H2abc}).

We define now the Lagrangian
\begin{equation}
\label{eq:lagr_tot_trunc}
  \mathcal{L}(\boldsymbol{z},\boldsymbol{v},\boldsymbol{a},Q^L,\dot{Q}^L,S^L)=
  \mathcal{L}_{\mathrm{orb}}(\boldsymbol{z},\boldsymbol{v},\boldsymbol{a},Q^L,\dot{Q}^L,S^L) +  \mathcal{L}_2^{\mathrm{int}}(Q^L,S^L) \,.
\end{equation}
We have split the total Lagrangian of the system in two contributions. The internal Lagrangian $\mathcal{L}_2^{\mathrm{int}}$ depends only on the internal degrees of freedom $Q^{ab}$, $Q^{abc}$, $S^{ab}$, $S^{abc}$, and describes the internal dynamics of body $2$ (the only one tidally deformed), while the orbital Lagrangian $\mathcal{L}_{\mathrm{orb}}$ depends both on the orbital degrees of freedom and on the moments $Q^L$, $S^L$. This means that adding the internal Lagrangian would not change the orbital equation of motion~\eqref{eq:orbitaleom}, because its derivatives with respect to the orbital degrees of freedom vanish. Note that $\mathcal{L}_{Q2}$ also depends on $E_2^{\mathrm{int}}/c^2$ (see Eq.~\eqref{eq:LQ2}). Thus, in order to write the Euler-Lagrange equations for the internal degrees of freedom, we need to know the explicit form of $E_2^{\mathrm{int}}$ as a function of the moments $Q^L$, $S^L$, at 0PN order.

We know that to fully determine the dynamics of the system at Newtonian order, we need only the evolution equations for the mass multipole moments (cf. section~\ref{sec:newtgrav}, Eq.~\eqref{eq:adiabnewt}) 
\begin{equation}
\label{eq:adiabnewt2}
\begin{split}
Q^{ab}  &=   \lambda_{2} G_2^{ab}  \\
Q^{abc} &=   \lambda_{3} G_2^{abc}  \\
\end{split} \ .
\end{equation}
Replacing the above adiabatic relations in Eq.~\eqref{eq:tidheat}, we find, at leading order,
\begin{align}
\label{eq:tidheat1}
  {\dot E}_2^{\mathrm{int}}=&\frac{1}{2}G_{g,2}^{ab}{\dot Q}^{ab}+\frac{1}{6}G_{g,2}^{abc}{\dot Q}^{abc}+
  O(c^{-2})\nonumber\\
  =&\frac{1}{4\lambda_2}\frac{d}{dt}\left(Q^{ab}Q^{ab}\right)+\frac{1}{12\lambda_3}\frac{d}{dt}\left(Q^{abc}Q^{abc}\right)  +\mathcal{O}(c^{-2})\,.
\end{align}
Therefore, up to a constant term, the internal energy (at Newtonian level) has the form
\begin{equation}
\label{EintN}
  E_2^{\mathrm{int}}=\frac{1}{4\lambda_2}Q^{ab}Q^{ab}+\frac{1}{12\lambda_3}Q^{abc}Q^{abc}+\mathcal{O}(c^{-2})\,, 
\end{equation}
or, in other words, the internal Lagrangian which correctly reproduces the Newtonian adiabatic relations~\eqref{eq:adiabnewt2}) is $\mathcal{L}_2^{\mathrm{int}}=-E_2^{\mathrm{int}}$.

Then, we look for an expression which reduces to Eq.~\eqref{EintN} at 0PN order, and which yields the correct adiabatic relations~\eqref{eq:adiabatic2} at 1PN order. We find that the correct Lagrangian (as shown below) is given by
\begin{align}
\label{EintPN}
  \mathcal{L}_2^{\mathrm{int}}=&-E_2^{\mathrm{int}}=-\frac{1}{4\lambda_2}Q^{ab}Q^{ab}-\frac{1}{12\lambda_3}Q^{abc}Q^{abc}-\frac{1}{6\sigma_2}S^{ab}S^{ab}-\frac{1}{16\sigma_3}S^{abc}S^{abc}\nonumber\\
  &+\alpha J_2^a Q^{bc}S^{abc}+ \beta J_2^a S^{bc}Q^{abc}\,,
\end{align}
where $\alpha$ and $\beta$ are two coupling constants that will turn out to be proportional to the rotational tidal Love numbers. Note that this is the most general Lagrangian function, which can be built from the multipole moments of our truncation by the requirement that $\mathcal{L}_2^{\mathrm{int}}$ is scalar, parity invariant, at most quadratic in the internal degrees of freedom and linear in the spin. 

An important point should be remarked. The expression~\eqref{EintPN} does not contain any $1/c^2$ factor. This means that it is valid in the same form also at Newtonian order. This seems to violate the requirement that $E_2^{\mathrm{int}}$ should reduce to Eq.~\eqref{EintN} in the Newtonian limit. In other words, it seems that we are changing the Newtonian internal energy, violating as a consequence Eq.~\eqref{eq:tidheat1}. On the other hand, the adiabatic relations~\eqref{eq:adiabatic2} state that current multipole moments are induced, through the spin, by the electric tidal moments also at Newtonian order. In the 0PN order case we have not considered the current moments, because even if present, they do not gravitate at Newtonian level, and therefore they can not affect the dynamics. However, if current moments do exist, they must contribute somehow to the internal energy of the body, also at Newtonian order. The point is that the contribution of the terms depending on current moments in Eq.~\eqref{EintPN} is a constant at Newtonian order. Therefore, its time derivative vanishes and Eq.~\eqref{EintN} is satisfied~\footnote{This can be checked explicitly taking the time derivative of Eq.~\eqref{EintPN}, replacing the adiabatic relations~\eqref{eq:adiabatic2}, and keeping only the 0PN order terms. Working to linear order in the spin, we get that the contribution to $E_2^\mathrm{int}$, coming from the current moments at Newtonian order, is quadratic in the spin, and therefore we should neglect it for consistency. Dropping the linear approximation, and considering the full contribution in the spin, we find anyway that the contribution of the current moments cancels out, and ${\dot E}_2^{\mathrm{int}}$ is given by Eq.~\eqref{eq:tidheat1}.}.

Now we apply the Euler-Lagrange equations (for the internal degrees of freedom) to the total Lagrangian~\eqref{eq:lagr_tot_trunc}, with $ \mathcal{L}_2^{\mathrm{int}}$ given by Eq.~\eqref{EintPN}. For the mass quadrupole we obtain
\begin{equation}
\label{eq:adiabaticquad}
\begin{gathered}
\left( \frac{\partial}{\partial Q^{ab}} -\frac{d}{dt} \frac{\partial}{\partial \dot{Q}^{ab}} \right) \mathcal{L} =0 \\
U^{ab}-\dot{V}^{ab} + \left(-\frac{1}{2 \lambda_2}Q^{ab}+ \alpha J_2^c S^{abc} \right)\left(1-W \right)=  \mathcal{O} \left( c^{-4} \right) \\
Q^{ab} = \lambda_2 2\left(1+ W \right) \left( U^{ab} -\dot{V}^{ab} \right) + 2 \alpha \lambda_2 J_2^c S^{abc} + \mathcal{O} \left( c^{-4} \right) \,,
\end{gathered}
\end{equation}
where we have used the fact that $W= \mathcal{O}(1/c^2)$. Remarkably, it can be shown that~\cite{Vines:2010ca} (cf. Eqs.~\eqref{eq:G2ab} and~\eqref{def:U}--\eqref{def:W})
\begin{equation}
G_2^{ab} = 2 \left(1+ W \right) \left( U^{ \langle ab \rangle } -\dot{V}^{\langle ab \rangle } \right)   + \mathcal{O} \left( c^{-4} \right) \,,
\end{equation}
which yields~\footnote{\label{foot}Note that we have replaced the tensors $U^{ab}$, $V^{ab}$ by their STF parts $U^{\langle ab \rangle}$, $V^{\langle ab \rangle}$. We are allowed to do this, because the mass quadrupole moment $Q^{ab}$ is a STF tensor. Using either $U^{ab}$, $V^{ab}$ or their STF parts into the Lagrangian $\mathcal{L}_{Q2}$ gives the same result, since for a generic tensor $T^{ab}$, $Q^{ab}T^{\langle ab \rangle}= Q^{ab} T^{ab}$.}
\begin{equation}
\label{eq:finalquadmass}
Q^{ab} = \lambda_2 G_2^{ab} + 2 \alpha \lambda_2 J_2^c S^{abc} + \mathcal{O} \left( c^{-4} \right) \,.
\end{equation}
In the mass octupole case we get
\begin{equation}
\begin{gathered}
\left( \frac{\partial}{\partial Q^{abc}} -\frac{d}{dt} \frac{\partial}{\partial \dot{Q}^{abc}} \right) \mathcal{L} =0 \\
 \left(\frac{1}{6} G_2^{abc} + \mathcal{O} \left( c^{-2} \right) \right) + \left(-\frac{1}{6 \lambda_3}Q^{abc}+ \beta J_2^c S^{ab} \right)\left(1-W \right)= \mathcal{O} \left( c^{-4} \right) \\
Q^{abc} = \lambda_3 \left(1+ W \right) \left(G_2^{abc} + \mathcal{O} \left( c^{-2} \right) \right) + 6\beta \lambda_3 J_2^c S^{ab} + \mathcal{O} \left( c^{-4} \right) \,.
\end{gathered}
\end{equation}
Neglecting higher PN order contributions, we obtain~\footnote{Note that we have replaced $J_2^c S^{ab}$ by $J_2^{\langle c} S^{ab \rangle}$, using the same argument explained in footnote~\ref{foot}.}
\begin{equation}
\label{eq:finaloctumass}
Q^{abc} = \lambda_3  \left(G_2^{abc} + \mathcal{O} \left( c^{-2} \right) \right) + 6\beta \lambda_3 J_2^{\langle c} S^{ab \rangle} + \mathcal{O} \left( c^{-4} \right) \,.
\end{equation}
Similarly, for the current quadrupole and octupole we obtain
\begin{equation}
\begin{gathered}
\left( \frac{\partial}{\partial S^{ab}} -\frac{d}{dt} \frac{\partial}{\partial \dot{S}^{ab}} \right) \mathcal{L} =0 \\
 \frac{1}{3c^2} H_2^{ab} + \left(-\frac{1}{3 \sigma_2}S^{ab}+ \beta J_2^c Q^{abc} \right)\left(1-W \right)= \mathcal{O} \left( c^{-4} \right) \\
S^{ab} =  \frac{\sigma_2}{c^2}H_2^{ab} + 3\beta \sigma_2 J_2^c Q^{abc} + \mathcal{O} \left( c^{-4} \right)
\end{gathered}
\end{equation}
and
\begin{equation}
\label{eq:adiabaticoctu}
\begin{gathered}
\left( \frac{\partial}{\partial S^{abc}} -\frac{d}{dt} \frac{\partial}{\partial \dot{S}^{abc}} \right) \mathcal{L} =0 \\
 \frac{1}{8c^2} H_2^{abc} + \left(-\frac{1}{8 \sigma_3}S^{abc}+ \alpha J_2^c Q^{ab} \right)\left(1-W \right)= \mathcal{O} \left( c^{-4} \right) \\
S^{abc} =  \frac{\sigma_3}{c^2}H_2^{abc} + 8 \alpha \sigma_3 J_2^{\langle c} Q^{ab \rangle} + \mathcal{O} \left( c^{-4} \right) \,.
\end{gathered}
\end{equation}

Gathering these results, the equations of motion for multipole moments read
\begin{align}
  Q^{ab}  =  &  \lambda_2G_2^{ab}+ 2\lambda_{2} \alpha   J_2^c S^{abc}+\mathcal{O}(c^{-4})\,,\nonumber\\
  Q^{abc} =  & \lambda_{3} \left(G_2^{abc}+\mathcal{O}(c^{-2})\right) +6\lambda_3 \beta
  J_2^{\langle c} S^{ab \rangle}+\mathcal{O}(c^{-4}) \,,\nonumber\\
  S^{ab}  =  & \frac{\sigma_{2}}{c^2} H_{2}^{ab}   + 3\sigma_2 \beta  J_2^c Q^{abc} +\mathcal{O}(c^{-4})\,, \nonumber\\
  S^{abc} =  & \frac{\sigma_{3}}{c^2} H_{2}^{abc}  + 8\sigma_3 \alpha
  J_2^{\langle c} Q^{ab \rangle}+\mathcal{O}(c^{-4})\,.
\end{align}
Replacing recursively $Q^L$ and $S^L$ in the above expressions, and truncating the result to linear order in the spin (consistently with our approximation), we obtain
\begin{align}
\label{eq:adiabatic1}
  Q^{ab}  =  &  \lambda_2G_2^{ab}+ \frac{2\lambda_{2}\sigma_3 \alpha}{c^2}  J_2^c H_{2}^{abc} +\mathcal{O}\left(J_2^a J_2^b \right)+\mathcal{O}(c^{-4})\,,\nonumber\\
  Q^{abc} =  & \lambda_{3} \left(G_{2}^{abc}+\mathcal{O}(c^{-2})\right) +\frac{6\lambda_3\sigma_2 \beta}{c^2}
  J_2^{\langle c} H_{2}^{ab \rangle} +\mathcal{O}\left(J_2^a J_2^b \right)+\mathcal{O}(c^{-4}) \,,\nonumber\\
  S^{ab}  =  & \frac{\sigma_{2}}{c^2} H_{2}^{ab}   + 3\lambda_3\sigma_2 \beta  J_2^c G_{2}^{abc}+\mathcal{O}\left(J_2^a J_2^b \right) +\mathcal{O}(c^{-4})\,, \nonumber\\
  S^{abc} =  & \frac{\sigma_{3}}{c^2} H_{2}^{abc}  + 8\lambda_2 \sigma_3 \alpha
  J_2^{\langle c} G_{2}^{ab \rangle}+\mathcal{O}\left(J_2^a J_2^b \right)+\mathcal{O}(c^{-4})\,.
\end{align}
These expressions coincide with the adiabatic relations~\eqref{eq:adiabatic2}, if we make the replacement
\begin{equation}
\label{eq:coupling}
\begin{aligned}
\lambda_{23} = & 2\lambda_2 \sigma_3 \alpha  \\
\lambda_{32} = & 6\lambda_3 \sigma_2 \beta  \\
\sigma_{23} = & 3\lambda_3 \sigma_2 \beta   \\
\sigma_{32} = & 8\lambda_2 \sigma_3 \alpha   
\end{aligned} \ .
\end{equation}
We have shown that the equations of motion for the $l \geq 2$ multipole moments in the adiabatic approximation can be obtained from an action principle. Surprisingly, the internal Lagrangian~\eqref{EintPN} enforces the relations~\eqref{eq:coupling}, which imply that only two out of four rotational tidal Love numbers are independent, while the other two should be proportional to the first ones. In particular, $\sigma_{32} \propto \lambda_{23} $ and $\sigma_{23} \propto \lambda_{32} $. This is unexpected, because such a behavior does not emerge from the perturbative approach (see section~\ref{sec:RTLN}), and should hold regardless of the internal composition of the object (i.e., it should be independent of the neutron star equation of state). We comment more on this issue in section~\ref{sec:issue}.

\subsection{Gravitational radiation}
\label{subsec:waveform}
In this section we derive the phase of the waveform of the gravitational radiation emitted by the binary system. We assume that the binary is slowly inspiralling in quasi-circular orbit, and that the spins of its components are orthogonal to the plane of the orbit (which means that the binary is non-precessing).

Replacing the adiabatic relations~\eqref{eq:adiabatic2} into the equation of motion~\eqref{eq:orbitaleom}, and using the expression~\eqref{EintPN} for the internal energy~\footnote{We point out that only the mass quadrupole actually contributes to the internal energy in Eq.~\eqref{eq:orbitaleomq}. All the other moments in $E_2^{\mathrm{int}}$ give rise to subleading contributions with respect to the other terms where they appear.}, we find
\begin{equation}
\label{eq:eomred}
a^i = a^i_M +a^i_J + a^i_{\lambda 2} + a^i_{\lambda 3} + a^i_{\sigma 2}+ a^i_{\sigma 3} + a^i_{\mathrm{RTLN}} \,.
\end{equation}
The mass and spin contributions are given by Eqs.~\eqref{eq:orbitaleomm}
\begin{align}
  a^i_M=&
  -\frac{M}{r^2} n^i
  -\frac{1}{c^2} \frac{M}{r^2} \left\{n^i \left[ (1+3 \nu) v^2-
    \frac{3 \nu}{2} \dot{r}^2-2(2+\nu) \frac{M}{r} \right]
  -2(2-\nu)\dot{r} v^i \right\} \nonumber \\
  &+\mathcal{O}\left(c^{-4}\right)
  \end{align}
and~\eqref{eq:orbitaleomj} 
  \begin{align}
  a^i_J=&
  \frac{\epsilon^{abc} J_2^c}{c^2 \eta_2 r^3} 
  \left[(3+\eta_2) v^a \delta^{bi}
    -3 (1+\eta_2) \dot{r} n^a \delta^{bi} + 6 n^{ai} v^b\right] \nonumber \\
   & +\frac{\epsilon^{abc} J_1^c}{c^2 \eta_1 r^3} \left[(3+\eta_1) v^a \delta^{bi} -3
    (1+\eta_1) \dot{r} n^a \delta^{bi} + 6 n^{ai} v^b \right]+\mathcal{O}\left(c^{-4}\right)\,,
\end{align}
respectively. The term proportional to the quadrupolar electric Love number $\lambda_2$ is
\begin{equation}
\begin{aligned}
a^i_{\lambda 2} = & \lambda_2 \bigg(- \frac{9 \eta_1}{\eta_2} \frac{M}{r^7}n^i + \frac{M}{c^2 r^7} \bigg\{\bigg[ -\frac{9 \eta_1}{2 \eta_2} \left( 2-\eta_2\right) \left(1+6 \eta_2 \right) v^2  \\
&+ \frac{36 \eta_1}{\eta_2} \left(1-6 \eta_2+ \eta_2^2 \right) \dot{r}^2 + \frac{3 \eta_1}{2 \eta_2} \left(66 + 9 \eta_2- 19 \eta_2^2 \right) \frac{M}{r} \bigg] n^i   \\
&+ \frac{9 \eta_1}{ \eta_2} \left(2-\eta_2 \right) \left(3-2 \eta_2 \right) \dot{r}v^i \bigg\} +\frac{9 \eta_1^2}{c^2\eta_2^2} \frac{\epsilon^{i ab} J_2^a}{r^8} \left(8 n^b \dot{r}-v^b \right) \\
&+ \frac{1}{c^2 r^8} \bigg\{ \frac{9+57\eta_1}{\eta_2}  \epsilon^{iab}   J^a_1 v^b + \epsilon^{ibc} n^{ab} \bigg[120 \frac{\eta_1}{\eta_2} J^c_1 v^a + 12\left(3 - \frac{3+\eta_1}{\eta_2} \right) J^a_1 v^c \bigg] \\
& -  \left(36 + \frac{36+84\eta_1}{\eta_2} \right)\epsilon^{abc} n^{ai} J_1^b v^c \bigg\} \bigg)
+\mathcal{O}\left(c^{-4}\right)  \,.
\end{aligned}
\end{equation}
The contribution proportional to the octupolar electric Love number $\lambda_3$ reads
\begin{equation}
a^i_{\lambda 3} = \lambda_3 \left(-\frac{60 \eta_1}{\eta_2} \frac{M}{r^9} n^i \right) +\mathcal{O}\left(c^{-2}\right) \,.
\end{equation}
The contribution proportional to the quadrupolar magnetic Love number $\sigma_2$ is
\begin{equation}
\begin{aligned}
a^i_{\sigma2} =& \sigma_2 \bigg(\frac{48\eta_1}{c^4 \eta_2}\frac{M}{r^7} \left[-\left(v^2 +2 \dot{r}^2 \right)n^i+3 \dot{r} v^i \right]+ \frac{1}{c^4 r^8} \bigg\{192 \left(\epsilon^{ibc} n^{ab} J_1^c v^a- \epsilon^{abc}n^{ai} J_1^b v^c \right)  \\
&+48 \epsilon^{iab} J_1^a v^b+ \frac{\eta_1}{\eta_2} \left[ 160 \left(\epsilon^{ibc} n^{ab} J_1^c v^a- \epsilon^{abc} n^{ai} J_1^b v^c \right)\right] +32\epsilon^{ibc}n^{ab}J_1^a v^c\\
&+ 16 \epsilon^{iab} J_1^a v^b \bigg\} \bigg) +\mathcal{O}\left(c^{-6}\right) \,.
\end{aligned}
\end{equation}
The term proportional to the octupolar magnetic Love number $\sigma_3$ reads
\begin{equation}
\begin{aligned}
a^i_{\sigma3} =&  \sigma_3 \bigg(\frac{120\eta_1}{c^4 \eta_2}\frac{M}{r^9} \left[-\left(3v^2 +5 \dot{r}^2 \right)n^i+8 \dot{r} v^i \right] \\
&+ \frac{1}{c^4 r^{10}} \bigg\{1800 \left(\epsilon^{ibc} n^{ab} J_1^c v^a- \epsilon^{abc}n^{ai} J_1^b v^c \right) +360 \epsilon^{iab} J_1^a v^b \\
&+ \frac{\eta_1}{\eta_2} \left[ 1575 \left(\epsilon^{ibc} n^{ab} J_1^c v^a- \epsilon^{abc} n^{ai} J_1^b v^c \right)\right] +225\epsilon^{ibc}n^{ab}J_1^a v^c+ 135 \epsilon^{iab} J_1^a v^b \bigg\} \bigg)\\
& +\mathcal{O}\left(c^{-6}\right) \,.
\end{aligned}
\end{equation}
The contribution coming from the rotational tidal Love numbers is (we recall that we work to linear order in the spin)
\begin{equation}
\begin{aligned}
a^i_{\mathrm{RTLN}} = & -\frac{\lambda_{23}}{c^2} \frac{24 \eta_1 M}{\eta_2 r^8} \left( 2 \epsilon^{ibc} n^{ab} v^c J_2^a +3 \epsilon^{abc} n^{ai} v^c J_2^b \right) \\
& -\frac{\lambda_{32}}{c^2} \frac{24 \eta_1 M}{\eta_2 r^8} \left(  \epsilon^{ibc} n^{ab} v^c J_2^a - \epsilon^{abc} n^{ai} v^c J_2^b \right) \\
&-\frac{\sigma_{23}}{c^2} \frac{24 \eta_1 M}{\eta_2 r^8} \left( 4 \dot{r} \epsilon^{ibc} n^b J_2^c +  \epsilon^{iab} v^bJ_2^a  - 2 \epsilon^{ibc} n^{ab} v^c J_2^a - 2 \epsilon^{abc} n^{ai} v^c J_2^b \right) \\
&+\frac{\sigma_{32}}{c^2} \frac{6 \eta_1 M}{\eta_2 r^8} \left( 8 \dot{r} \epsilon^{ibc} n^b J_2^c + 2 \epsilon^{iab} v^bJ_2^a  + 2 \epsilon^{ibc} n^{ab} v^c J_2^a - 5 \epsilon^{abc} n^{ai} v^c J_2^b \right) \\
&+\mathcal{O}\left(c^{-4}\right) \,.
\end{aligned}
\end{equation}
Note that the contributions coming from the quadrupolar and octupolar magnetic Love numbers $\sigma_2$, $\sigma_3$ are proportional to $1/c^4$. Since we are using the 1PN approximation, this may seem inconsistent (i.e., we should drop these terms). However, this arises from the fact that, even if we have defined the current multipole moments only at Newtonian level, they are tidally induced at 1PN order (cf. with Eqs.~\eqref{eq:adiabaticrelspin2} and the discussion below it). In other words, the leading order contribution of the magnetic Love numbers appears inevitably with that power of $c$. Going to higher PN order would give rise only to subleading terms.

Furthermore, we are treating the rotational tidal Love numbers as if they were independent, though they are actually related by the Eqs.~\eqref{eq:coupling} (at least according to the Lagrangian formulation). We ignore this problem for the moment (see section~\ref{sec:issue}).

\subsubsection{Radius-frequency relation}
Now we focus on non-precessing circular orbits with angular frequency $\omega/(2\pi)$. Without loss of generality, we can assume that the binary lies in the plane $z^3 = 0$. In this case, $n^i=(\cos(\omega t),\sin(\omega t),0)$ (we remind that $n^2=1$), and
\begin{equation}
v^i =  r \omega  \phi^i \qquad a^i =  - r \omega^2  n^i \,,
\end{equation}
where $\phi^i = \{ - \sin{(\omega t)}, \cos{(\omega t)}, 0 \}$, and it satisfies the relations
\begin{equation}
\phi^2 = 1 \qquad n^i \phi^i = 0 \qquad \dot{n}^i =   \omega  \phi^i \qquad \dot{\phi}^i =  -\omega  n^i \qquad \dot{r} = n^i v^i = 0 \,.
\end{equation}
Furthermore, we assume the spins of the two bodies to be parallel to the orbital angular momentum (consistently with the non-precessing requirement). Note that the spins are constant in the adiabatic approximation, see Eqs.~\eqref{eq:spin_newt3}, \eqref{eq:spin_tr} and~\eqref{eq:spinconserv}. Therefore, we can write
\begin{equation}
J^i_1 = J_1 s^i \qquad J_2^i = J_2 s^i \qquad s^i = \epsilon^{ijk} n^j \phi^k \,,
\end{equation}
where $J_1,\ J_2$ are the spin magnitudes, and $s^i = ( 0,0,1 )$. Also, we define the dimensionless spin variables 
\begin{equation}
\chi_1 = \frac{c J_1}{M_1^2} =\frac{c J_1}{(\eta_1 M)^2} \qquad \chi_2 = \frac{c J_2}{\ ^n M_2^2} =\frac{c J_2}{(\eta_2 M)^2} \,.
\end{equation}

The orbital equation of motion admits a solution of this form. Replacing the above ansatz in Eq.~\eqref{eq:eomred}, we obtain
\begin{equation}
\label{eq:radfreq}
\begin{aligned}
- r \omega^2 n^i = & \bigg(-\frac{M}{r^2}  -\frac{1}{c^2} \frac{M}{r^2} \left[ (1+3 \nu) \omega^2 r^2 -2(2+\nu) \frac{M}{r} \right]  \\
& + \frac{ \omega M^2}{c^3r^2}\left[  \chi_2\left(3- \eta_2\right)\eta_2 +  \chi_1\left(3- \eta_1\right)\eta_1 \right]  + \lambda_2 \bigg\{- \frac{9 \eta_1}{\eta_2} \frac{M}{r^7}\\
&  + \frac{M}{c^2 r^7} \bigg[ -\frac{9 \eta_1}{2 \eta_2} \left( 2-\eta_2\right) \left(1+6 \eta_2 \right) \omega^2 r^2  + \frac{3 \eta_1}{2 \eta_2} \left(66 + 9 \eta_2- 19 \eta_2^2 \right) \frac{M}{r} \bigg]    \\
&  +9 \eta_1^2 \frac{\omega M^2}{c^3r^7} \left[\chi_2+ \chi_1 \left(1 + \frac{6}{\eta_2} \right) \right] \bigg\} + \lambda_3 \left(-\frac{60 \eta_1}{\eta_2} \frac{M}{r^9}  \right) \\
& + \sigma_2  \left(-\frac{48 \eta_1}{\eta_2} \frac{\omega^2 M}{c^4 r^5} +\frac{144 \eta_1^2 \chi_1}{\eta_2} \frac{\omega M^2}{c^5 r^7} \right) + \sigma_3  \left(-\frac{360 \eta_1}{\eta_2} \frac{\omega^2 M}{c^4 r^7} \right.  \\
& \left. +\frac{1440 \eta_1^2 \chi_1}{\eta_2} \frac{\omega M^2}{c^5 r^9} \right)  + \left(72 \lambda_{23}-24 \lambda_{32}-24 \sigma_{23} +18 \sigma_{32} \right) \frac{ \omega \nu M^3 \chi_2}{c^3 r^7} \bigg) n^i \,.
\end{aligned}
\end{equation}
Now we define the quantity
\begin{equation}
x= \frac{(\omega M)^{2/3}}{c^2} \,,
\end{equation}
which is the true physical small parameter of the PN expansion (indeed, $x= v^2/c^2 + \mathcal{O} \left( c^{-4} \right)$). We solve Eq.~\eqref{eq:radfreq} for $r$ as a function of $\omega$, determining the relation between the radius of the orbit and the orbital frequency~\cite{Vines:2010ca}. Working perturbatively in the PN parameter $x$, in the spins and in the tidal Love numbers, we find
\begin{align}
  \label{eq:romega}
  r =&  \frac{M^{1/3}}{ \omega^{2/3}} \bigg{\{}  1  + \frac{\nu-3}{3} x +
  \left[ \frac{(\eta_1-3)\eta_1}{3} \chi_1 + \frac{(\eta_2-3) \eta_2}{3} \chi_2 \right]x^{1.5}+  \mathcal{O}\left(x^2\right)\nonumber\\
  &+\frac{3 \eta_1}{\eta_2}
  \frac{c^{10}}{M^5} \lambda_2 x^5  + \left[-\frac{\eta_1}{2\eta_2} \left(6- 26
    \eta_2 + \eta_2^2 \right) \frac{c^{10}}{M^5} \lambda_2   + \frac{16 \eta_1}{\eta_2} \frac{c^8}{M^5} \sigma_2 \right]x^6 \\
    &+\left[ \left( 2\eta_1 \left(9 -2 \eta_2 \right)
    \chi_2 -\frac{4 \eta_1^3}{\eta_2}
    \chi_1 \right) \frac{c^{10}}{M^5} \lambda_2- \frac{48 \eta_1^2}{\eta_2} \chi_1
    \frac{c^8}{M^5} \sigma_2 \right. \nonumber\\
&\left. + \frac{\nu \chi_2 c^{10}}{M^4} \left( - 24   \lambda_{23}  
+ 8   \lambda_{32} + 8   \sigma_{23} - 6   \sigma_{32} \right) \right] x^{6.5} + \mathcal{O} \left(x^7\right) \nonumber \\
&  + \frac{20 \eta_1}{\eta_2}  \frac{c^{14}}{M^7} \lambda_3 x^7
  +\frac{120 \eta_1}{\eta_2} \frac{c^{12}}{M^7} \sigma_3 x^8 - \frac{480 \eta_1^2}{\eta_2} \chi_1 \frac{c^{12}}{M^7} \sigma_3 x^{8.5} \bigg\}  \,.
\end{align}
The PN order of the tidal effects is already clear from this expression. The first line in the above equation refers to the point-particle (black hole) terms. The following three lines include \emph{all} the tidal terms derived consistently to linear order in the spin and up to 6.5PN order. The last line refers to the leading order contributions of the octupolar Love numbers $\lambda_3$ and $\sigma_3$, resulting as a by-product of this computation. The above relation is needed to derive the gravitational waveform phase.

\subsubsection{Total energy}
Replacing the adiabatic relations~\eqref{eq:adiabatic1} in the total Lagrangian~\eqref{eq:lagr_tot_trunc}, yields the \emph{reduced Lagrangian} 
\begin{align}
\label{eq:redlag}
  \mathcal{L}(\boldsymbol{z},\boldsymbol{v},\boldsymbol{a}) = &
  \frac{\mu v^2}{2}+\frac{\mu M}{r}\left(1+\frac{3\eta_1}{2\eta_2}
  \frac{\lambda_2}{r^5} + \frac{15 \eta_1}{2 \eta_2} \frac{\lambda_3}{r^7}\right) +\frac{\mu}{c^2}\left\{\frac{1-3\nu}{8} v^4 \right. \nonumber\\
  & \left. +\frac{M}{r}
  \left[v^2\left(\frac{3+\nu}{2}+\frac{3\eta_1^2(5+\eta_2) }{4 \eta_2}\frac{\lambda_2}{r^5}\right)+{\dot r}^2\left(\frac{\nu}{2} -\frac{9 \eta_1(1-6\eta_2+
      \eta_2^2)}{2 \eta_2}\frac{\lambda_2}{r^5}\right) \right. \right. \nonumber \\
    & \left.\left. +\frac{M}{r} \left(-\frac{1}{2}+\frac{3 \eta_1(-7+
      5\eta_2)}{2\eta_2}\frac{\lambda_2}{r^5}\right)\right]\right\} +\frac{\epsilon^{abc}}{c^2}   v^b \left[ \left( \eta_2 J_1^a + \eta_1
    J_2^a \right)  \frac{2M}{r^2} n^c  \right. \nonumber \\
  & \left. + \left( \eta_2^2 J_1^a + \eta_1^2 J_2^a  \right) \frac{a^c}{2}  \right]    + \frac{\lambda_2}{c^2}\frac{9 \eta_1 M}{r^7} \epsilon^{abc} n^a J_1^b v^c  + \frac{\sigma_2}{c^4} \left[ \frac{12 \eta_1^2 M^2}{r^6} \left(v^2-\dot{r}^2
    \right) \right. \nonumber \\
  & \left. + \frac{24 \eta_1 M}{r^7} \epsilon^{abc} n^a J_1^b v^c\right] + \frac{\sigma_3}{c^4} \left[ \frac{60 \eta_1^2 M^2}{r^8} \left(v^2-\dot{r}^2
    \right) + \frac{180 \eta_1 M}{r^9} \epsilon^{abc} n^a J_1^b v^c\right] \nonumber \\
  & + \frac{\eta_1^2 M^2}{c^2 r^7}\left( 48 \lambda_2 \sigma_3 {\alpha}  -36 \lambda_3
  \sigma_2 {\beta}  \right) \epsilon^{abc} n^a J_2^b v^c \,,
\end{align}
which depends only on the orbital degrees of freedom. Note that the contributions from the rotational tidal Love numbers in the above equation only enter through the terms proportional to $\alpha$ and $\beta$. For consistency, the variation of this action should give the equation of motion~\eqref{eq:eomred}. We have checked that this is indeed the case, once one has replaced the rotational tidal Love numbers using the conditions~\eqref{eq:coupling}. 

From the above reduced Lagrangian, the conserved energy of our truncation is given by~\cite{Mikoczi:2016fiy} (see Eqs.~\eqref{eq:euleren} and~\eqref{eq:euleren2})
\begin{equation}
E = \left( p_1^i \dot{z}^i + p_2^i \ddot{z}^i  \right) -\mathcal{L} \,,
\end{equation}
where $p_1^i$, $p_2^i$ are the first and second momentum, respectively,
\begin{equation}
p_1^i =  \frac{\partial \mathcal{L}}{\partial \dot{z}^i} - \frac{d}{dt} \frac{\partial \mathcal{L}}{\partial \ddot{z}^i}  \qquad
p_2^i = \frac{\partial \mathcal{L}}{\partial \ddot{z}^i} \,.
\end{equation}
The result is
\begin{align}
\label{cons_en}
  E=& v^i \left(\frac{\partial\mathcal{L}}{\partial v^i} -\frac{d}{dt}\frac{\partial\mathcal{L}}{\partial a^i} \right)+ 
  a^i\frac{\partial\mathcal{L}}{\partial a^i} -\mathcal{L}\nonumber\\
  =&\frac{\mu v^2}{2}-\frac{\mu M}{r}\left(1+
  \frac{3\eta_1}{2\eta_2} \frac{\lambda_2}{r^5} +  \frac{15 \eta_1}{2 \eta_2} \frac{\lambda_3}{r^7}\right)
 +\frac{\mu}{c^2}\left\{ \frac{3(1-3\nu)}{8} v^4\right.\nonumber\\
 &+\frac{M}{r}\left[v^2\left(\frac{3+\nu}{2}+\frac{3\eta_1^2(5+\eta_2)}{4 \eta_2} \frac{\lambda_2}{r^5}\right) +{\dot r}^2\left(\frac{\nu}{2} -\frac{9 \eta_1(1-6\eta_2+
     \eta_2^2)}{2 \eta_2}\frac{\lambda_2}{r^5}\right) \right.\nonumber \\
   & \left.\left. -\frac{M}{r} \left(-\frac{1}{2}+\frac{3 \eta_1(-7+5\eta_2)}{2\eta_2}
   \frac{\lambda_2}{r^5}\right)\right]\right\}  +\frac{\epsilon^{abc}}{c^2}   v^b a^c \left( \eta_2^2 J_1^a + \eta_1^2 J_2^a  \right) \nonumber \\
 & + \frac{\sigma_2}{c^4}  \frac{12 \eta_1^2 M^2}{r^6} \left(v^2-\dot{r}^2 \right)+\frac{\sigma_3}{c^4}  \frac{60 \eta_1^2 M^2}{r^8} \left(v^2-\dot{r}^2 \right) \,.
\end{align}
Note that the terms proportional to $\alpha$ and $\beta$ (as well as other terms) in the reduced Lagrangian do not contribute to the conserved energy (and then to the gravitational waveform phase), because these terms are linear in the velocity, and therefore cancel out. As we show below, the rotational tidal Love numbers will appear through the radius-frequency relation. For circular orbits we get
\begin{align}
  E=&\frac{\mu \omega^2 r^2}{2}-\frac{\mu M}{r}\left(1+
  \frac{3\eta_1}{2\eta_2} \frac{\lambda_2}{r^5} +  \frac{15 \eta_1}{2 \eta_2} \frac{\lambda_3}{r^7}\right)
 +\frac{\mu}{c^2}\left\{ \frac{3(1-3\nu)}{8} \omega^4 r^4\right.\nonumber\\
 &\left.+\frac{M}{r}\left[\omega^2 r^2\left(\frac{3+\nu}{2}+\frac{3\eta_1^2(5+\eta_2)}{4 \eta_2} \frac{\lambda_2}{r^5}\right)  -\frac{M}{r} \left(-\frac{1}{2}+\frac{3 \eta_1(-7+5\eta_2)}{2\eta_2}
   \frac{\lambda_2}{r^5}\right)\right]\right\}\nonumber \\
   &   +\frac{r^2 \omega^3 \mu^2}{c^3}  \left( \chi_1+ \chi_2 \right)  + \frac{\sigma_2}{c^4}  \frac{12 \eta_1^2 M^2 \omega^2}{r^4} +\frac{\sigma_3}{c^4}  \frac{60 \eta_1^2 M^2 \omega^2}{r^6}  \,,
\end{align}
and replacing the radius-frequency relation~\eqref{eq:romega},
\begin{align}
  E =& - \frac{ \mu}{2} (M \omega)^{2/3} \Bigg{\{}  1 - \frac{9+ \nu}{12} x  +\left[ \frac{2\eta_2(\eta_2+3)}{3 }  \chi_2+\frac{2\eta_1(\eta_1+3)}{3 } \chi_1\right]x^{1.5}+\mathcal{O}\left(x^2\right)\nonumber\\
&  - \frac{9 \eta_1}{\eta_2} \frac{c^{10}}{M^5} \lambda_2 x^5  -\left[ \frac{11 \eta_1}{2 \eta_2}\left(3+2\eta_2+3\eta_2^2 \right)  \frac{c^{10}}{M^5}
  \lambda_2 + \frac{88 \eta_1}{\eta_2} \frac{c^{8}}{M^5} \sigma_2 \right]x^6 \nonumber\\
&  + \left\{ \left[24 \eta_1 (\eta_2-3) \chi_2 +   \frac{24 \eta_1^3 }{\eta_2}  \chi_1
  \right] \frac{c^{10}}{M^5}  \lambda_2  +\frac{192\eta_1^2 }{\eta_2} \chi_1 \frac{c^{8}}{M^5} \sigma_2 \right.   \nonumber\\
& \left. + \frac{c^{10} \nu \chi_2 }{M^4} \left(96 \lambda_{23} -32 \lambda_{32} -32 \sigma_{23}
+24 \sigma_{32} \right) \right\} x^{6.5} +\mathcal{O}\left(x^7\right) \nonumber\\
& -  \frac{65 \eta_1}{ \eta_2} \frac{c^{14}}{M^7} \lambda_3 x^7 - \frac{600 \eta_1}{\eta_2}
\frac{c^{12}}{M^7} \sigma_3 x^8 + \frac{1920 \eta_1^2 }{\eta_2}\chi_1 \frac{c^{12}}{M^7} \sigma_3 x^{8.5}
\Bigg{\}} \,.
\end{align}
Note that in this equation the rotational tidal Love numbers ($\lambda_{23}$, $\lambda_{32}$, $\sigma_{23}$, $\sigma_{32}$) appear explicitly, since the adiabatic relations have been used to obtain Eq.~\eqref{eq:romega}. Using the radius-frequency is then possible to express every quantity as of function of $\omega$ (or equivalently, $x$). This is important, because the orbital frequency is gauge-invariant, while the orbital radius is not.

\subsubsection{Gravitational flux}
The last ingredient, that we need to calculate the gravitational waveform phase, is the energy loss by gravitational wave emission. The gravitational wave flux (at 1.5PN order) is given in Eq.~\eqref{eq:flux}, whereas the multipole moments of system are given by Eqs.~\eqref{sysM}-\eqref{sysZ}. Within our truncation they read~\footnote{We recall that to get the gravitational wave phase up to 6.5PN order, we need to include the mass octupole moment $Q^{ijk}$ of the body $2$ only at the leading order. Since this term enters the gravitational wave flux at the next-to-leading order, we can safely neglect its contribution to the system multipole moments.}
\begin{align}
\label{eq:sysq2}
  M^{ij}_{sys} = & Q^{ij}+  \mu r^2 n^{\langle ij \rangle} +  \frac{1}{c^2}
  \bigg{\{}    \mu r^2 \left[ \left(  \frac{29(1-3\nu) v^2}{42}  + \frac{(8\nu -5) M }{7r} \right)  n^{\langle ij \rangle}\right. \nonumber\\
    & \left.    +  \frac{11(1-3\nu) }{21}  v^{\langle ij \rangle} +\frac{4\left(3\nu-1 \right)}{7} \dot{r} n^{i \langle} v^{j \rangle}\right] \nonumber\\
  & + \frac{4 r}{3} \left( 2 v^a n^{\langle i } - n^a v^{\langle i } \right)
  \epsilon^{j \rangle a b} \left(\eta_1^2 J_2^b + \eta_2^2 J_1^b \right) \nonumber\\
& + \bigg[ \left(E_2^{int} + 3 U_{Q2}  \right) \eta_1^2 r^2 n^{\langle ij \rangle} -\frac{\eta_1 M}{42 r} \bigg( 2 \left(46 \eta_1^2 + 109 \eta_1 \eta_2 +63 \eta_2^2 \right)Q^{ij} \nonumber \\
& -3  \left(52 \eta_1^2 + 4 \eta_1 \eta_2 -25 \eta_2^2 \right) n^{\langle ij \rangle ab} Q^{ab}  -6 \left(15 \eta_1^2 + 21 \eta_1 \eta_2 +11 \eta_2^2 \right) n^{a \langle i} Q^{j \rangle a}  \bigg) \nonumber\\
 & + \frac{\eta_1^2 }{42}  \left( 29v^2 Q^{ij} -66 v^{a \langle i} Q^{ j \rangle a} \right) + \frac{2 \eta_1^2 r}{21} \left( n^{\langle i } \dot{Q}^{j \rangle a} v^a
    + 8 v^{\langle i } \dot{Q}^{j \rangle a} n^a \right) - \frac{2 \eta_1^2 }{7} \dot{r} \dot{Q}^{ij} \nonumber\\
& + \frac{\eta_1^2 r^2}{42} \left(11 \ddot{Q}^{ij}  -12 n^{a \langle i } \ddot{Q}^{j \rangle a} \right) \bigg]  +  \frac{8 \eta_1}{9} \left( 2 \epsilon^{ab \langle i } S^{j \rangle b}  v^a -  r \epsilon^{ab \langle i }
  \dot{S}^{j \rangle b}  n^a \right)\bigg{\}} \nonumber\\
  &+\mathcal{O} \left(c^{-2} Q^{ijk} \right) +\mathcal{O} \left(c^{-4} \right) \,,
  \end{align}
  \begin{align}
  \label{eq:sysq3}
  M^{ijk}_{sys} = &  \mu r^3 (\eta_1 - \eta_2)  n^{\langle ijk \rangle}
  + 3 \eta_1 r Q^{\langle ij} n^{k \rangle} +\mathcal{O} \left(Q^{ijk} \right)+ \mathcal{O} \left(c^{-2} \right) \,, \\
  \label{eq:syss2}
  J^{ij}_{sys} = & S^{ij} +  \mu  r^2 (\eta_1 -\eta_2)
  \epsilon^{ab \langle i} n^{j \rangle a} v^b +\frac{3r}{2}
  \left(\eta_1 J_2^{\langle i }- \eta_2 J_1^{\langle i } \right) n^{ j \rangle} \nonumber \\
 & +\frac{\eta_1 }{2} \left(-2  \epsilon^{ ab \langle i } Q^{j \rangle b}v^a+ r \epsilon^{ ab \langle i } \dot{Q}^{j \rangle b} n^a  \right) +\mathcal{O} \left(c^{-2} \right) \,,
\end{align}
where we have used the relations~\eqref{partitioning} and~\eqref{eq:newtconv}, when needed. The tail term in Eq.~\eqref{eq:tail} is needed only at leading order, to derive the tail-tidal coupling due to the quadrupolar electric Love number $\lambda_2$. Other contributions from it would be subleading with respect to the terms coming from the instantaneous part. Thus, using Eq.~\eqref{eq:sysq2} and~\eqref{eq:G2ab}, we obtain
\begin{equation}
\label{eq:tailint}
\begin{aligned}
  U^{ij}_{tail}(U) =& 2M \int_0^{\infty} \left(\mu r^2 \ddddot{n}^{\langle ij \rangle }+\ddddot{Q}^{ij} \right) \left[ \log{\left(\frac{c \tau}{2 r_0} \right)+ \frac{11}{12}} \right] d\tau \,, \\
  =&   2M \int_0^{\infty} \left(\mu r^2 \ddddot{n}^{\langle ij \rangle }+ \lambda_2 \ddddot{G}_{g,2}^{ij} \right) \left[ \log{\left(\frac{c \tau}{2 r_0} \right)+ \frac{11}{12}} \right] d\tau \,, \\ 
  =&    2M \left(\mu r^2+ \lambda_2 \frac{3 \eta_1M}{r^3}  \right) \int_0^{\infty}  \ddddot{n}^{\langle ij \rangle }\left(U-\tau \right) \left[ \log{\left(\frac{c \tau}{2 r_0} \right)+ \frac{11}{12}} \right] d\tau \,, 
 \end{aligned}
\end{equation}
where $n^i(U-\tau) =\left (\cos{\left[\omega(U-\tau)\right]},\sin{\left[\omega(U-\tau)\right]},0\right)$, and the time derivatives are computed with respect to $U$~\footnote{We note that also the system multipole moments in the instantaneous flux should be evaluated in the radiative coordinate $U$, rather than in the harmonic coordinate $t$. However, the explicit time dependence cancels out in the result, whatever coordinate we use.}. We recall that $U$ is the retarded time in radiative coordinates, and $r_0$ a gauge-dependent constant which cancels out in the final result. The integral in Eq.~\eqref{eq:tailint} can be evaluated using the formula~\cite{Blanchet:2013haa}
\begin{equation}
\begin{gathered}
\int_0^{\infty} \log{\left(\frac{\tau}{2 \tau_0} \right)  \mathrm{e}^{-\mathrm{i} \omega \tau} \, d \tau} = \frac{\mathrm{i}}{\omega} \left( \log{\left( 2 \omega \tau_0\right)} + \gamma_{\mathrm{E}} + \mathrm{i} \frac{\pi}{2} \right) \,,
\end{gathered}
\end{equation} 
where $\omega >0$ is the orbital frequency, $\tau_0=r_0/c$ and $\gamma_{\mathrm{E}}$ is the Euler-Mascheroni constant.

Replacing Eqs.~\eqref{eq:sysq2}-\eqref{eq:tailint} into Eq.~\eqref{eq:flux}, and using the adiabatic relations~\eqref{eq:adiabatic1} and the radius-frequency relation~\eqref{eq:romega}, the gravitational wave flux can be written as
\begin{align}
F =&  \frac{32}{5} \nu^2 c^5 x^5 \Bigg{\{}  1 -\left(\frac{1247}{336}+ \frac{35}{12}\nu \right)x  \nonumber \\
&+ \left[4\pi- \frac{\eta_2(5+6\eta_2)}{4} \chi_2  -\frac{\eta_1(5+6\eta_1)}{4} \chi_1  \right] x^{1.5} +\mathcal{O} \left( x^2 \right) \nonumber\\
&  + \frac{6(3-2 \eta_2)}{\eta_2}\frac{c^{10}}{M^5} \lambda_2 x^5  \nonumber\\
&+ \left[ \frac{\left(-704-1803 \eta_2+4501 \eta_2^2 -2170 \eta_2^3 \right)}{28 \eta_2}
  \frac{c^{10}}{M^5} \lambda_2  + \frac{2(113 -114\eta_2)}{3 \eta_2}  \frac{c^{8}}{M^5} \sigma_2  \right] x^6 \nonumber\\
& + \left\{ \left[  \frac{24\pi(3-2 \eta_2)}{\eta_2} +  \frac{(667- 939 \eta_2 + 304 \eta_2^2)}{8}
  \chi_2  \right. \right. \nonumber\\ 
  & + \left. \frac{(-395 +1110 \eta_2 -1019 \eta_2^2 + 304 \eta_2^3)}{ 8 \eta_2}   \chi_1 \right]
\frac{c^{10}}{M^5} \lambda_2 \nonumber\\ 
& + \left[ \chi_2   +\frac{(-613+1225
    \eta_2-612 \eta_2^2)}{3 \eta_2}  \chi_1 \right] \frac{c^8}{M^5} \sigma_2   \nonumber\\
& \left. \left. + \frac{c^{10} \chi_2}{M^4} \bigg[ 8\eta_2(12\eta_2-17)  \lambda_{23}   + 32 \nu  \lambda_{32} 
 +\frac{ \eta_2 (113-114 \eta_2)}{3}   \sigma_{23}   - 24 \nu  \sigma_{32}
  \right] \right\} x^{6.5} \nonumber\\
  &  +\mathcal{O} \left( x^7 \right) + \frac{80 \eta_1}{\eta_2} \frac{c^{14}}{M^7} \lambda_3 x^7 + \frac{480 \eta_1}{\eta_2} \frac{c^{12}}{M^7} \sigma_3 x^8 - \frac{1920 \eta_1^2 \chi_1}{\eta_2} \frac{c^{12}}{M^7} \sigma_3 x^{8.5}\Bigg{\}}\,.
\end{align}

\subsubsection{TaylorF2 approximant}
\label{sec:phase}
Finally, integrating twice Eq.~\eqref{eq:wav}, we obtain the phase of gravitational waveform in the frequency domain. The TaylorF2 approximant (see section~\ref{sec:bns}) reads
\begin{align}
\label{phase}
  \psi = & \frac{3}{128 \nu x^{5/2}} \Bigg\{ 1 + \left(\frac{3715}{756}+\frac{55}{9} \nu \right) x \nonumber\\
  &+ \left(\frac{113}{3}(\eta_1 \chi_1 + \eta_2 \chi_2)  -\frac{38}{3} \nu (\chi_1 + \chi_2 )   -16\pi \right) x^{1.5}+ \mathcal{O} \left( x^2 \right) \nonumber \\
  & + \left(264 -\frac{288}{\eta_2}\right) \frac{c^{10}\lambda_2}{M^5}  x^5+ \left[ \left(  \frac{4595}{28}- \frac{15895}{28 \eta_2}  + \frac{5715 \eta_2}{14} -
    \frac{325 \eta_2^2}{7}   \right)\frac{c^{10}\lambda_2}{M^5} \right. \nonumber\\
 & \left.+ \left( \frac{6920}{7} - \frac{20740}{21 \eta_2} \right) \frac{c^{8}\sigma_2}{M^5} \right] x^6 +  \left\{ \left[ \left( \frac{593}{4} - \frac{1105}{8 \eta_2} +\frac{567 \eta_2}{8}
    -81 \eta_2^2 \right) \chi_1\right. \right. \nonumber\\
    & \left. + \left( -\frac{6607}{8} +\frac{6639 \eta_2}{8} -81 \eta_2^2 \right) \chi_2  -  \pi\left(264 -\frac{288}{\eta_2}\right) \right] \frac{c^{10}\lambda_2}{M^5} \nonumber\\
  &  + \left[ \left(-\frac{9865}{3} + \frac{4933}{3 \eta_2} + 1644 \eta_2 \right) \chi_1
    -\chi_2 \right]   \frac{c^{8}\sigma_2}{M^5} \nonumber\\
  & +\frac{c^{10} \chi_2}{M^4}\bigg[ \left(   856 \eta_2 -  816 \eta_2^2 \right){\lambda_{23}}   - \left(\frac{833 \eta_2}{3} - 278 \eta_2^2  \right) {\sigma_{23}} \nonumber\\
    &- \nu \left(272 {\lambda_{32}} -204 {\sigma_{32} }\right)  \bigg] \bigg\}x^{6.5}+ \mathcal{O} \left( x^7 \right) + \left(\frac{4000}{9} - \frac{4000}{9\eta_1} \right)
   \frac{c^{14}\lambda_3}{M^7}   x^7 \nonumber\\
   &+\left( \frac{29400}{11}  -
   \frac{29400}{11 \eta_1} \right) \frac{c^{12}\sigma_3}{M^7}  x^8  + \left( \frac{22400}{3 \eta_1}
    + \frac{22400 \eta_1}{3 }-\frac{44800}{3} \right)  \chi_2  \frac{c^{12}\sigma_3}{M^7}  x^{8.5} \Bigg\}.
\end{align} 
The first two lines are the point-particle (black hole) contributions up to 1.5PN order. Note the term proportional to $\pi$, coming from the tail part of the gravitational flux. The other lines are the terms due to tidal deformations. The 5PN order term is the leading-order contribution of the quadrupolar electric Love number, first derived by Flanagan and Hinderer~\cite{Flanagan:2007ix}. At 6PN order there are the next-to-leading contribution of $\lambda_2$, derived by Vines, Flanagan and Hinderer~\cite{Vines:2010ca,Vines:2011ud}, and the leading-order contribution of the quadrupolar magnetic Love number, first derived by Yagi~\cite{Yagi:2013sva} (see also~\cite{Banihashemi:2018xfb}). The tail-tidal coupling appears at leading-order at 6.5PN order, and depends on $\lambda_2$ (the term proportional to $\pi$, independent of the spins). It was first derived by Damour, Nagar and Villain~\cite{Damour:2012yf}. Finally, the other terms appearing at 6.5PN order are the new contributions computed in this thesis, due to the interaction between the rotation of the object and the tidal field. Note that they are linear in the spins, and therefore vanish for non-rotating objects. These terms arise from two different effects. They depend on: (i) the quadrupolar Love numbers $\lambda_2$ and $\sigma_2$, and (ii) the rotational tidal Love numbers $\lambda_{23}$, $\lambda_{32}$, $\sigma_{23}$ and $\sigma_{32}$. The first contribution is due to the spin-tidal coupling (similarly to the tail-tidal coupling) between the mass and current quadrupole moments and the external tidal field (which depends on the spin of the other body, see Eq.~\eqref{eq:G2ab}--\eqref{eq:H2abc}). In other words, these terms are the next-to-next-to-leading order contribution of $\lambda_2$ and the next-to-leading order contribution of $\sigma_2$. The second contribution is instead the leading order contribution arising from the rotational tidal Love numbers. Note that all the four rotational tidal Love numbers enter at the same order, both the quadrupolar and octupolar ones. In the next section we clarify this point. The result obtained in this thesis completes the tidal part of the gravitational waveform phase up to 6.5PN order. The higher-order terms appearing at 7PN order and beyond are the leading-order contributions of the octupolar Love numbers $\lambda_3$ and $\sigma_3$.

In Table~\ref{tab:tidal} we summarize the different PN orders of the contributions of the tidal Love numbers. We stress that we are not including the tidal terms derived within the effective-one-body (EOB) approach~\cite{Buonanno:1998gg}, where the contribution of the electric quadrupolar Love number has been partially derived up to 7.5PN order (neglecting spins, magnetic Love numbers, and higher-order electric multipoles)~\cite{Damour:2012yf}. Indeed, the EOB formalism, extended by including tidal effects and fitting to numerical relativity simulations, provides an accurate description of inspiralling binary neutron stars up to the merger~\cite{Bernuzzi:2014owa,Dietrich:2017aum,Nagar:2018zoe}. The EOB resummation improves the PN results giving information on higher-order terms beyond current analytical knowledge (for instance, tail effects emerge naturally in the EOB formalism~\cite{Damour:2012yf}). On this subject, our results (Eq.~\eqref{phase}) can be used as a starting point to improve current EOB templates, including the spin-tidal contributions. The functional form of the Lagrangian~\eqref{EintPN} that we have derived can be used as an \emph{ansatz} to find the correct Hamiltonian to be included in the EOB action~\cite{Damour:2009wj,Bini:2012gu} in order to take into account the rotational tidal Love numbers.

\begin{table}
\centering
 \begin{tabular}{c|c|c|c|c|c}
\toprule
 PN order & $\lambda_{2}$ & $\sigma_{2}$ & $\lambda_{23}$, $\lambda_{32}$, $\sigma_{23}$, $\sigma_{32}$  & $\lambda_{3}$ & $\sigma_{3}$ \\ 
 \midrule
 $5$ 	& LO   &      &     	  &     &     \\
 $6$ 	& NLO  & LO   &     	  &     &     \\
 $\bm{6.5}$  & {\bf NNLO} (tail + {\bf spin})    & {\bf NLO (spin)} 
 & {\bf LO (spin)}   &    &     \\
 $7$ 	& $\dots$ & $\dots$  &  $\dots$   	  & LO  &     \\
 $8$ 	& $\dots$ & $\dots$  &  $\dots$   	  & $\dots$  &  LO   \\
\bottomrule
 \end{tabular}
 \caption{\textsl{Schematic representation of the PN contributions of the Love numbers to the gravitational wave phase of a binary system, to linear order in the spin. ``LO'', ``NLO'', and ``NNLO'' stand for leading order, next-to-leading order, etc. The entries in boldface are the new 6.5PN order terms computed in this thesis. They are all proportional to the spins of the binary components, and are zero in the non-spinning case.}}
\captionsetup{format=hang,labelfont={sf,bf}}
   \label{tab:tidal}
\end{table}

Equation~\eqref{phase} is the gravitational waveform phase for a binary system where only one of the two objects is tidally deformed (the body $2$). As previously explained, since we are neglecting quadratic terms in the tidally induced multipole moments, to obtain the full gravitational wave phase up to octupole mass and current moments for both bodies, it is sufficient to add to Eq.~\eqref{phase} the same expression (for the tidal part) obtained by exchanging the indices $1$ and $2$ of the two bodies. The result is
\begin{align}
  \psi = & \frac{3}{128 \nu x^{5/2}} \Bigg\{ 1 + \left(\frac{3715}{756}+\frac{55}{9} \nu \right) x \nonumber\\
  &+ \left(\frac{113}{3}(\eta_1 \chi_1 + \eta_2 \chi_2)  -\frac{38}{3} \nu (\chi_1 + \chi_2 )   -16\pi \right) x^{1.5}+ \mathcal{O} \left( x^2 \right) \nonumber \\
  & + \Bigg[ \left(264 -\frac{288}{\eta_2}\right) \frac{c^{10}\lambda_2^{(2)}}{M^5}  x^5+ \left[ \left(  \frac{4595}{28}- \frac{15895}{28 \eta_2}  + \frac{5715 \eta_2}{14} -
    \frac{325 \eta_2^2}{7}   \right)\frac{c^{10}\lambda_2^{(2)}}{M^5} \right. \nonumber\\
 & \left.+ \left( \frac{6920}{7} - \frac{20740}{21 \eta_2} \right) \frac{c^{8}\sigma_2^{(2)}}{M^5} \right] x^6 +  \left\{ \left[ \left( \frac{593}{4} - \frac{1105}{8 \eta_2} +\frac{567 \eta_2}{8}
    -81 \eta_2^2 \right) \chi_1\right. \right. \nonumber\\
    & \left. + \left( -\frac{6607}{8} +\frac{6639 \eta_2}{8} -81 \eta_2^2 \right) \chi_2  -  \pi\left(264 -\frac{288}{\eta_2}\right) \right] \frac{c^{10}\lambda_2^{(2)}}{M^5} \nonumber\\
  &  + \left[ \left(-\frac{9865}{3} + \frac{4933}{3 \eta_2} + 1644 \eta_2 \right) \chi_1
    -\chi_2 \right]   \frac{c^{8}\sigma_2^{(2)}}{M^5} \nonumber\\
  & +\frac{c^{10} \chi_2}{M^4}\bigg[ \left(   856 \eta_2 -  816 \eta_2^2 \right){\lambda_{23}^{(2)}}   - \left(\frac{833 \eta_2}{3} - 278 \eta_2^2  \right) {\sigma_{23}^{(2)}} \nonumber\\
    &- \nu \left(272 {\lambda_{32}^{(2)}} -204 {\sigma_{32}^{(2)} }\right)  \bigg] \bigg\}x^{6.5}+ \mathcal{O} \left( x^7 \right) + \left(\frac{4000}{9} - \frac{4000}{9\eta_1} \right)
   \frac{c^{14}\lambda_3^{(2)}}{M^7}   x^7 \nonumber\\
   &+\left( \frac{29400}{11}  -
   \frac{29400}{11 \eta_1} \right) \frac{c^{12}\sigma_3^{(2)}}{M^7}  x^8  + \left( \frac{22400}{3 \eta_1}
    + \frac{22400 \eta_1}{3 }-\frac{44800}{3} \right)  \chi_2  \frac{c^{12}\sigma_3^{(2)}}{M^7}  x^{8.5} \nonumber \\
    & + \left( 1 \leftrightarrow 2\right) \Bigg]    \Bigg\} \ ,
\end{align} 
where we have restored the superscript $(2)$ in the Love numbers of body $2$. Note that the coefficients of the tidal terms are not of order $\mathcal{O}(1)$. For instance, the 5PN oder term is magnified by a factor $\lambda_2^{(A)}/M_A^5 \sim k_2^{E\, (A)} (R_A/M_A)^5 \ (A=1,2)$ (see Eq.~\eqref{eq:lovetidal}), where $k_2^{E}$ is the dimensionless electric quadrupolar Love number and $R$ and $M$ the radius and the mass of the object, respectively. For a neutron star $(R/M)^5 \sim 10^3 \div 10^4$, thus the tidal terms are comparable with the 3.5PN order point-particle terms~\cite{Vines:2011ud} (not shown here), cf. with the discussion below Eq.~\eqref{eq:pnimpact}.

\ 

After this work was completed, we have been informed of a related work by Landry~\cite{Landry:2018bil}. Our work differs from Ref.~\cite{Landry:2018bil}, because it includes also the spin-tidal terms proportional to the ordinary tidal Love numbers, while Landry computed only the corrections due to the rotational tidal Love numbers. Furthermore, our result for the gravitational wave phase does not agree with that derived in~\cite{Landry:2018bil}. The source of this discrepancy is a different definition of the energy of the binary system, when terms proportional to the relative velocity $v^i$ are involved. This happens for the magnetic standard tidal Love numbers, as well as for the rotational tidal Love numbers. We also note that, neglecting the spin effects, our result for the 6PN order term proportional to the magnetic, quadrupolar tidal Love number $\sigma_2$ agrees with those of Refs.~\cite{Yagi:2013sva,Banihashemi:2018xfb}, while that of Ref.~\cite{Landry:2018bil} does not.

\subsection{Post-Newtonian order counting of the spin-tidal terms}
\label{subsec:counting}
In this section, we show how it can be easily understood why the spin-tidal coupling computed affects the gravitational wave phase~\eqref{phase} at 6.5PN order, and we generalize this counting to multipole moments and tidal moments of generic order $l \geq 2$. Henceforth, we set $c=1$, thus the PN order is given by the small parameter $v^2 \sim M/r$.

First, we notice that the mass moments $Q^L$ enter the waveform at $l$PN order, whereas the current moments $S^L$ at ($l+1/2$)PN order~\cite{Blanchet:2013haa}. Indeed, the contribution of the multipole moments to the radial acceleration
in a binary system is of order
\begin{equation}
 |a^i|\sim \frac{M}{r^2} \left( \frac{Q^L}{r^{l}} + \frac{v S^L}{r^{l}} \right)\,, 
\end{equation}
(cf. Eqs.~\eqref{eq:orbitaleomm}--\eqref{eq:finaleqf}). Now we assume that the multipole moments are tidally induced. Let us start from the non-spinning case. The tidal moments $G^L$ and $H^L$ enter at ($l+1$) and ($l+3/2$) leading PN order, respectively (cf. Eq.~\eqref{eq:G2ab}--\eqref{eq:H2abc}),
\begin{equation}
 G^L \sim \frac{M}{r^{l+1}} \qquad H^L \sim \frac{vM}{r^{l+1}} \,.
\end{equation}
Therefore, the leading-order contribution of the standard electric ($Q^L \sim \lambda_l G^L$) and magnetic ($S^L \sim \sigma_l H^L$) tidal Love numbers is, respectively, ($2l+1$)PN and ($2l+2$)PN. Indeed, replacing the adiabatic relations in the radial acceleration gives
\begin{equation}
 |a^i|\sim \frac{M}{r^2} \left(  \frac{\lambda_l M}{r^{2l+1}} + \frac{ \sigma_l v^2M }{r^{2l+1}} \right)\,.
\end{equation}

Now we focus on the spinning case. The first effect is that the tidal moments depends also on the spin. The leading contribution to $G^L$ and $H^L$ coming from the spin is of order ($l+5/2$)PN and ($l+2$)PN, respectively (cf. Eq.~\eqref{eq:G2ab}--\eqref{eq:H2abc}),
\begin{equation}
 G^L \sim \frac{vJ}{r^{l+2}} \qquad H^L \sim \frac{J}{r^{l+2}} \,,
\end{equation}
which gives
\begin{equation}
 |a^i|\sim \frac{M}{r^2} \left(  \frac{\lambda_lvJ}{r^{2l+2}} + \frac{ \sigma_lvJ}{r^{2l+2}} \right)\,.
\end{equation}
The leading contribution due to the coupling between the spin and the standard tidal Love numbers is thus of ($2l+5/2$)PN order, for both the electric and magnetic sectors. The spin-tidal contribution of standard quadrupolar Love numbers is then of 6.5PN order, consistently with our computation.

Let us now consider the rotational tidal Love numbers. To linear order in the spin, the multipole moments with a given parity and order $l$ are induced by the tidal moments with opposite parity and order $l\pm 1$, i.e., $Q^L \sim \lambda_{l\, l\pm1} J H^{L\pm1}$ and $S^L \sim \sigma_{l\, l\pm1} J G^{L\pm1}$. Using the above formulas, we get
\begin{equation}
 |a^i|\sim \frac{M}{r^2} \left(  \frac{\lambda_{l\, l\pm1} v JM}{r^{2l+1 \pm1}} + \frac{ \sigma_{l\, l\pm1}vJM}{r^{2l+1\pm1}} \right)\,.
\end{equation}
Therefore, the PN order of the correction proportional to the rotational tidal Love numbers is $2l+3/2 \pm1$, where the plus and minus signs refer to the coupling between a multipole moment of order $l$, and the tidal moment with $l+1$ and $l-1$, respectively. We stress that:
\begin{itemize}
 \item[1)] When $l\geq3$, the minus sign provides the lowest PN correction, namely $(2l+1/2)$. For example, in the octupolar case $l=3$, we obtain a 6.5PN order correction induced by the quadrupolar tidal moments, as computed in the previous sections.
 \item[2)] On the other hand, for $l=2$ there is no dipolar tidal moment which can induce a quadrupole moment, and we have to select the plus sign. This gives again a 6.5PN order term (induced by the octupolar tidal moments), consistently with our computation. This explains why all the rotational tidal Love numbers considered in this thesis enter at 6.5PN order.
 \item[3)] For both signs, the PN order of the rotational tidal Love numbers is the \emph{average} between the PN order of an ordinary tidal term of order $l$ and the tidal term of opposite parity and with $l\pm1$. For example, $\lambda_{23} \sim 6.5$PN is the average of $\lambda_2 \sim 5$PN and $\sigma_3 \sim 8$PN. This is reminiscent of the fact that the rotational tidal Love numbers are the coupling constants between moments of different parity and order.
\end{itemize}

Summarizing, the leading order corrections due to a multipole moment of order $l$ are
\begin{align}
{\rm PN~order}_{\rm electric~TLNs}=&\,2l+1\,,\nonumber\\
{\rm PN~order}_{\rm magnetic~TLNs}=&\,2l+2\,,\nonumber\\
 {\rm PN~order}_{\rm spin-TLNs}=&\,2l+\frac{5}{2}\,,\nonumber\\
 {\rm PN~order}_{\rm RTLNs}=&\,2l+\frac{1}{2}+2\delta_{l2}\,.
\end{align}
Note that for $l \geq 3$, the contribution of the rotational tidal Love numbers enters always at \emph{lower} PN order than the usual tidal Love numbers (both electric and magnetic) in the non-spinning case.

\subsection{Lagrangian formulation of the rotational tidal Love numbers} 
\label{sec:issue}
Within our truncation we consider the $l=2,3$ mass and current multipole moments of a spinning object. In a perturbative approach, to linear order in the spin, \emph{four} rotational tidal Love numbers are introduced to describe the coupling between these (four) multipole moments and the $l=2,3$ tidal moments~\cite{Pani:2015hfa,Pani:2015nua}. Respectively, $\lambda_{23}$ describes how a mass quadrupole moment is induced by an octupolar magnetic tidal moment, $\lambda_{32}$ describes how a mass octupole moment is induced by a quadrupolar magnetic tidal moment, $\sigma_{23}$ describes how a current quadrupole moment is induced by an octupolar electric tidal moment and $\sigma_{32}$ describes how a current octupole moment is
induced by a quadrupolar electric tidal moment.

The Lagrangian~\eqref{EintPN} implies the relations~\eqref{eq:coupling}, i.e., $\lambda_{23} \propto \sigma_{32}$ and $\lambda_{32} \propto \sigma_{23}$. However, these rotational tidal Love numbers have been computed numerically in~\cite{Pani:2015nua} for different equations of state (see section~\ref{sec:RTLN}). Their results show explicitly that the above putative relations are violated. Even supposing that there is some error in their computation, there is no reason in general to think that these four constants are not independent. This because the internal composition of the object (i.e., the neutron star equation of state) breaks eventual symmetries arising between different sectors. In other words, it would be very unlikely that some \emph{true} (not approximately as in section~\ref{sec:universal}) EOS independent relation exists among the rotational tidal Love numbers~\footnote{For instance, the axial and polar quasi-normal modes of a non-rotating object are isospectral only for the Schwarzschild black hole~\cite{Chandra:1983}. There is no EOS independent relation between them in the case of a neutron star.}. 

Yet, as we showed in the previous sections, such universal relations arise from our Lagrangian formulation. The reason for this is that our internal Lagrangian contains only \emph{two} coupling constants, $\alpha$ and $\beta$, which are responsible for the coupling, proportional to the spin, between multipole moments and tidal moments with opposite parity and $l\leftrightarrow l\pm1$. In other words, a Lagrangian formulation seems to predict \emph{two} rotational Love numbers, rather than four. This fact seems intrinsically related with the Lagrangian formulation, which introduces the same coupling constant in two different Euler-Lagrange equations (two fields are coupled to each other through one constant). For instance, the coupling $\alpha J_2^a Q^{bc} S^{abc}$ in $\mathcal{L}_2^{\mathrm{int}}$ contributes to the Euler-Lagrange equations for both $Q^{ab}$ and $S^{abc}$. In the former case it gives a term $\sim \alpha J_2^a S^{abc}$ (see Eq.~\eqref{eq:adiabaticquad}), whereas in the latter one it gives $\sim\alpha J_2^a Q^{bc}$ (see Eq.~\eqref{eq:adiabaticoctu}). In both cases the terms depend on the same coupling constant, $\alpha$.

Unfortunately, we have not been able to solve this apparent inconsistency. Possible explanations could be: (i) the Lagrangian formulation fails to reproduce the full couplings that arise in perturbation theory, which however we think unlikely. (ii) The numerical computation of the Love numbers carried out in~\cite{Pani:2015nua} is wrong, and some hidden symmetry of the perturbative equations effectively implies a relation among the Love numbers. (iii) There is a non-trivial stationary limit of the dynamical action describing the time evolution of the induced multipole moments~\cite{Steinhoff:2016rfi,Hinderer:2016eia}, which reduces the degrees of freedom. We will investigate this issue in future work. However, we stress that the expression for the gravitational wave phase in Eq.~\eqref{phase} can also accommodate putative relations among the rotational tidal Love numbers. Indeed, we have derived it assuming that they are independent, but this does not change the result. The only explicit contribution from the coupling constants $\alpha$ and $\beta$ appears in the reduced Lagrangian~\eqref{eq:redlag}, but it cancels out in the conserved energy (see the discussion below Eq.~\eqref{cons_en}), not affecting the final expression of the gravitational wave phase.

\section{Impact of spin-tidal effects on parameter estimation}
\label{sec:impact}
In this section we estimate the impact of the new spin-tidal terms computed in this thesis on the parameter estimation of binary neutron stars, in particular on the measurement of the electric, quadrupolar tidal deformabilities. We focus on the tidal part of the gravitational waveform up to 6.5PN order, i.e., we neglect higher-order PN terms due to octupolar deformations. Moreover, we neglect the rotational tidal Love numbers, because of the unresolved issue with the Lagrangian formulation discussed in the previous sections. However, we estimate that the contribution of the latter is roughly of the same order of the other spin-tidal terms (see the quasi-universal relations for the rotational tidal Love numbers in~\cite{Gagnon-Bischoff:2017tnz}). Thus, the tidal terms in the gravitational wave phase considered are (henceforth we set $c=1$)
 \begin{equation}
 \label{eq:tidal_phase}
 \psi_{T}(x) = \frac{3}{128 \nu x^{5/2}} \left\{ -\frac{39}{2}\tilde\Lambda x^5   + \left(\delta\Lambda+\tilde\Sigma \right) x^6 + \left(\frac{39}{2} \pi \tilde\Lambda+  \hat\Lambda+\hat\Sigma \right) x^{6.5} \right\}\,, 
\end{equation}
where
\begin{align}
\label{defLambda}
 \tilde\Lambda =&\frac{16}{13}\left(\frac{12}{\eta_1}-11\right) \eta_1^5\Lambda_1+
  (1\leftrightarrow 2)\,, \\
 \delta\Lambda=&\left( \frac{4595}{28}- \frac{15895}{28 \eta_1} + \frac{5715 
\eta_1}{14} -
  \frac{325 \eta_1^2}{7}  \right)\eta_1^5\Lambda_1 +(1\leftrightarrow 2) \,,\\
 \tilde\Sigma =& 
  \left( \frac{6920}{7} - \frac{20740}{21 \eta_1} \right) 
  \eta_1^5\Sigma_1 +(1\leftrightarrow 2)\,, \\
\label{hatLambdaE}
\hat\Lambda =&
  \left[ \left( \frac{593}{4} - \frac{1105}{8 \eta_1} +\frac{567 \eta_1}{8} 
-81 
\eta_1^2 \right) \chi_2+ \left( -\frac{6607}{8} +\frac{6639 \eta_1}{8} -81 \eta_1^2 \right) 
\chi_1 \right] \eta_1^5\Lambda_1 \nonumber \\
& +(1\leftrightarrow 2)\,, \\
\label{hatLambdaM}
 \hat\Sigma =&\left[\left(-\frac{9865}{3} + \frac{4933}{3 \eta_1} + 1644 
\eta_1 \right) \chi_2 -\chi_1 \right]
 \eta_1^5\Sigma_1 +(1\leftrightarrow 2) \,,
 \end{align}
whereas $\Lambda_A=\lambda_2^{(A)}/(\eta_A M)^5$ and $\Sigma_A=\sigma_2^{(A)}/(\eta_A M)^5$ ($A=1,2$). Note that with the above notation, $\tilde \Lambda$ coincides with the (quadrupolar, electric) average-weighted tidal deformability, recently constrained by the LIGO/Virgo collaboration to be $\tilde \Lambda = 300^{+ 500}_{-190}$ ($90\%$ symmetric credible interval), for the binary neutron star merger GW170817~\cite{Abbott:2018wiz}. $\Lambda_1$, $\Lambda_2$ are the dimensionless (quadrupolar, electric) tidal deformabilities of the single stars. For equal masses, and assuming that the neutron stars are described by the same equation of state~\cite{Abbott:2018exr}, $\Lambda_1 = \Lambda_2 = \tilde \Lambda$.

\begin{figure}[]
\centering
\includegraphics[width=0.99\textwidth]{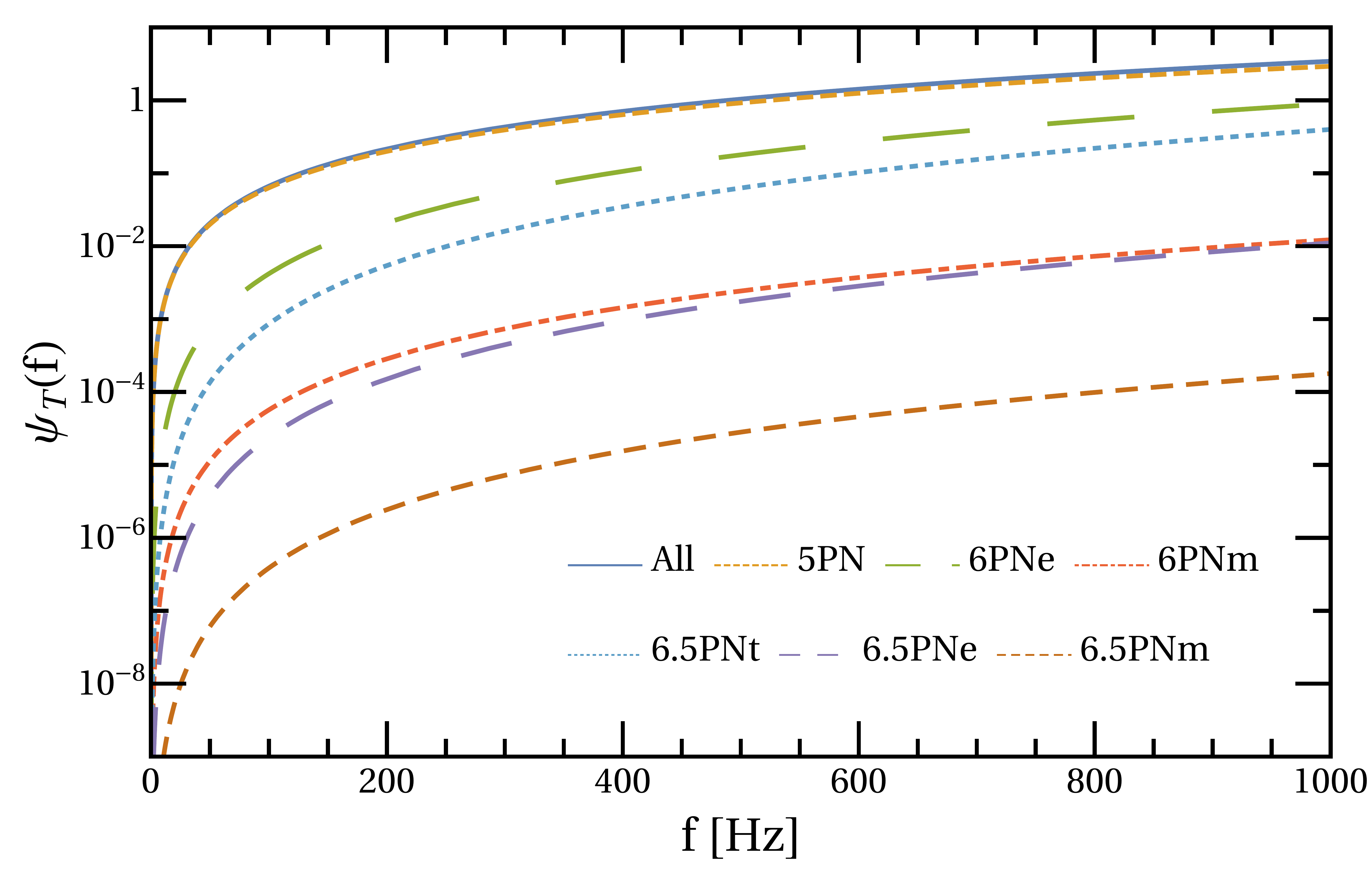}
\caption{\textsl{Contribution (in absolute value) of each tidal term in Eq.~\eqref{eq:tidal_phase} to the gravitational wave phase as a function of the frequency. We considered a $1.4 M_{\odot} $ equal-mass binary, with spins $\chi_1= \chi_2 = 0.05$ and tidal deformabilities $\Lambda_1 = \Lambda_2 =300$, and computed the magnetic tidal Love numbers using the quasi-universal relation~\eqref{eq:fit} in the irrotational case. See the text for the description of the various terms.}}
\captionsetup{format=hang,labelfont={sf,bf}}
\label{fig:phase_tidal}
\end{figure}

First of all, we show qualitatively the contribution of each individual term in Eq.~\eqref{eq:tidal_phase} to the overall tidal part of the gravitational wave phase. In Fig.~\ref{fig:phase_tidal} we plot the absolute value of the different PN tidal terms as a function of the gravitational wave frequency $f$ (which is related to the PN parameter by $x= (\pi M f)^{2/3}$), for an equal-mass ($\eta_1= \eta_2 = 1/2$, $\nu = 1/4$), GW170817-like, binary neutron star. The parameters of the source are: total mass $M =2.8 M_{\odot}$, dimensionless spins $\chi_1 = \chi_2 = 0.05$ and tidal deformabilities $\Lambda_1 = \Lambda_2 =300$. We choose the value of the electric tidal deformability to be compatible with the median value reported by the LIGO/Virgo collaboration for a similar system~\cite{Abbott:2018wiz}. Moreover, we set the values of the spins equal to the upper limit of the low-spin prior used by the collaboration (see the discussion in section~\ref{sec:LIGOspin}), to enhance the effect of the $6.5$PN order spin-tidal terms (which scale linearly with the spins). The range of frequencies includes that relevant for second-generation detectors before the merger of the binary (cf. section~\ref{sec:res}). We compute the magnetic tidal Love numbers using the quasi-universal relation~\eqref{eq:fit}, for irrotational fluids (which is the case more physical). We stress that $\Lambda_A/\Sigma_A \sim 100$.

We can see that the leading 5PN order term (proportional to $\tilde \Lambda$) dominates by far the tidal part of the gravitational wave phase, representing almost the entire contribution to the waveform~\footnote{Note that in Fig.~\ref{fig:phase_tidal} we have plotted the absolute value of each term, while in general different contributions can sum up incoherently (for instance, the 5PN order coefficient and the tail-tidal term have opposite sign).}. The next term in order of importance is the 6PN order correction to the electric Love number, $\delta \Lambda$ (6PNe in the plot), which contributes on average about the $20\%$ of the total tidal phase evolution ($\sim 17\%$ at $f=500$ Hz). Currently, the gravitational wave templates used by the LIGO/Virgo collaboration include only up to this term. Other sub-leading PN terms are not accounted for in the analysis~\cite{Abbott:2018wiz}. The next term in order of relevance is the 6.5PN order tail-tidal coefficient (proportional to $\pi$ in Eq.~\eqref{eq:tidal_phase}, 6.5PNt in the plot). The spin-tidal term $\hat \Lambda$ (6.5PNe in the plot) is significantly less dominant: its relative contribution is smaller than the total tidal phase by about two orders of magnitude, and contributes about a $3\%$ with respect to the total 6.5PN order coefficient. This suggests that it might be neglected for binaries with $\chi_A \lesssim 0.05$. On the other hand, this term grows linearly with the spin, therefore it might become important if highly-spinning neutron star binaries exist in Nature (see the discussion in section~\ref{sec:LIGOspin}). We quantify its impact in section~\ref{sec:res}. Finally, the lowest contributions come from the magnetic tidal Love numbers. This is a consequence of the small ratio between the magnetic and the electric tidal Love numbers shown in section~\ref{sec:universal}. We can see that the contributions of the magnetic Love numbers $\tilde \Sigma$ and $\hat \Sigma$ (6PNm and 6.5PNm in the plot, respectively) are smaller than the respective electric terms by about two orders of magnitude. Furthermore, the leading 6PN order magnetic term is comparable to 6.5PN order spin-tidal electric coefficient, suggesting that also the impact of the magnetic tidal Love numbers is small~\cite{Jimenez-Forteza:2018buh} (and moreover it does not grow larger with the spin). Considering static fluids, rather than irrotational ones, would not change the situation (see section~\ref{sec:universal}). In general, the parameter space of binary neutron stars is not large, therefore these results do not differ sensibly for binary systems with different parameters.

\subsection{Statistical framework}
\label{sec:stat}
In this section we briefly review the statistical tools needed for our analysis. The output of a gravitational wave detector is a data stream $d(t)$ (i.e., a time series), containing a given realization of the detector noise $n(t)$, and possibly a time-domain gravitational wave signal $h(t)$:
\begin{equation}
d(t) = h(t) + n(t) \,.
\end{equation}
We assume that the detector noise is described by a stochastic process that is:
\begin{description}
\item[stationary] Stationarity means that the probability distribution of the process is invariant under time translations, $P_n(t) = P_n(t+ \tau)$. This automatically implies that the average (over the ensemble) of the process is constant, $\langle n(t) \rangle = const $, and the autocorrelation function $R(t,t') = \langle n(t) n(t') \rangle$ depends only on the time difference $\tau=t'-t$, $R(\tau) = \langle n(t) n(t+\tau) \rangle$. Since the mean value is constant, without loss of generality it can be assumed $\langle n(t) \rangle=0$.
\item[ergodic] Ergodicity means that a single realization allows us to determine the properties of the stochastic process. In other words, the ensemble averages can be replaced by averages over time, $ \langle  \dots  \rangle = \lim_{T \to \infty} \frac{1}{T} \int_0^T \dots \, dt$.
\item[Gaussian] Gaussian processes have normal probability distribution and are completely specified by the average $\langle n(t) \rangle$, and the autocorrelation function $R(t,t')$. 
\end{description}
A stochastic process which satisfies all the above properties can then be characterized in terms of its autocorrelation function $R(\tau)$ only, or, going to the frequency domain, by the \emph{one-side power spectral density} $S_n(f)$, defined by
\begin{equation}
\frac{1}{2}  S_n(f) = \int_{-\infty}^{+\infty} R(\tau) \mathrm{e}^{\mathrm{i} 2 \pi f \tau} \, d\tau \,.
\end{equation}
Note that, since $R(\tau)$ is real and invariant under time translation, $S_n(f)$ is real and symmetric for $f \to -f$.

Generally, in gravitational wave data analysis one has to dig out a gravitational signal buried in a noise, which is comparable or much larger (in amplitude) than the signal itself, $|n(t)| \gg |h(t)|$. If one knows the functional form $h(t)$ of the signal which (s)he is looking for, then it is possible to apply the \emph{matched filtering} technique~\cite{Maggiore:2008}. More specifically, we define the linear functional
\begin{equation}
\hat{K}[f(t)] = \int_{-\infty}^{+\infty} f(t) K(t) dt \,,
\end{equation}
and we ask for the filter function $K(t)$ which maximizes the signal-to-noise ratio (SNR) $\rho \equiv S/N$, where $S$ is the expectation value of $\hat K$ when the signal is present, and $N$ the root mean square value of $\hat K$ when there is only noise:
\begin{equation}
\rho \equiv \frac{S}{N} = \frac{\left \langle \hat K[d(t)] \right \rangle}{\sqrt{\left  \langle  \hat K^2[n(t)]  \right \rangle }} \,.
\end{equation} 
Going to the frequency domain, it can be shown that the filter function which gives the optimal value of $\rho$ is
\begin{equation}
\label{eq:wiener}
\tilde K(f) = c \frac{\tilde h(f)}{S_n(f)} \,,
\end{equation} 
where $\tilde K(f)$ and $\tilde h(f)$ are the Fourier transforms of the filter function and of the gravitational signal, respectively, and $c$ is an unimportant overall constant which factors out in the SNR. The filter function in Eq.~\eqref{eq:wiener} is called \emph{matched filter}, or \emph{Wiener filter}. Using the optimal filter, it is straightforward to show that the SNR can be written in the frequency domain as
\begin{equation}
\label{eq:optimal}
\rho^2 = 4 \int_{0}^{\infty} \frac{|\tilde h(f)|^2}{S_n(f)} \, df \,,
\end{equation}
where we have used the fact that $\tilde h(f) =\tilde h^*(-f) $ and $\tilde K(f) =\tilde K^*(-f) $, being $h(t)$ and $K(t)$ real functions.

However, in practical situations, one does not know the exact functional form of the gravitational signal (assuming that a signal is actually present in the data stream). Therefore, one constructs a bank of \emph{waveform templates} $h_{T}(t; \vec{\theta}_T)$, depending on a set of parameters $\vec{\theta}_T$, and look for the combination of parameters whose corresponding optimal filter $\tilde K_T(f; \vec{\theta}_T) =  \tilde h_T(f; \vec{\theta}_T)/S_n(f)$ makes the SNR exceeding a given threshold. In other words, the combination of parameters, which maximizes the SNR, gives an estimate of the \emph{true} parameters $\vec{\theta}_0$ of the gravitational signal $h(t, \vec{\theta}_0)$. In practice, the SNR is given by the scalar product between two waveforms
\begin{equation}
\label{eq:snr}
\rho^2 = (h|h_T) = 4 \, \text{Re} \int_{f_{min}}^{f_{max}} \frac{\tilde h(f) \, \tilde h_T^*(f)}{S_n(f)}\, df \,,
\end{equation}
where $f_{min}$ and $f_{max}$ delimit the range of frequencies where the detector is actually sensitive. Note that when the template matches perfectly the signal, Eq.~\eqref{eq:snr} reduces to Eq.~\eqref{eq:optimal}.

The crucial issue in waveform modeling is providing a waveform template whose functional dependence on time $h_T(t)$ (or, equivalently, on frequency $\tilde h_T(f)$) follows closely that one of the gravitational signal detected. Indeed, if $h(t) \neq h_T(t)$, the combination of parameters $\vec{\theta}_T$ which maximizes the SNR is different from the true parameters of the source $\vec{\theta}_0$. In other words, a poor approximation of the true gravitational waveform can introduce a non-negligible bias in the parameter estimation of the source, as we discuss in the following.

We can distinguish the waveform parameters in \emph{intrinsic} and \emph{extrinsic} ones. The intrinsic parameters $\vec{\gamma}$ are the physical parameters which characterize the source of the gravitational radiation. For instance, in binary systems they are the masses, spins, tidal deformabilities, etc. The extrinsic parameters $\vec{\xi}$ account instead for sky-position, wave polarizations, distance to the source, etc., and for compact binaries include also the unphysical waveform parameters $\phi_c$ and $t_c$, namely the phase and the time at the coalescence (see Eq.~\eqref{eq:phasefreq}). 

The extrinsic parameters are irrelevant for waveform modeling purposes, since they can be naturally factored out. For binary systems, the dependence of the SNR on them is removed fixing or averaging over the sky-locations, normalizing to remove the amplitude scaling with the distance, and maximizing over the unphysical parameters $\phi_c$, $t_c$. This procedure defines the \emph{match} between two waveforms, which replaces the SNR in many waveform modeling computations:
\begin{equation}
\label{eq:match}
\mathcal{M}[h(\vec{\gamma}_0),h_T(\vec{\gamma}_T)]=\underset{\phi_c,t_c}{max}\frac{\left( h | h_T 
\right)}{\sqrt{\left( h | h \right) \left( h_T | h_T 
\right)}} \,.
\end{equation}
The match is a useful tool to measure the metric distance between two waveform representations. Indeed, since the scalar product $(\cdot | \cdot)$ is positive definite, $\mathcal{M}\in[0,1]$, with $\mathcal{M}=1,0$ being perfect and zero match, respectively. In general, $\mathcal{M}$ is used as an indicator of the performance of waveform models. For high SNR and Gaussian noise, the match may be used to provide an estimate of the systematic errors produced by the different waveform representations, as we show below.

The parameter estimation of gravitational wave signals is based on the application of Bayesian probability theory (see, e.g.,~\cite{Dagos:2003}) to the observed data streams~\cite{Cutler:1994ys,Maggiore:2008}. For Gaussian noise, the probability distribution of a given noise realization reads
\begin{equation}
\mathcal{P}(n) \propto \mathrm{exp} \left[ -\frac{1}{2} (n|n)\right] \,.
\end{equation}
Focusing only on the intrinsic parameters, replacing $n = d- h(\vec{\gamma}_0)$ in the above equation, and using Bayes theorem, we get the posterior probability distribution of the true parameters of the gravitational waveform given the data:
\begin{equation}
\mathcal{P}(\vec{\gamma}_0|d) \propto \mathrm{exp} \left[ -\frac{1}{2} (d-h(\vec{\gamma}_0)|d-h(\vec{\gamma}_0))\right] \mathcal{P}_0 (\vec{\gamma}_0) \,,
\end{equation}
where $\mathcal{P}_0 (\vec{\gamma}_0)$ is the prior information on the parameters.
Since we reconstruct the source parameters using a waveform template, the last equation is actually written in terms of the template parameters $\vec{\gamma}_T$,
\begin{equation}
\label{eq:wave_bayes}
\mathcal{P}(\vec{\gamma}_T|d) \propto  \mathrm{exp} \left[ -\frac{1}{2} (d-h_T(\vec{\gamma}_T)|d-h_T(\vec{\gamma}_T))\right] \mathcal{P}_0 (\vec{\gamma}_T) \,.
\end{equation}
For high SNR and assuming flat priors, the above equation can be simplified neglecting the noise-related factors~\cite{Chatziioannou:2017tdw}, reducing to
\begin{equation}
\label{eq:probFMmr}
\mathcal{P}(\vec{\gamma}_T) \propto \exp\left[-\rho^2\left (1-\mathcal{M}[h(\vec{\gamma}_0),h_T(\vec{\gamma}_T)] \right)\right]\,,
\end{equation}
where $h(\vec{\gamma}_0)$ is the true gravitational wave signal embedded in $d$. The above equation allows us to describe completely the statistics in terms of the SNR and of the match $\mathcal{M}[h,h_T]$. In other words, for a given SNR $\rho$ and a given template $h_T$, the mismatch $1-\mathcal{M}$ determines the probability distribution around the true values $\vec{\gamma}_0$. Note that the true parameters given by $\vec{\gamma}_0$ do not correspond to the recovered ones $\vec{\gamma}_T$, unless the real (or injected) waveform and the template bank used are equal, $h=h_T$. This may insert non-negligible systematic errors, that in some cases may compete in significance with the statistical ones. Thus, if we replace $h$ by a given waveform template, Eq.~\eqref{eq:probFMmr} allows us to estimate the impact of using one or another waveform template in our parameter estimation. In the next section we evaluate these effects for the spin-tidal PN corrections in Eq.~\eqref{eq:tidal_phase}.

On the other hand, an estimate of the statistical errors is provided by the \emph{Fisher matrix} approach~\cite{Cutler:1994ys,Vallisneri:2007ev}. For high SNR, the gravitational signal in Eq.~\eqref{eq:wave_bayes} can be consider a linear function of its parameters. Neglecting the possible discrepancy between the real waveform and the template bank (i.e., assuming $h=h_T$), and expanding $h_T(\vec{\gamma}_T)$ around the true values $\vec{\gamma}_0$,
\begin{equation}
h_T(\vec{\gamma}_T) \sim h_T(\vec{\gamma}_0) + \left(\gamma_T^i-\gamma_0^i \right) \frac{\partial h_T}{\partial \gamma_T^i}\bigg{|}_{\vec{\gamma}_T=\vec{\gamma}_0}  + \dots \,,
\end{equation}
Eq.~\eqref{eq:wave_bayes} reduces to the Gaussian
\begin{equation}
\label{eq:probFMs}
\mathcal{P}(\vec{\Delta \gamma})\propto \mathrm{exp}\left[-\frac{1}{2}\Gamma_{ij} \Delta \gamma^i \Delta \gamma^j\right] \,,
\end{equation}
where
\begin{equation}
\label{eq:fishermatrix}
\Gamma_{ij}=\left(\frac{\partial h_{T}}{\partial\gamma_{T}^i} \bigg{|}
 \frac{\partial h_{T}}{\partial\gamma_{T}^j}\right)\bigg{|}_{\vec{\gamma}_T=\vec{\gamma}_0}
\end{equation}
is the Fisher information Matrix (FIM), and $\vec{\Delta \gamma}=(\vec\gamma_T-\vec{\gamma}_0)$. Note that the linear term in Eq.~\eqref{eq:probFMs} vanishes, because we are expanding the posterior $\mathcal{P}(\vec{\gamma}_T)$ around its maximum $\vec{\gamma}_T=\vec{\gamma}_0$. Thus, Eq.~\eqref{eq:probFMs} gives the probability of having each of the reconstructed parameters shifted by $\Delta \gamma^i=(\gamma_T^i-\gamma_0^i)$, from the real values. The covariance matrix on the reconstructed parameters is given by the inverse of the Fisher matrix: $\Sigma_{ij} = \Gamma^{-1}_{ij}$.

Then, we can compute the value of the $D$-dimensional posterior when each of the reconstructed parameters is $n\,\sigma$ away from the maximum-likelihood ones, $|\Delta \gamma^i| =n \sigma_{\gamma^i}$, where $\sigma_{\gamma^i} = \sqrt{\Gamma^{-1}_{ii}}$ is the statistical error on the $i$-th parameter $\gamma_T^i$. The result is
\begin{equation}
\label{eq:probFMs2}
\mathcal{P}(\vec{\gamma}_{n\, \sigma}) \propto \mathrm{exp}\left[-\frac{1}{2}\Gamma_{ij} n^2 \sigma_{\gamma^i} \sigma_{\gamma^j}\right] \,,
\end{equation}
where we have omitted the plus/minus sign in $\Delta \gamma^i$ because it cancels out in the following~\footnote{A weaker requirement would have been looking for the \emph{global} $n\, \sigma$ confidence level hypersurface. Since in the FIM approximation the posterior distribution is Gaussian, this surface is a $D$-dimensional ellipsoid, and
\begin{equation}
\mathcal{P}(\vec{\gamma}_{n\,\sigma})\propto \exp\left({-\frac{1}{2} r^2 }\right) \,,
\end{equation}
where $r$ is the Mahalanobis distance
\begin{equation}
r^2 = \phi(c(n),D) \,,
\end{equation}
with $\phi$ the inverse of the cumulative distribution of the $\chi^2$-distribution with $D$ degrees of freedom, and $c(n)$ the probability of falling inside the $n\, \sigma$ confidence region ($c(1) \sim 0.68, \, c(2) \sim 0.95, \, \text{etc}.$).}. Neglecting the correlations among the parameters~\footnote{This assumption is justified by the fact that we are interested mainly in the weighted-tidal deformability $\tilde{\Lambda}$ parameter, which at high SNR is weakly correlated to the other parameters (cf. section~\ref{sec:res}).}, Eq.~\eqref{eq:probFMs2} reduces to
\begin{equation}
\label{eq:probFMs3}
\mathcal{P}(\vec{\gamma}_{n\, \sigma})\propto \mathrm{e}^{-\frac{D}{2}n^2} \,.
\end{equation}

We can now compare systematic and statistical uncertainties in gravitational wave data analysis. By equating Eqs.~\eqref{eq:probFMmr} and~\eqref{eq:probFMs3}, one gets~\cite{Chatziioannou:2017tdw}
\begin{equation}
\label{eq:ind}
2 \rho^2 (1-\mathcal{M}) = D \,n^2  \,.
\end{equation}
The above expression allows one to define the \emph{distinguishability criterion} between two waveform models, previously derived in terms of the waveform amplitude and phase in~\cite{Lindblom:2008cm}. In other words, it allows us to estimate the minimum SNR required to distinguish two waveform models within a certain ${n=\Delta \gamma}/{\sigma}$ significance, with the latter ratio equal to unity to distinguish two models with $1\sigma$ significance, for instance. In the next section, we use this definition as a quantitative indicator of the impact of the new spin-tidal terms.

\subsection{Results of the parameter estimation analysis}
\label{sec:res}
Systematic uncertainties on gravitational wave parameter estimation are induced by the incompleteness of the waveform template banks. This may produce an artificial bias with respect to the true parameters, that in some cases may overtake the statistical uncertainties driven by the detector noise~\cite{Favata:2013rwa}. In this section, we evaluate the impact of neglecting the 6.5PN order spin-tidal terms in Eq.~\eqref{eq:tidal_phase}, comparing different analytic waveform models. In particular, we want to estimate how much this would affect the measurement of the average tidal deformability $\tilde \Lambda$ parameter.

We stress that these new terms modify the waveform at high PN order. This implies that their effects gain importance as the signal approaches the high-frequency regime, possibly probing a region where current gravitational detectors are less sensitive. In general, the impact of these terms depends on the source parameters and on the merger frequency relative to the detector sensitivity. Since the parameter range of neutron star mergers is not large (total mass $M \in [2 \div 4]M_{\odot}$ and mass ratio $\eta_1/\eta_2 \in [1 \div 2] $), different configurations do not affect significantly the results. Therefore, as explained in the following, we restrict our analysis to a prototype $1.4M_{\odot}$ equal-mass binary neutron star, compatible with the only event detected so far by the LIGO/Virgo collaboration, GW170817~\cite{TheLIGOScientific:2017qsa,Abbott:2018wiz}. 

Furthermore, in order to maximize the effect of the tidal terms, we consider second-generation detectors at design sensitivity and third-generation detectors. More specifically, we consider: (i) the LIGO interferometer in its zero-detuned, high-power configuration~\cite{zerodet} (see also the new updated LIGO sensitivity curve~\cite{zerodet2}), and (ii) the planned Einstein Telescope (ET) interferometer in the so-called ET-D xylophone configuration\cite{Hild:2010id,eintel}. The current prospects for these two interferometers predict a sensitivity gain of a factor $\sim 3$ and $\sim 45$ (see below), for LIGO and the ET, respectively, compared to current second-generation detectors.

\begin{figure}[]
\centering
\includegraphics[width=0.99\textwidth]{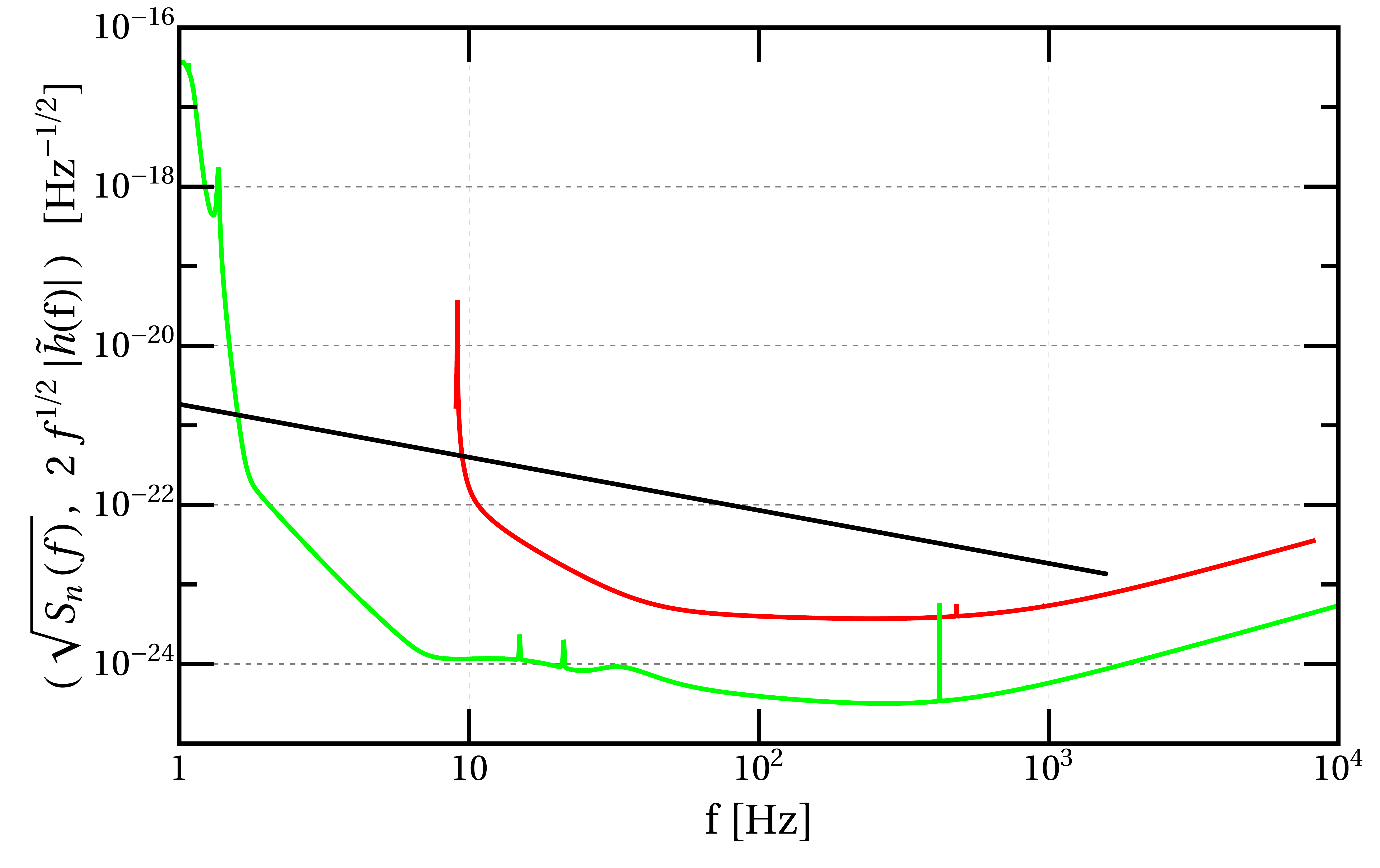}
\caption{\textsl{Power spectral density curves for LIGO at design sensitivity (in red) and ET-D (in green). The black line is the strain of the gravitational signal emitted by a $1.4M_{\odot}$ equal-mass binary neutron star at $100\,$Mpc, during the inspiral phase. The ending frequency corresponds to the innermost stable circular orbit (ISCO), $r=6M $, of a Schwarzschild black hole with mass equal to the total mass of the system, $f=(6^{3/2} \pi M )^{-1} \sim 1570\,$Hz. Roughly speaking, signals which lie above each curve can be detected by the corresponding interferometer.}}
\captionsetup{format=hang,labelfont={sf,bf}}
\label{fig:sensitivity}
\end{figure}

In Fig.~\ref{fig:sensitivity} we show the (square root of the) power spectral density sensitivity curve $\sqrt{S_n(f)}$ of the two detectors, for comparison. We plotted also the strain amplitude $2 f^{1/2}|\tilde h(f)|$~\footnote{The normalization of the gravitational signal is chosen in such a way that it can be directly compared to the detector sensitivity in the definition of the SNR. Indeed, Eq.~\eqref{eq:optimal} can be rewritten as
\begin{equation}
\rho^2 = 4 \int_{0}^{\infty} \frac{|\tilde h(f)|^2}{S_n(f)} \, df = \int_{-\infty}^{+\infty} \left(\frac{2 f^{1/2} |\tilde h(f)|}{\sqrt{S_n(f)}} \right)^2 \, d(\log{f}) \,.
\end{equation}
} of the gravitational signal emitted by a $1.4M_{\odot}$ equal-mass binary neutron star in the inspiral phase, at the prototype distance of $d=100\,$Mpc. Note that at the leading PN order (quadrupole approximation) the amplitude $\mathcal{A}(f)$ of the gravitational waveform (see Eq.~\eqref{eq:amplitudefreq}) is $\mathcal{A}(f) \sim \mathcal{M}_c^{5/6} f^{-7/6}/d $, where $\mathcal{M}_c= M \nu^{3/5}$ is the chirp mass.

Regarding the gravitational wave templates, we model the waveform phase adding the tidal phase in Eq.~\eqref{eq:tidal_phase} to the standard PN point-particle TaylorF2 phase up to 3.5PN order and to linear order in the spin~\cite{Khan:2015jqa}. Consistently with the previous sections, we neglect quadratic and higher-order spin corrections. With this choice, the point-particle phase does not depend on the spin-induced quadrupole moments of the binary components, which are quadratic in the spin and depend on the equation of state (cf. section~\ref{sec:universal}). For the waveform amplitude we consider only the Newtonian order, since amplitude corrections are negligible with respect to phase deviations in parameter estimation analysis~\cite{Cutler:1994ys}. We notice that since all the spin-dependent terms in the entire PN phase are linear in the spin, our results will be symmetric under spin inversion, $\chi_i\to -\chi_i$. Furthermore, we include the magnetic tidal Love numbers through the quasi-universal relations~\eqref{eq:fit}, though we have shown in section~\ref{sec:impact} that their contribution is negligible, compared to the electric Love numbers. Since they are more realistic (see section~\ref{sec:axial}), we consider only the magnetic tidal Love numbers arising from an irrotational fluid. However, considering the magnetic Love numbers for static fluids does not affect the results, because, except for the sign, the magnitude of the Love numbers in the two cases is comparable (see section~\ref{sec:universal}).

To quantify the effect of the spin-tidal 6.5PN order contributions, we make an analysis based on the match/FIM distinguishability described in the previous section. In particular, our analysis is valid
for high SNR and Gaussian noise. We explore the possibility of detecting a (injected) gravitational wave signal $h$, and to reconstruct its parameters, through the matched filter procedure, with a waveform template bank $h_T$. For the sake of clarity, the two waveform phases considered are the following:
\begin{itemize}
\item{\textbf{The gravitational signal $\boldsymbol{h}$:} point-particle TaylorF2 waveform phase plus all the terms present in tidal phase~\eqref{eq:tidal_phase}, containing in particular the new spin-tidal terms $\hat \Lambda$ and $\hat \Sigma$.}
\item{\textbf{The waveform template $\boldsymbol{h_T}$:} point-particle TaylorF2 waveform phase plus the tidal phase~\eqref{eq:tidal_phase}, setting to zero the spin-tidal terms, $\hat \Lambda= \hat \Sigma=0$.}
\end{itemize}
In the following we show our results for LIGO and the ET.

\subsubsection{LIGO}
\label{sec:LIGOspin}
We take the only binary neutron star event observed so far by the LIGO/Virgo collaboration, GW170817, as a reference~\cite{TheLIGOScientific:2017qsa,Abbott:2018wiz}. This event, observed with a SNR of $\rho = 32.4$~\footnote{For a network of $N$ detectors the SNR scales approximately as $\sqrt{N}$~\cite{Cutler:1994ys}}, was consistent with a binary neutron star system with masses compatible to 1.4 solar masses, $M_1 \sim M_2 \sim 1.4 M_\odot$, and with spins compatible to zero, $\chi_1 \sim \chi_2 \sim 0$. Moreover, the $90\%$ symmetric credible interval on the average-weighted (electric, quadrupolar) tidal deformability $\tilde\Lambda$ has been recently constrained to lie within $\tilde\Lambda \in [110,800]$, with the median value being $\tilde{\Lambda}=300$~\cite{Abbott:2018wiz}.

\begin{figure}[]
\centering
\includegraphics[width=0.99\textwidth]{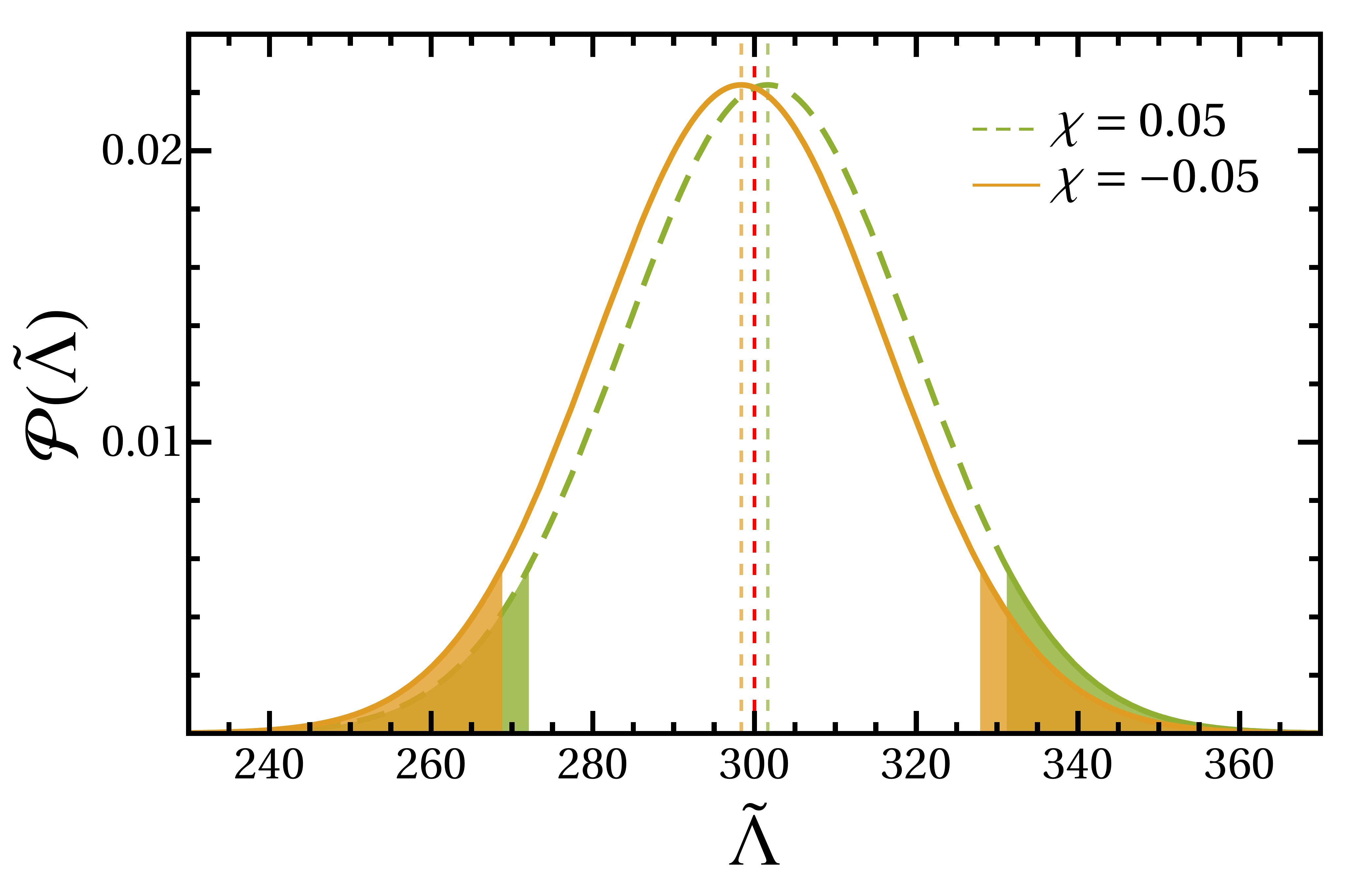}
\caption{\textsl{Probability distributions obtained for a $1.4M_{\odot}$ equal-mass binary spinning at $\chi=0.05$ (dashed-green) and $\chi=-0.05$ (orange), with the zero-detuned LIGO noise sensitivity curve and a SNR $\rho =100$. The vertical dashed lines define the maximum-posterior values, while the solid areas define the $90\%$ credible intervals. The red dashed vertical line defines the injected value $\tilde\Lambda_0=300$.}}
\captionsetup{format=hang,labelfont={sf,bf}}
\label{fig:LIGOlowspin}
\end{figure}

Based on the above discussion, we examine a \emph{standard} scenario where the physical parameters of the system are taken to be those compatible with GW170817. We consider an equal-mass binary neutron star with masses $M_1=M_2=1.4M_\odot$, and tidal deformabilities (assuming the neutron stars described by the same equation of state~\cite{Abbott:2018exr}) $\Lambda_1 = \Lambda_2 =\tilde{\Lambda}=300$.

The component spins in binary neutron stars are expected to be small, $|\chi_{1,2}| \lesssim 0.05$. Indeed, though the distribution of neutron star spins is uncertain, old neutron stars in the late stages of a binary inspiral are expected to rotate rather slowly. The most fastly-spinning neutron star observed so far in a compact system is the most massive component of the double pulsar system PSR J0737-3039A~\cite{Burgay:2003jj}, with a spin period of $\sim 23\,{\rm ms}$, which corresponds to $\chi\sim0.02\div0.05$, depending on the equation of state~\cite{Brown:2012qf,Kastaun:2013mv}. Such rotation rate is not expected to decrease substantially as this system approaches the merger~\cite{Dietrich:2015pxa}. On the other hand, the observation of numerous isolated millisecond pulsars (a spin period of 1 ms corresponds roughly to a dimensionless spin $\chi \sim 0.4$) suggests that spin rates as high as $\chi\sim0.1$ might be found also in binary neutron star systems~\cite{Dietrich:2015pxa}.

To maximize the effect of the spin-tidal coupling we set $\chi_1 = \chi_2 = \chi=\pm 0.05 $, consistently with the upper limit of the low-spin prior used by the LIGO/Virgo collaboration~\cite{TheLIGOScientific:2017qsa,Abbott:2018wiz}. Finally, we assume that the injected gravitational wave signal would have been detected by current second-generation detectors, with a SNR equal to that of the GW170817 event, $\rho=32.4$. Since at design sensitivity one expects a gain factor $\sim 3$, we set $\rho = 100$ for the zero-detuned configuration of LIGO.

We are interested in the measurement of the $\tilde \Lambda$ parameter, therefore we restrict our analysis to one dimension. We fix masses and spins at the true values chosen above, and vary only $\tilde\Lambda$ in the computation of the match in Eq.~\eqref{eq:match} (we recall that $\Lambda_1=\Lambda_2 = \tilde \Lambda$ for equal masses and the same equation of state). Varying only $\tilde \Lambda$ we compute the match $\mathcal{M}[ h(\tilde \Lambda_0),h_{T}(\tilde \Lambda)]$, where $\tilde \Lambda_0=300$ is the injected value. In the computation of $\mathcal{M}$, we fix the frequency sensitivity range for LIGO to $\left\lbrace f_{min},f_{max}\right\rbrace= \left\lbrace 9\, \mathrm{Hz},f_{ISCO}\right\rbrace$, where $f_{ISCO}\sim 1570\,$Hz is the frequency of the ISCO (see the caption of Fig.~\ref{fig:sensitivity}). Then, we translate the match to the probability distribution of the average tidal deformability, $\mathcal{P}(\tilde \Lambda)$, through Eq.~\eqref{eq:probFMmr}. The result for aligned ($\chi=0.05$) and anti-aligned ($\chi=-0.05$) spins is shown in Fig.~\ref{fig:LIGOlowspin}.

We can see that the probability distributions are Gaussian (consistently with the FIM approximation), and match almost perfectly the predictions described by the template $h_T$, though the spin-tidal effects tend to induce a minimal shift on $\mathcal{P}(\tilde\Lambda)$ that depends on the sign of the spin. The impact on the recovery of $\tilde\Lambda$ is absolutely negligible. The sign of the offset tends to overestimate and underestimate $\tilde\Lambda$, for positive (dashed and green) and negative (orange) spins, respectively. This can be explained observing the relation between the electric $6.5$PN order spin-tidal coefficient $\hat\Lambda$, and the leading $5$PN order $\tilde \Lambda$ one. We observe that for positive spins the two terms contribute with the same sign to the gravitational phase. Therefore, the lack of the spin-tidal terms in $h_T$ is compensated increasing the value of $\tilde \Lambda$. Conversely, for negative spins they are in counter-phase, and then a lower value of $\tilde \Lambda$ in $h_T$ provides a better match. Moreover, as anticipated, we notice that the results are perfectly symmetric under spin inversion $\chi_i \to - \chi_i$, which is reminiscent of the fact that the waveform is linear in the spin. 

Thus, Fig.~\ref{fig:LIGOlowspin} shows that the spin-tidal coefficients for an event fully compatible with GW170817 are negligible, when assuming a detection with LIGO at design sensitivity. A more refined multi-dimensional analysis, including all the parameters and the correlations among them, could only strengthen this (negative) result.

\subsubsection{ET}
\begin{figure}[]
\centering
\includegraphics[width=0.99\textwidth]{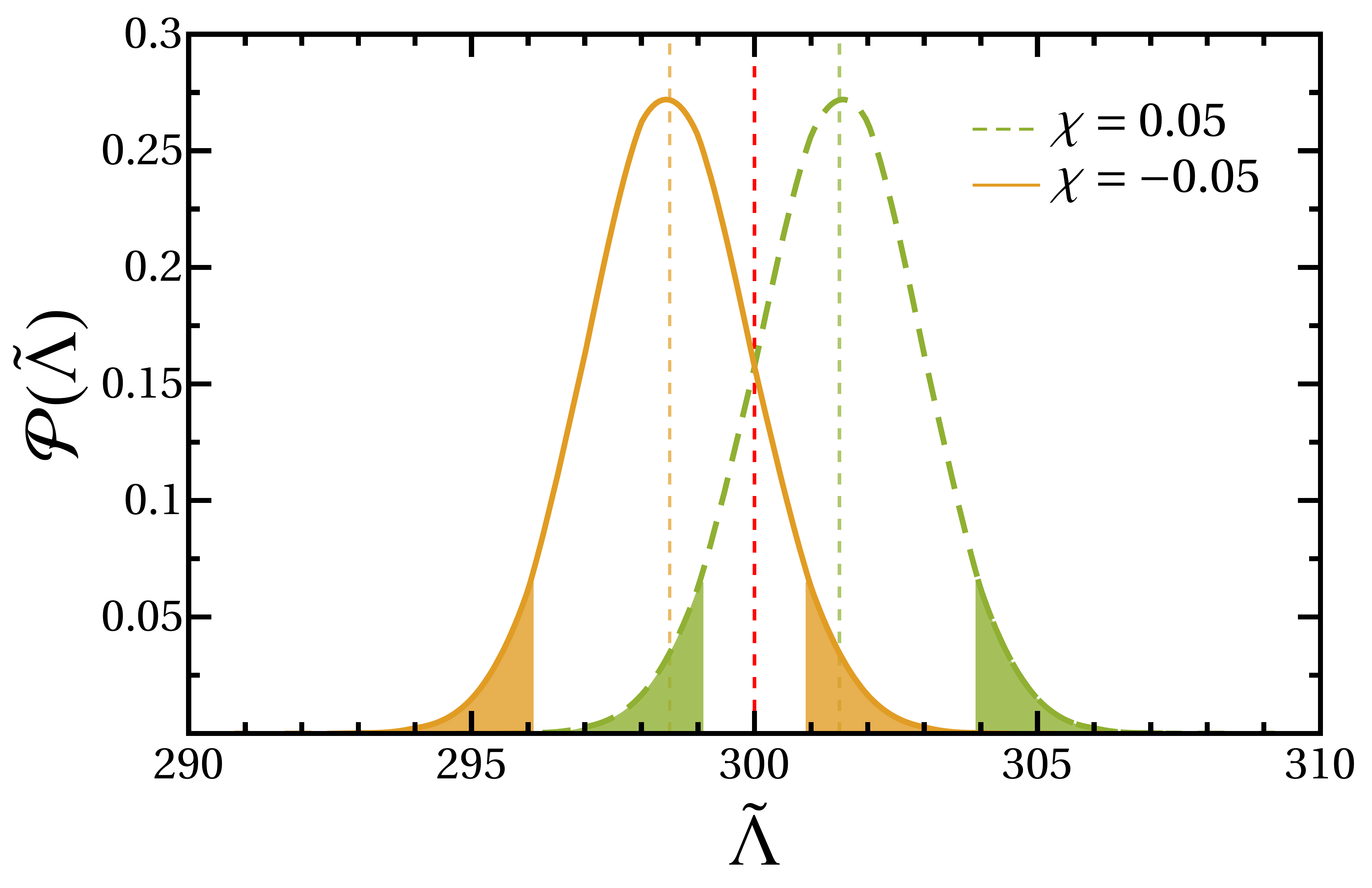}
\caption{\textsl{Probability distributions obtained for a $1.4M_{\odot}$ equal-mass binary spinning at $\chi=0.05$ (dashed-green) and $\chi=-0.05$ (orange), with the ET noise sensitivity curve and a SNR $\rho=1500$. The vertical dashed lines define the maximum-posterior values, while the solid areas define the $90\%$ credible intervals. The red dashed vertical line defines the injected value $\tilde\Lambda_0=300$.}}
\captionsetup{format=hang,labelfont={sf,bf}}
\label{fig:ETlowspin}
\end{figure}
\begin{figure}[]
\captionsetup[subfigure]{labelformat=empty}
\centering
{\includegraphics[width=0.49\textwidth]{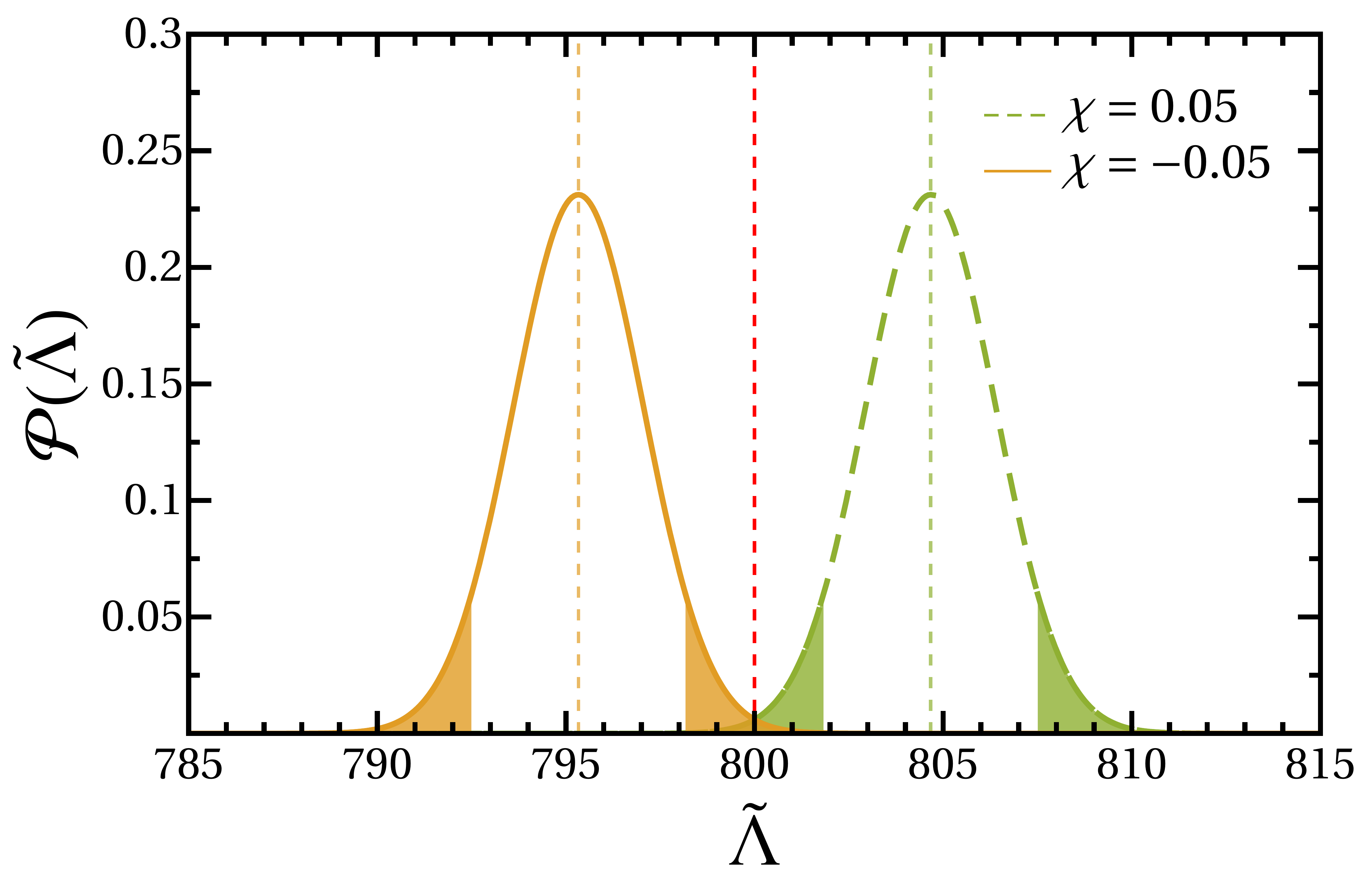}}  
{\includegraphics[width=0.49\textwidth]{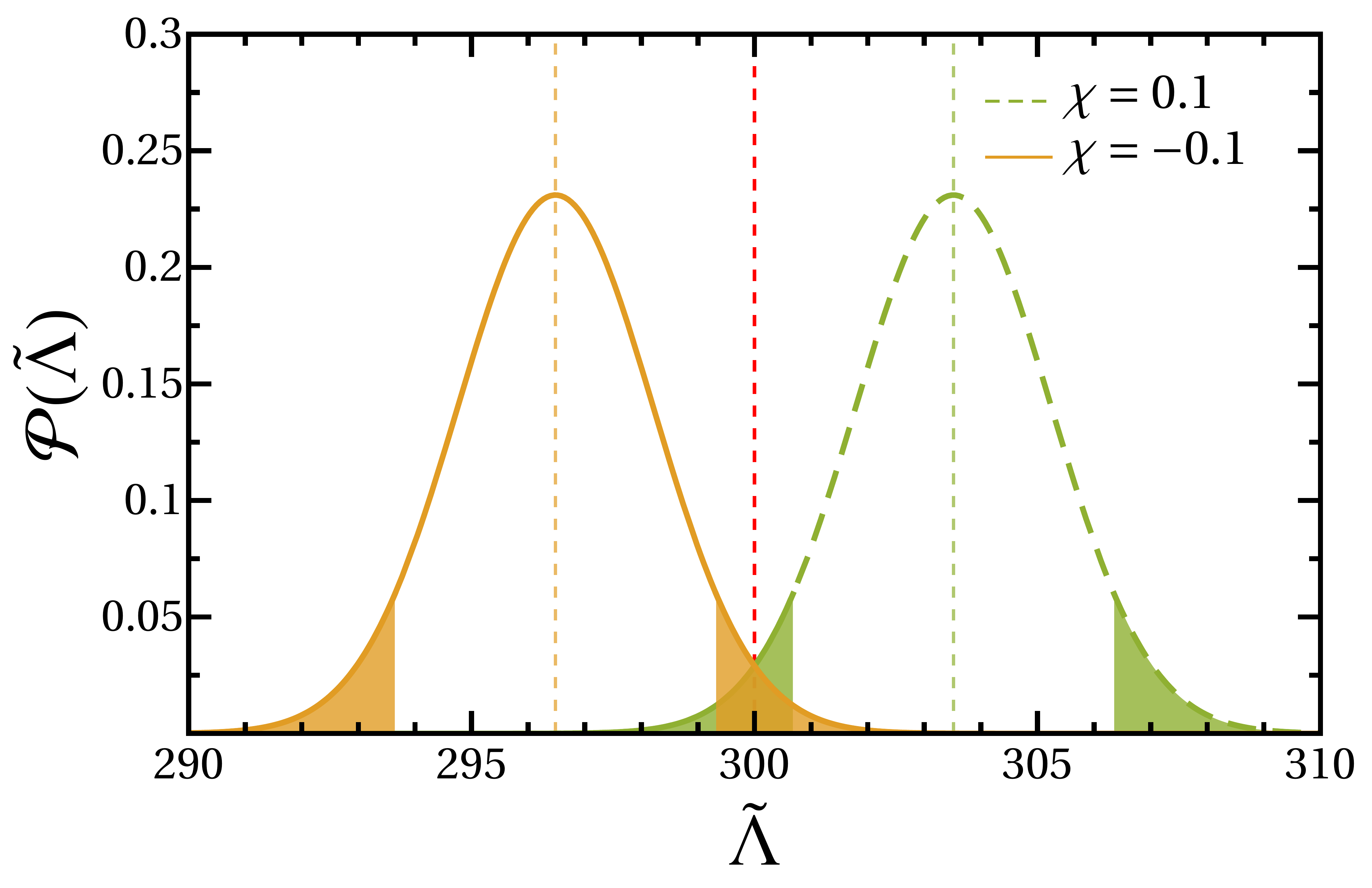}}
\caption{\textsl{(Left) Probability distributions obtained for a $1.4M_{\odot}$ equal-mass binary spinning at $\chi=0.05$ (dashed-green) and $\chi=-0.05$ (orange), with the ET noise sensitivity curve and a SNR $\rho=1700$. The vertical dashed lines define the maximum-posterior values, while the solid areas define the $90\%$ credible intervals. The red dashed vertical line defines the injected value $\tilde\Lambda_0=800$. (Right) Probability distributions obtained for a $1.4M_{\odot}$ equal-mass binary spinning at $\chi=0.1$ (dashed-green) and $\chi=-0.1$ (orange), with the ET noise sensitivity curve and a SNR $\rho=1700$. The vertical dashed lines define the maximum-posterior values, while the solid areas define the $90\%$ credible intervals. The red dashed vertical line defines the injected value $\tilde\Lambda_0=300$.}}
\captionsetup{format=hang,labelfont={sf,bf}}
\label{fig:ETdouble}
\end{figure}
We repeat the analysis for the same system as before, but this time using the sensitivity curve of the ET. The latter is expected to increase the sensitivity about a factor $15$, with respect to second-generation detectors at design sensitivity, and then of a factor $\sim 45$ with respect to current detectors. We said before that the SNR of GW170817 was $\rho=32.4$. Though the observation of such high SNR event (luminosity distance $D_L \sim 40 \,$Mpc) was rather unlikely considering the previous event rate predictions~\cite{Abbott:2016ymx}, the inclination reported tends to favor an off-axis orientation with respect to the Earth observation line (inclination angle $ \theta \sim 30º$)~\cite{Abbott:2018wiz}. For this configuration, the triangular shape of the ET (which will be actually composed of three detectors) increases the SNR by a factor $\sim 1.5$, relative to a single L-shaped interferometer~\cite{Hild:2010id}. Taking into account also that the SNR of GW170817 was increased by approximately a factor $\sqrt{2}$ with respect to a single-detector observation~\footnote{Only the two LIGO sites contributed to the total SNR of GW170817, due to the unlucky sky-position of the source relative to the Virgo interferometer orientation.}, the same event would have been seen by the ET with an SNR of $\sim (32.4/\sqrt{2}) \times 45 \times 1.5 $. Therefore, we set $\rho=1500$.

In the computation of the match, we fix the frequency sensitivity range for the ET to $\left\lbrace f_{min},f_{max}\right\rbrace= \left\lbrace 1\, \mathrm{Hz},f_{ISCO}\right\rbrace$. The results are shown in Fig.~\ref{fig:ETlowspin}. We notice that, due to the very large SNR, the distributions are a lot more peaked, compared to those obtained with LIGO in Fig.~\ref{fig:LIGOlowspin}. Instead, the induced shift with respect to the injected value, due to the imperfect modeling of the signal, is the same as before. This is expected, because the observed offset is independent of the SNR, which cancels out in the computation of the match. It depends only on the the difference between the signal $h$ and the template $h_T$, and the relative shape of the sensitivity curves of LIGO and the ET, which are anyway similar in the high-frequency region, where the tidal effects are relevant. This can be easily checked, translating the LIGO curve onto the ET one in Fig.~\ref{fig:sensitivity}. However, the impact on the recovery of $\tilde\Lambda$ is still not sufficient to distinguish the spin-tidal effects, being the induced bias completely contained inside the $90\%$ credible interval, and therefore overtaken by the statistical uncertainties. This means that the spin-tidal terms seem negligible, even when assuming a detection with the ET. 

\begin{figure}[]
\centering
\includegraphics[width=0.99\textwidth]{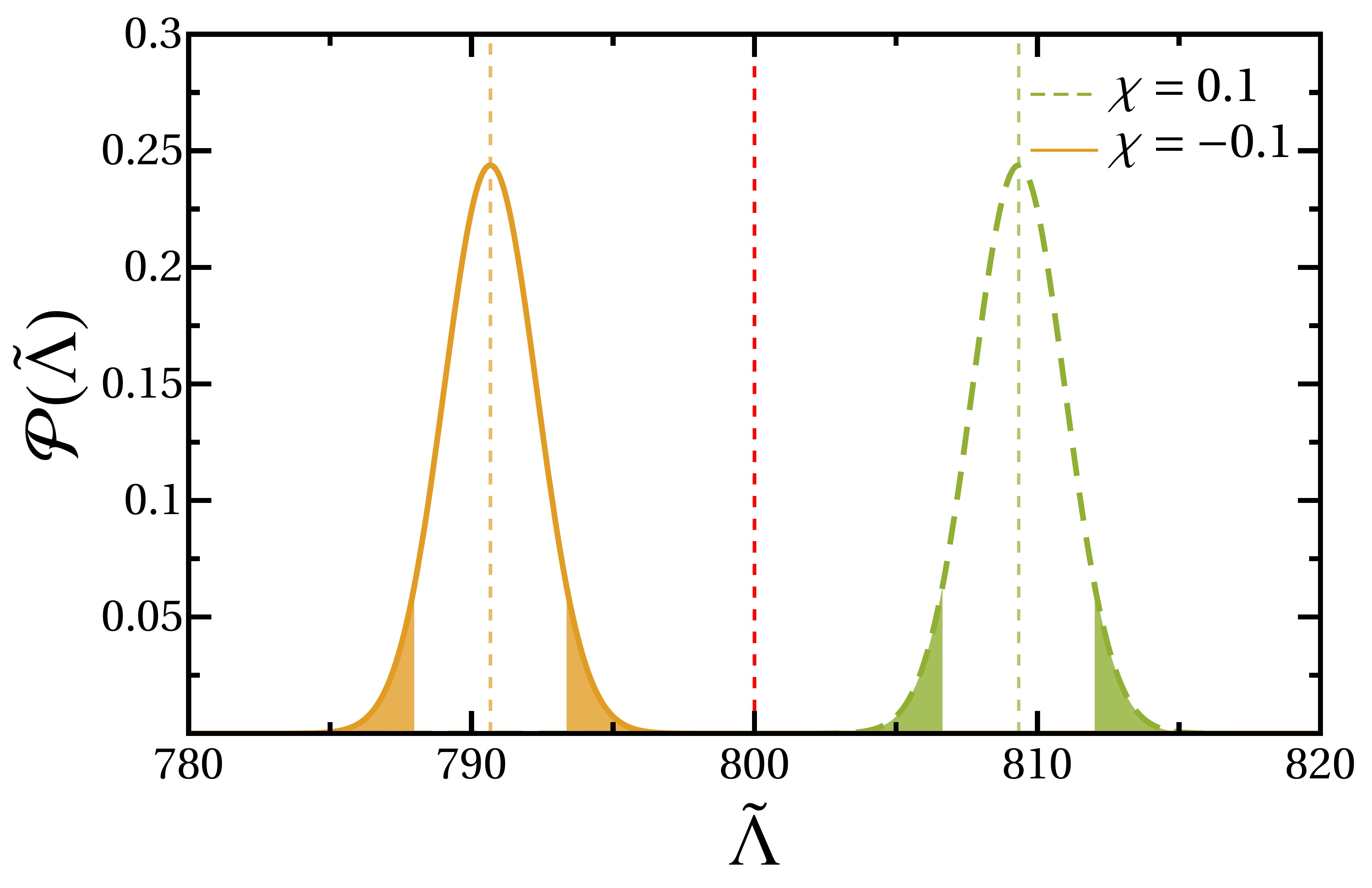}
\caption{\textsl{Probability distributions obtained for a $1.4M_{\odot}$ equal-mass binary spinning at $\chi=0.1$ (dashed-green) and $\chi=-0.1$ (orange), with the ET noise sensitivity curve and a SNR $\rho=1700$. The vertical dashed lines define the maximum-posterior values, while the solid areas define the $90\%$ credible intervals. The red dashed vertical line defines the injected value $\tilde\Lambda_0=800$.}}
\captionsetup{format=hang,labelfont={sf,bf}}
\label{fig:EThighspin}
\end{figure}

We further investigate this problem by considering a more \emph{optimistic} scenario, consisting in the hypothetical case of observing an event with the physical parameters compatible with GW170817 ($1.4M_{\odot}$ equal-mass binary), but in a face-on orientation. In this configuration, the SNR of the event detected by the ET is increased by another factor $\sim 1.15$, with respect to the previous one (the total gain of the triangular interferometer is then $\sim 1.5 \times 1.15 \sim 1.7$, relative to a single L-shaped detector). Furthermore, to maximize the detectability of the spin-tidal terms, we consider three cases:
\begin{itemize}
\item[1)] a \emph{large tidal-deformability} case, where we set the injected value of the average tidal deformability equal to the upper limit of the $90\%$ credible interval reported by the LIGO/Virgo collaboration for GW170817, $\tilde{\Lambda}_0=800$, keeping the spins equal to $\chi_1=\chi_2=\chi= \pm 0.05$;
\item[2)] a \emph{high-spin} case, where we fix the star spins to $\chi_1=\chi_2=\chi= \pm 0.1$, which is allowed by the high-spin prior used by the collaboration, that imposes $|\chi_{1,2}| \leq 0.79$~\cite{TheLIGOScientific:2017qsa,Abbott:2018wiz}, and we keep the tidal deformability equal to $\tilde{\Lambda}_0=300$;
\item[3)] we combine the above cases, increasing both the spins and the tidal deformability, $\chi_1=\chi_2=\chi= \pm 0.1$ and $\tilde{\Lambda}_0=800$.
\end{itemize}

We run the analysis on the above three configurations assuming to detect the event with the ET with a SNR of $\rho =1700$ (1.15 times larger than the $\rho=1500$ assumed before). The results for the first two cases are shown in Fig.~\ref{fig:ETdouble}. In the left panel we show the large-tidal deformability case, whereas in the right panel the high-spin case. We notice that the picture slightly improves in these scenarios. The bias relative to the injected value is increased, with respect to the \emph{standard} scenario in Fig.~\ref{fig:ETlowspin}, thanks to the larger values of tidal deformability and spin in the first and second case, respectively. This reflects the linear dependence of the gravitational phase on these parameters. As a consequence, and thanks also to the higher SNR, the injected value $\tilde \Lambda_0$ lies on the tails of the distributions, slightly outside of the $90\%$ credible intervals delimited by the solid areas in both cases. This implies that the gravitational signal $h$, which includes the spin-tidal effects in the waveform, is marginally distinguishable from the template $h_T$. However, even in these optimistic scenarios, and due to the simplifications we have taken into account, the offset is only at $\sim 2\sigma$, suggesting that it might be very challenging to measure this effect. 

\begin{figure}[]
\captionsetup[subfigure]{labelformat=empty}
\centering
{\includegraphics[width=0.7\textwidth]{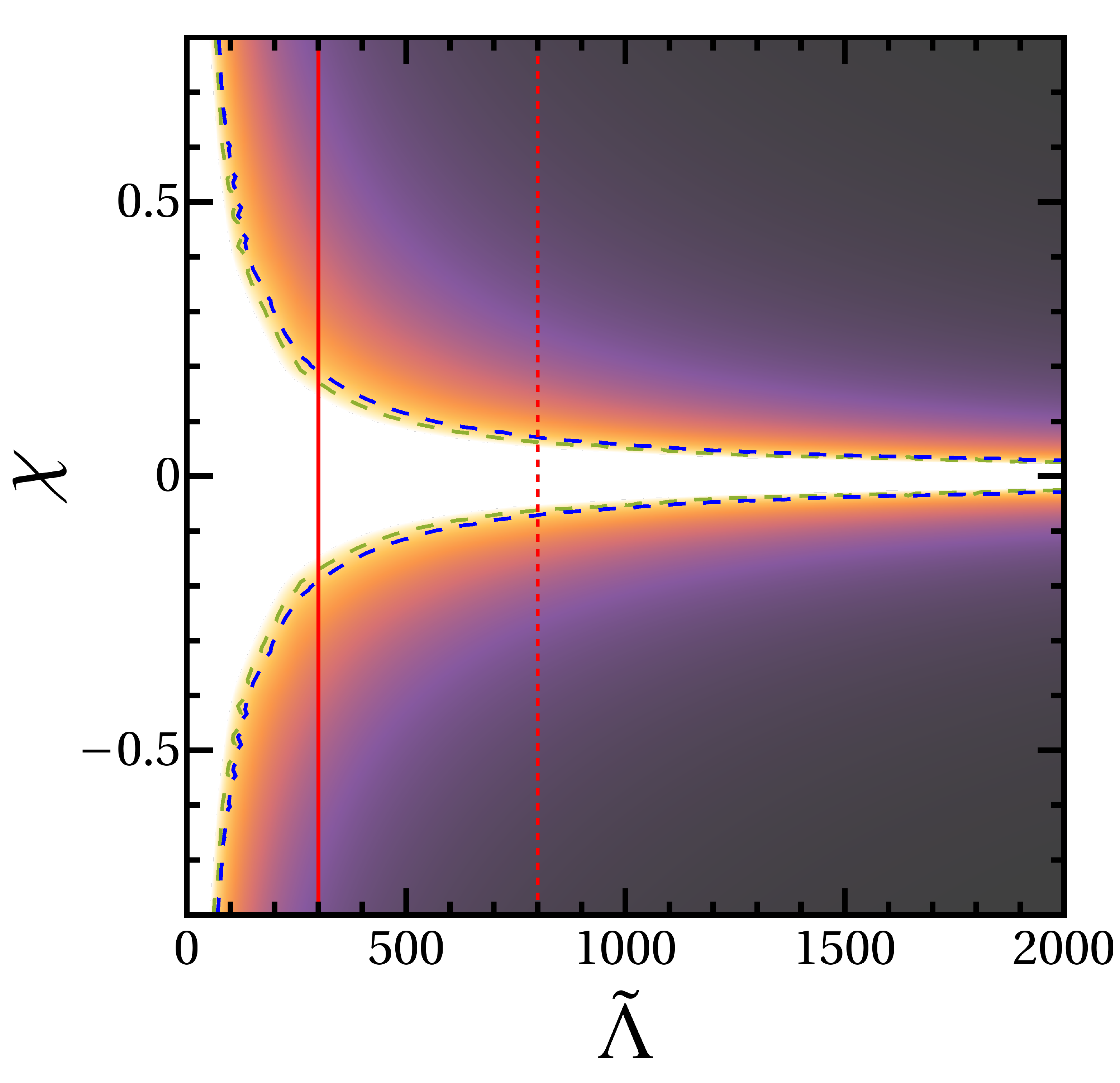}} \ 
{\includegraphics[width=0.09\textwidth]{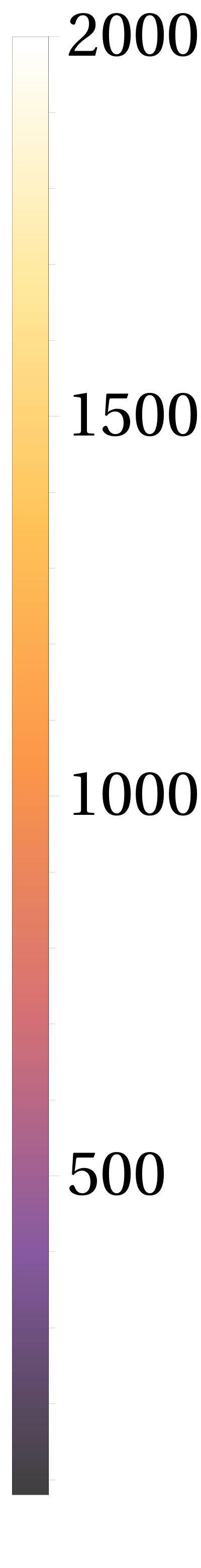}}
\caption{\textsl{Estimation of the SNR required to distinguish the effects of the spin-tidal terms considering the ET noise sensitivity curve. The vertical red grid lines fix the tidal deformabilities consistent with the median (solid) and $90\%$ upper limit (dashed) provided by LIGO/Virgo collaboration~\cite{Abbott:2018wiz}. The blue and green contour lines correspond to the SNR of our \textit{standard}, $\rho=1500$, and \textit{optimistic}, $\rho=1700$, scenarios, respectively.}}
\captionsetup{format=hang,labelfont={sf,bf}}
\label{fig:spincontour}
\end{figure}

If we consider instead the third case, the situation is quite different. As shown in Fig.~\ref{fig:EThighspin}, by assuming a spin $\chi=0.1$ and an average tidal deformability $\tilde{\Lambda}_0 = 800$, we sum up the two effects, introducing a bias which shifts the distribution completely off from the true value (more than $5 \sigma$ away). In this scenario the systematic uncertainties are much larger than the statistical ones, implying that not accounting for the spin-tidal terms would result in an incorrect estimate of the tidal deformability. However, this would be the case only if neutron stars with large tidal deformabilities and moderately high spins exist in Nature.
 
Finally, we provide an estimate of the minimum values for the triplet ${\rho-\chi-\tilde\Lambda}$ (with $\chi_1=\chi_2=\chi$) required to distinguish the effects of the spin-tidal terms for a GW170817-like event detected with the ET. To do so, we compute the match of the gravitational signal $h(\chi,\tilde\Lambda)$ against the template $h_T(\chi,\tilde\Lambda)$, for $\tilde\Lambda \in[0,2000]$ and $\chi\in[-0.79,0.79]$, for a $1.4M_{\odot}$ equal-mass binary. The results of the match are translated to the SNR, through Eq.~\eqref{eq:ind}, for a ${D=6: \left\lbrace M_1, M_2, \chi_1,\chi_2, \Lambda_1, \Lambda_2 \right \rbrace } $ parameter space, where we require to estimate all the parameters at $90\%$ credible level, i.e., setting ${n=1.64}$ in Eq.~\eqref{eq:ind}. The results of this analysis are shown in Fig.~\ref{fig:spincontour}. The density plot represents the minimum SNR needed to observe some characteristic combination of $\tilde\Lambda$ and ${\chi_1=\chi_2=\chi}$. The solid and dashed vertical red grid lines ${\tilde\Lambda=\left\lbrace 300,800\right\rbrace}$, set the median and $90\%$ upper limit provided by Ref.~\cite{Abbott:2018wiz}, respectively. The blue and green contour lines correspond to the SNR of our \textit{standard}, $\rho=1500$, and \textit{optimistic}, $\rho=1700$, scenarios, respectively. 
 
We can see that for low spins and/or small tidal deformabilities the contour lines get closer. This is expected, since the spin-tidal corrections are harder to detect in this limit, implying a fast increasing of the required SNR. On other hand, this means that small variations of SNR are less significant in this regime. Indeed, there is small difference between the $\rho=1500$ and the $\rho= 1700$ contour lines. In general, larger spins are required to attain the same SNR as $\tilde\Lambda$ decreases.

The intersection of ${\tilde\Lambda=800}$ with the ${\rho=\left\lbrace 1500,1700\right\rbrace}$ contours shows that the minimum spin required to distinguish the spin-tidal effects from a $h_T$ template, at the $90\%$ level, is ${\chi \sim \left\lbrace \pm 0.07,\pm 0.06\right\rbrace}$, respectively. Notice that the intersection of the $\tilde\Lambda=800$ line with the green contour line ${(\rho=1700)}$ corresponds to the particular cases shown in the left panel of Fig.~\ref{fig:ETdouble}, and in Fig.~\ref{fig:EThighspin}. We stress that the parallelism between the results in Fig.~\ref{fig:spincontour} and those shown in Figs.~\ref{fig:ETlowspin}--\ref{fig:EThighspin} is not perfect. The reason for this is that in Fig.~\ref{fig:spincontour} we are accounting for $D=6$ parameters through Eq.~\eqref{eq:ind}, while the previous plots are the result of a single-parameter analysis. Indeed, for $\tilde\Lambda=300$ and ${\rho=1700}$, the intersection occurs at ${\chi \sim \pm 0.17}$, to be compared to the right panel of Fig.~\ref{fig:ETdouble}.

In conclusion, spin-tidal couplings are only expected to affect significantly the signal for putative binary neutron star events with the SNR of GW170817 (as seen by current second-generation detectors), observed with third-generation detectors, and for moderately high spins. On the other hand, the calibration of these effects on current waveform templates would have a non-negligible impact only if binaries with $\chi_i \gtrsim 0.1$ evolve and merge in the local Universe.

\chapter{The relativistic inverse stellar problem}
\label{sec:inverse}
\begin{figure}
\captionsetup[subfigure]{labelformat=empty}
\centering
{\begin{tikzpicture}[baseline]
\begin{loglogaxis}[xmin=10^14,xmax=3*10^15,
axis x line=bottom,
axis y line=left,
enlargelimits,
legend style={draw=none,font=\footnotesize},
legend style={at={(0.81,0.55)},anchor=north},
width=0.47\textwidth,
height=0.47\textwidth,
minor tick style={draw=none},
ytick={1e33,1e34,1e35,1e36},xtick={1e14,1e15},
xlabel=$\epsilon/c^2 \, \text{[g/cm}^3\text{]}$,ylabel=$p \, \text{[dyn/cm}^2\text{]}$]
\addplot[smooth,thick,solid,red] table[x index=2,y index=3] {chapter_3/figures/fps.d};
\addplot[smooth,thick,dashed,orange] table[x index=2,y index=3] {chapter_3/figures/apr.d};
\addplot[smooth,thick,dotted,green] table[x index=2,y index=3] {chapter_3/figures/sly4.d};
\addplot[smooth,thick,dashdotted,cyan] table[x index=2,y index=3] {chapter_3/figures/bgn1h1.d};
\addplot[smooth,thick,densely dashed,blue] table[x index=2,y index=3] {chapter_3/figures/gm1.d};
\addplot[smooth,thick,densely dotted,magenta] table[x index=2,y index=3] {chapter_3/figures/gnh3.d};
\addlegendentry{FPS};
\addlegendentry{APR4};
\addlegendentry{SLY};
\addlegendentry{BGN1H1};
\addlegendentry{GM1};
\addlegendentry{GNH3};
\end{loglogaxis}
\end{tikzpicture}
}
{
\begin{tikzpicture}[baseline]
\begin{axis}[xmin=9.5,xmax=16,
axis x line=bottom,
axis y line=left,
enlargelimits,
width=0.47\textwidth,
height=0.47\textwidth,
scaled y ticks = false,
xlabel=$R \, \text{[}   \text{km} \text{]} $,ylabel=$M \, \text{[} M_{\odot} \text{]}$]
\addplot[smooth,thick,red,solid] table {chapter_3/figures/fps.dat};
\addplot[smooth,thick,orange,dashed] table {chapter_3/figures/apr.dat};
\addplot[smooth,thick,green,dotted] table {chapter_3/figures/sly4.dat};
\addplot[smooth,thick,cyan,dashdotted] table {chapter_3/figures/bgn1h1.dat};
\addplot[smooth,thick,blue,densely dashed] table {chapter_3/figures/gm1.dat};
\addplot[smooth,thick,magenta,densely dotted] table {chapter_3/figures/gnh3.dat};
\addplot[name path = A,domain=9:16,samples=100,smooth,thick,yellow]{2.05};
\addplot[name path = B,domain=9:16,samples=100,smooth,thick,yellow]{1.97};
\addplot[yellow!20] fill between[of=A and B];
\end{axis}
\end{tikzpicture}
}
\caption{\textsl{(Left) Pressure-energy profiles predicted by various models of EOS, in the core region of neutron stars. For the description of the micro-physical properties the EOS models see, e.g.,~\cite{Lattimer:2000nx}. (Right) Mass-radius profiles obtained by integrating the TOV equations with the EOSs shown in the left panel. The yellow horizontal band represents the astrophysical constraint imposed by the observation of a neutron star with mass $M=(2.01 \pm 0.04) M_{\odot}$~\cite{Antoniadis:2013pzd}. Roughly speaking, all EOSs whose corresponding mass-radius curves lie below the band are ruled out.}}
\captionsetup{format=hang,labelfont={sf,bf}}
\label{fig:eostab}
\end{figure}
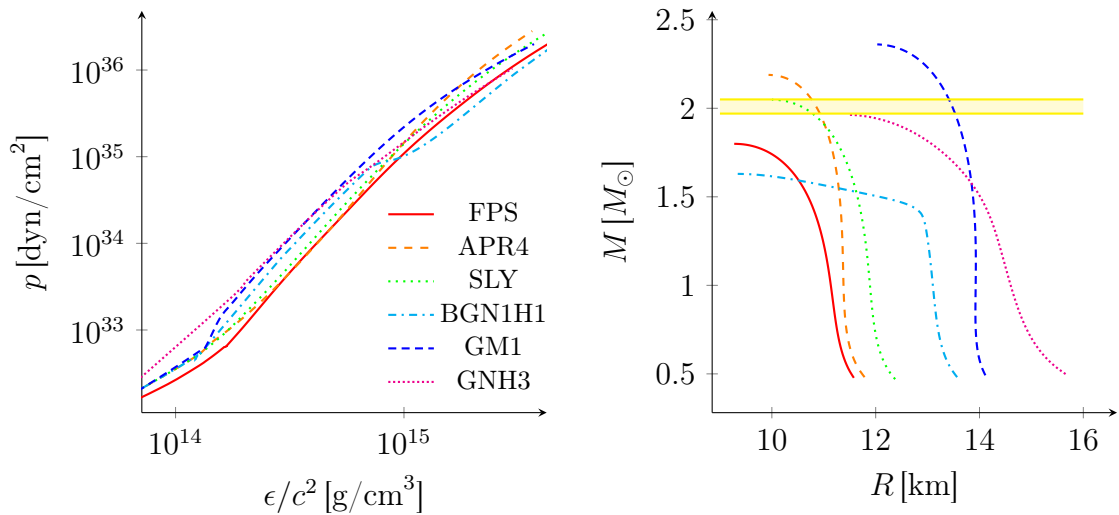

The lack of knowledge of the behavior of matter at supranuclear densities has led to large uncertainties on the equation of state (EOS) inside the core of neutron stars (cf. section~\ref{sec:tabeos}). Various theoretical approaches to model the microphysical interactions among hadrons have been developed, predicting different scenarios for the composition of cold nuclear matter at densities above the nuclear saturation point, $\rho_{nu} \sim 2.7 \times 10^{14}\ \mathrm{g/cm}^3$~\cite{Lattimer:2000nx}. This gave rise to many models of the EOS, i.e., different predictions on the relation between the pressure and the energy density, $p= p(\epsilon)$ (we recall that the cold nuclear matter inside neutron stars is modeled through a barotropic EOS, see section~\ref{sec:eos}). In the left panel of Fig.~\ref{fig:eostab}, we show several models of EOS in the energy density--pressure plane.

The EOS is an essential ingredient to determine the structure of neutron stars and to make predictions on the macroscopic observables, such as mass $M$, radius $R$, etc. (see section~\ref{sec:tov}). The integration of the TOV equations, using various models of EOS, allows us to find the mass-radius profiles, which can sensibly differ from each other. We show this feature in the right panel of Fig.~\ref{fig:eostab}. We can see that different EOSs predict different values of the neutron star maximum mass, and that for a given mass they predict different values of the neutron star radius.

This naturally suggests that astrophysical observations can be exploited to constrain the neutron star EOS, ruling out those proposed models which are incompatible with the measurements. For instance, the observation of two-solar-mass neutron stars~\cite{Demorest:2010bx,Antoniadis:2013pzd} (colored band in Fig.~\ref{fig:eostab}) has ruled out all the EOS models which can not support such a large mass. In general, the simultaneous measurement of both the mass and radius of a neutron star (or, as we will see, of any other pair of independent observables) fixes a configuration in the mass--radius plane, imposing a constraint on the EOS.

We stress that besides observational constraints, experimental and theoretical bounds can further be imposed on the EOS. Data from laboratory experiments~\cite{Chen:2010qx,Abrahamyan:2012gp,Tsang:2008fd,Tamii:2011pv,Piekarewicz:2012pp,Trippa:2008gr} and nuclear-physics calculations~\cite{Gandolfi:2011xu,Tews:2012fj,Hebeler:2013nza,Drischler:2016djf} constrain the low-density regime, whereas perturbative QCD computations~\cite{Kurkela:2009gj,Fraga:2013qra} bind the ultra-high density region, $ \rho \gg \rho_{nu}$. Another theoretical bound is the causality constraint, which requires that the speed of sound, $c_s=c \sqrt{dp/d\epsilon}$, does not exceed the speed of light $c$. Several models of EOS obtained through non-relativistic computations do not satisfy this requirement at large densities.

\begin{figure}[]
\centering
\includegraphics[width=0.99\textwidth]{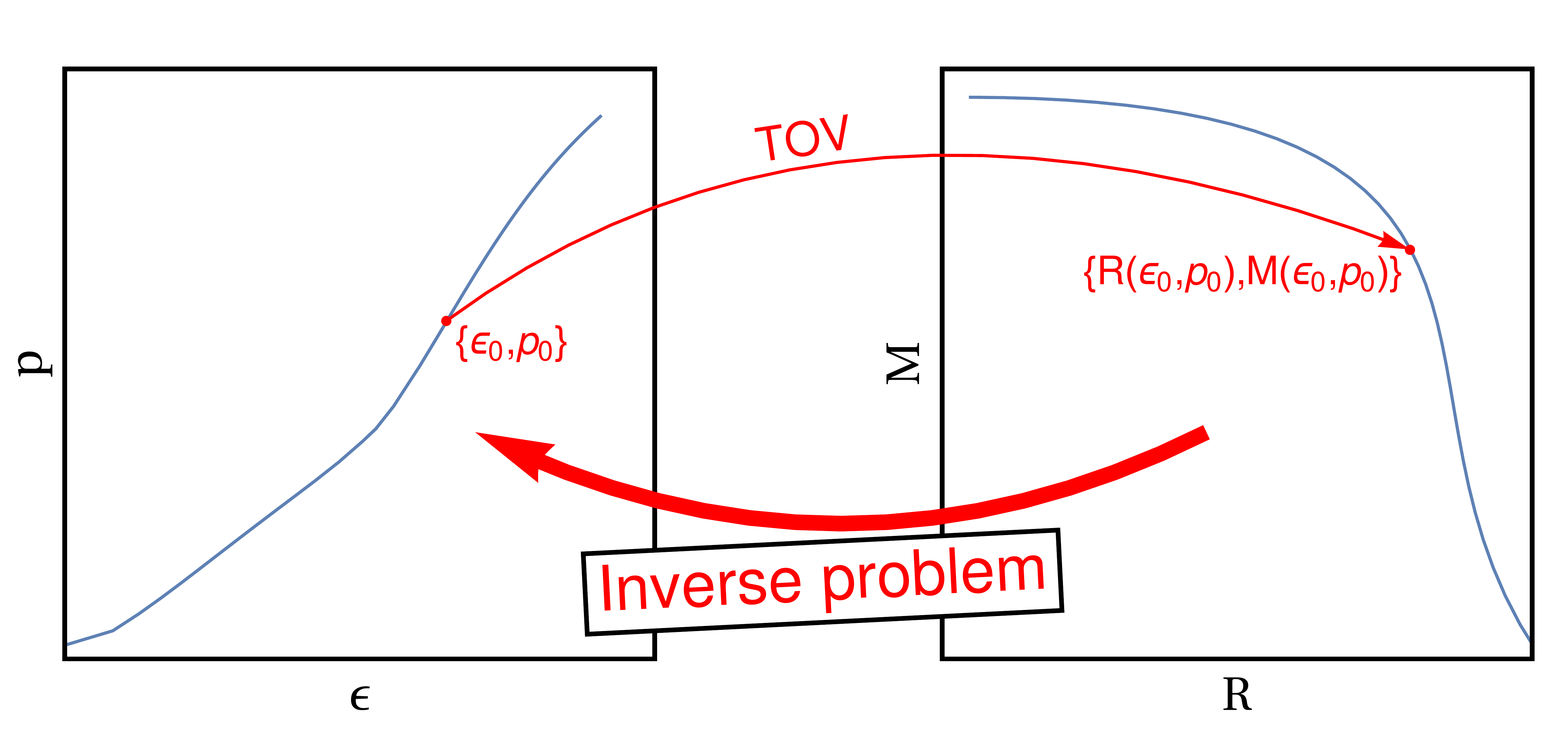}
\caption{\textsl{The unique mapping between the EOS and the mass-radius relation, provided by the TOV equations. The $\{R(\{\epsilon_0,p_0\}),M(\{\epsilon_0,p_0\})\}$ point is the equilibrium configuration with central energy density and pressure given by $\{\epsilon_0,p_0\}$. The knowledge of the whole mass-radius curve is needed, in principle, to invert the mapping and obtain the energy-pressure relation.}}
\captionsetup{format=hang,labelfont={sf,bf}}
\label{fig:inverse}
\end{figure}

In 1992, Lindblom showed that the TOV equations provide a unique mapping between the energy-pressure relation (i.e., the EOS) and the mass-radius profile~\cite{1992ApJ...398..569L}. In other words, the equations of stellar structure link the microscopical properties of matter to the macroscopical characteristics of the neutron star, as shown in Fig.~\ref{fig:inverse}. In principle, the complete knowledge of the $M(R)$ relation could be used to invert the mapping and fully determine the EOS. This requires a collection of simultaneous measurements of neutron star masses and radii, dense and accurate enough to make possible the inversion and to reconstruct the EOS. This procedure is known as the \emph{relativistic inverse stellar problem}: constraining the neutron star EOS through the observations of macroscopic quantities of neutron stars.

The main issues in solving the inverse stellar problem are: (i) a limited number of simultaneous measurements of neutron star observables~\footnote{The available neutron star observations are not uniformly distributed over the range of masses. The total population of observed neutron stars is peaked around the Chandrasekhar mass, $M \sim 1.4 M_{\odot}$~\cite{Lattimer:2012nd,Ozel:2016oaf}.}, and (ii), the large uncertainties affecting the measurements of these quantities (in particular the radius). Despite these difficulties, measurements of masses and radii, obtained through neutron star observations in the electromagnetic band~\footnote{The estimate of neutron star radii, obtained from electromagnetic observations through X-ray spectral modeling, is affected by systematic uncertainties, which led to some disagreement among the inferences of different groups~\cite{Watts:2016uzu}.}, have allowed us to constrain the EOS, ruling out extreme models which predict very large radii ($R \gtrsim 13$ km)~\cite{Ozel:2008kb,Ozel:2010fw,Guver:2010td,Guver:2008gc,Guver:2013xa,Steiner:2010fz,Steiner:2012xt,Lattimer:2013hma,Lattimer:2014sga,Guillot:2013wu,Guillot:2014lla,Ozel:2015fia}. 

The pair of observables mass-radius is not the only one which can provide information on the neutron star internal structure. The measurement of any macroscopic quantity, such as compactness, moment of inertia, etc., can in principle constrain the EOS. In the field of gravitational wave physics, an important candidate to probe the interior of neutron stars is the (electric quadrupolar) tidal deformability $\lambda_2$~\cite{Hinderer:2009ca,Lackey:2011vz,Lackey:2013axa,Maselli:2013rza,Wade:2014vqa,Read:2013zra}. Indeed, as extensively discussed in the previous chapters, the tidal deformability carries information on the EOS, and leave an imprint on the gravitational waveform emitted by inspiralling binary neutron stars (or neutron star-black hole systems)~\footnote{Tidal deformations of neutron stars affect the emitted gravitational waveform through the quadrupolar electric tidal deformabilities, at leading-order (see section~\ref{sec:phase}). Current templates used by the LIGO/Virgo collaboration model only the main contribution given by $\lambda_2$~\cite{TheLIGOScientific:2017qsa,Abbott:2018wiz}. Other terms, such as magnetic tidal Love numbers, spin corrections and higher-order multipole moments, have not been included yet (on the other hand, we have shown in section~\ref{sec:impact} that their impact is negligible for second-generation detectors).}. Therefore, the inverse stellar problem can be solved as well using the pair of observables mass-tidal deformability, which can be measured through gravitational wave detections. In other words, the measurement of the tidal deformability can constrain the high-density regime of the neutron star EOS.

This is indeed what happened after the first gravitational wave detection of a binary neutron star merger, GW170817\cite{TheLIGOScientific:2017qsa}. The LIGO/Virgo collaboration exploited the constraints on the star tidal deformabilities (see section~\ref{sec:impact}) to rule out some proposed models of EOS, which predict more deformable matter and then large tidal deformabilities~\cite{Abbott:2018wiz}. The LIGO/Virgo collaboration~\cite{Abbott:2018exr} and many other works~\cite{De:2018uhw,Annala:2017llu,Most:2018hfd,Bauswein:2017vtn,Raithel:2018ncd} translated this result on the neutron star radius, constraining it in the range $\sim 9\div 13 $ km, in agreement with electromagnetic astrophysical observations.

In this second part of the thesis, we demonstrate the feasibility of using gravitational wave signals emitted by coalescing neutron star binaries to solve the relativistic inverse stellar problem, i.e., to infer the parameters of a phenomenological representation of the EOS from measurements of the stellar mass and tidal deformability, performing a model selection among the EOSs proposed in the literature. In section~\ref{sec:parametrize} we review the piecewise polytropic phenomenologically parametrized model of the EOS. In section~\ref{sec:MCMC} we describe the statistical approach used, and, finally, in section~\ref{sec:setup} we present our results (Abdelsalhin \emph{et al.}~\cite{Abdelsalhin:2017cih}).

Since throughout this chapter we refer only to the quadrupolar electric tidal deformability $\lambda_2$, henceforth we omit the subscript $2$ in the $\lambda$ symbol.

\section{Phenomenological representations of the equation of state}
\label{sec:parametrize}
The models of EOS proposed in the literature depend on several parameters arising from the way hadron interactions are modeled, and on the particle content (see section~\ref{sec:tabeos}). On the other hand, measurements of mass and, especially, radius/tidal deformability can be affected by large uncertainties, which may make difficult to solve the inverse stellar problem mapping the correct EOS, if too many parameters are involved. Furthermore, constraining the parameters within one given realistic model does not rule out automatically other models based on different theoretical approaches.

\begin{figure}[]
\centering
\includegraphics[width=0.99\textwidth]{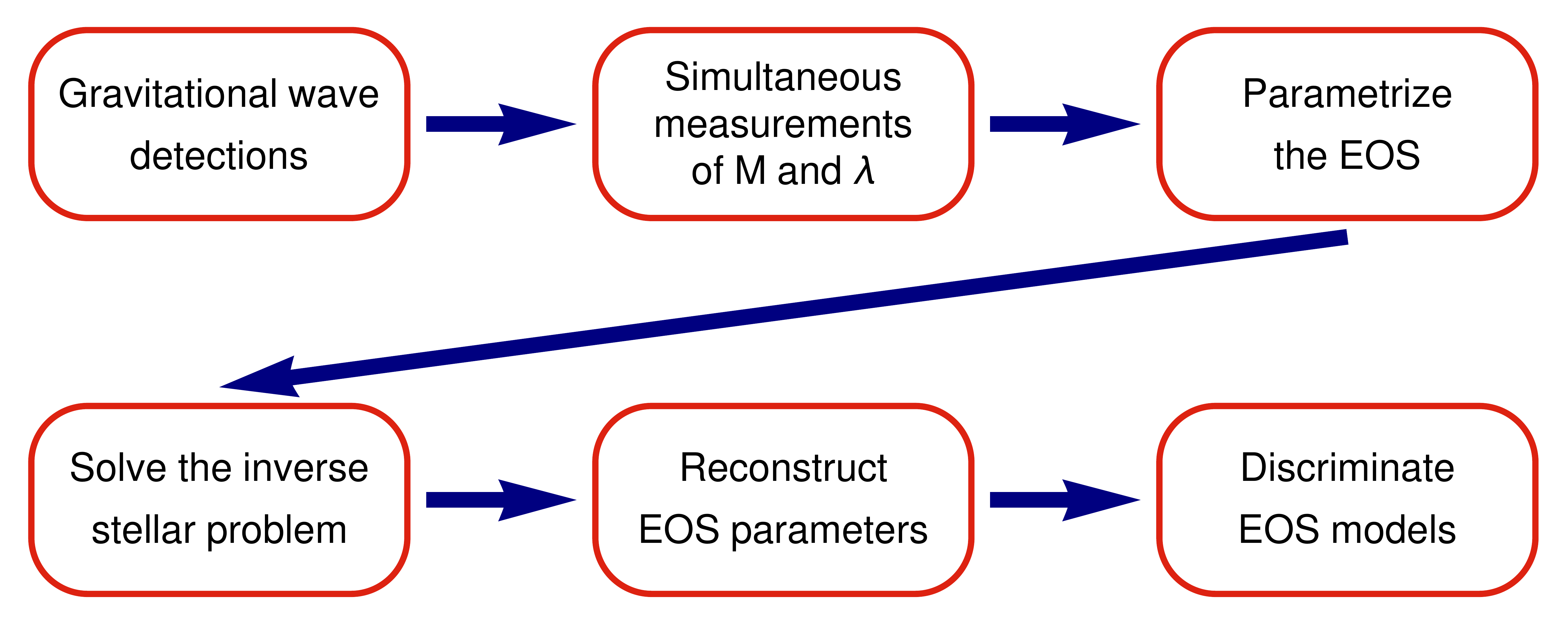}
\caption{\textsl{Flowchart summarizing the steps of the inverse stellar problem procedure that we use to constrain the EOS with gravitational wave observations.}}
\captionsetup{format=hang,labelfont={sf,bf}}
\label{fig:inverse2}
\end{figure}

Phenomenological parametrizations of the EOS of neutron stars are a way to overcome this limitation. Indeed, phenomenological models provide an effective approach to solve the inverse stellar problem, since they allow one to describe a large class of realistic EOSs through a relatively small set of coefficients, to be constrained by observational data~\cite{Ozel:2009da,Raithel:2017ity,Lindblom:2012zi,Lindblom:2013kra,Lindblom:2018ntw,Read:2009yp,Shibata:2010zz}. These EOSs can be used to combine measurements of various neutron star parameters, even coming from different channels. In other words, phenomenological representations can be exploited to combine the results of gravitational and electromagnetic observations, leading to multimessenger constraints on the EOS. Moreover, it may be possible that the true EOS differs from the models proposed in literature so far. In this case, a phenomenological approach would be extremely
useful to constrain the main features of the correct EOS.

Phenomenological models developed so far include:
\begin{itemize}
\item[1)] the piecewise polytropic EOS developed by Read et al.~\cite{Read:2008iy} (a variant of this model was proposed by Ozel and collaborators~\cite{Ozel:2009da,Raithel:2016bux}).
\item[2)] the spectral representation proposed by Lindblom~\cite{Lindblom:2010bb}, based on an expansion of the adiabatic index in terms of the pressure/enthalpy (see also the causal version of the model based on the speed of sound~\cite{Lindblom:2018rfr}).
\item[3)] the semi-phenomenological model described by Steiner et al.~\cite{Steiner:2010fz}, where a pressure-energy relation depending on nuclear physics parameters (such as symmetry energy, compressibility, etc.), for $\rho \lesssim 2 \rho_{nu}$, is matched to a two-piece polytropic relation at larger densities, which fits the inner core.  
\end{itemize}
Ozel and collaborators have shown that the piecewise polytropic model allows us to discriminate among realistic EOSs using electromagnetic measurements of neutron star radii~\cite{Ozel:2015fia}. Both the model by Read et al. and the spectral representation by Lindblom have been directly included in the gravitational wave templates, to parametrize the dependence of the tidal deformability on the EOS. Lackey, Wade and collaborators have shown that this way of parametrizing (alternative to the parametrization of the tidal deformability in terms of neutron star mass~\cite{DelPozzo:2013ala}) allows us to constrain the EOS from gravitational wave detections\cite{Lackey:2014fwa,Carney:2018sdv}. Indeed, the spectral model has been used by the LIGO/Virgo collaboration to infer the radii of the detected neutron stars of the GW170817 event\cite{Abbott:2018exr}.

In this thesis we use the piecewise polytropic model by Read et al.~\cite{Read:2008iy}. In Fig.~\ref{fig:inverse2} we show a flowchart illustrating the various steps of our inverse stellar problem procedure. Henceforth we use $G=c=1$ units.

\subsection{Polytropic equations of state}
\label{sec:polytropicstate}
In this section we describe the general form of a polytropic EOS. We recall that for a barotropic EOS, $p= p(\epsilon)$, the first law of thermodynamics reads (see section~\ref{sec:eos})
\begin{equation}
\label{eq:thermoeos}
d\epsilon = \frac{\epsilon+p}{n} dn \,,
\end{equation}
where $\epsilon$ is the energy density, $p$ the pressure and $n$ the baryon number density. The adiabatic index $\Gamma$ is defined as
\begin{equation}
\label{eq:adiabatic}
\Gamma  = \frac{d \log p}{d(\log n)} = \frac{\epsilon+p}{p}\frac{d p}{d \epsilon} \,,
\end{equation}
where in the last equality we have made use of Eq.~\eqref{eq:thermoeos}. The adiabatic index is a useful parameter to quantify the \emph{stiffness} of an EOS. 

The stiffness is a property of the EOS which indicates how much matter can be stored in a given volume. If the matter is more compressible, more of it can be stored in the same volume, the average density is larger, and the EOS is said \emph{soft}. Viceversa, EOSs characterized by less compressible matter correspond to lower average densities, and are called \emph{stiff}. Since the structure and composition of matter change inside a neutron star, the stiffness generally depends on the density. From a different but complementary point of view, the stiffness can be also seen as a measure of the pressure that the matter is subjected to, at a given density. Soft matter (being more compressible) exhibits lower pressures, whereas stiff matter is characterized by higher pressures. In relation to the adiabatic index, soft EOSs show generally a larger $\Gamma$, while stiff EOSs a lower one. Note that the adiabatic index is a function of the density, reflecting that the stiffness in general is not constant. 

Thus, soft EOSs lead to more compact neutron stars (larger masses for a fixed radius, or, equivalently, smaller radii for a fixed mass), which are less affected by tidal forces, and then less deformable, i.e., the tidal deformability $\lambda$ is smaller. Viceversa, stiff EOSs lead to less compact objects (smaller masses for the same radius, larger radii for the same mass), which are more deformable, and have larger tidal deformabilities. Measurements of neutron star radii from the electromagnetic band~\cite{Guillot:2014lla,Lattimer:2014sga,Ozel:2015fia}, and of the average tidal deformability (see section~\ref{sec:impact}) from the gravitational wave event GW170817\cite{TheLIGOScientific:2017qsa,Abbott:2018wiz,Abbott:2018exr} favor soft EOSs, ruling out extreme stiff models.

The stiffness is also related to the speed of sound $c_s$, which is defined by
\begin{equation}
c_s^2  = \frac{d p}{d \epsilon} = \frac{p \Gamma}{\epsilon+p} \,,
\end{equation}
where in the last equality we have used Eq.~\eqref{eq:adiabatic}. Then, a higher speed of sound corresponds to stiffer EOSs, whereas a lower speed of sound to a softer one.

Assuming that all baryons have the same mass, typically the neutron mass $m_n$, we can write $\rho = m_n n$, where $\rho$ is the rest-mass density~\footnote{This assumption is justified by the fact that: (i) for plain nuclear matter inside a neutron star, the neutron fraction is larger than the $90\%$ (see section~\ref{sec:tabeos}), and the masses of the proton and the neutron are very similar. (ii) We are going to use a phenomenologically parametrized model of the EOS, not one coming from nuclear physics calculations. Therefore, we can assume that the neutron star matter is composed of a single-component fluid.}. This allows us to rewrite Eq.~\eqref{eq:thermoeos} as
\begin{equation}
\label{eq:thermoeos2}
d\epsilon = \frac{\epsilon+p}{\rho} d\rho \,.
\end{equation}
Polytropic EOSs are power law relations between the pressure and the energy/rest-mass density. The thermodynamics of fully degenerate Fermi gases in the non-relativistic/ultra-relativistic limit leads to polytropic EOSs~\cite{Chandrasekhar:1931ih,Chandrasekhar:1931ftj,Chandrasekhar:1935zz}. Moreover, polytropic models approximate the EOS arising from the strong interacting matter present in the neutron star cores (cf. the next section below).

There are two kinds of polytropic relations: energy polytropes and rest-mass polytropes. The first one is a relativistic version, which has the form 
\begin{equation}
p(\epsilon)=K \epsilon^{\gamma} \,,
\end{equation}
where $K$ is the polytropic constant and $\gamma$ the polytropic index. The second type has instead the form
\begin{equation}
p(\rho)=K \rho^{\gamma} \,.
\end{equation}
Note that in the latter case, the polytropic index coincides with the adiabatic index, $\gamma = \Gamma$. This means that the adiabatic index (and then the stiffness) of a polytrope is constant. Thus, we can write
\begin{equation}
\label{eq:eos_def}
p(\rho)=K \rho^{\Gamma} \,,
\end{equation}
which is the form of the polytrope that we will use henceforth.

The expression for the energy density can be derived using the first law of thermodynamics. Replacing Eq.~\eqref{eq:eos_def} in Eq.~\eqref{eq:thermoeos2}, one obtains the differential equation
\begin{equation}
\frac{d\epsilon}{d\rho}=\frac{\epsilon+K \rho^{\Gamma}}{\rho} \,,
\end{equation}
whose solution is
\begin{equation}
\begin{aligned}
\epsilon(\rho)& =(1+a)\rho+\frac{K \rho^{\Gamma}}{\Gamma-1} \qquad \Gamma \neq 1 \\
\epsilon(\rho)& =(1+b)\rho + K \rho \log{\rho} \qquad \Gamma = 1 \,,
\end{aligned}
\end{equation}
where $a$ and $b$ are integration constants. Imposing that in the non-relativistic limit the energy density reduces to the rest-mass density,
\begin{equation}
\label{eq:constraint}
\lim_{\rho \to 0} \frac{\epsilon(\rho)}{\rho} = 1 \,,
\end{equation}
requires $\Gamma > 1$ and $a=0$ (cf. the next section). The result is
\begin{equation}
\epsilon(\rho)=\rho+\frac{K \rho^{\Gamma}}{\Gamma-1} \qquad \Gamma >1 \,.
\end{equation}

The speed of sound is given by
\begin{equation}
c^2_s(\rho)= \frac{K \Gamma (\Gamma-1) \rho^{\Gamma}}{(\Gamma-1) \rho + K \Gamma  \rho^{\Gamma}} \qquad \Gamma >1 \,.
\end{equation}
Note that $c_s$ is monotonically increasing with the rest-mass density. Thus, polytropic EOSs can violate causality: for large enough densities $c_s > 1$, i.e., the speed of sound exceeds the speed of light.

\subsection{Piecewise polytropic equations of state}
\label{sec:piecewise}
In this section we describe the piecewise polytropic model, developed by Read et al.~\cite{Read:2008iy}, that we use to solve the inverse stellar problem. We said that the adiabatic index of a polytrope is constant. Thus, simple polytropic EOSs can not be a good approximation to ``realistic'' EOSs which undergo great changes of stiffness. Piecewise polytropic models provide a solution to this problem. 

Read et al. showed that piecewise polytropes accurately fit the pressure-density profiles of a large variety of EOSs based on realistic nuclear physics calculations. These include models with plain $\mathrm{npe}\mu$ nuclear matter, hyperons, meson condensates and phase transitions to deconfined quarks, obtained within both non-relativistic many-body methods and relativistic mean-field approaches (cf. section~\ref{sec:tabeos}). The neutron star macroscopic observables, like masses and radii, are accurately reproduced within $\lesssim 1\%$ of the corresponding values predicted by the realistic models.

\begin{figure}[]
\centering
\includegraphics[width=0.6\textwidth]{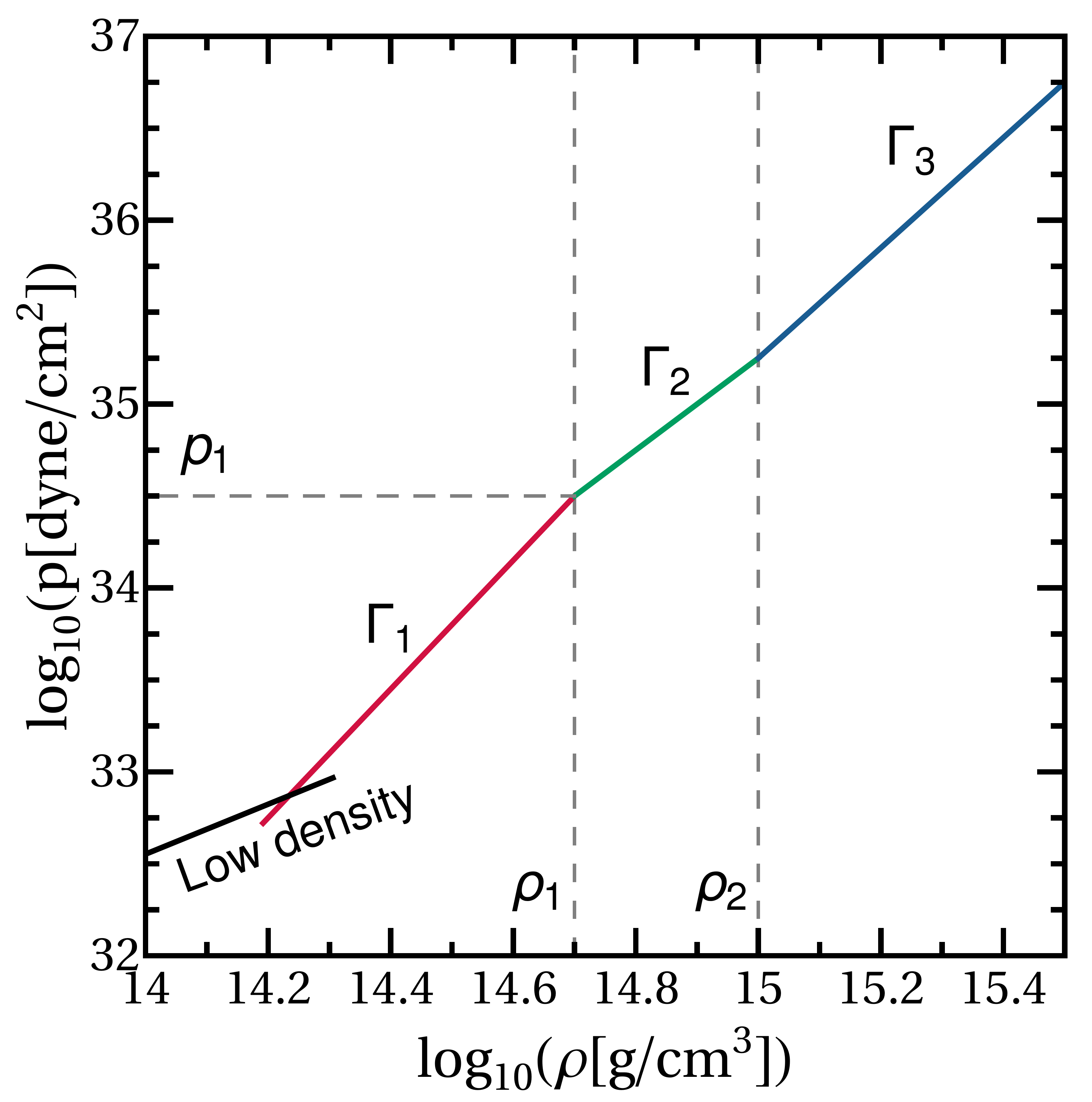}
\caption{\textsl{Schematic representation of the regions of the piecewise polytropic model in the neutron star core. See the text for details.}}
\captionsetup{format=hang,labelfont={sf,bf}}
\label{fig:piececore}
\end{figure}

A piecewise polytropic EOS is defined for $\rho \geq \rho_0$ (where $\rho_0$ can be freely chosen) by
\begin{equation}
p(\rho)=K_i \rho^{\Gamma_i} \qquad \rho_{i-1} \leq \rho \leq \rho_i \qquad i =1, \dots, N \,,
\end{equation}
where $\rho_0<\cdots<\rho_i<\cdots<\rho_N$ are $N+1$ dividing densities which partition the range of densities $\rho \geq \rho_0$ in $N+1$ regions. Imposing the continuity of the pressure across each boundary gives
\begin{equation}
\label{eq:contk}
K_{i+1} = \frac{p(\rho_ {i})}{\rho_{i}^{\Gamma_{i+1}}} \qquad i =1, \dots, N \,.
\end{equation}
The polytropic constant $K_1$ of the first region is determined specifying the value of the pressure $p_0$ at $\rho = \rho_0$,
\begin{equation}
K_1 = \frac{p_0}{\rho_{0}^{\Gamma_{1}}}  \qquad p_0 = p(\rho_0) \,.
\end{equation}
The integral of the first law of thermodynamics~\eqref{eq:thermoeos2} in each region gives
\begin{equation}
\begin{aligned}
\epsilon(\rho)& =(1+a_i)\rho+\frac{K_i }{\Gamma_i-1}\rho^{\Gamma_i} \qquad \Gamma_i \neq 1\\
\epsilon(\rho)& =(1+b_i)\rho+ K_i \rho \log{\rho} \qquad \Gamma_i= 1
\end{aligned} \qquad \rho_{i-1} \leq \rho \leq \rho_i \qquad i =1, \dots, N \,,
\end{equation}
where
\begin{table}
\centering
\begin{tabular}{ccc}
\toprule
{$\Gamma_i $ }     &   { $K_i/c^2 \ \left[\text{g}^{1-\Gamma_i} \ \text{cm}^{3(\Gamma_i-1)}\right] $}    &   {  $\rho_i \ \left[\text{g/cm}^3 \right] $  }    \\
\midrule
1.58425       &        6.80110 $\times 10^{-9}  $     &      2.44034 $\times 10^{7} $   \\[0.5ex]
1.28733       &        1.06186 $\times 10^{-6} $      &      3.78358 $ \times 10^{11}$     \\[0.5ex]
0.62223       &        5.32697 $\times 10^{1} $     &      $\rho_{\mathrm{L}} \equiv  $ 2.62780 $ \times  10^{12} $   \\[0.5ex]
$\Gamma_{\mathrm{SLy}} \equiv   $ 1.35692       &       $K_{\mathrm{SLy}} \equiv  $ 3.99874 $ \times 10^{-8}   $    &     {$\quad \ \  \rho_{M}   $  }        \\
\bottomrule
\end{tabular}
\caption{\textsl{Parameters of the piecewise polytropic version of the SLy EOS used in the neutron star crust region. We restored the value of the speed of light $c$. See the text for details on the highlighted constants.}}
\captionsetup{format=hang,labelfont={sf,bf}}
\label{tab:sly}
\end{table}
\begin{equation}
\begin{aligned}
a_i & =\frac{\epsilon(\rho_{i-1})}{\rho_{i-1}}-1-\frac{K_i }{\Gamma_i-1}\rho_{i-1}^{\Gamma_i-1} \\
b_i & =\frac{\epsilon(\rho_{i-1})}{\rho_{i-1}}-1- K_i \log{\rho_{i-1}}
\end{aligned}  \qquad i =1, \dots, N 
\end{equation}
to make the energy density continuous. In the first region, $i=1$, this requires to specify the value of the energy density $\epsilon_0 = \epsilon(\rho_0)$. Note that the above solution is continuous in $\Gamma_i$. Indeed, replacing $a_i$ and $b_i$, we get
\begin{equation}
\begin{aligned}
\epsilon(\rho) & =\frac{\epsilon(\rho_{i-1})}{\rho_{i-1}} \rho+ K_i \rho  \frac{\rho^{\Gamma_i-1}- \rho_{i-1}^{\Gamma_i-1}}{\Gamma_i-1} \quad \Gamma_i \neq 1\\
\epsilon(\rho)& =\frac{\epsilon(\rho_{i-1})}{\rho_{i-1}} \rho+ K_i \rho \log{\left(\frac{\rho}{\rho_{i-1}} \right)} \quad \Gamma_i= 1
\end{aligned} \quad \rho_{i-1} \leq \rho \leq \rho_i \qquad i =1, \dots, N \,,
\end{equation}
and we recognizes that
\begin{equation}
\lim_{\Gamma_i \to 1} \frac{\rho^{\Gamma_i-1}- \rho_{i-1}^{\Gamma_i-1}}{\Gamma_i-1} = \log{\left(\frac{\rho}{\rho_{i-1}} \right)} \,.
\end{equation}
Finally, the speed of sound reads
\begin{equation}
c^2_s(\rho)=\frac{\Gamma_i p}{\epsilon+p}  \qquad \rho_{i-1} \leq \rho \leq \rho_i \qquad i =1, \dots, N 
\end{equation}
and it is discontinuous at each $\rho_i$, because of the change of the adiabatic index.

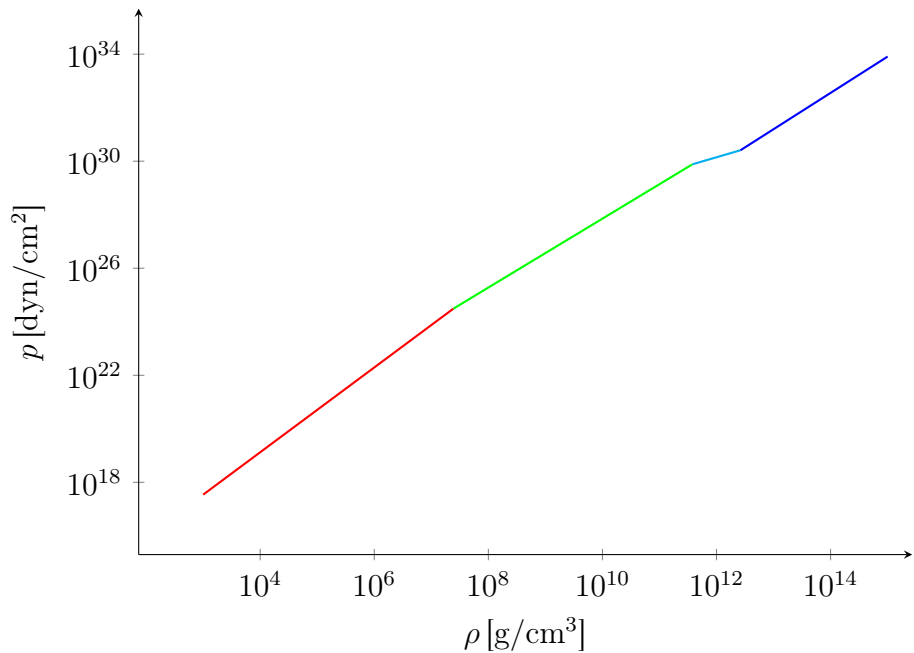
\begin{figure}[]
\centering
\begin{tikzpicture}
\begin{loglogaxis}[xmin=1e3,xmax=2e14,ymin=1e17,ymax=1e34,
axis x line=bottom,
axis y line=left,
enlargelimits,
width=0.8\textwidth,
height=0.6\textwidth,
xtick={1e4,1e6,1e8,1e10,1e12,1e14},ytick={1e18,1e22,1e26,1e30,1e34},
xlabel=$\rho \, {\text{[g$/$cm$^3$]}}$,ylabel=$p \, {\text{[dyn$/$cm$^2$]}}$]
\addplot
[domain=1e3:2.44034e7,samples=100,smooth,
thick,red]
{(6.80110e-9*(2.998e10)^2)*x^1.58425};
\addplot
[domain=2.44034e7:3.78358e11,samples=100,smooth,
thick,green]
{(1.06186e-6*(2.998e10)^2)*x^1.28733};
\addplot
[domain=3.78358e11:2.62780e12,samples=100,smooth,
thick,cyan]
{(5.32697e1*(2.998e10)^2)*x^0.62223};
\addplot
[domain=2.62780e12:10e14,samples=100,smooth,
thick,blue]
{(3.99874e-8*(2.998e10)^2)*x^1.35692};
\end{loglogaxis}
\end{tikzpicture}
\caption{\textsl{Plot of the analytical version of the SLy EOS. The four regions of the piecewise polytropic representation correspond approximately to: a non-relativistic electron gas, a relativistic electron gas, the neutron drip and the inner crust.}}
\captionsetup{format=hang,labelfont={sf,bf}}
\label{fig:crust}
\end{figure}

The microscopic stability condition, $dp/d\epsilon \geq 0$, imposes $\Gamma_i \geq 0$. Values of the adiabatic index in the range $0 \leq \Gamma_i \leq 1$ are allowed in piecewise polytropic EOSs, provided that $\rho_0 \neq 0$ (see the previous section, Eq.~\eqref{eq:constraint}). In particular, $\Gamma_i = 0$ (which means a constant pressure across the interval) allows us to take into account first-order phase transitions in the neutron star matter.

Read et al. found that modeling the neutron star high-density core with three polytropic segments accurately reproduce a large set of realistic EOSs. The values of the dividing densities which minimize the discrepancy with respect to the tabulated EOSs correspond to  $\rho_1 = \text{10}^{\text{14.7}} \, \text{g/cm}^{\text{3}}$  and $\rho_2 = \text{10}^{\text{15}} \, \text{g/cm}^{\text{3}}$. This model has four independent parameters: the adiabatic indices of the three regions and the value of the pressure at the first dividing density, namely $\{p_1,\Gamma_1,\Gamma_2,\Gamma_3\}$, where $p_1 = p(\rho_1)$. The polytropic constants $K_i$ are given by Eq.~\eqref{eq:contk}, with $K_1 = p_1/\rho_1^{\Gamma_1}$. A schematic picture of this model is shown in Fig.~\ref{fig:piececore} (note that the polytropic branches are straight lines in log-log scale, with the slope given by the adiabatic index).

At low densities, the outermost polytropic segment is matched dynamically to a fixed crust, which is chosen to be a parametrized four-piece polytropic version of the SLy EOS~\cite{1958NucPh...9..615S,Douchin:2001sv}. Its parameters are summarized in Table~\ref{tab:sly}. The pressure-density profile of the crust EOS is shown in Fig.~\ref{fig:crust}.

\begin{figure}[]
\captionsetup[subfigure]{labelformat=empty}
\centering
{\includegraphics[width=0.45\textwidth]{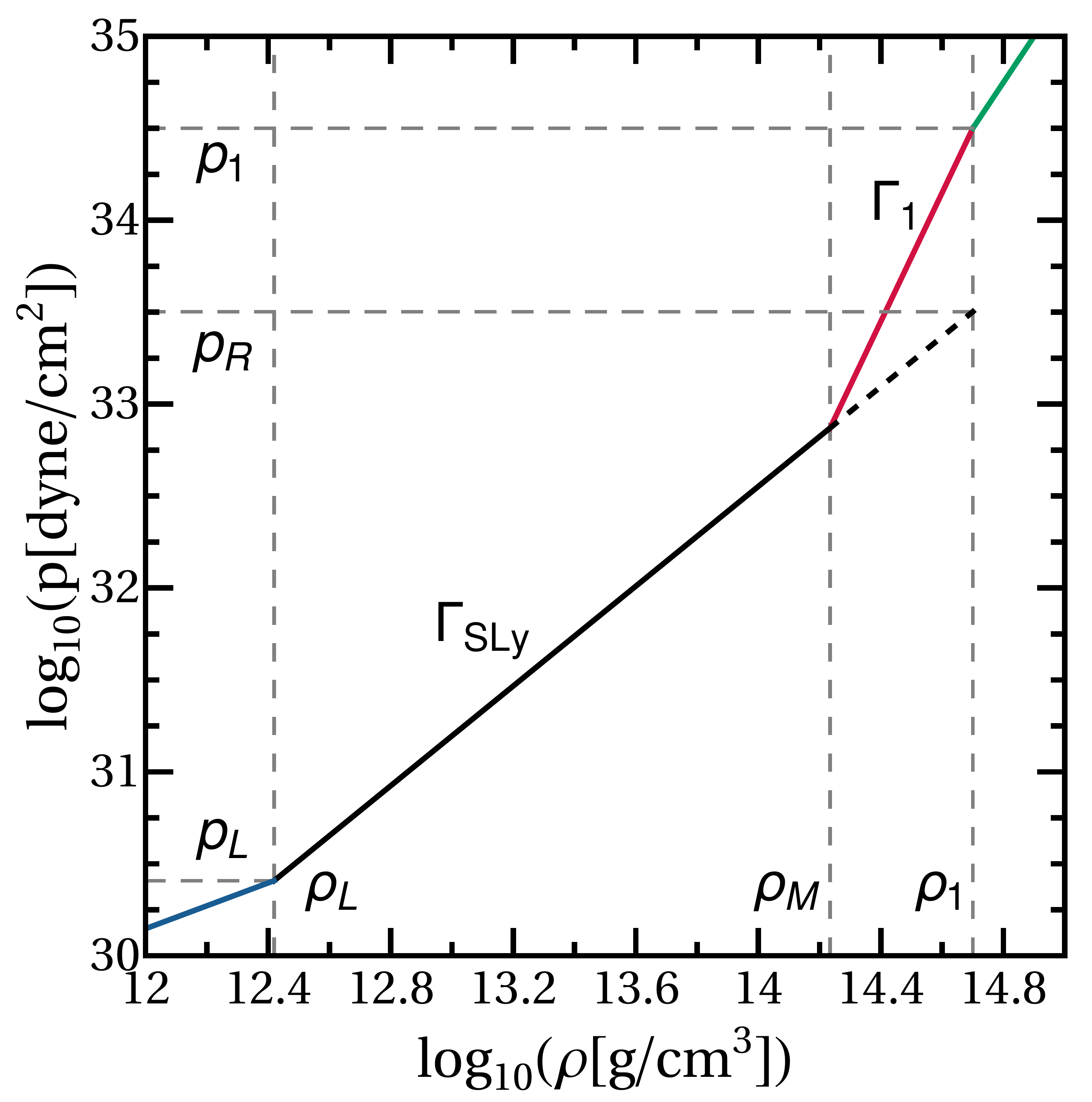}} \ \ \ 
{\includegraphics[width=0.45\textwidth]{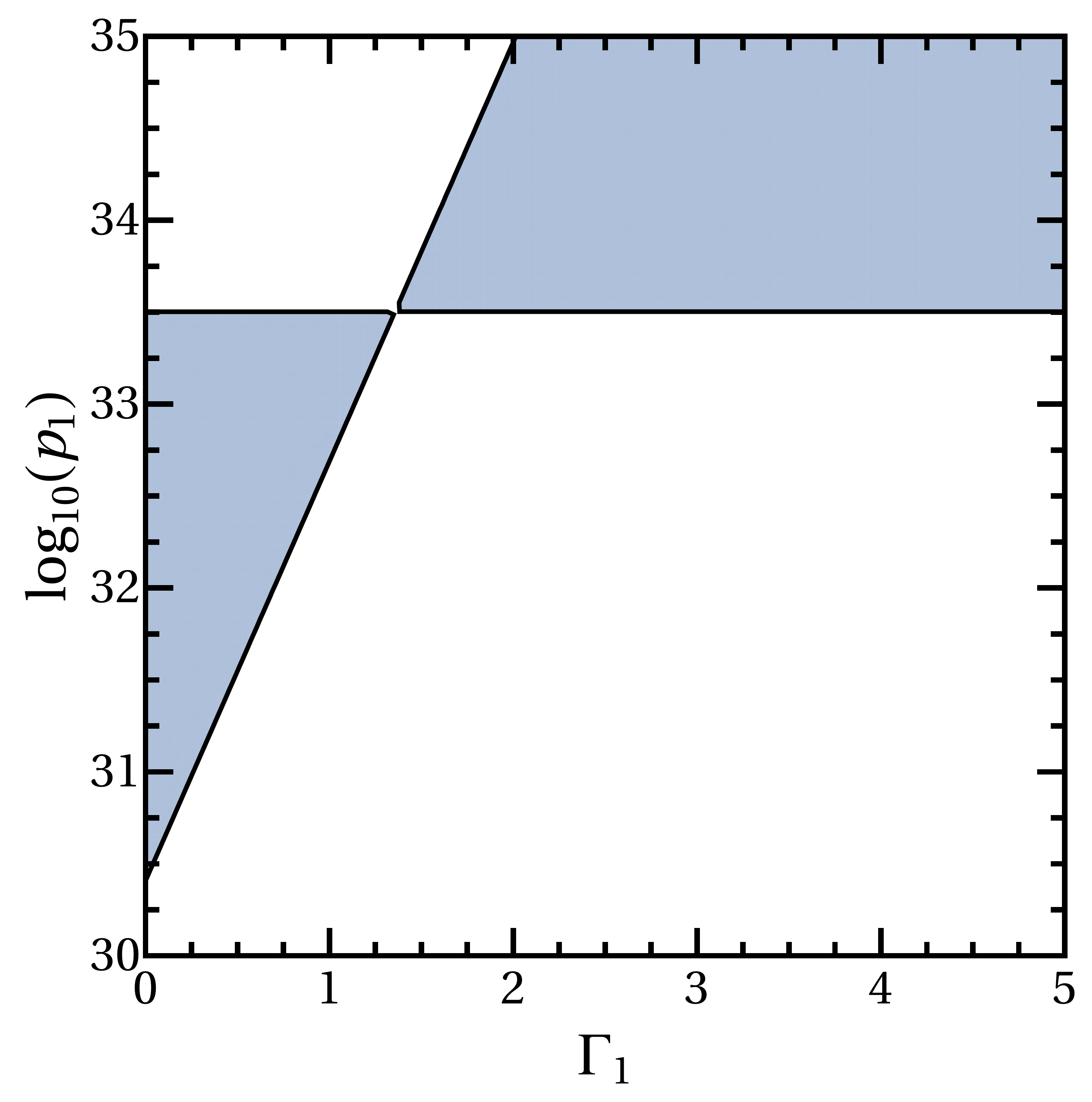}}
\caption{\textsl{(Left) Schematic representation of the match between the crust and the core EOSs. See the text for details. (Right) Constraint imposed on the parameters $p_1$ and $\Gamma_1$ by the matching procedure. The allowed region is shown in blue.}}
\captionsetup{format=hang,labelfont={sf,bf}}
\label{fig:match}
\end{figure}

The matching point between crust and core is simply given by the value of density where the low and high-density EOSs intersect each other, and depends only on $p_1$ and $\Gamma_1$. It reads
\begin{equation}
\rho_{M}=\left(\frac{K_1}{K_{\mathrm{SLy}}}\right)^{1/(\Gamma_{\mathrm{SLy}}-\Gamma_1)} \,,
\end{equation}
where $K_{\mathrm{SLy}}$ and $\Gamma_{\mathrm{SLy}}$ are the polytropic parameters of the innermost crust region, given in Table~\ref{tab:sly}. This choice naturally implies a constraint on $p_1$ and $ \Gamma_1$, since specific combinations of them do exist, which yield no intersection between the crust and the core EOSs, and are therefore incompatible. The allowed region can be found analytically, and satisfies the following relation
\begin{equation}
\label{constraint}
\begin{gathered}
\left\{ \bigg[p_1 \geq p_R \bigg] \bigcap \left[ \Gamma_1 \geq \frac{\log{\left( p_1/p_{L} \right)}}{\log{\left( \rho_1/\rho_{L} \right)}} \right] \right\} 
 \bigcup
\left\{ \bigg[p_1 \leq p_R \bigg] \bigcap \left[ \Gamma_1 \leq \frac{\log{\left( p_1/p_{L} \right)}}{\log{\left( \rho_1/\rho_{L} \right)}} \right] \right\} \,,
\end{gathered}
\end{equation}
where $\rho_L$ and $p_L = K_{\mathrm{Sly}} \rho_L^{\Gamma_{\mathrm{Sly}}}$ are, respectively, the density and the pressure at the interface of the two innermost crust regions, and $p_R = K_{\mathrm{Sly}} \rho_1^{\Gamma_{\mathrm{Sly}}}$ is the value of the pressure assumed by the last low-density polytropic segment if extended up to $\rho_1$ (see the left panel of Fig.~\ref{fig:match}). The values of $\rho_L$ is given in Table~\ref{tab:sly}. We show the above region, in the $p_1$--$\Gamma_1$ plane, in the right panel of Fig.~\ref{fig:match}. Note that we are implicitly requiring that the match occurs in the innermost crust region, i.e., at density $\rho_M \geq \rho_L$.

\begin{figure}[]
\captionsetup[subfigure]{labelformat=empty}
\centering
{\includegraphics[width=0.45\textwidth]{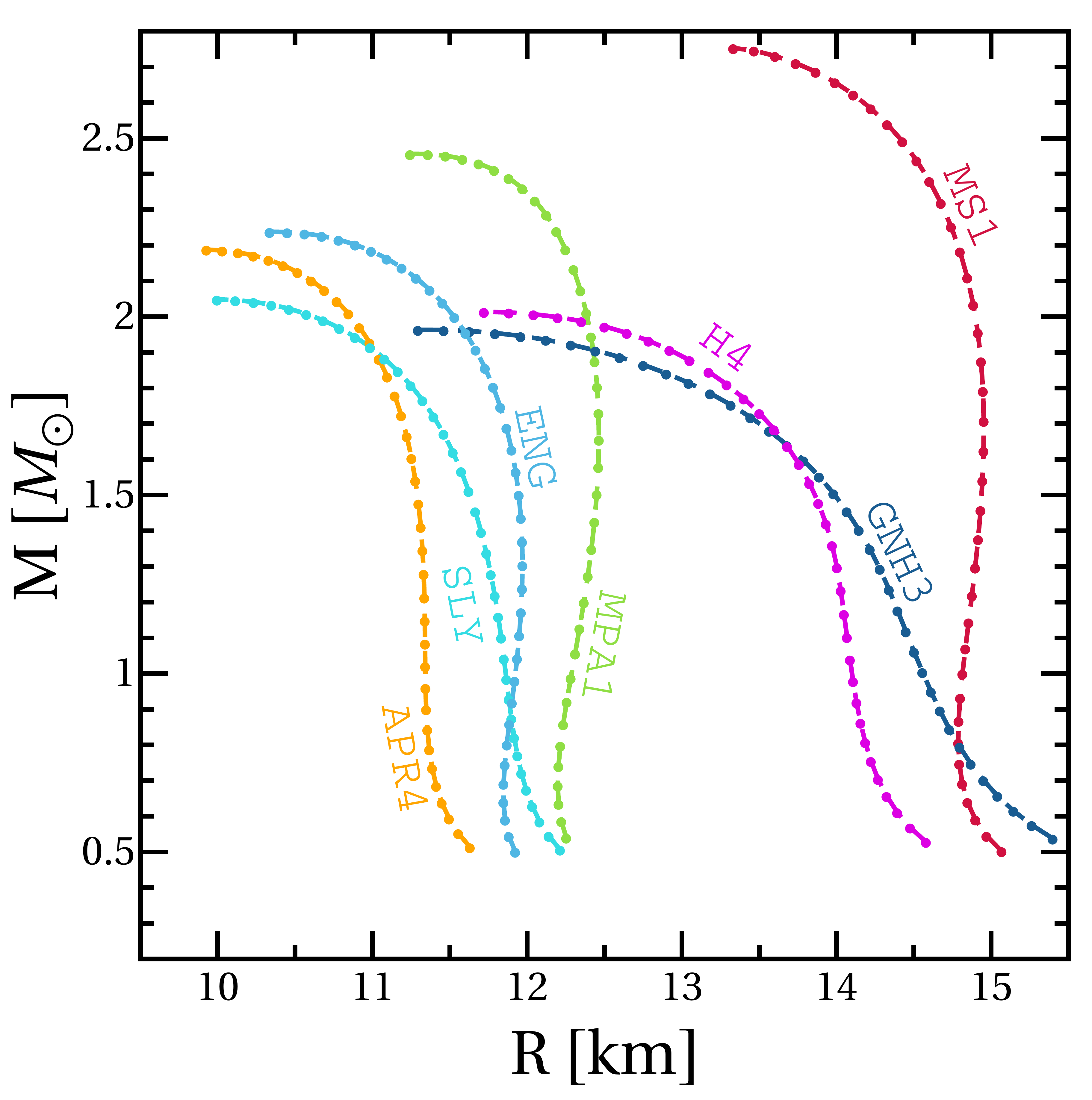}} \ \ \ 
{\includegraphics[width=0.45\textwidth]{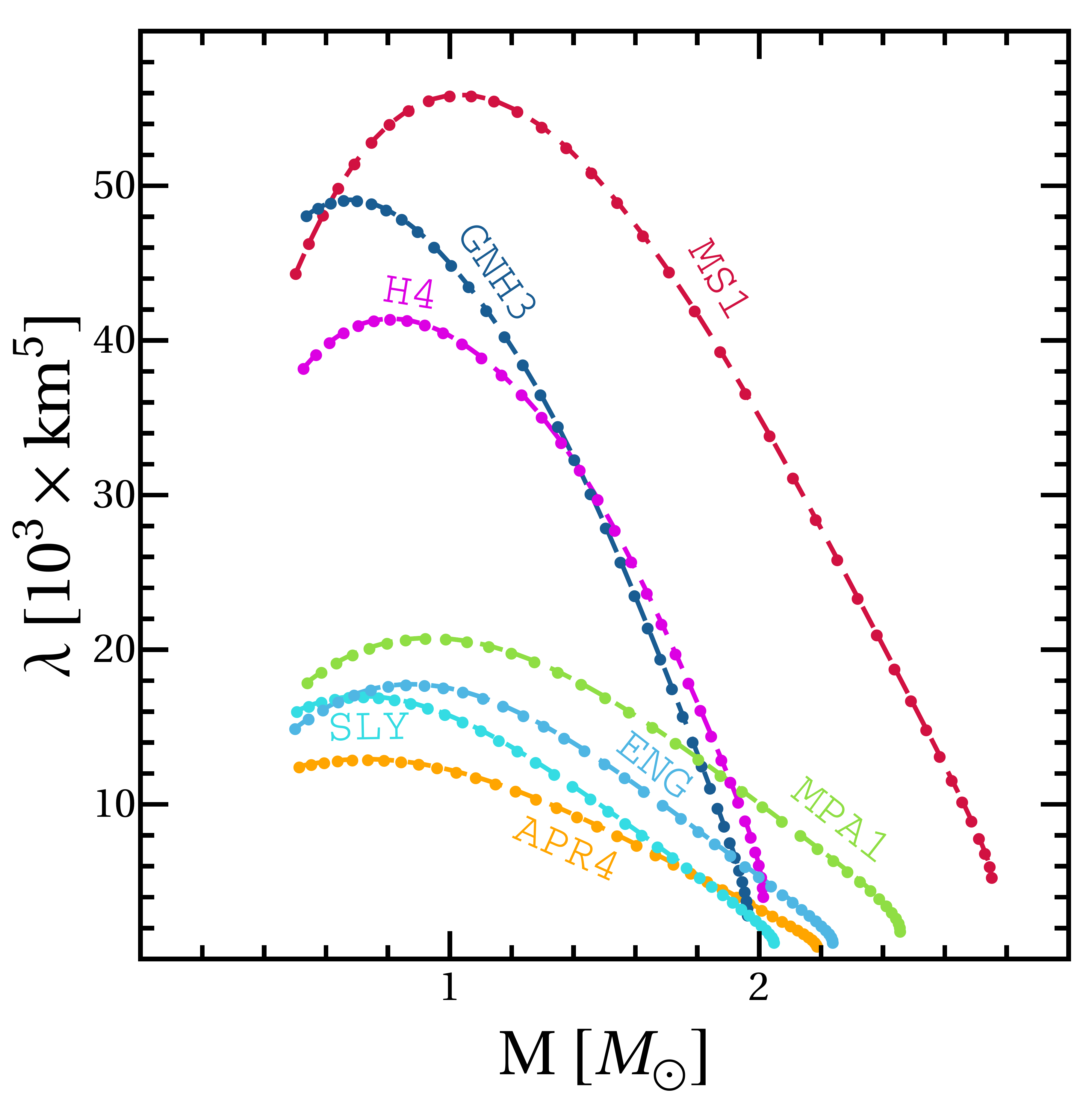}}
\caption{\textsl{(Left) Mass-radius relations for some realistic EOSs modeled through the piecewise polytropic representation. The values of the parameters $\{p_1, \Gamma_1, \Gamma_2, \Gamma_3\}$ which specify the EOSs, as well as the details on the microscopic composition of each realistic model, can be found in~\cite{Read:2008iy}. (Right) Tidal deformability as a function of the neutron star mass for the same EOSs considered in the left panel.}}
\captionsetup{format=hang,labelfont={sf,bf}}
\label{fig:mrlambda}
\end{figure}

Finally, in Fig.~\ref{fig:mrlambda} we show the mass-radius and mass-tidal deformability diagrams obtained solving the relativistic equations of stellar structure for different EOSs (see sections~\ref{sec:tov} and~\ref{sec:polar} and the Appendix~\ref{sec:appA}), modeled through the piecewise polytropic parametrization. The EOS models which give large radii and tidal deformabilities are the stiff ones, whereas the EOSs that lead to more compact, less deformable objects are the soft ones.

With a slight abuse of notation, henceforth we define $p_1$ to be the logarithm (to base ten) of the pressure evaluated at $\rho_1$, $p_1 \equiv  \log_{10}{p(\rho_1)}$.

\section{The Bayesian framework}
\label{sec:MCMC}
In this section we describe the approach that we use to estimate the EOS parameters $\{p_1,\Gamma_1,\Gamma_2,\Gamma_3\}$ of the piecewise polytropic representation, starting from the macroscopic observables provided by gravitational wave observations, namely the mass $M$ and the tidal deformability $\lambda$ of the detected neutron stars. We stress that our method is completely general, and can be applied also using different neutron star observables, obtained either with electromagnetic or gravitational wave observations, leading to a multimessenger framework. 

In general, for a given set of $N$ observed stars, we have $m+N$ free parameters to determine, i.e., $m$ parameters of the EOS model, and $N$ central pressures $p^c_{i} \ (i=1,\dots,N)$. We assume that any detected neutron star provides $n=2$ observables, which in our case are the mass and the tidal deformability. Therefore, to fully characterize the parametrized EOS, we need at least $N=m$ observations~\footnote{\label{foot2}The counting of the observations needed to characterize the EOS derives from the idealistic inverse stellar problem, where the uncertainties on the measurements are not taken into account. In the latter case, at least $N$ pairs of independent observables are needed to constrain $N$ EOS parameters, otherwise the solution is not unique. An insufficient number of observations would constrain the EOS parameters anyway, but it would not be able to remove completely the degeneracy.}. 

As discussed in the previous section, piecewise polytropes are characterized by $m=4$ parameters, which lead to $8$ unknown parameters to be found:
\begin{equation}
\vec{\theta}=\{p_1,\Gamma_1,\Gamma_2,\Gamma_3,p^c_1,p^c_2,p^c_3,p^c_4\} \,.
\end{equation}
Therefore, we need at least $N=4$ observations, which provide the required set of $8$ measured quantities:
\begin{equation}
\vec{d}=\{M_1,\lambda_1,M_2,\lambda_2,M_3,\lambda_3,M_4,\lambda_4\} \,.
\end{equation}

Within the Bayesian scheme of inference (see, e.g.,~\cite{Dagos:2003}), we are interested in determining the posterior probability density function (PDF) of the EOS parameters given the experimental data, $\mathcal{P} (\vec{\theta}\vert\vec{d})$~\footnote{When realistic data are used, the deterministic solution of the idealistic inverse problem (see footnote~\ref{foot2}) is spoiled by the intrinsic probabilistic nature of the experimental measurements, and transformed into a probability distribution.}. Using Bayes theorem, we can write the joint posterior PDF as 
\begin{equation} \mathcal{P} (\vec{\theta}\vert\vec{d})  \propto
 \mathcal{L} (\vec{d}\vert\vec{\theta})  \, \mathcal{P}_0(\vec{\theta}) \,,
\label{eq:PDF} 
\end{equation} 
where $\mathcal{L}(\vec{d}\vert\vec{\theta})$ is the likelihood function, i.e., the PDF of the experimental data given the EOS parameters, and $\mathcal{P}_0(\vec{\theta})$ the prior PDF, which describes the former information on the parameters. The probability distribution of $l$ parameters is given by marginalizing over the remaining $8-l$ variables, i.e.,
\begin{equation} 
\mathcal{P}(\theta_1,\dots,\theta_{l}\vert\vec{d})=\int \mathcal{P}(\vec{\theta}\vert\vec{d}) \, d\theta_{l+1}\dots d\theta_{8} \,.  
\end{equation}

In our analysis we assume that the set of data $\vec{d}$ obtained from gravitational wave detections are independent and Gaussian distributed, with the values of each observable $M_i\ (\lambda_i)$ being affected by an experimental uncertainty $\sigma_{M_i}\ (\sigma_{\lambda_i})$~\footnote{\label{foot3}This analysis was started and almost completed before the first gravitational wave detection of a binary neutron star. Therefore, no real data were available at the time, which is the reason why we have used mock data. Even after the GW170817 event, the observed neutron stars would have not been enough to fully constrain the EOS parameters. We stress that our goal is to show the feasibility of constraining the EOS using gravitational wave observations, and not to exploit the information coming from the binary neutron star detection to actually infer the EOS.}. Under these assumptions, the likelihood can be written as
\begin{equation}
{\cal L}\propto\mathrm{e}^{-\chi^2} \,,
\end{equation}
where the chi-square variable reads
\begin{equation} \chi^2 = \sum_{i=1}^4 \left\{\frac{\left[ M \left(p_1,\Gamma_1,\Gamma_2,\Gamma_3,p^c_i \right)-M_i \right]^2}{2\sigma_{M_i}^2}  +\frac{\left[\lambda \left(p_1,\Gamma_1,\Gamma_2,\Gamma_3,p^c_i \right)-\lambda_i \right]^2}{2\sigma_{\lambda_i}^2}\right\} . 
\end{equation}

It is straightforward to generalize the above formalism to an EOS representation with an arbitrary number of parameters $m$ and/or to an arbitrary number of neutron star observations $N$, each of them providing $n$ independent observables.

\subsection{The Markov chain Monte Carlo}
We sample the posterior probability distribution $\mathcal{P}(\vec{\theta}|\vec{d})$ in Eq.~\eqref{eq:PDF} using Markov chain Monte Carlo (MCMC) simulations based on the Metropolis-Hastings algorithm (see, e.g.,~\cite{Gilks:1996,Brooks:2011}). The procedure of this framework can be summarized with the following steps.

Given an initial point $\vec{\theta}_1=\left\{p_1,\Gamma_1,\Gamma_2,\Gamma_3,p^c_1,p^c_2,p^c_3,p^c_4\right\}$, randomly chosen within the parameter space, we propose a jump to a new state, $\vec{\theta}_2$, with probability specified by the proposal function $f=f(\vec{\theta}_1,\vec{\theta}_2)$. The latter is chosen to be a multivariate Gaussian distribution centered in the current state $\vec{\theta}_1$,
\begin{equation}
f(\vec{\theta}_1,\vec{\theta}_2)  = \mathrm{exp} \left[ -\frac{1}{2} \left( \vec{\theta}_2-\vec{\theta}_1\right)^{\mathrm{T}} \mathbf{\Sigma}^{-1} \left( \vec{\theta}_2-\vec{\theta}_1\right) \right] \,,
\end{equation}
where $\mathbf{\Sigma}$ is the covariance matrix (see below) and $\null^{\mathrm{T}}$ denotes the transpose operator. Note that with this choice $f$ is symmetric, i.e., $f(\vec{\theta}_2,\vec{\theta}_1)=f(\vec{\theta}_1,\vec{\theta}_2)$. Then, we compute the ratio
\begin{equation}
r(\vec{\theta}_1,\vec{\theta}_2) = \frac{\mathcal{P}(\vec{\theta}_2|\vec{d})}{\mathcal{P}(\vec{\theta}_1|\vec{d})}\,,
\end{equation}
and accept the proposed move with probability
\begin{equation}
a(\vec{\theta}_1,\vec{\theta}_2)=\mathrm{min}\left\{1,r(\vec{\theta}_1,\vec{\theta}_2)\right\} \,.
\end{equation}
In this way, the chain is updated to the state $\vec{\theta}_2$ with probability $a(\vec{\theta}_1,\vec{\theta}_2)$, or remains fixed in $\vec{\theta}_1$ with probability $1-a(\vec{\theta}_1,\vec{\theta}_2)$. If $\mathcal{P}(\vec{\theta}_2|\vec{d}) \geq \mathcal{P}(\vec{\theta}_1|\vec{d})$ the jump is always accepted, while if $\mathcal{P}(\vec{\theta}_2|\vec{d}) < \mathcal{P}(\vec{\theta}_1|\vec{d})$ it is accepted with probability $r(\vec{\theta}_1,\vec{\theta}_2)$. The previous steps are then iterated $n$ times, allowing the chain to explore the parameter space of the model (see Algorithm~\ref{mh} below).

The MCMC theory guarantees that, from any initial state and proposal function, the system evolves towards the desired target distribution $\mathcal{P}(\vec{\theta}|\vec{d})$. However, in practical situations the convergence of the chain is strongly affected by the choice of the proposal function. In this thesis we adopt an adaptive framework, in which the covariance matrix $\mathbf{\Sigma}$ of $f(\vec{\theta}_1,\vec{\theta}_2)$ is continuously updated through a \emph{Gaussian adaptation} (GaA) algorithm~\cite{1085030,5586491}. A remarkable feature of this approach is that the acceptance probability $P$ of the proposed jump can be fixed a priori. In the following section we describe in detail the features of the algorithm.

\subsection{The Gaussian adaptation algorithm}
\label{sec:gaussian}
According to the GaA algorithm, the covariance matrix $\mathbf{\Sigma}$ of the proposal distribution
$f(\vec{\theta}_1,\vec{\theta}_2)$ is defined as
\begin{equation}
\mathbf{\Sigma} = \big(\rho \mathbf{Q}\ \big) \big( \rho \mathbf{Q}^{\mathrm{T}} \big) \,, 
\end{equation} 
where $\rho$ is the step size of the algorithm and $\mathbf{Q}$ the square root of the covariance matrix, normalized such that $\mathrm{det}(\mathbf{Q})=1$. We compute $\mathbf{Q}$ from $\mathbf{\Sigma}$ using the Cholesky decomposition.

The structure of the adaptive Metropolis-Hastings algorithm used in the MCMC is the following: we start from an initial state $\vec{\theta}_1$, setting $\rho=1$ and $\mathbf{\Sigma} = \mathbf{Q} = \mathbbm{1}$, where $\mathbbm{1}$ is the identity matrix. Then, at each step a new point is sampled as
\begin{equation} 
\vec{\theta}_{i+1} = \vec{\theta}_{i} + \rho \, \mathbf{Q} \cdot \vec{\eta}\,, 
\end{equation}
where $\vec{\eta}$ is a vector drawn from a Gaussian distribution with zero mean and unit variance, $\vec{\eta} \sim \mathcal{N}(\vec{0},\mathbbm{1})$. If the proposed move $\vec{\theta}_{i+1}$ is accepted, the step size and the covariance matrix are updated according to the following rules:
\begin{equation} 
\begin{gathered} 
\rho  \to f_e \, \rho \,,\\
\mathbf{\Sigma}  \to \left(1-\frac{1}{N_\tn{C}} \right) \mathbf{\Sigma} + \frac{1}{N_\tn{C}} \big(\Delta \vec{\theta} \big) \big( \Delta \vec{\theta} \big)^{\mathrm{T}} \,, 
\end{gathered}
\end{equation} 
where $f_e>1$ is called {\it expansion factor}, $N_\tn{C}$ is a free parameter of the GaA algorithm and
$\Delta \vec{\theta} =\vec{\theta}_{i+1}-\vec{\theta}_i$. Conversely, if the proposed jump is rejected, the covariance matrix is not updated and the step size is reduced by a {\it contraction factor} $f_c<1$:
\begin{equation}
\begin{gathered}
\rho  \to f_c \, \rho \,, \\
\mathbf{\Sigma} \to  \mathbf{\Sigma} \,.  
\end{gathered}
\end{equation} 
A workflow of this procedure is shown in Algorithm~\ref{mh}.

The GaA algorithm relies on some free parameters, which following~\cite{5586491}, we have fixed to the following values:
\begin{equation} 
\begin{aligned} 
f_e & = 1+\beta (1-P) \\
 f_c & = 1- \beta P \\ 
\beta & = 1/N_\tn{C} \\ 
N_C & =(D+1)^2 / \log{(D+1)} \,,
\end{aligned}
\end{equation} 
where $D$ is the dimension of the MCMC parameter space and $P$ is the acceptance probability of the proposed move. For our simulations we found that an optimal value of such probability, which guarantees an efficient mixing of the chains~\footnote{We stress that the GaA algorithm just described comes with a flaw. The MCMC theory guarantees that the Markov chain converges asymptotically to the desired target distribution $\mathcal{P}(\vec{\theta}|\vec{d})$, for any given proposal distribution $f(\vec{\theta_1},\vec{\theta_2})$, if such function is stationary (i.e., it does not change at each step) or if it has a \emph{diminishing} adaptation~\cite{haario2001,10.2307/27595854,doi:10.1198/jcgs.2009.06134}. The latter property means that asymptotically the \emph{local} adaptation of $f(\vec{\theta_1},\vec{\theta_2})$ from the step $n$ to the step $n+1$ must be infinitely small (with $n$). Note that this requirement allows anyway a \emph{global} residual finite adaptation even at large $n$. However, the above GaA algorithm is not in this class of functions, since the local adaptation for $n \to \infty$ can be arbitrarily large. From this follows that the GaA algorithm \emph{could} break the ergodicity of the MCMC, which means that the Markov chain could not converge to the desired target distribution, or it could not converge to a stationary distribution at all. We checked if this is the case for our simulations simply switching off the adaptation at large times, preserving in this way the ergodicity of the system.}, corresponds to $P=0.25$. 

\renewcommand\algorithmicthen{}
\renewcommand\algorithmicdo{}
\begin{algorithm}
\caption{Adaptive Metropolis-Hastings}\label{mh}
\begin{algorithmic}[0]
\State \textbf{Start}: $\vec{\theta}_1,\rho=1,\mathbf{\Sigma} = \mathbbm{1}$
\For{$i=1,\dots,n$}
\State \emph{evaluate} $\mathbf{Q}$ by Cholesky decomposition of $\mathbf{\Sigma}$
\State \emph{normalize} $\mathbf{Q} \to \mathbf{Q}/\mathrm{det}(\mathbf{Q})^{1/D}$ 
\State \emph{propose move} $\vec{y} = \vec{\theta}_i+ \rho \, \mathbf{Q} \cdot \vec{\eta}$ with $\vec{\eta}\sim\mathcal{N}(\vec{0},\mathbbm{1})$
\State \emph{evaluate ratio} $\mathcal{P}(\vec{y}|\vec{d})/\mathcal{P}(\vec{\theta}_i|\vec{d})$
\If{\textbf{accepted}}
\State $\vec{\theta}_{i+1}= \vec{y}$
\State $\rho \to f_e \, \rho$
\State $\mathbf{\Sigma}  \to \left(1-\frac{1}{N_\tn{C}} \right) \mathbf{\Sigma} 
+ \frac{1}{N_\tn{C}} \big(\vec{\theta}_{i+1} -\vec{\theta}_i \big) \big( \vec{\theta}_{i+1} -\vec{\theta}_i \big)^{\mathrm{T}}$
\EndIf
\If{\textbf{rejected}}
\State $\vec{\theta}_{i+1}= \vec{\theta}_i$
\State $\rho \to f_c \, \rho$
\State $\mathbf{\Sigma}  \to  \mathbf{\Sigma} $
\EndIf
\State \ 
\EndFor
\end{algorithmic}
\end{algorithm}

\section{Results of the numerical simulations}
\label{sec:setup}
To test the ability of our approach to reconstruct the parameters of the piecewise polytropes, we analyze different possible scenarios. We consider non-spinning neutron stars (we recall that spin-tidal effects are negligible for second generation interferometers, see section~\ref{sec:LIGOspin}) with mass $M$ in the range $(1.1 \div 1.6)M_\odot$, which covers most of the mass range determined so far by electromagnetic observations of double neutron stars~\cite{Lattimer:2012nd,Ozel:2016oaf}. Also, it includes the observed masses of the first gravitational wave detection from a binary neutron star~\cite{TheLIGOScientific:2017qsa,Abbott:2018wiz}. Moreover, we focus on two EOSs, \texttt{apr4}~\cite{Akmal:1998cf} and \texttt{h4}~\cite{Lackey:2005tk}. As shown in Fig.~\ref{fig:mrlambda}, these models span a wide range of mass-radius/tidal deformability configurations. Furthermore, they fit within the $90\%$ credible interval estimated by the LIGO/Virgo collaboration after the gravitational wave event GW170817~\cite{TheLIGOScientific:2017qsa}~\footnote{We notice that in a following, more refined, analysis, performed after our work was completed, the LIGO/Virgo collaboration ruled out also the \texttt{h4} EOS~\cite{Abbott:2018wiz}}. Therefore, \texttt{apr4} and \texttt{h4} are the best candidates to represent extreme cases of {\it soft} and {\it stiff} nuclear matter, compatible with astrophysical observations. For both EOSs, we compare the features of a canonical $1.4M_\odot$ neutron star in Table~\ref{table:massradius}, which also shows that the tidal deformability of the two EOSs differs by a factor $\gtrsim 3$. We recall that large values of $\lambda$ yield stronger changes in the gravitational wave signal (see Eq.~\eqref{phase}), and therefore lead to tighter constraints.

\begin{table}
  \centering
  \begin{tabular}{c|ccc}
\toprule
\multicolumn{1}{c}{} &EOS &  $R$ [km] & $\lambda$ [$10^3 \times$km$^5$]\\
\midrule
\rotatebox[origin=c]{90}{\ soft\ } &\texttt{apr4} &  11.34  & 9.502 \\ 
\midrule
\rotatebox[origin=c]{90}{\ stiff\ } &\texttt{h4} & 13.99 & 32.86 \\
\bottomrule
\end{tabular}
\caption{\textsl{Radius and tidal deformability of prototype $1.4M_\odot$ neutron star modeled with the EOSs \texttt{apr4} and \texttt{h4}.}}
\captionsetup{format=hang,labelfont={sf,bf}}
\label{table:massradius}
\end{table}
 
The uncertainties on the observables are computed for the advanced generation of detectors. More specifically, we assume that the gravitational wave events are detected by a network of four interferometers, composed by the two LIGO sites, Virgo and the Japanese KAGRA, which is going to join the next observation run (O3) by the end of 2019~\cite{Akutsu:2018axf}. For all the measurements we consider the detector configurations at design sensitivity~\cite{zerodet} (see also the new updated LIGO sensitivity~\cite{zerodet2}). Henceforth, we refer to such network as HLVK.

Following~\cite{Rodriguez:2013oaa}, we fix the uncertainty on the neutron star mass $\sigma_M$ to $10\%$ of the measured value for HLVK, in agreement with the uncertainties reported by the LIGO/Virgo collaboration for the component masses of the observed neutron star binary system. We compute the uncertainty on the tidal deformability $\sigma_\lambda$ using a Fisher matrix approach (see section~\ref{sec:stat}, Eq.~\eqref{eq:fishermatrix}), assuming equal-mass binary neutron stars at a prototype distance of $100$ Mpc. Note that for $N$ independent interferometers the error on the tidal deformability is roughly reduced by a factor $\sim1/\sqrt{N}$, with respect to the single detector analysis~\cite{Cutler:1994ys}. We find $\sigma_\lambda$ of order $\sim (10 \div 30) \% $, depending on the mass and EOS considered, in agreement with the simulations in~\cite{Wade:2014vqa,Read:2013zra}. As expected, softer (stiffer) EOSs, corresponding to smaller (larger) tidal deformability, lead to larger (smaller) uncertainties. We remark that the gravitational wave event GW170817 has not put very strong bounds on the individual tidal deformabilities of the neutron stars. The best constraint reported by the LIGO/Virgo collaboration is that on the average weighted tidal deformability of the two stars~\cite{Abbott:2018wiz,TheLIGOScientific:2017qsa} (see section~\ref{sec:impact}).

It is important to stress that for $M\lesssim 1.6M_\odot$ the adiabatic index $\Gamma_3$ does not affect the structure of the star for both \texttt{apr4} and \texttt{h4}, because the central densities of such stars are smaller than $\rho_2$ (cf. Fig.~\ref{fig:radius} below). Therefore, we can safely neglect this coefficient within the analysis, reducing the parameter space volume to $\vec{\theta}=\{p_1,\Gamma_1,\Gamma_2,p^c_1,p^c_2,p^c_3\}$. Note that the EOS is now fully specified by only three variables, and as a consequence we only need six observables, which correspond to three observed neutron stars (i.e., two binary neutron star coalescences, or a binary neutron star and a black hole-neutron star system).

We choose flat prior distributions $\mathcal{P}_0(\vec{\theta})$ for all the parameters, within the ranges: $p_1\in[33,35]$ (where the pressure is measured in $\text{dyn/cm}^2$), $\Gamma_{i}\in [1,4]$ and $p^c_i\in[10^{-6},10^{-3}]\ \text{km}^{-2}$ (in geometric units)~\footnote{The conversion factor for the pressure between CGS and geometric units is: $1\, \text{km}^{-2} = 10^{10} (G/c^4)  \,\text{dyn/cm}^2$ (for instance, $10^{35} \,\text{dyn/cm}^2$ correspond to $\sim 0.827 \times 10^{-4} \, \text{km}^{-2} $).}. The range of the EOS parameters is large enough to include all the EOS models considered by Read et al in~\cite{Read:2008iy}. The parameters of the outer core, $(p_1,\Gamma_1)$, are also constrained by the theoretical bound given in Eq.~\eqref{constraint}. 

\begin{table}
\centering
\begin{tabular}{c|c|cc|cc}
\toprule
\multicolumn{2}{c|}{}& \multicolumn{2}{c|}{\texttt{apr4}} & \multicolumn{2}{c}{\texttt{h4}}\\
\midrule
\multicolumn{1}{c}{} &parameter & injected & $1\sigma$ &  injected  & $1\sigma$\\
\midrule
\multirow{6}{*}{\rotatebox[origin=c]{90}{\texttt{m246}}} &$p_1 $ & 34.269 & [34.205 - 34.427] & 34.669 & [34.611 - 34.738]\\ 
&$\Gamma_1$ & 2.830 &   [2.700 - 3.896]  &  2.909  &   [2.479 - 3.401] \\
&$\Gamma_2$ & 3.445 &  [2.415 - 3.907] & 2.246 &  [1.732 - 3.518]\\
&$p_1^c \ [10^{-4}\times \text{km}^{-2}]$ & 0.862 & [0.750 - 1.15] & 0.372 & [0.310 - 0.446]\\
&$p_2^c \ [10^{-4}\times \text{km}^{-2}]$ & 1.22 &  [1.06 - 1.58] & 0.533  & [0.486 - 0.614] \\
&$p_3^c \ [10^{-4}\times \text{km}^{-2}]$ & 1.74 &  [1.39 - 2.58]& 0.804  &  [0.721 - 0.930] \\
\midrule
\multirow{6}{*}{\rotatebox[origin=c]{90}{\texttt{m456}}} &$p_1 $ & 34.269 & [34.247 - 34.582] & 34.669 & [34.628 - 34.742]\\ 
&$\Gamma_1$ & 2.830 &  [2.212 - 3.846] & 2.909 &  [1.956 - 3.906]\\
&$\Gamma_2$ & 3.445 & [1.817 - 3.599] & 2.246 & [1.056 - 2.383]\\
&$p_1^c \ [ 10^{-4}\times \text{km}^{-2}]$ & 1.22 & [1.09 - 1.76] & 0.533 & [0.423 - 0.643]\\
&$p_2^c \ [10^{-4}\times \text{km}^{-2}]$ & 1.45 & [1.29 - 2.12] & 0.650 & [0.556 - 0.773]\\
&$p_3^c \ [10^{-4}\times \text{km}^{-2}]$ & 1.74 & [1.46 - 2.70] & 0.804 & [0.706 - 0.957]\\
\midrule
\multirow{6}{*}{\rotatebox[origin=c]{90}{\texttt{m123}}} &$p_1 $  & 34.269 &[34.209 - 34.367]  & 34.669 & [34.644 - 34.771] \\ 
&$\Gamma_1$  & 2.830 & [2.458 - 3.898]  & 2.909 & [2.752 - 3.520] \\
&$\Gamma_2$ & 3.445 & [2.691 - 3.952] & 2.246 & [1.055 - 3.596] \\
&$p_1^c \ [10^{-4}\times \text{km}^{-2}]$  &  0.722 &  [0.623 - 0.919]  &  0.311 &  [0.260 - 0.355] \\
&$p_2^c \ [10^{-4}\times \text{km}^{-2}]$ &  0.862 &  [0.752 - 1.07]   &  0.372 &  [0.330 - 0.427] \\
&$p_3^c \ [10^{-4}\times \text{km}^{-2}]$  &   1.03 &  [0.893 - 1.26] &   0.443 &  [0.407 - 0.512]  \\
\bottomrule
\end{tabular}
\caption{\textsl{Comparison between injected and reconstructed values of the \texttt{apr4} and \texttt{h4} parameters, for the three models analyzed. For each parameter of the piecewise polytropic EOS we show the $1\sigma$ ($\sim 68 \%$) credible interval of the marginalized posterior distribution.}}
\captionsetup{format=hang,labelfont={sf,bf}}
\label{table:injected}
\end{table}

Finally, for each set of data, we run four parallel processes of $n=5\times 10^5$ samples, starting from different, random initial points of the parameter space. We assess the convergence of the MCMC simulations to the target distribution by:
\begin{itemize}
\item[1)] analyzing the autocorrelation of each chain, defined as a function of the lag variable $k$ (for single-parameter simulations) by
\begin{equation}
C(k) = \frac{\sum_{i=1}^{n-k}\left(\theta_i-\mu\right)\left(\theta_{i+k}-\mu\right)}{\sum_{i=1}^{n}\left(\theta_i-\mu\right)\left(\theta_i-\mu\right)} \,,
\end{equation}
where $\theta_i$ is the $i$-th state, $\mu$ the mean value and $n$ the number of steps of the time series. The autocorrelation function gives an estimate of the time scale $\bar{k}$ (i.e., the number of steps) that is needed to obtain effectively independent samples~\cite{Brooks:2011} ($C(k) \sim 0 $ for $k \sim \bar{k}$). Using this information, one chooses $n \gg \bar{k}$. 
\item[2)] performing the Gelman-Rubin convergence diagnostic~\cite{10.2307/2246093,10.2307/1390675}, which allows one to check if multiple MCMC chains are converging to the same target distribution, and estimates if longer simulations can improve the results. The Gelman-Rubin test compares the variances of different chains through the \emph{potential scale reduction factor} $R$, defined for a single-parameter MCMC simulation by
\begin{equation}
R = \frac{V}{W} \qquad V= \sigma^2 + \frac{B}{m n} \qquad \sigma^2 = \frac{n-1}{n} W + \frac{B}{n} \,,
\end{equation}
where $B/n$ and $W$ are the between-sequence variance and within-sequence variance, respectively,
\begin{equation}
\begin{aligned}
\frac{B}{n} & = \frac{1}{m-1} \sum_{j=1}^{m} \left(\mu_j - \bar{\mu} \right)^2\\
W &= \frac{1}{m} \sum_{j=1}^m \frac{1}{n-1} \sum_{i=1}^n \left(\theta_{ij}- \mu_j \right)^2 \,,
\end{aligned}
\end{equation}
whereas $\theta_{ij}$ is the $i$-th state of the $j$-th chain, $\mu_j$ is the mean value of the $j$-th chain and $\bar{\mu}$ the mean value over all chains. $n$ and $m$ are the numbers of steps of each chain and the number of chains, respectively. For large enough $n$, $R \to 1$ from above, assessing the convergence of the simulations.
\end{itemize} 
We obtain the final distributions summing up the four individual chains of each set of data, after discarding the first $10\%$ of them, as a burn-in procedure (namely, the points of the chains for which the convergence has not been reached yet). In the Appendix~\ref{sec:appC} we report some examples of the chains generated by the MCMC simulations, for the models that we discuss in the next section.

\subsection{Reconstruct the parameters of the equation of state}
\label{sec:results}
\begin{figure}[]
\captionsetup[subfigure]{labelformat=empty}
\centering
{\includegraphics[width=0.32\textwidth]{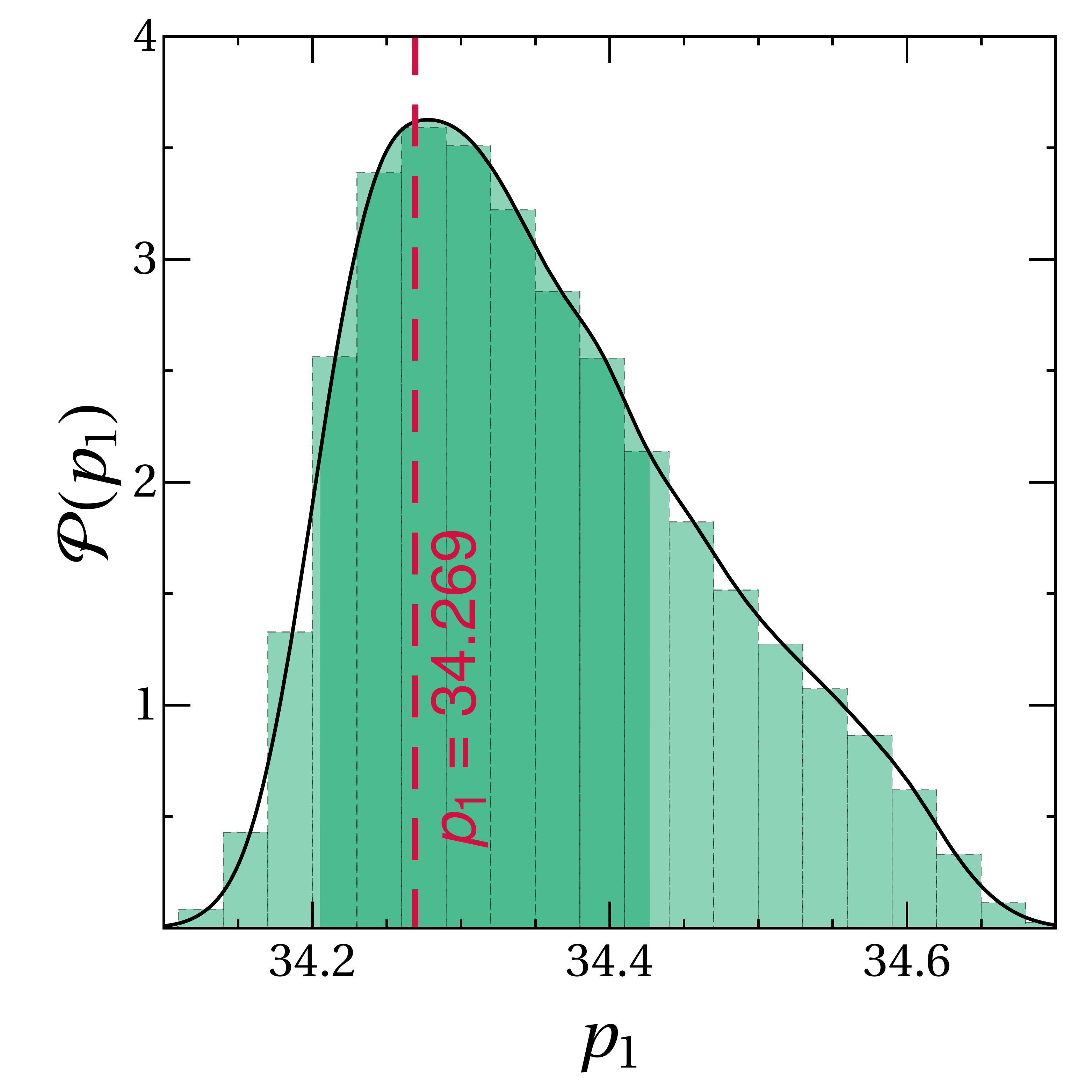}}  
{\includegraphics[width=0.32\textwidth]{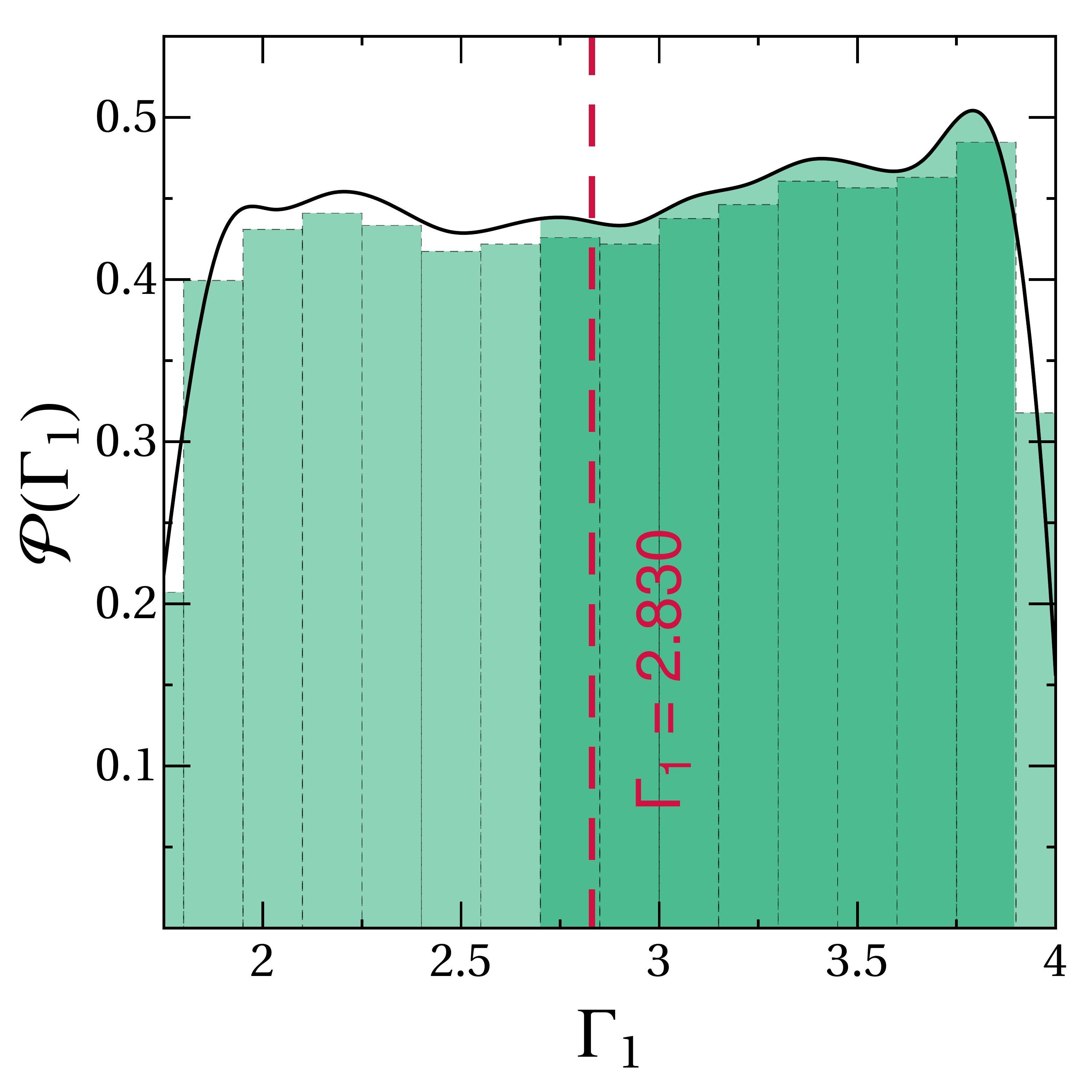}}
{\includegraphics[width=0.32\textwidth]{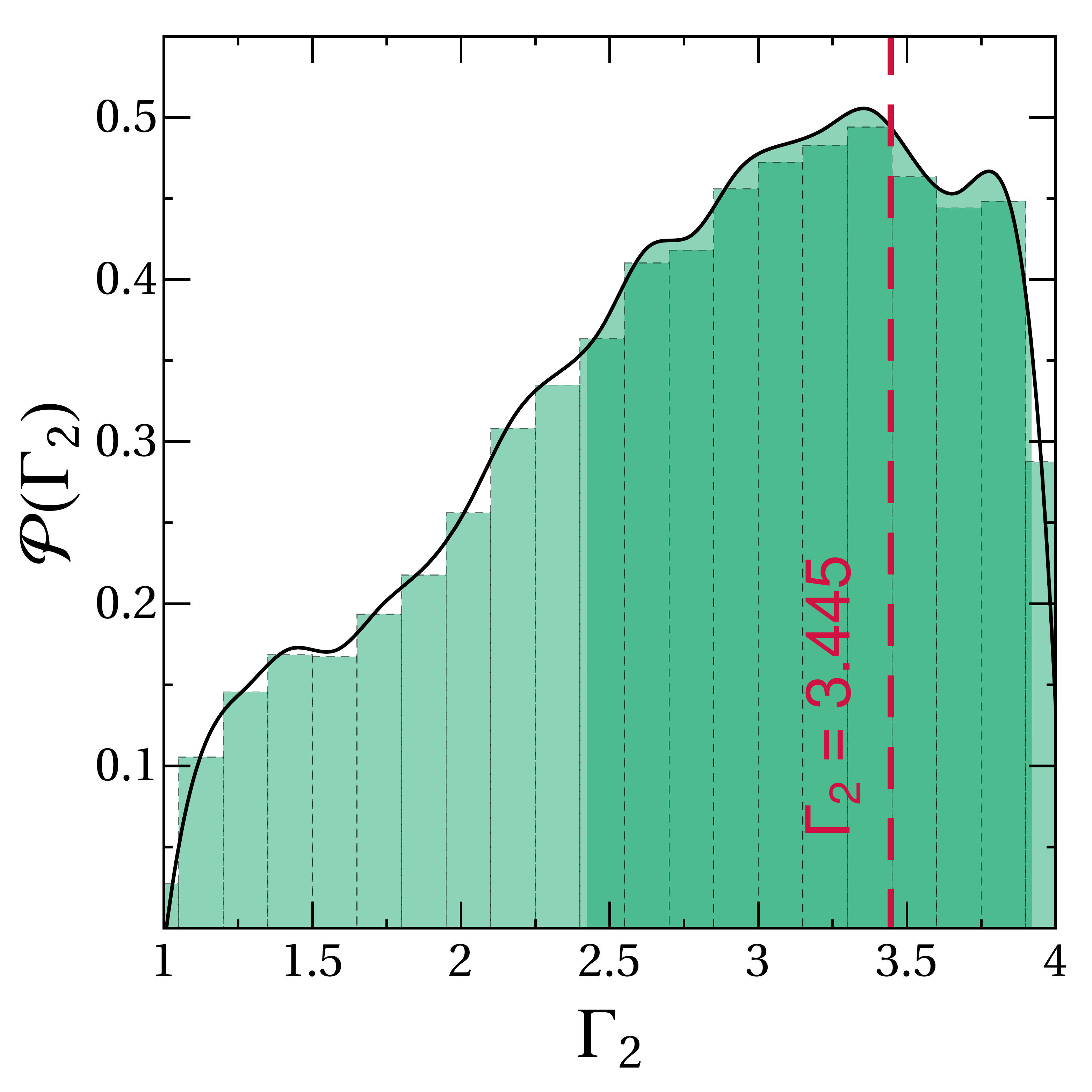}}
{\includegraphics[width=0.32\textwidth]{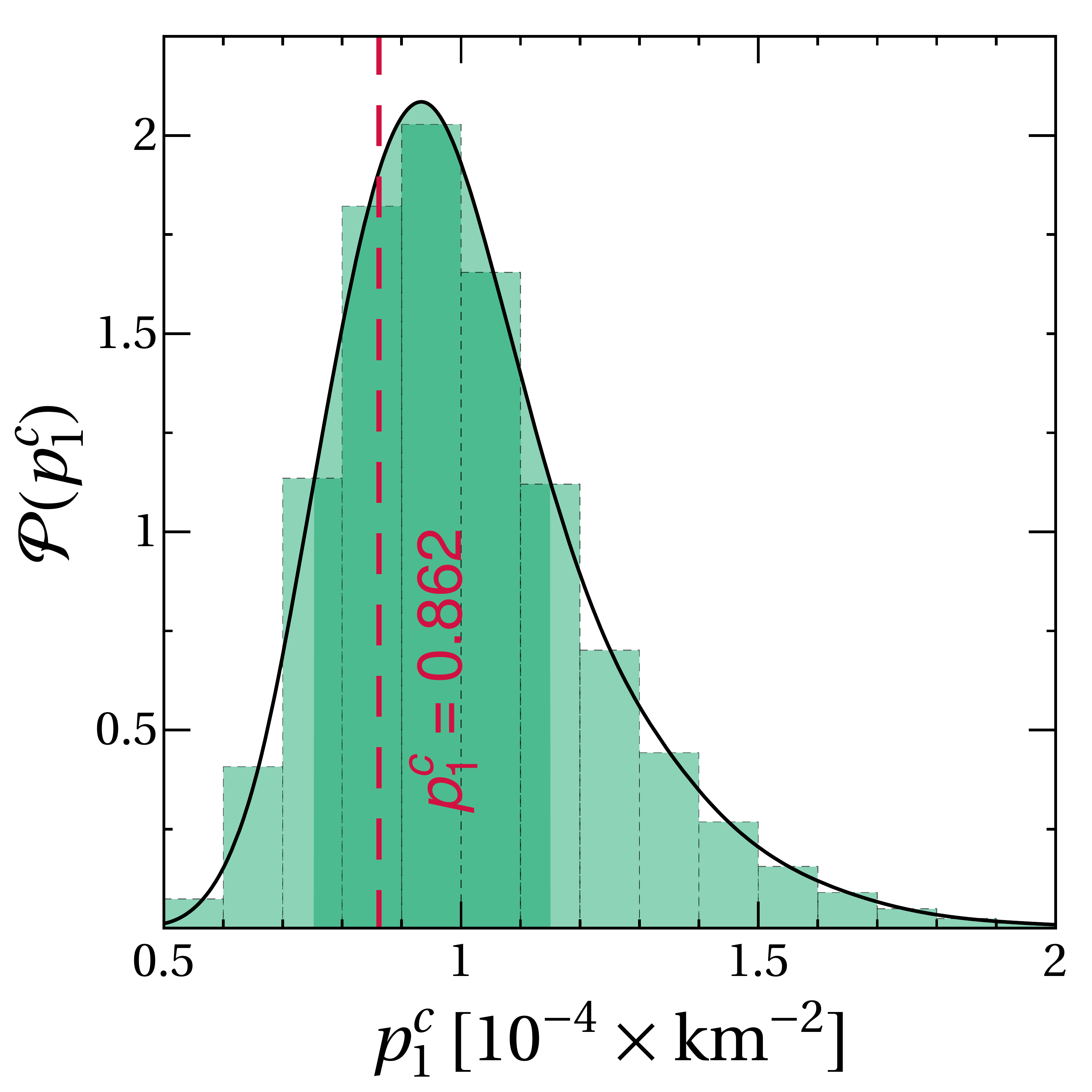}}  
{\includegraphics[width=0.32\textwidth]{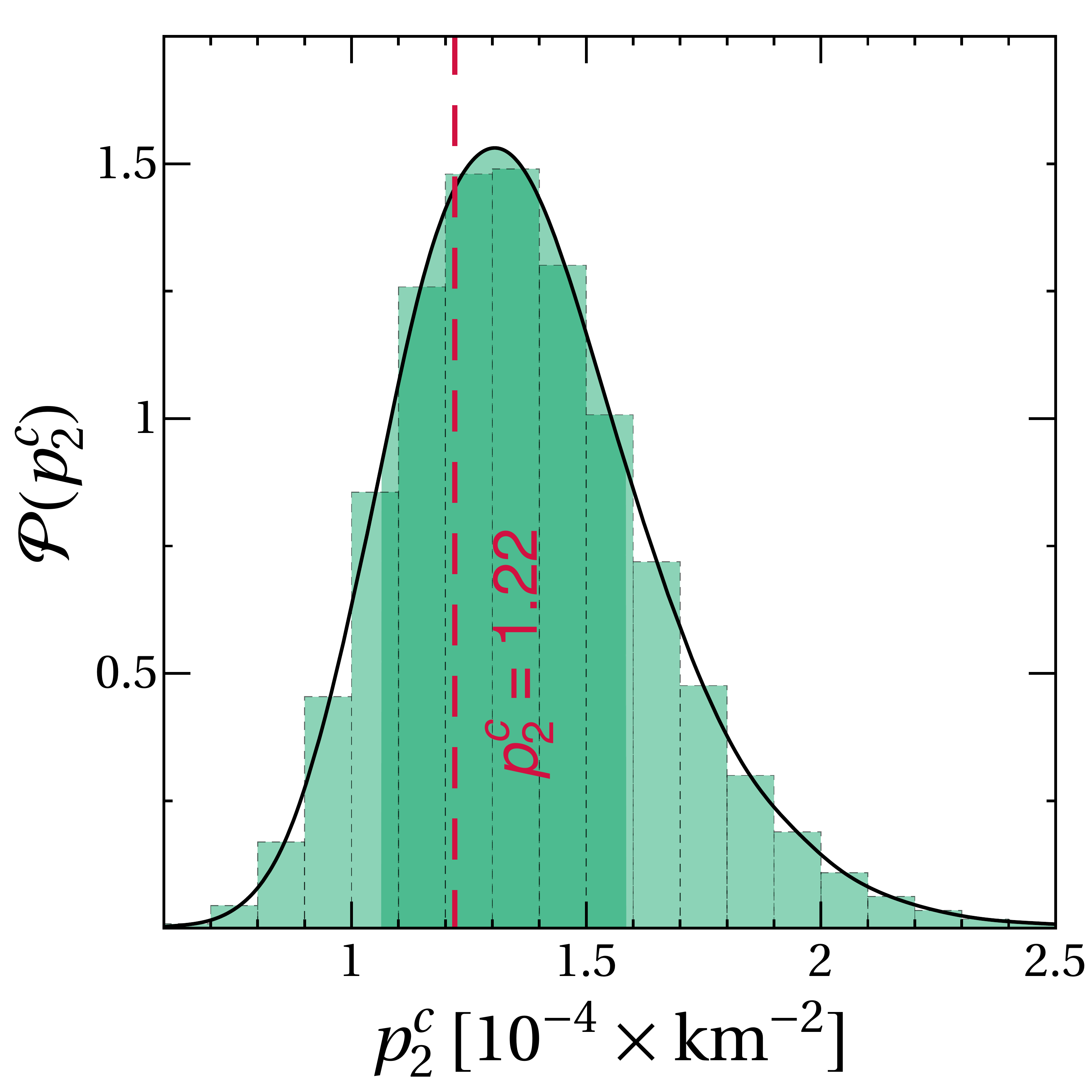}}
{\includegraphics[width=0.32\textwidth]{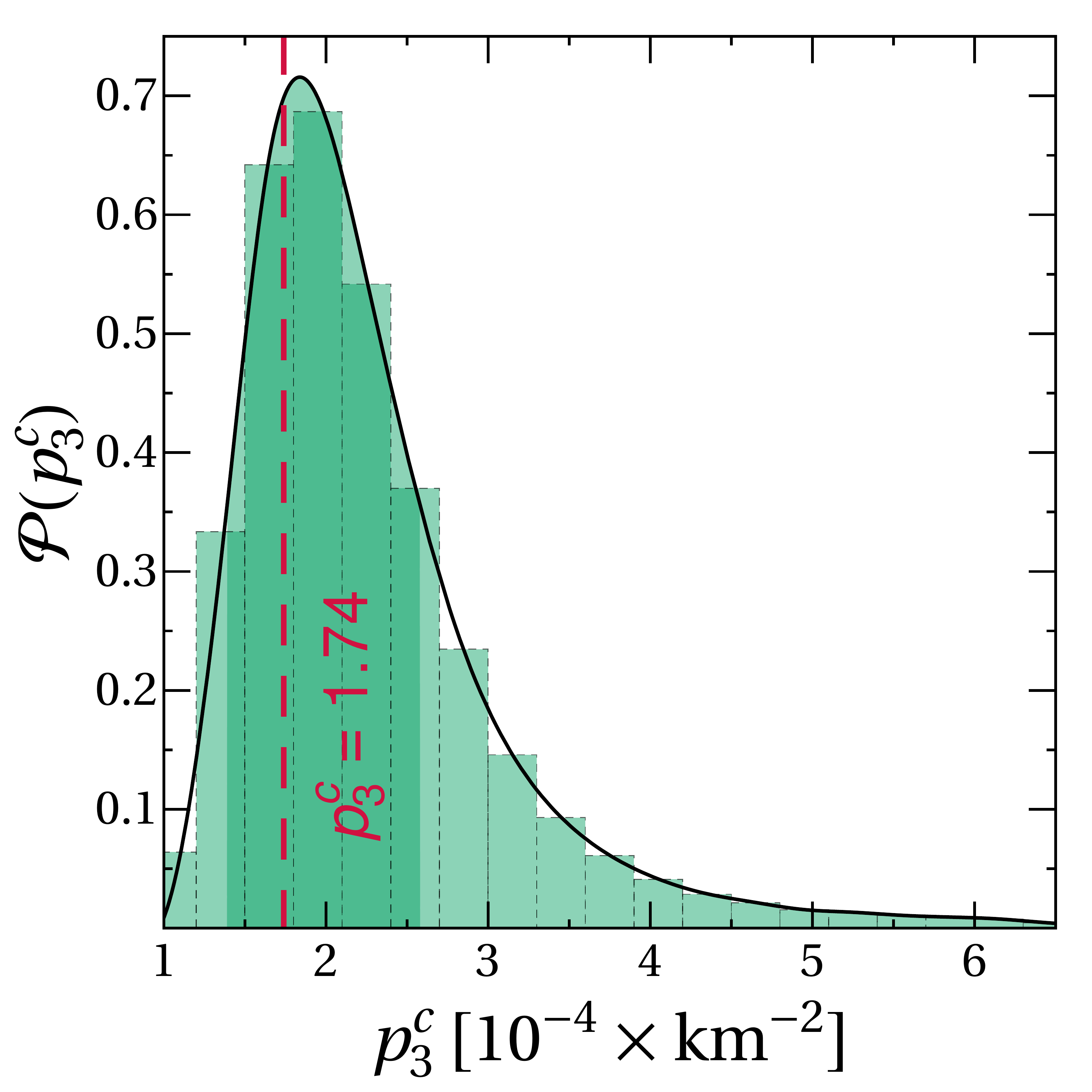}}
\caption{\textsl{Marginalized posterior PDF for the parameters of the \texttt{apr4} EOS, derived for the \texttt{m246} model with neutron stars masses $(1.2,1.4,1.6)M_\odot$. The histograms of the sampled points are shown below each function. The red, dashed vertical lines identify the injected true values, while the shaded bands correspond to the $1\sigma$ credible regions of each parameter.}}
\captionsetup{format=hang,labelfont={sf,bf}}
\label{fig:m246apr4}
\end{figure}

\begin{figure}[]
\captionsetup[subfigure]{labelformat=empty}
\centering
{\includegraphics[width=0.32\textwidth]{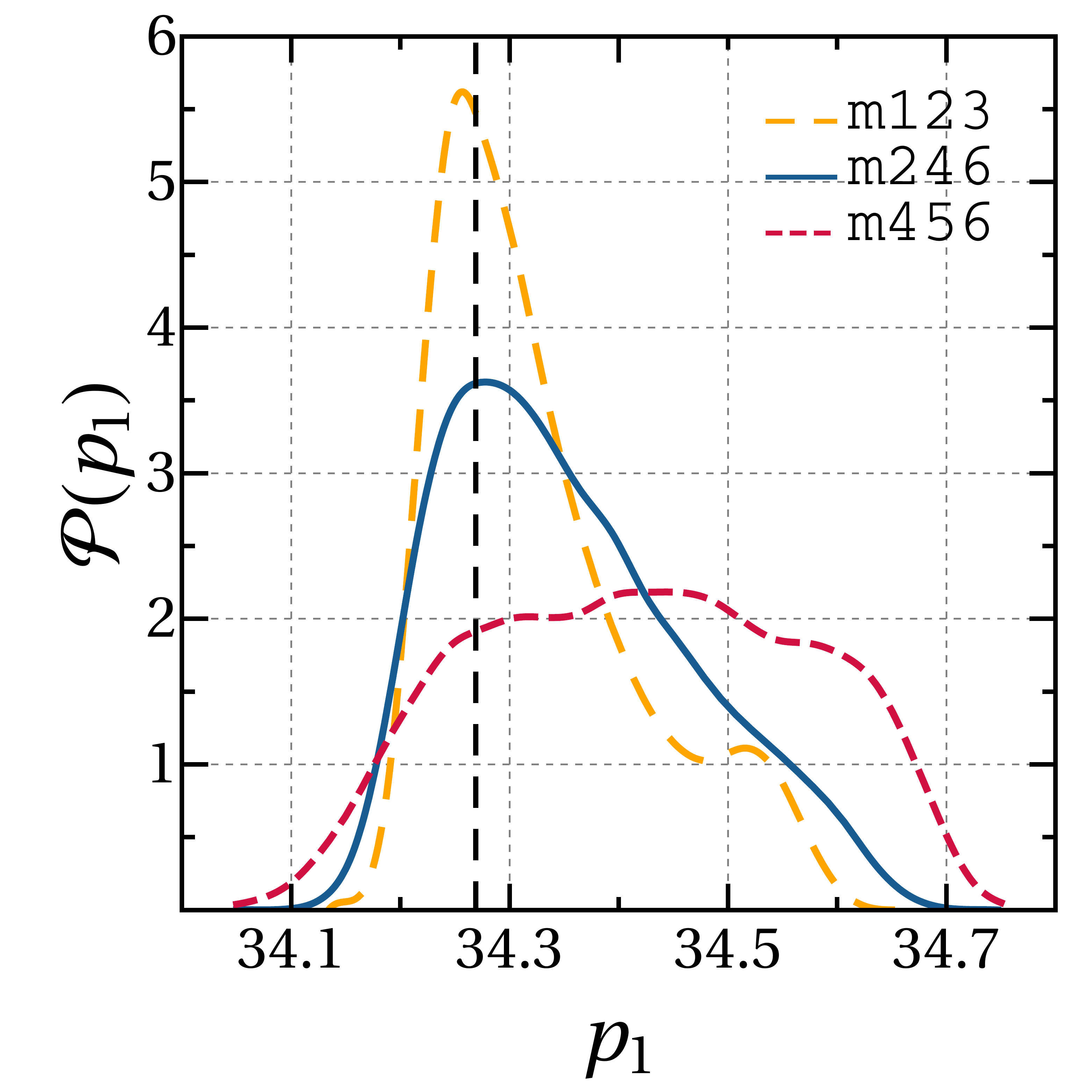}}  
{\includegraphics[width=0.32\textwidth]{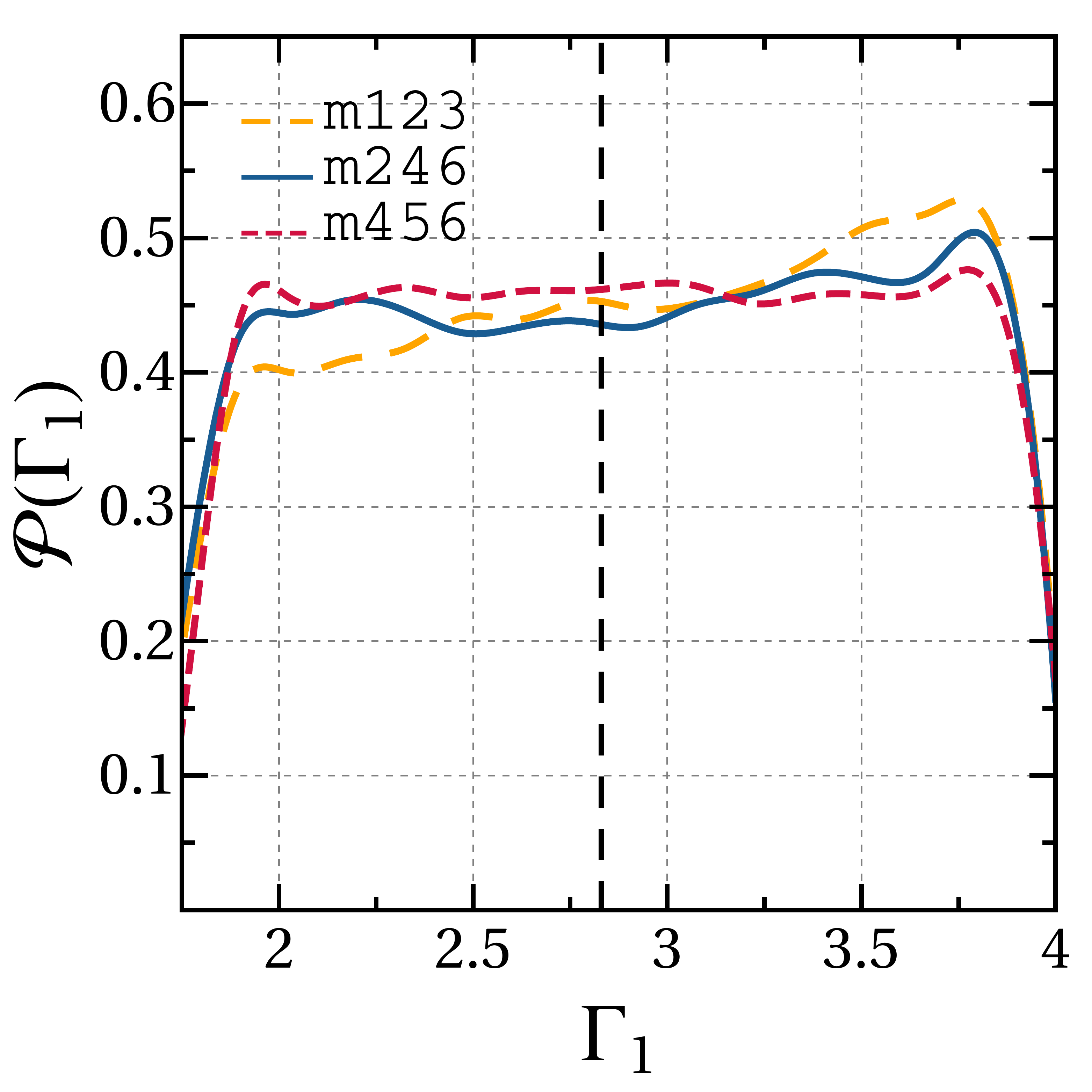}}
{\includegraphics[width=0.32\textwidth]{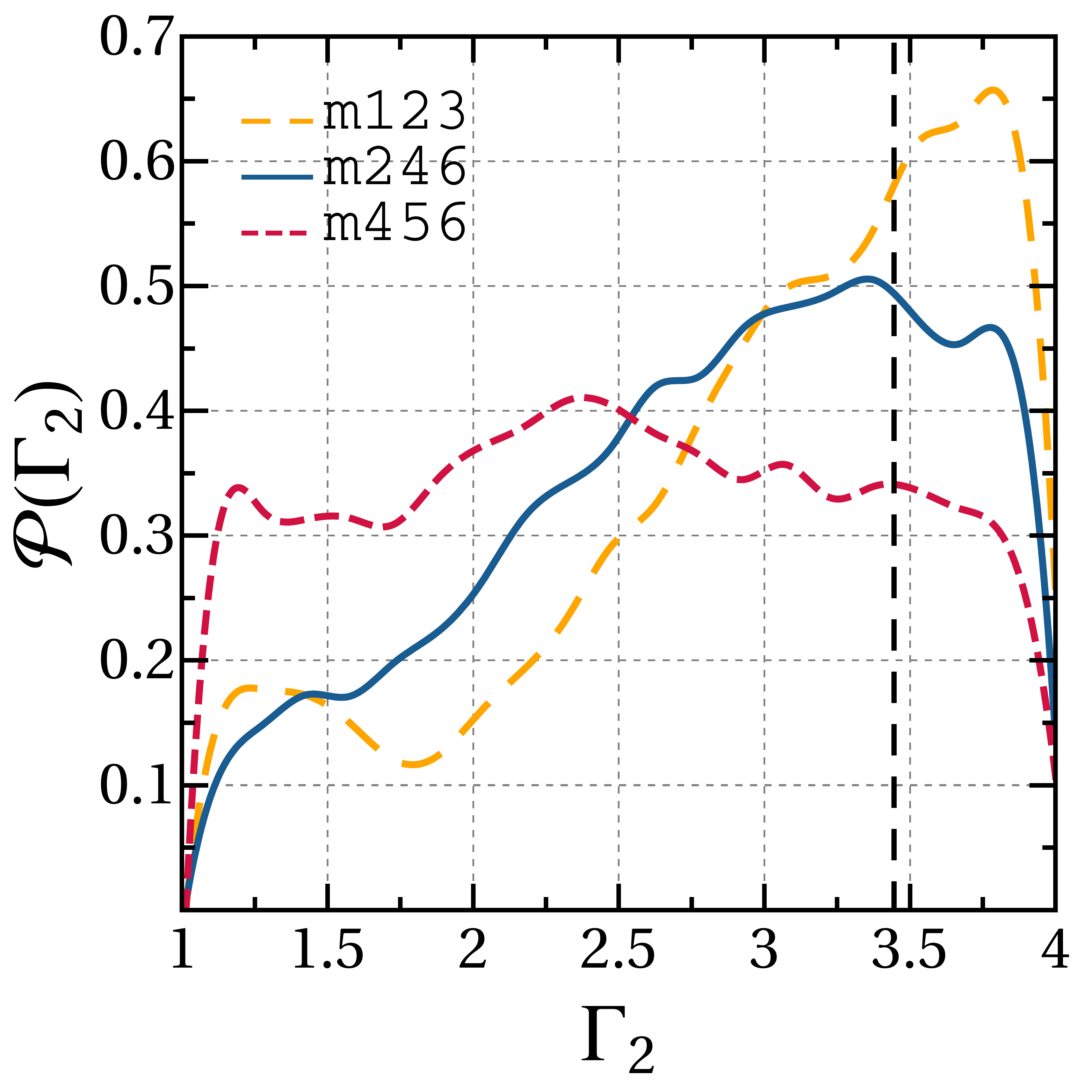}}
\caption{\textsl{Comparison among the marginalized posteriors of $p_1$, $\Gamma_1$ and $\Gamma_2$ for the \texttt{apr4} EOS, derived for the models \texttt{m246}, \texttt{m456} and \texttt{m123}. The dashed vertical lines correspond to the true values of the parameters.}}
\captionsetup{format=hang,labelfont={sf,bf}}
\label{fig:massapr4}
\end{figure}

\begin{figure}[]
\captionsetup[subfigure]{labelformat=empty}
\centering
{\includegraphics[width=0.32\textwidth]{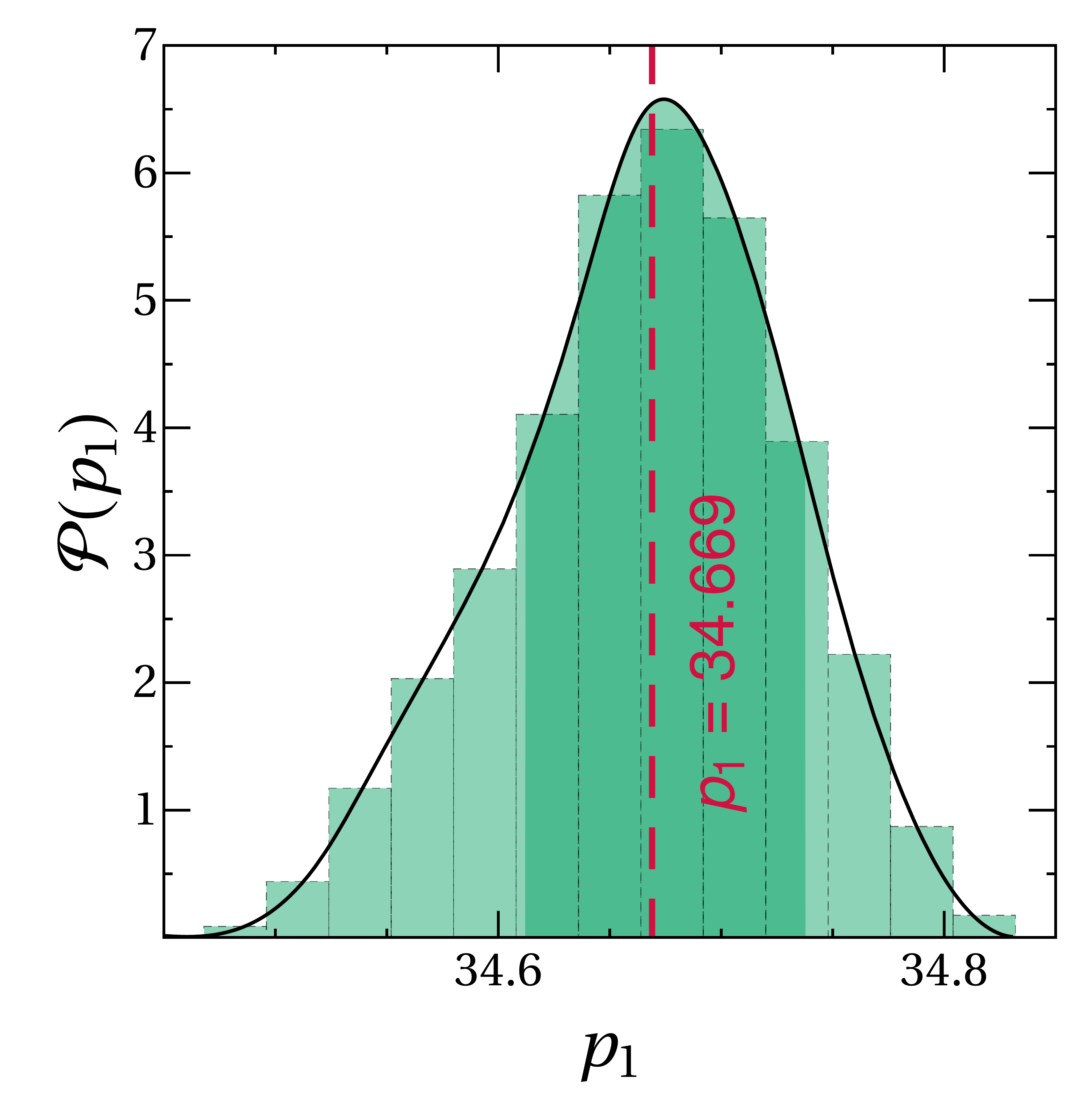}}  
{\includegraphics[width=0.32\textwidth]{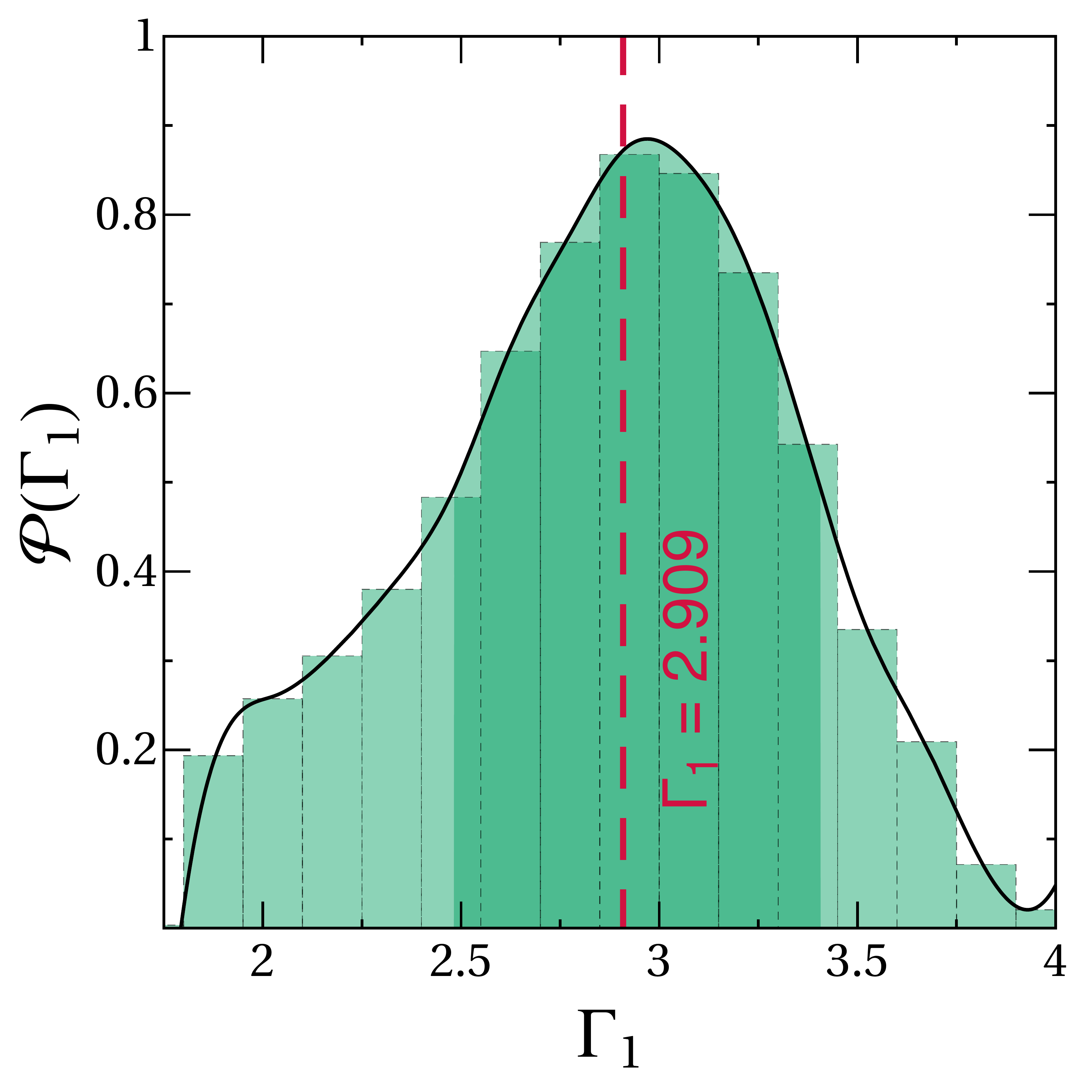}}
{\includegraphics[width=0.32\textwidth]{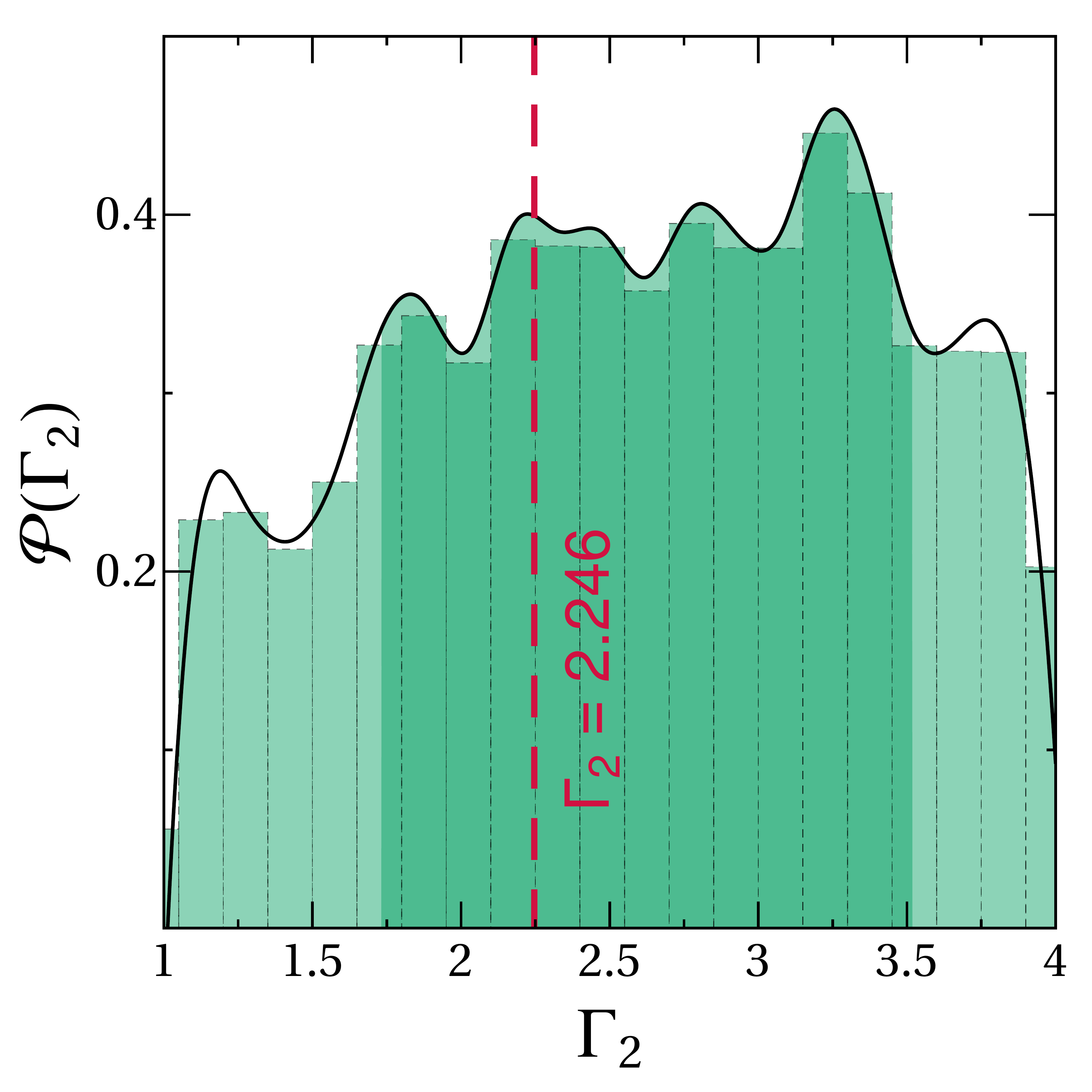}}
{\includegraphics[width=0.32\textwidth]{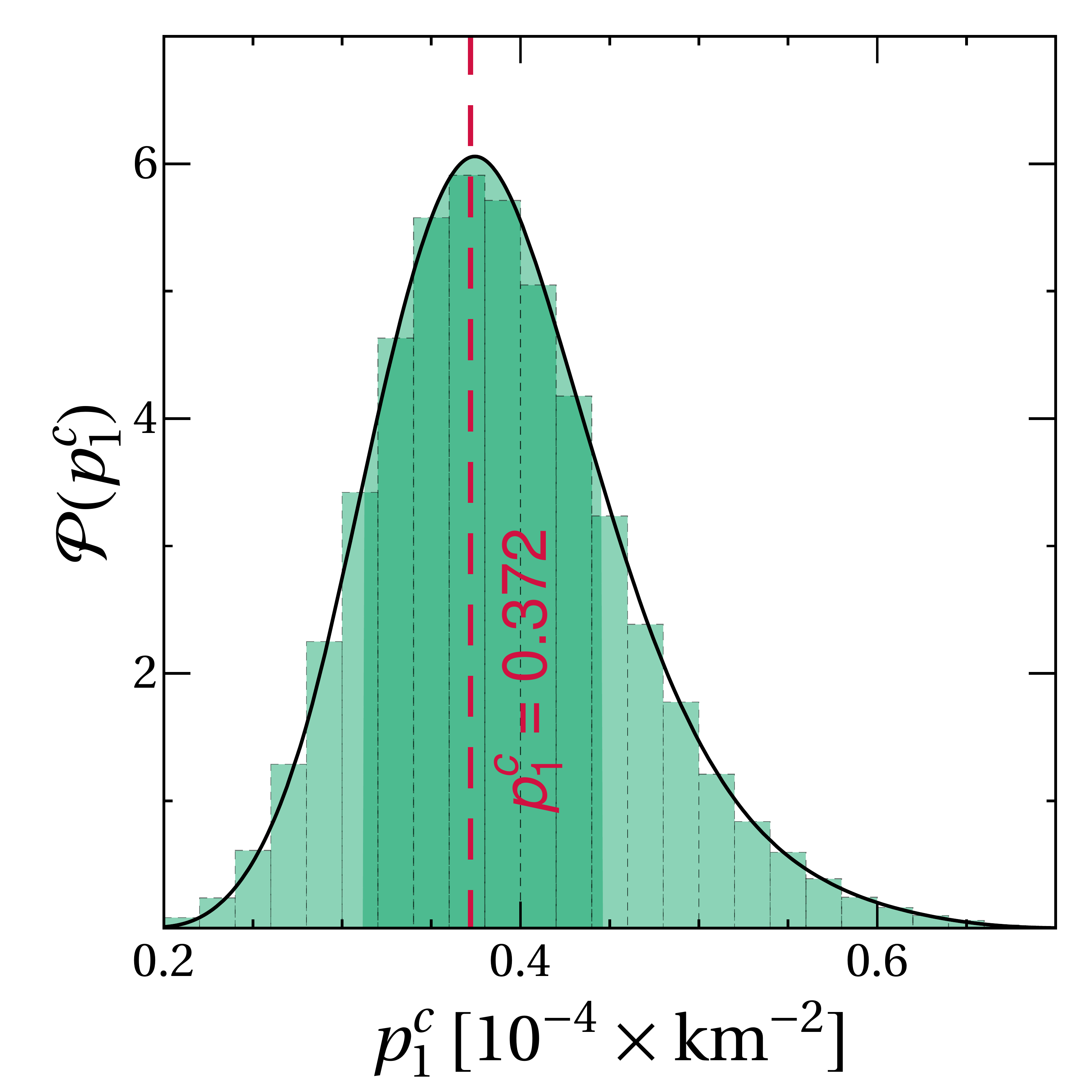}}  
{\includegraphics[width=0.32\textwidth]{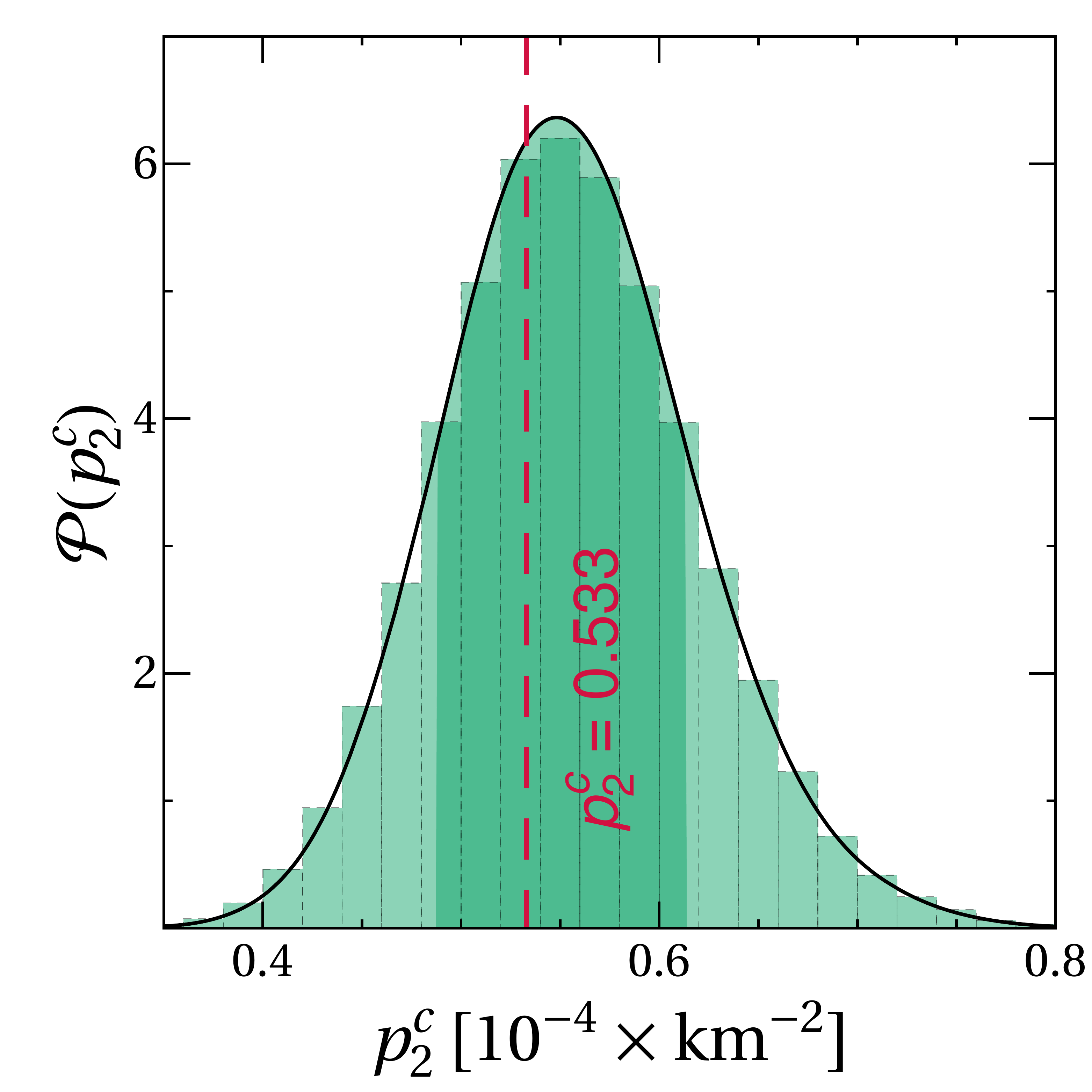}}
{\includegraphics[width=0.32\textwidth]{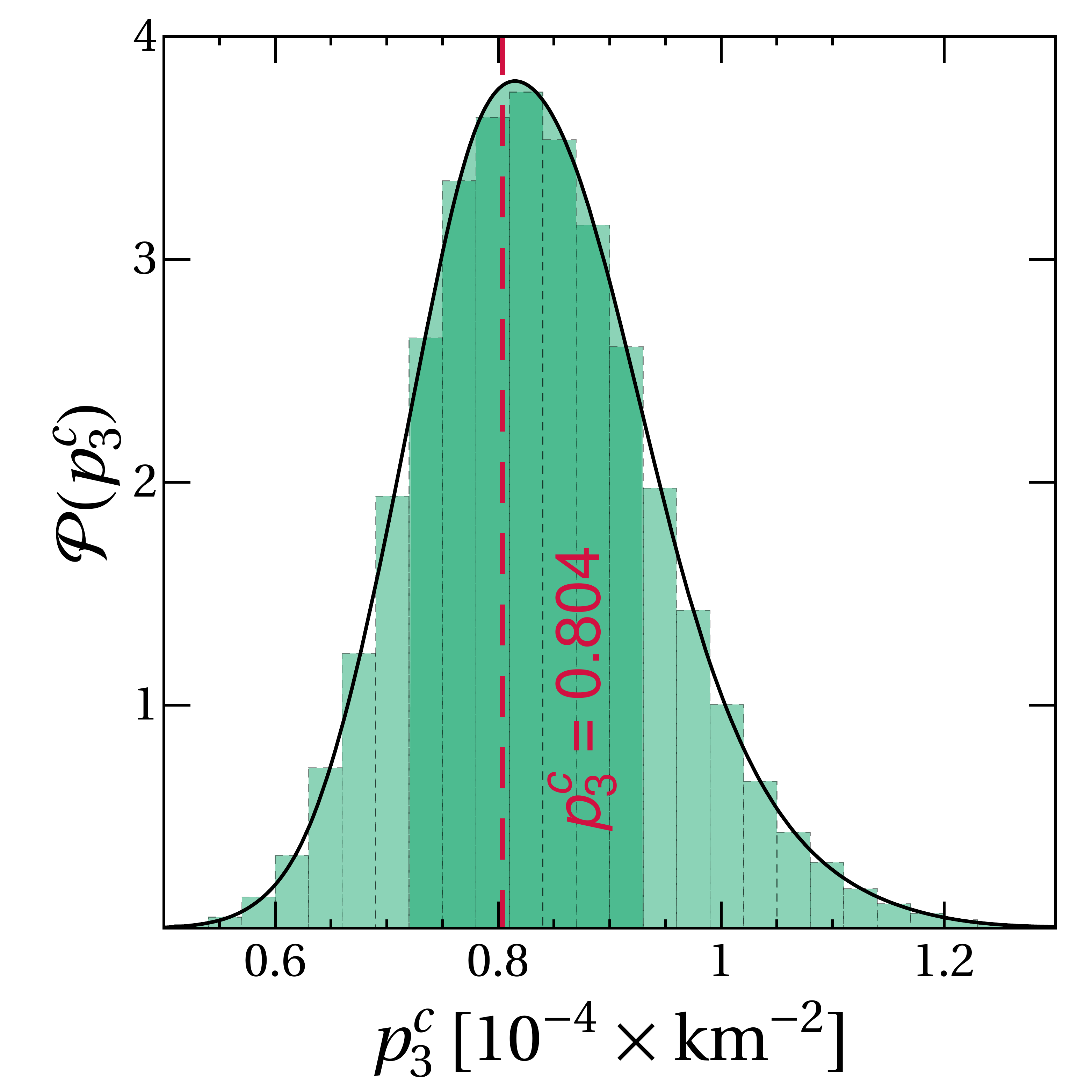}}
\caption{\textsl{Marginalized posterior PDF for the parameters of the \texttt{h4} EOS, derived for the \texttt{m246} model with neutron stars masses $(1.2,1.4,1.6)M_\odot$. The histograms of the sampled points are shown below each function. The red, dashed vertical lines identify the injected true values, while the shaded bands correspond to the $1\sigma$ credible regions of each parameter.}}
\captionsetup{format=hang,labelfont={sf,bf}}
\label{fig:m246h4}
\end{figure}

\begin{figure}[]
\centering
\includegraphics[width=0.6\textwidth]{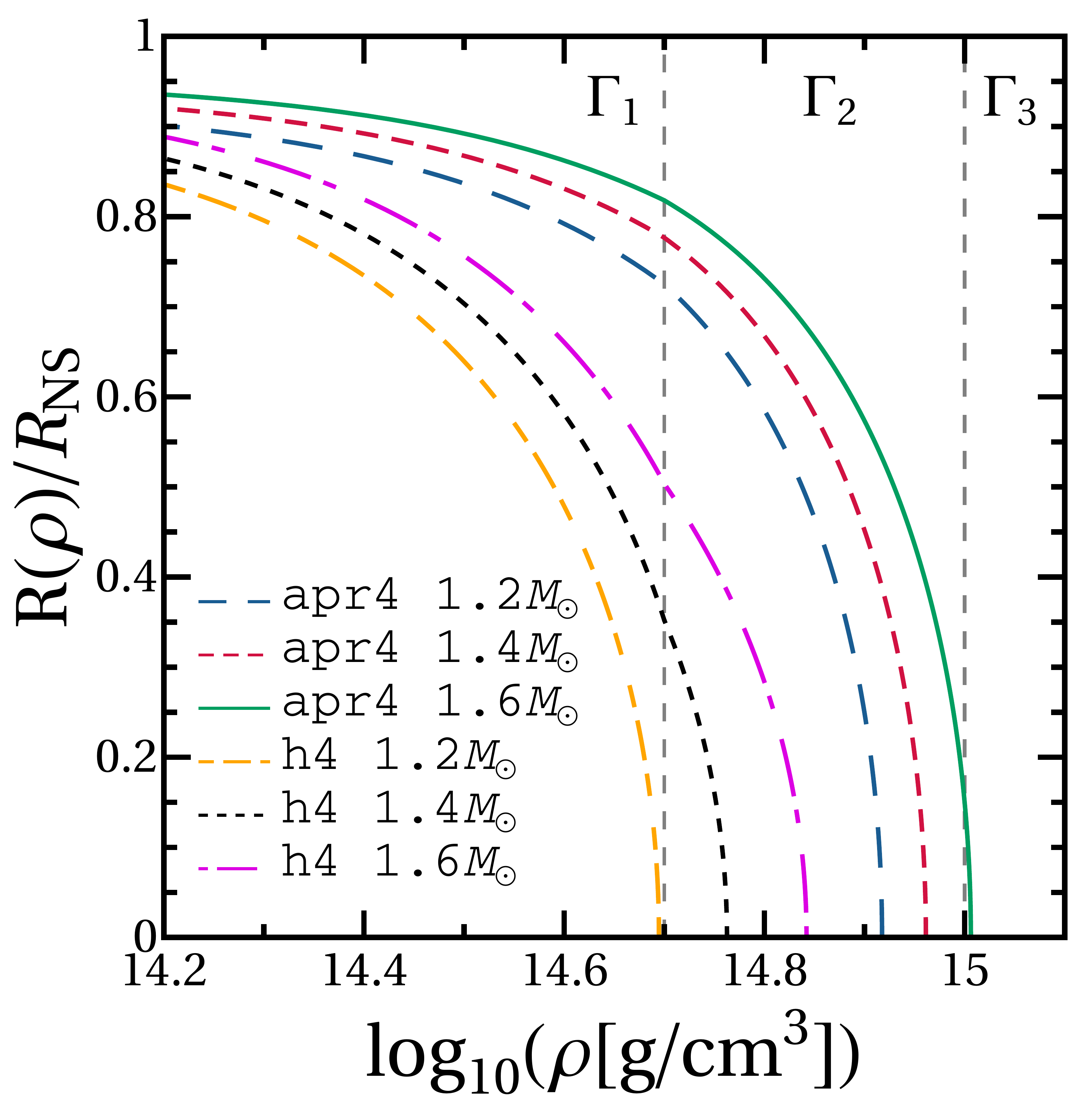}
\caption{\textsl{The radial distance from the center of the star, normalized to its radius, is plotted as a function of the density. The different curves correspond to the masses and EOSs analyzed in the \texttt{m246} configuration. The vertical lines separate the three regions of the piecewise polytropic parametrization.}}
\captionsetup{format=hang,labelfont={sf,bf}}
\label{fig:radius}
\end{figure}

\begin{figure}[]
\captionsetup[subfigure]{labelformat=empty}
\centering
{\includegraphics[width=0.32\textwidth]{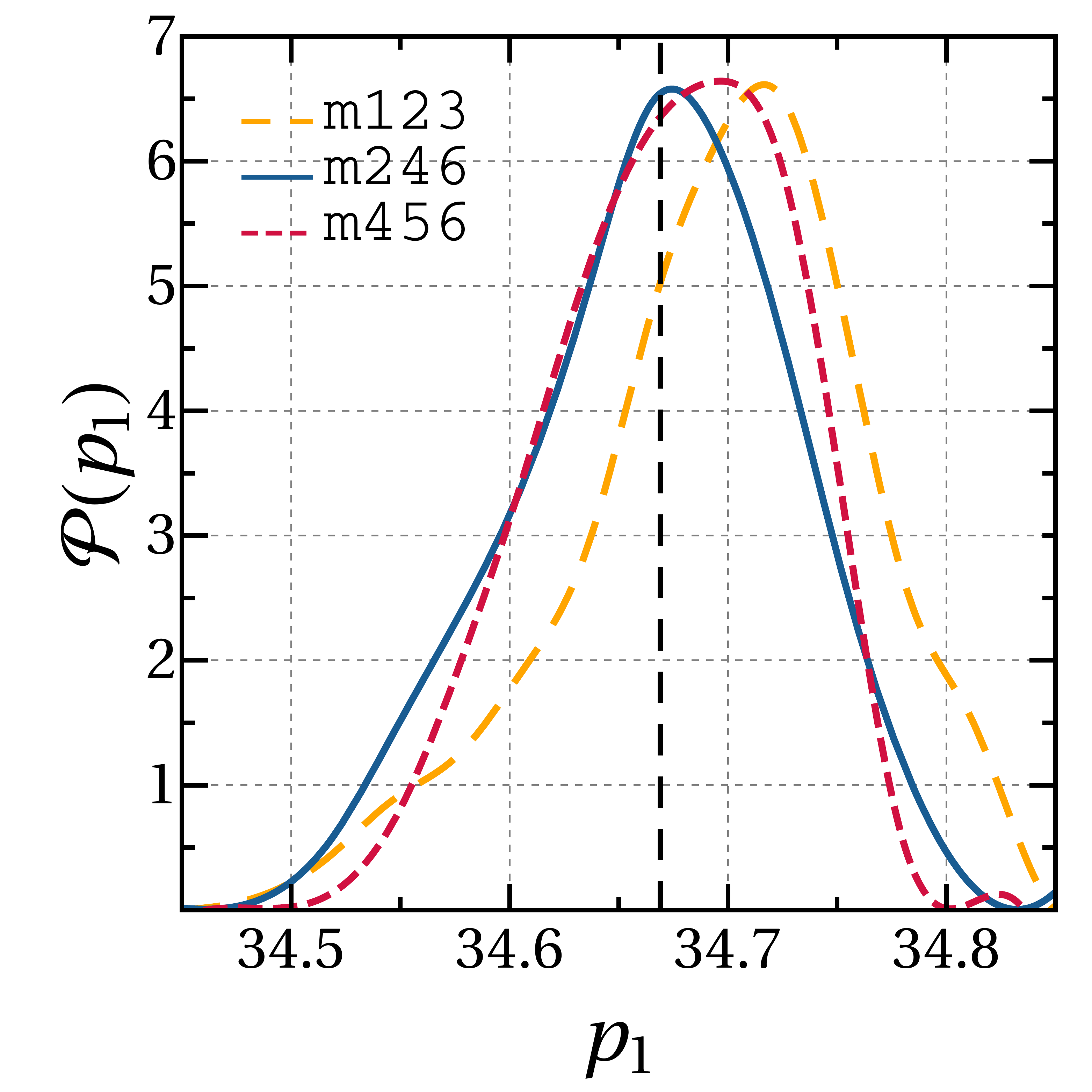}}  
{\includegraphics[width=0.32\textwidth]{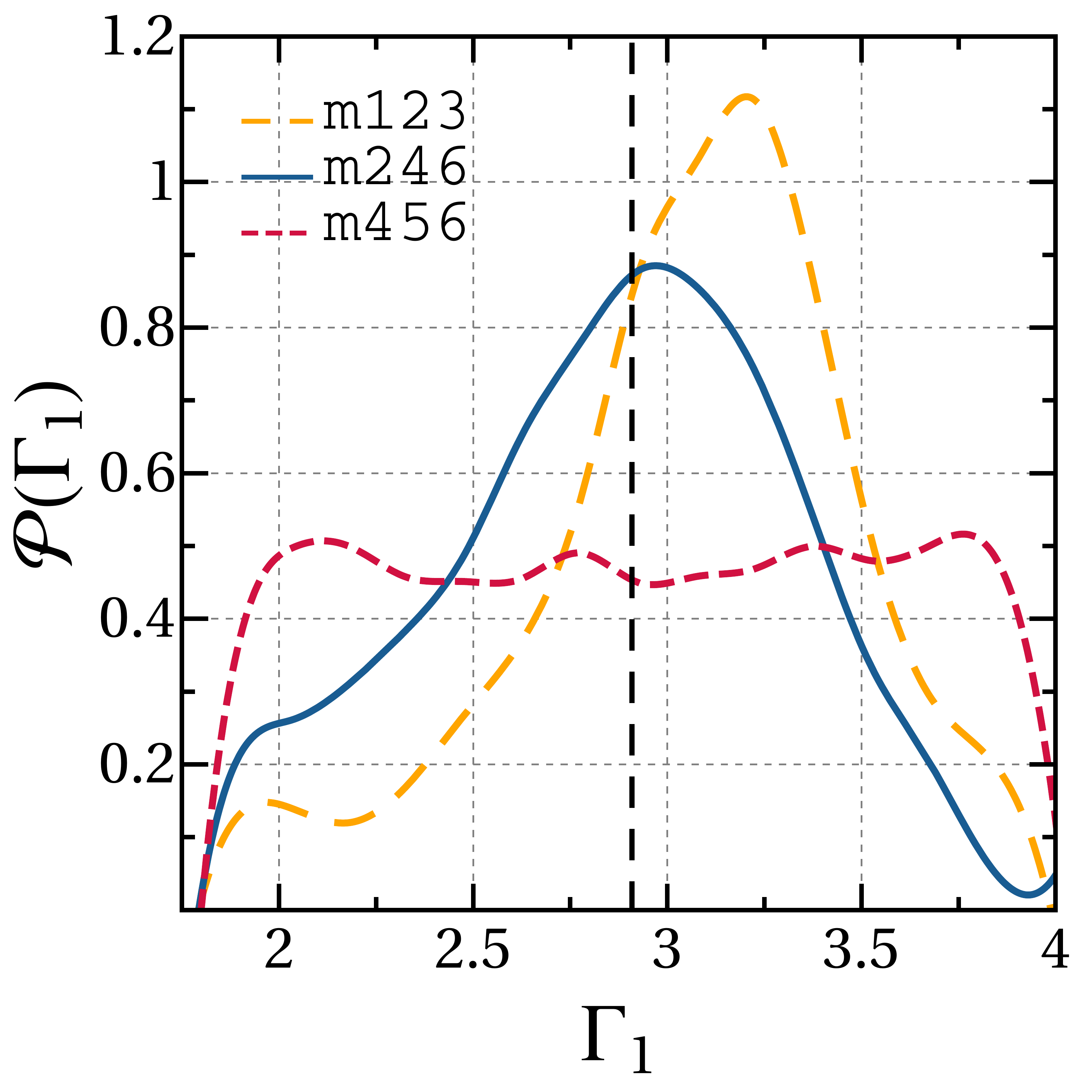}}
{\includegraphics[width=0.32\textwidth]{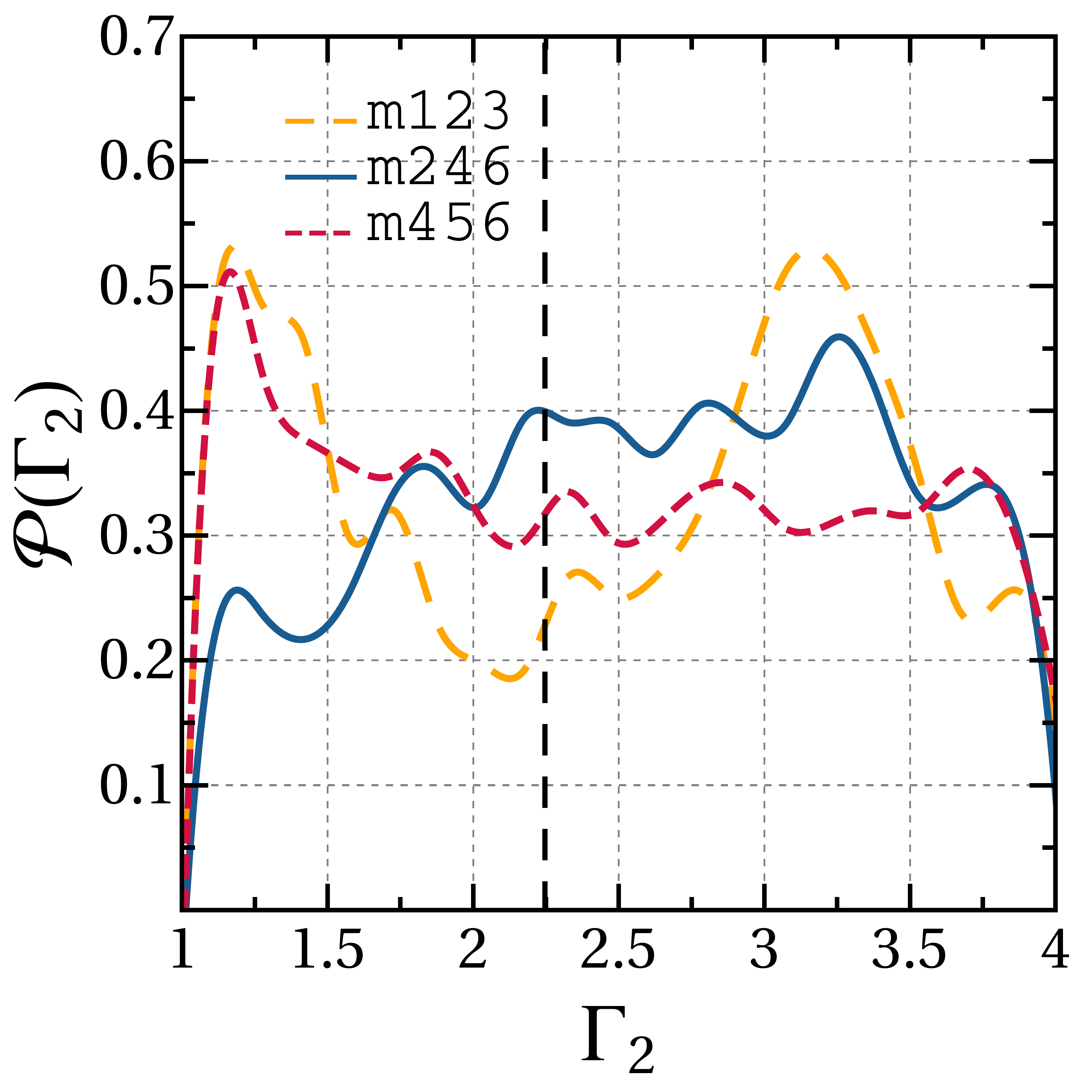}}
\caption{\textsl{Comparison among the marginalized posteriors of $p_1$, $\Gamma_1$ and $\Gamma_2$ for the \texttt{h4} EOS, derived for the models \texttt{m246}, \texttt{m456} and \texttt{m123}. The dashed vertical lines correspond to the true values of the parameters.}}
\captionsetup{format=hang,labelfont={sf,bf}}
\label{fig:massh4}
\end{figure}

The first goal of our approach is to determine the parameters of the piecewise polytropic EOS. As said before, we have six unknown variables to constrain, i.e., $\vec{\theta}=\{p_1,\Gamma_1,\Gamma_2,p^c_1,p^c_2,p^c_3\}$, which require three neutron star observations. We test our method on the following prototype configurations: (i) the model \texttt{m246} with three objects of mass $(1.2,1.4,1.6)M_\odot$, (ii) a \textit{heavier} one \texttt{m456} composed of stars of $(1.4,1.5,1.6)M_\odot$, (iii) a \textit{lighter} system \texttt{m123} with masses $(1.1,1.2,1.3)M_\odot$. The numerical values of injected and reconstructed parameters are listed in Table~\ref{table:injected}, for the considered configurations, and for the EOSs \texttt{apr4} and \texttt{h4}.

In Fig.~\ref{fig:m246apr4} we show the marginalized posterior distributions of the parameters corresponding to the \texttt{apr4} EOS, derived for the configuration \texttt{m246}. The dashed vertical line in each panel indicates the true, injected value of the parameter, while the darker bands correspond to the $1\sigma$ credible intervals. The PDF is constructed from the sample histograms using a Gaussian kernel density estimator. We can see that the true values of all the parameters are always reconstructed within the $1\sigma$ confidence level. The posteriors of the neutrons star central pressures are always peaked around the injected values with nearly symmetrical distributions. The pressure $p_1$ of the outer core region is also well measured, with the relative difference between the injected valued and the reconstructed median being below $1\%$. 

In general, the adiabatic indices of the piecewise polytropic representation are determined with less accuracy, although some differences do exist between the various polytropic segments. The top panels of Fig.~\ref{fig:m246apr4} show indeed that $\Gamma_1$ is unconstrained, with an almost flat posterior within the allowed range of values. Conversely, the second index $\Gamma_2$ provides better results, with a median close to the true quantity, and a probability distribution that tends to favor larger values. Analyzing the joint distribution between various pairs of parameters we find that $p_1$-$\Gamma_2$ is the only one that shows a significant correlation, which is, otherwise, small (see Fig.~\ref{fig:select} below and the Appendix~\ref{sec:appC}).

Most of the features described for the \texttt{m246} configuration do not change qualitatively if we analyze the other two models \texttt{m456} and \texttt{m123}, for the same EOS \texttt{apr4}. Smaller masses lead in general to stronger constraints. This is expected, since, for a fixed EOS, lighter neutron stars yield larger tidal deformabilities (see the right panel of Fig.~\ref{fig:mrlambda}), which enhance the tidal contribution to the gravitational wave signal, and therefore provide smaller (relative) errors $\sigma_\lambda$. For completeness, the full marginalized posterior distributions of the other two configurations, \texttt{m456} and \texttt{m123}, can be found in the Appendix~\ref{sec:appC}.

A direct comparison among the posterior distributions of the EOS parameters $p_1$, $\Gamma_1$ and $\Gamma_2$, obtained for the three considered configurations, is shown in Fig.~\ref{fig:massapr4}. We can see that the best results for $p_1$ and $\Gamma_2$ occur for the model \texttt{m123}, which is composed of three neutron stars with masses $(1.1,1.2,1.3)M_{\odot}$. Conversely, for \texttt{m456} which considers a collection of data with heavier objects, $(1.4,1.5,1.6)M_{\odot}$, the posterior distributions of both $p_1$ and $\Gamma_2$ broaden significantly (approaching a flat distribution) and the $1\sigma$ level becomes much looser. In all cases the index $\Gamma_1$ is instead unconstrained.

The picture described above changes qualitatively when we consider neutron stars made of a stiffer EOS, which leads to more deformable objects. In Fig.~\ref{fig:m246h4} we show the posterior probability distributions of the parameters for the model \texttt{m246}, assuming \texttt{h4} as the underlying equation of state. We can see that the star central pressures $p^c_i$ are found with an accuracy comparable to that shown in Fig.~\ref{fig:m246apr4}, for the \texttt{apr4} EOS. The top left panel of the figure shows that the pressure $p_1$ at the first dividing density is, again, the EOS parameter which is constrained with the largest precision, the posterior distribution being nearly Gaussian and symmetric around the true value. However, a direct comparison with Fig.~\ref{fig:m246apr4} shows that the role of the adiabatic indices $\Gamma_1$ and $\Gamma_2$ seems now to be reverted. Indeed, for the EOS \texttt{h4} it is $\Gamma_1$ which is very well estimated, with a relative difference of the median with respect to the true value smaller than $1\%$. The parameter $\Gamma_2$ is essentially unbounded, with a slightly noisy distribution not much dissimilar from a flat one. Moreover, the pair of parameters $p_1$-$\Gamma_1$ shows now correlation (see Fig.~\ref{fig:select} and the Appendix~\ref{sec:appC}).

The different features of the results for the two EOSs can be understood looking at Fig.~\ref{fig:radius}, where we plot, for each neutron star and EOS considered for the model \texttt{m246}, the radial distance $R(\rho)$ normalized to the radius of the star $R_{\text{NS}}$, as a function of the density $\rho$. The major difference between the two EOSs is that the radial profiles of the \texttt{apr4} stars extend to larger values of $\rho$, well inside the region of the second branch of the piecewise polytropic specified by $\Gamma_2$. Conversely, the  \texttt{h4} stars are mainly dominated by the first branch specified by $\Gamma_1$. For this EOS, stars with masses below $1.2M_\odot$ have a central pressure smaller than $p_1$ (or, equivalently, central density smaller than $\rho_1$), and therefore are outside the $\Gamma_2$ interval.

Furthermore, Fig.~\ref{fig:radius} shows that at the boundary between the first two regions, the function $R(\rho)$ of the \texttt{apr4} stars is already about the $80\%$ of its overall value. Therefore, it seems quite natural that for this EOS the tidal deformability, which is proportional to $R^5$ (see Eq.~\eqref{eq:lovetidal}), is more sensible to variations of $ \Gamma_2$. Conversely, the radius of the \texttt{h4} stars is almost completely determined by the integration of the stellar equations within the density region belonging to the first polytropic branch, and this is why the inverse stellar problem constrains $\Gamma_1$ with a larger accuracy. 

Like before, we report the posterior distributions for the configurations \texttt{m456} and \texttt{m123} in the Appendix~\ref{sec:appC}, whereas in Fig.~\ref{fig:massh4} we compare the EOS parameters of three models for the \texttt{h4} EOS. We notice that for the lightest configuration \texttt{m123}, the reconstructed value of $p_1$ shows an offset with respect to the injected parameter. This is the opposite behavior with respect to the \texttt{apr4} EOS configurations, where lighter neutron stars provide better results.

This feature is probably due to a non-negligible contribution coming from the low density part of the EOS, which reduces the impact of the first polytropic region on the tidal deformability, for neutron stars with low masses (see again Fig.~\ref{fig:radius}). In particular, sampling the parameter space, we have found that the subspace $p_1$-$\Gamma_1$ is characterized by a large region in which the posterior distribution assumes values only slightly lower than the absolute maximum, making extremely difficult to resolve it through the Monte Carlo simulation. As a consequence, the marginalized distributions are shifted with respect to the injected values.

Alike the \texttt{apr4} EOS, instead, the configuration with larger masses, \texttt{m456}, (corresponding to smaller tidal deformabilities) shows the worst result, with the distribution of the index $\Gamma_1$ which is essentially flat. In all cases the parameter $\Gamma_2$ is unconstrained, with noisy and flat-like distribution.

\subsection{Discriminate among realistic models of equation of state}
\begin{figure}[]
\captionsetup[subfigure]{labelformat=empty}
\centering
{\includegraphics[width=0.41\textwidth]{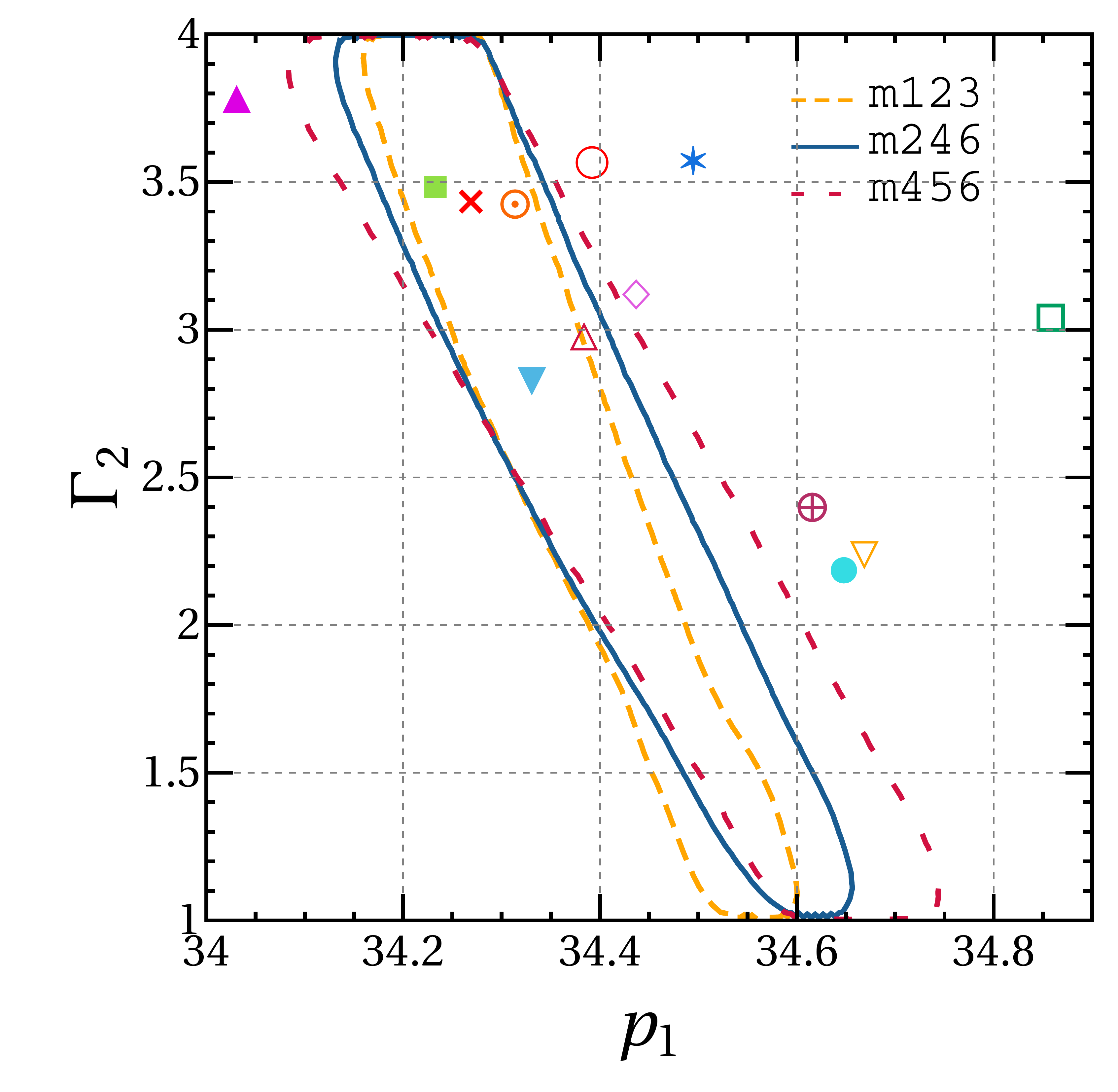}}  
{\includegraphics[width=0.41\textwidth]{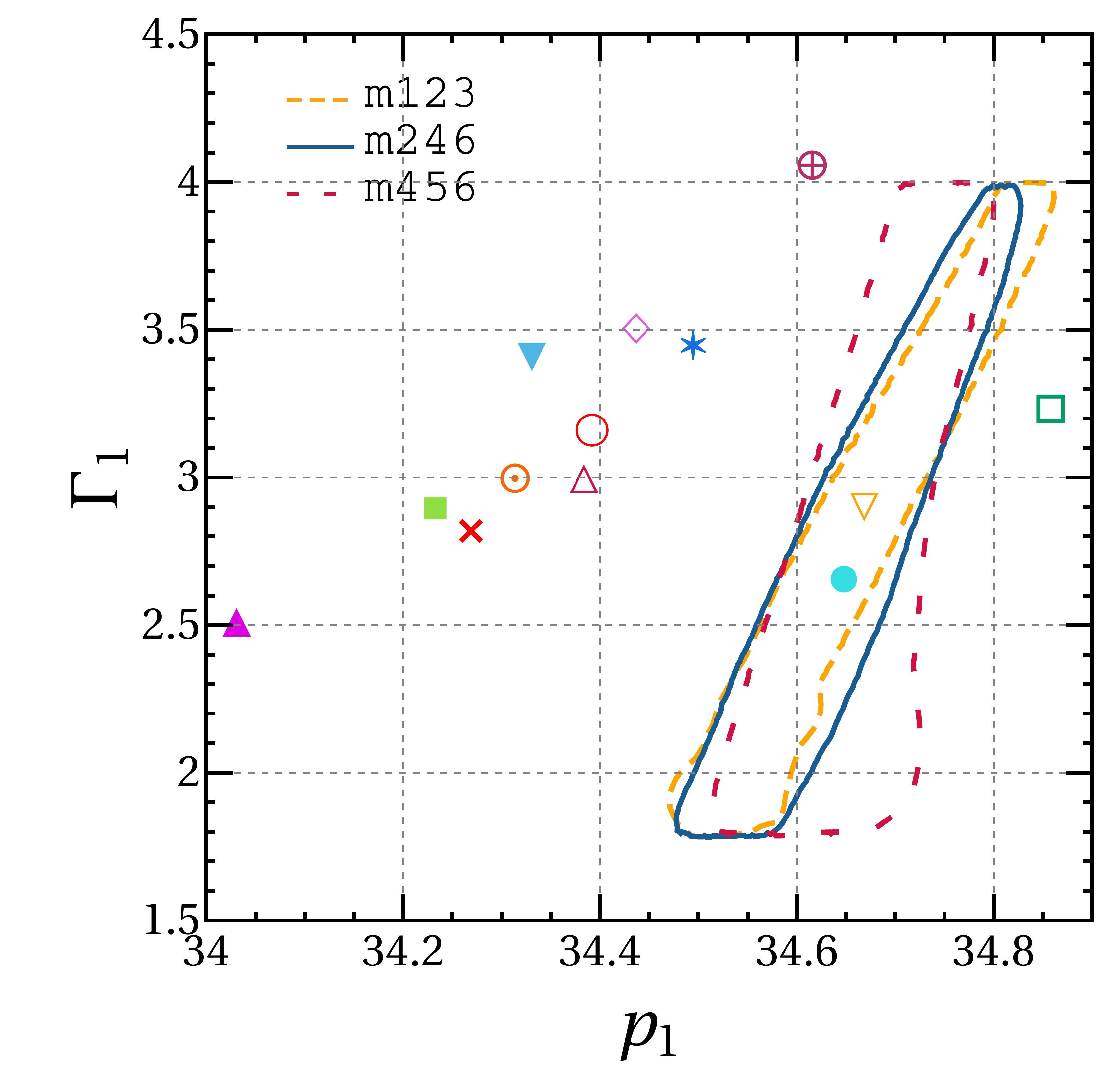}}
{\includegraphics[width=0.15\textwidth]{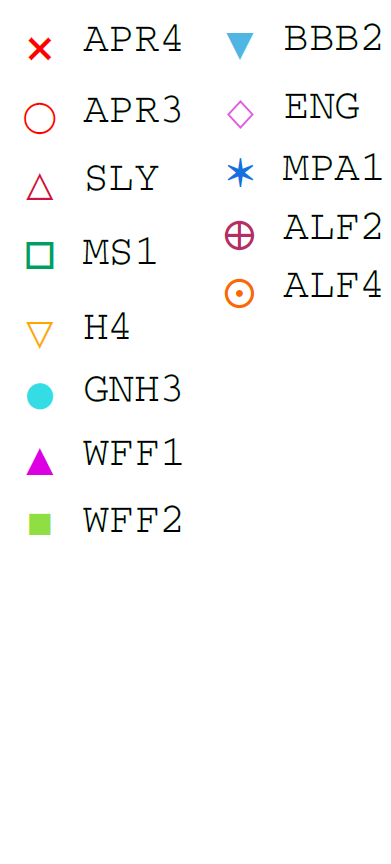}}
\caption{\textsl{(Left) 2D credible regions at 2$\sigma$ level for the joint probability distribution $\mathcal{P}(p_1,\Gamma_2)$, computed assuming \texttt{apr4} as the true equation of state (red cross), for the three models considered. Different markers correspond to the values of $p_1$ and $\Gamma_2$ for various EOSs. (Right) 2D credible regions at 2$\sigma$ level for the joint probability distribution $\mathcal{P}(p_1,\Gamma_1)$, computed assuming \texttt{h4} as the true equation of state (yellow reversed triangle), for the three models considered. Different markers correspond to the values of $p_1$ and $\Gamma_1$ for various EOSs.}}
\captionsetup{format=hang,labelfont={sf,bf}}
\label{fig:select}
\end{figure}

The relativistic inverse stellar problem, which relies on a parametrized representation, provides a powerful framework to perform EOS selection, i.e., to rule out models which are incompatible with astrophysical observations. Remarkably, it provides a straightforward method to combine measurements coming from different neutron stars. Our study shows that for soft (stiff) matter, the joint probability distribution of $p_1$--$\Gamma_2$ ($p_1$--$\Gamma_1$) offers the best prospects for EOS selection. We show this in Fig.~\ref{fig:select}. In the left (right) panel we plot the $2\sigma$ ($\sim 95\%$) credible regions, obtained from the posterior distributions of the parameters $p_1$--$\Gamma_2$ ($p_1$--$\Gamma_1$) for the \texttt{apr4} (\texttt{h4}) EOS and the three models considered. The red cross (\texttt{apr4}) and the yellow reversed triangle (\texttt{h4}) indicate the injected values, whereas the different markers are the values of the parameters corresponding to various EOSs, which have been mapped on the piecewise polytropic model by Read et al. in~\cite{Read:2008iy}. 

For both EOSs, the joint distributions seem quite effective in selecting the correct EOS, constraining a  portion of the parameter space. In both cases we are able to rule out the EOS models with stiffness different from that of the injected one, with an accuracy larger than the $90\%$. If the true EOS of supranuclear matter were stiff, measuring the tidal deformability with sufficient accuracy would allow us to rule out many known EOSs. However, since electromagnetic and gravitational wave observations suggest instead that true EOS is soft\cite{Ozel:2015fia,Abbott:2018exr}, we can rule out only a limited number of models. We notice how these bounds do not depend strongly on the neutron stars masses of the various configurations which we have analyzed.

\chapter*{Conclusions and outlook}
\chaptermark{Conclusions and outlook}
\addcontentsline{toc}{chapter}{Conclusions and outlook}
\label{sec:conclu}
In this thesis I have studied the tidal deformations of neutron stars in binary systems, and the corresponding gravitational radiation emitted, under two main lines of research.
\begin{itemize}
\item[1)] Within the first line of research, I have computed the spin-tidal interactions which affect the dynamics of two orbiting bodies in General Relativity, at the leading PN order and to linear order in the spin. These corrections belong to two classes. The first ones depend on the coupling between the standard tidal Love numbers and the spins of the compact objects. The latter ones rely instead on the rotational tidal Love numbers. Both of them depend linearly on the spins of the two bodies. 

I have computed the spin-tidal corrections to the waveform phase of the gravitational radiation emitted by binary systems in circular orbit with spins orthogonal to the orbital plane. At leading PN order, these new spin-tidal terms depend on the quadrupolar, both electric and magnetic, ordinary tidal Love numbers, and on the quadrupolar and octupolar rotational tidal Love numbers. All these terms modify the gravitational wave phase at 6.5PN order, i.e., at 1.5PN order relative to leading order, electric, quadrupolar tidal term. Thus, at linear order in the spin, the terms computed here should include \emph{all} the tidal terms up to 6.5PN order.

I stress that the spin-tidal terms computed enter the gravitational wave phase (and in general the orbital dynamics) at a lower PN order, relative to the standard electric, octupolar tidal term (which enters at 7PN order). Using simple arguments, I have derived a general rule to evaluate at which PN order the spin-tidal couplings, due to higher-order multipole moments, affect the dynamics. By means of this, I have shown that any rotational tidal Love number with $l \geq 3$ enters always at lower PN order with respect to the corresponding ordinary tidal Love numbers.

I have encountered a conceptual problem related to the inclusion of the rotational tidal Love numbers in the Lagrangian formulation, that I have not been able to solve. However, I remark that this issue could eventually affect only the numerical coefficients in the PN expansion, but not the correct identification of the PN order.

Furthermore, I have estimated the impact of the new 6.5PN order spin-tidal corrections computed on the analysis of gravitational wave signals emitted by neutron star binaries, with physical parameters consistent with the GW170817 event. I have quantified the impact of these terms by means of the bias produced on the measurement of the average weighted tidal deformability $\tilde\Lambda $, that arises from neglecting these terms in our waveform templates.

I have performed a simple analysis based on the FIM approximation. I have found that the spin-tidal effects are significant for GW170817-like binary neutron star events, detected by third-generation interferometers, if the component spins of the binaries are $\chi_i \gtrsim 0.1$. Therefore, these corrections could be relevant for binary neutron star waveform approximants only if binaries with moderately high spins merge in our local Universe. 

\item[2)] In the second line of research I have studied the feasibility of solving the relativistic inverse stellar problem with gravitational wave observations of binary neutron star coalescences. I have presented a Bayesian approach to reconstruct the phenomenological parameters which characterize the EOS in the neutron star core, using masses and tidal deformabilities obtained from gravitational wave detections. 

In my analysis I have adopted a piecewise polytropic representation for the EOS, and I have generated mock data using two candidates of the EOS, APR4 and H4, which represent the prototypes of soft and stiff nuclear matter, respectively, encompassing a wide range of admissible models of the EOS. My results show that few observations of coalescing neutron star binaries, by a network of advanced interferometers, would be sufficient to put interesting constraints on some of the parameters of the piecewise polytropic model, depending on the stiffness of the EOS. In particular, I have found that if the EOS is soft (stiff) we are able to better constrain the parameter which characterizes the inner (outer) part of the core.

Furthermore, constraints on different parameters can be used to make EOS selection. I have found that the joint-2D posterior distributions on pairs of EOS parameters are the best tool to rule out EOSs not in agreement with gravitational wave observations. In all the cases analyzed, I have been able to discriminate among soft and stiff models of EOS, with an accuracy larger than the $95\%$.
\end{itemize} 

\

The work done in this thesis can be extended/improved in several ways, within both the lines of research.

\ 

\noindent \textbf{Spin-tidal interactions}
\begin{itemize}
\item First of all, regarding the PN modeling of the spin-tidal couplings, there is the unsolved issue about the inclusion of the rotational tidal Love numbers in the Lagrangian formulation. The latter predicts the existence of some \emph{truly} universal relations, which effectively reduce the number of the independent (quadrupolar and octupolar) rotational tidal Love numbers from 4 to 2. However, such relations do not emerge from perturbation theory, when the rotational tidal Love numbers are computed numerically~\cite{Pani:2015nua}. This issue deserves further investigation, which might be also useful to clarify some discrepancies found between the rotational tidal Love numbers computed in~\cite{Pani:2015nua} and in~\cite{Gagnon-Bischoff:2017tnz}. Moreover, until this problem is solved, it will not be possible to estimate the impact of the rotational tidal Love numbers on gravitational wave data analysis, as it has been done for the other spin-tidal terms in the gravitational waveform phase.

\item The parameter estimation carried on in this thesis to evaluate the impact of the new spin-tidal terms on binary neutron star gravitational wave templates relies on several approximations. A more accurate and refined analysis, involving the full multi-dimensional parameter space, and possibly based on Bayesian methods, is required to attest the detectability of these effects in a more robust way. Furthermore, I have focused on the ET detector, but slightly better results are expected for the Cosmic Explorer interferometer~\cite{Evans:2016mbw}, since the designed sensitivity of the latter is slightly larger than that of the ET. Another issue to take into account, in the possible inclusion of the spin-tidal terms in waveform approximants calibrated on numerical simulations, is the current accuracy of the Numerical Relativity codes, which might make the minimal variations produced by the spin-tidal couplings indistinguishable from the numerical noise~\cite{Dietrich:2017aum}.

\item Another extension/application of my work is related to the gravitational wave searches for exotic compact objects (ECO)~\cite{Cardoso:2017cqb,Cardoso:2017njb}. Since the tidal Love numbers of a black hole are exactly zero~\cite{Binnington:2009bb,Damour:2009vw}, measuring the effect of the tidal deformability in the waveform of a binary coalescence provides a way to distinguish black holes from other exotic compact alternatives, for which the tidal Love numbers do not vanish~\cite{Cardoso:2017cfl,Maselli:2017cmm,Sennett:2017etc,Johnson-McDaniel:2018uvs}. There is no reason to expect that such objects should be slowly spinning (this is particularly true for supermassive objects detectable by the third-generation space-based detector LISA~\cite{2017arXiv170200786A}, whose spin might grow through accretion or through subsequent mergers during the galaxy evolution). Therefore, for these exotic objects the spin-tidal effects are expected to be larger, and their inclusion will improve previous analysis~\cite{Maselli:2017cmm}.

\item Lastly, a possible generalization of this work could be the inclusion of a time-dependence in the definition of the tidal Love numbers. In this thesis I have extended the tidal deformations of binary systems to spinning objects, using the adiabatic approximation. In particular, though both the tidal fields and the body multipole moments slowly evolve in time during the binary inspiral, their ratios, i.e., the tidal Love numbers, do not: they are constants. This description is known to fail as the orbital separation reduces and non-linear effects start to gain importance~\cite{Maselli:2012zq}. It would be interesting to study the possibility of introducing a time-scaling parameter (for instance, the orbital frequency) in the definition of the Love numbers, in order to make them time-dependent.
\end{itemize}

\ 

\noindent \textbf{Inverse stellar problem}
\begin{itemize}
\item After the detection of a binary neutron star merger, GW170817~\cite{TheLIGOScientific:2017qsa,Abbott:2018wiz}, and with more events expected in the next future, constraining the EOS through gravitational wave detections has become an hot topic in astrophysics~\cite{Abbott:2018exr,De:2018uhw,Annala:2017llu,Most:2018hfd,Bauswein:2017vtn,Raithel:2018ncd}. This event gives us the opportunity to use real data instead of simulated ones. It would be very interesting to combine different neutron star observables to put multi-messenger constraints on the EOS. A possibility already under investigation~\cite{Fasano:2019zwm} is that of combining the gravitational wave data on the tidal deformabilities provided by the LIGO/Virgo collaboration, with the observations of neutron star radii obtained in the electromagnetic band~\cite{Ozel:2015fia}. Such multi-band analysis is straightforward using a phenomenological representation of the EOS.

\item Other possible extensions, that can be easily addressed within my approach, are: (i) compare the various parametrization of EOS available in the literature, in order to find the model which leads to the most accurate constraints~\cite{Carney:2018sdv}; (ii) test the capability of the third generation of detectors to constrain the EOS, through the analysis of simulated gravitational wave measurements of masses and tidal deformabilities; (iii) exploit the correlation between the star radius and the post-merger signal of binary neutron star coalescences~\cite{Bauswein:2011tp,Bauswein:2012ya} to infer the EOS~\cite{Chatziioannou:2017ixj}.
\end{itemize}

\appendix

\chapter{Numerical integration of the equations of stellar structure}
\label{sec:appA}
In this appendix we provide some useful relations to integrate numerically the TOV equations~\eqref{eq:tov} and the perturbative equations~\eqref{eq:polars} and~\eqref{eq:axials2}. First, we notice that to compute the equilibrium configuration of a neutron star (i.e., its mass and radius), we do not need the whole TOV system~\eqref{eq:tov}, but only the first two equations
\begin{equation}
\label{eq:tovred}
\left \{
\begin{aligned}
\frac{dm}{dr} & =4\pi r^2 \epsilon \\
\frac{dp}{dr}& = -\frac{(\epsilon+p)(m+4\pi r^3 p)}{r(r-2m)} 
\end{aligned}
\right.\,,
\end{equation}
together with the EOS, $p =p(\epsilon)$. Indeed, the ODE for the function $\nu(r)$,
\begin{equation}
\frac{d\nu}{dr} = \frac{2(m+4\pi r^3 p)}{r(r-2m)} \,,
\end{equation}
is necessary only if we are interested in computing the spacetime metric inside the star.

Expanding the system~\eqref{eq:tovred} near $r=0$, we get
\begin{equation}
\begin{aligned}
\label{eq:init}
m(r) & = m_3 r^3+ m_5 r^5 + \mathcal{O}(r^7) \\
p(r) & = p_0 + p_2 r^2+ \mathcal{O}(r^4) \,,
\end{aligned} 
\end{equation}
where
\begin{equation}
\begin{aligned}
m_3 & = \frac{4 \pi}{3}\epsilon_0 \\
m_5 & = \frac{4 \pi}{5}\frac{p_2}{c_{s,0}^2} \\
p_2 & =-\frac{2 \pi}{3}(\epsilon_0+p_0)(\epsilon_0+3p_0)
\end{aligned}
\end{equation}
and $p_0$, $\epsilon_0$ and $c_{s,0}^2$ are the values of pressure, energy density and speed of sound ($c_s^2 = dp/d\epsilon$) at the center of the star, respectively. We can use Eqs.~\eqref{eq:init} as initial conditions to numerically integrate the system~\eqref{eq:tovred}. 

To compute the electric, quadrupolar tidal deformability $\lambda_2$, we add Eq.~\eqref{eq:polars} with $l=2$,
\begin{equation}
\label{eq:H0app}
\begin{aligned}
\dfrac{d^2 H_0(r)}{dr^2} & + \left(\frac{2\left[r-m+2 \pi r^3(p-\epsilon) \right]}{r(r-2m)} \right) \frac{d H_0(r)}{dr} + \left( \frac{4 \pi r^2 \left[p (9-16 \pi r^2 p) + 5 \epsilon\right] }{(r-2m)^2 }\right.\\
&  + \frac{ 4 \pi r^3 \left[ (r-2m)(p+\epsilon)/c_s^2 -2m(13p + 5 \epsilon) \right]-4m^2 }{r^2(r-2m)^2}\\
& \left. -\frac{6}{r (r-2m)}\right) H_0(r) =0 \,,
\end{aligned}
\end{equation}
to the system~\eqref{eq:tovred}. We recall that to compute the tidal deformability, one needs actually only the quantity $y_e=(r/H_0)(dH_0/dr)$, evaluated at the star surface (cf. Eq.~\eqref{eq:lovenumber2el}). Thus, as pointed out in~\cite{Lindblom:2013kra}, it is numerically more efficient to transform the second-order, linear ODE for $H_0(r)$ in a first-order, non-linear ODE for $y_e(r)$:
\begin{equation}
\begin{aligned}
\label{eq:yeapp}
\frac{dy_e}{dr} =& - \frac{y_e^2}{r} - \frac{r+ 4 \pi r^3(p-\epsilon) }{r(r-2m)} y_e + \frac{4(m+4 \pi r^3 p)^2}{r(r-2m)^2} + \frac{6}{r-2m} \\
&- \frac{4 \pi r^2}{r-2m}\left(5 \epsilon + 9 p+\frac{\epsilon+p}{c_s^2} \right)\,.
\end{aligned}
\end{equation}
The initial condition at $r=0$ for the above equation is given by
\begin{equation}
y_e(r) =2 \left[ 1-\frac{2 \pi}{7} \left( \frac{\epsilon_0}{3}+11 p_0+ \frac{\epsilon_0+p_0}{c^2_{s,0}} \right) r^2 + \mathcal{O}(r^4)\right] \,.
\end{equation}
Note that in this form the dependence on the arbitrary constant $a_0$, arising from the boundary condition for $H_0(r)$ at the center of the star (see Eq.~\eqref{eq:H0init}), naturally disappears. Since $y_e(r)$ is the actually quantity which enters the tidal deformability, this proves that the Love number is independent of $a_0$ (cf. with the discussion below Eq.~\eqref{eq:H0init}). We notice also that the expansion of Eq.~\eqref{eq:yeapp} around $r=0$ gives actually two solutions, because the ODE is quadratic in $y_e(r)$. One then identifies the correct branch through a comparison with the initial condition for $H_0(r)$.

The case of the magnetic, quadrupolar tidal deformability is analog. For $l=2$, Eq.~\eqref{eq:axials2} reads
\begin{equation}
\frac{d^2h_0(r)}{dr^2} - \left( \frac{4 \pi r^2(p+\epsilon)}{r-2m} \right)  \frac{dh_0(r)}{dr} - \left(\frac{6r-4m \pm 8 \pi r^3(p+ \epsilon)}{r^2(r-2m)} \right)  h_0(r) =0 \,,
\end{equation}
which we can transform in an ODE for $y_m=(r/h_0)(dh_0/dr)$ (cf. Eq.~\eqref{eq:lovenumber2mag}):
\begin{equation}
\begin{aligned}
\frac{dy_m}{dr} =& - \frac{y_m^2}{r} - \frac{2m-r[1+4 \pi r^2 (p +\epsilon)] }{r(r-2m)} y_m - \frac{4m-r[6 \pm 8 \pi r^2(p+\epsilon)]}{r(r-2m)}\,,
\end{aligned}
\end{equation}
where the plus/minus sign refers to static/irrotational fluids, respectively. The initial condition at $r=0$ is
\begin{equation}
y_m(r)=3 \left[ 1+\frac{4 \pi}{63} \left( 15p_0+23 \epsilon_0 \right) r^2 + \mathcal{O}(r^4)\right]
\end{equation}
in the static case, and
\begin{equation}
y_m(r)=3 \left[ 1+\frac{4 \pi}{63} \left( 3p_0+11 \epsilon_0 \right) r^2 + \mathcal{O}(r^4)\right]
\end{equation}
in the irrotational one.

\chapter{Spherical harmonics}
\label{sec:appB}
In this appendix we recall some useful properties of the spherical harmonics. We define the (orbital) angular momentum operator as
\begin{equation}
\pmb{\mathbb{L}}= - \mathrm{i} \left(\boldsymbol{r} \times \boldsymbol{\nabla}\right) \,,
\end{equation}
which in spherical coordinates reads
\begin{equation}
\begin{aligned}
\mathbb{L}_x = & - \mathrm{i} \left( -\sin{\varphi} \frac{\partial}{\partial \theta}  -\frac{\cos{\theta}}{\sin{\theta}} \cos{\varphi} \frac{\partial}{\partial \varphi} \right) \\ 
\mathbb{L}_y = & - \mathrm{i}  \left(\cos{\varphi}  \frac{\partial}{\partial \theta}  - \frac{\cos{\theta}}{\sin{\theta}} \sin{\varphi} \frac{\partial}{\partial \varphi} \right) \\
\mathbb{L}_z = & - \mathrm{i} \frac{\partial}{\partial \varphi} \,,
\end{aligned}
\end{equation}
and its square is given by
\begin{equation}
\pmb{\mathbb{L}}^2 = - \left(\frac{\partial^2}{\partial \theta^2}+ \frac{\cos{\theta}}{\sin{\theta}} \frac{\partial}{\partial \theta} + \frac{1}{\sin^2{\theta}} \frac{\partial}{\partial \varphi^2} \right) \,.
\end{equation}
The \emph{spherical harmonics} $Y_{lm}(\theta, \varphi)$ are the eigenfunctions of the operator $\pmb{\mathbb{L}}^2$:
\begin{equation}
\pmb{\mathbb{L}}^2 Y_{lm}(\theta, \varphi)=l(l+1) Y_{lm}(\theta, \varphi).
\end{equation}
The index $l$ can assume non-negative integer values ($l = 0,1,2,\dots, \infty$), whereas the index $m$ varies on the integer values in the interval $m \leq |l|$ ($m = -l,-(l-1),\dots,-1,0,1,\dots,l-1,l$).

The explicit expression of $Y_{lm}(\theta, \varphi)$, for non-negative values of $m$, is
\begin{equation}
Y_{lm}(\theta, \varphi)= \sqrt{\frac{2l+1}{4\pi}\frac{(l-m)!}{ (l+m)!}} \,  \mathcal{P}^m_l(\cos{\theta}) \mathrm{e}^{\mathrm{i} m \varphi} \qquad m \ge 0 \,,
\end{equation}
where $\mathcal{P}^m_l(x)$ are the \emph{associated Legendre polynomials}
\begin{equation}
\mathcal{P}^m_l(x) =(-1)^m (1-x^2)^{m/2} \frac{d^m}{dx^m}\mathcal{P}_l(x) \,,
\end{equation}
and $\mathcal{P}_l(x)$ are the \emph{Legendre polynomials}
\begin{equation}
\mathcal{P}_l(x) =\frac{1}{2^l l!} \frac{d^l}{dx^l}(x^2-1)^l \,.
\end{equation}
The spherical harmonics for negative values of $m$ are obtained from the relation
\begin{equation}
Y_{lm}(\theta, \varphi)= (-1)^m Y^*_{l,-m}(\theta, \varphi) \,.
\end{equation}

The spherical harmonics form an orthonormal basis on the 2-sphere: 
\begin{equation}
\begin{aligned}
\int_0^{2\pi} \int_0^{\pi} Y^*_{l'm'}(\theta, \varphi) Y_{lm}(\theta, \varphi) \, \sin{\theta} d \theta d \varphi = \delta_{l'l} \delta_{m'm} \\
\sum_{l=0}^{\infty} \sum_{m=-l}^l Y^*_{lm}(\theta', \varphi') Y_{lm}(\theta, \varphi) = \frac{\delta(\theta-\theta') \delta(\varphi-\varphi')}{\sin{\theta}} \,.
\end{aligned}
\end{equation}
The above equations represent the orthonormality and completeness relations, respectively. Thus, any scalar function $f(\theta,\varphi)$ can be expanded in terms of the spherical harmonics
\begin{equation}
f(\theta,\varphi) = \sum_{l=0}^{\infty} \sum_{m=-l}^l f_{lm} Y_{lm}(\theta, \varphi) \,,
\end{equation}
where the coefficients $f_{lm}$ are given by
\begin{equation}
f_{lm} = \int_0^{2\pi} \int_0^{\pi} Y^*_{lm}(\theta, \varphi) f(\theta,\varphi) \, \sin{\theta} d \theta d \varphi  \,.
\end{equation}

Under a parity transformation, $\{\theta, \varphi\} \to \{\pi - \theta, \varphi + \pi \}$, the spherical harmonics transform as
\begin{equation}
\mathbb{P} Y_{lm}(\theta, \varphi) = (-1)^l Y_{lm}(\theta, \varphi) \,,
\end{equation}
and therefore, they are called \emph{even}, or \emph{polar}, or \emph{electric}.

The complete set of symmetric trace-free (STF) tensors $\mathcal{Y}^{lm}_L$ (see the~\nameref{sec:notation} for the multi-index definition) defined in~\cite{Thorne:1980ru} is intimately related to the spherical harmonics. Indeed, they satisfy the relations
\begin{equation}
\begin{aligned}
\mathcal{Y}^{lm}_L n_L =& Y_{lm} \\
\Phi_{i,lm} = & l \, \epsilon_{ijk} \, \mathcal{Y}^{lm}_{kL-1} \, n_{jL-1}
\end{aligned}
\end{equation}
where
\begin{equation}
\boldsymbol{\Phi}_{lm} = \mathrm{i} \, \pmb{\mathbb{L}} \, Y_{lm} =  \left(\boldsymbol{r} \times \boldsymbol{\nabla}\right) Y_{lm} \,,
\end{equation}
and $\boldsymbol{n}$ is the unit radial vector.

The spherical harmonics can be generalized to vector and tensor fields. Following the definition used in~\cite{Regge:1957td}, the \emph{vector spherical harmonics} are
\begin{equation}
\begin{aligned}
\left(  R_{\theta,lm} , R_{\varphi,lm}  \right)= & \left( \frac{\partial}{\partial \theta} Y_{lm}, \frac{\partial}{\partial \varphi} Y_{lm} \right) \\
\left(  S_{\theta,lm}  , S_{\varphi,lm}  \right) = & \left(  -\frac{1}{\sin{\theta}} \frac{\partial}{\partial \varphi} Y_{lm} ,\sin{\theta} \, \frac{\partial}{\partial \theta} Y_{lm} \right) \,.
\end{aligned}
\end{equation}
They are used to decompose the angular components of three and four-vectors. Note that $S_{i,lm}$ is the angular part of $\boldsymbol{\Phi}_{lm}$ in spherical coordinates (the radial part vanishes):
\begin{equation}
\begin{aligned}
S_{\theta,lm} = &\cos{\theta} \cos{\varphi} \, \Phi_{x,lm} + \cos{\theta} \sin{\varphi}\, \Phi_{y,lm} -\sin{\theta}\, \Phi_{z,lm} \\
S_{\varphi,lm} =& - \sin{\theta}\sin{\varphi} \,\Phi_{x,lm} +  \sin{\theta}\cos{\varphi} \,\Phi_{y,lm} \,.
\end{aligned}
\end{equation}

Under a parity transformation the vector spherical harmonics transform as
\begin{equation}
\begin{aligned}
\mathbb{P} R_{i,lm} & = (-1)^l R_{i,lm}  \\
\mathbb{P} S_{i,lm} & = (-1)^{l+1} S_{i,lm}
\end{aligned} \qquad i=\theta,\varphi \,.
\end{equation}
Therefore, $R_{i,lm}$ are even like the scalar harmonics, while $S_{i,lm}$ are called \emph{odd}, or \emph{axial}, or \emph{magnetic}.

The angular components of rank-2 symmetric tensors can be expanded in terms of \emph{tensor spherical harmonics}. They read~\cite{Regge:1957td}
\begin{equation}
\begin{aligned}
\gamma_{ij,lm} &= \begin{pmatrix}
 Y_{lm} & 0   \\
 * & \sin^2{\theta}Y_{lm}
\end{pmatrix} \\
\Psi_{ij,lm} &= \begin{pmatrix}
\frac{\partial^2}{\partial \theta^2}Y_{lm} & (\frac{\partial}{\partial \theta} -\frac{\cos{\theta}}{\sin{\theta}} )\frac{\partial}{\partial \varphi}Y_{lm}  \\
 * & (\sin{\theta} \cos{\theta}\frac{\partial}{\partial \theta} +\frac{\partial^2}{\partial \varphi^2} )Y_{lm}
\end{pmatrix} \\
\chi_{ij,lm} & = \begin{pmatrix}
 \frac{1}{\sin{\theta}}(\frac{\cos{\theta}}{\sin{\theta}} - \frac{\partial}{\partial \theta} )\frac{\partial}{\partial \varphi}Y_{lm} & \frac{1}{2} (\sin{\theta}\frac{\partial^2}{\partial \theta^2} - \cos{\theta} \frac{\partial}{\partial \theta} -\frac{1}{\sin{\theta}} \frac{\partial^2}{\partial \varphi^2})Y_{lm}  \\
 * & (\sin{\theta}\frac{\partial}{\partial \theta} - \cos{\theta} )\frac{\partial}{\partial \varphi}Y_{lm}
\end{pmatrix} 
\end{aligned} i,j=\theta,\varphi \,,
\end{equation}
where the star denotes the components obtained by symmetry.
Under a parity transformation, $\gamma_{ij,lm}$ and $\Psi_{ij,lm}$ are even, whereas $\chi_{ij,lm}$ is odd.

Lastly, we introduce the \emph{spin-weighted spherical harmonics} $\null_s Y_{lm}(\theta, \varphi)$. We say that a given function $f$, defined on the 2-sphere, has spin weight $s$, if under rotations around the unit radial vector $\boldsymbol{n}$, it transforms as~\cite{Newman:1966ub,Campbell:1971rm}
\begin{equation}
\label{eq:spinweightedhr}
f' = \mathrm{e}^{\mathrm{i} s \psi} f \,,
\end{equation}
where $\psi$ is the rotation angle. For instance, the unit vector $\boldsymbol{m} = (\boldsymbol{\mathrm{e}}_{\theta}+ \mathrm{i} \boldsymbol{\mathrm{e}}_{\varphi})/\sqrt{2}$ has $s=1$, whereas $\boldsymbol{m}^*$ has $s=-1$. $\boldsymbol{\mathrm{e}}_{\theta}$ and $\boldsymbol{\mathrm{e}}_{\varphi}$ are the unit vectors in the $\theta$ and $\varphi$ direction, respectively. Any function, which transforms as in Eq.~\eqref{eq:spinweightedhr} with spin weight $s$, can be expanded in terms of the spin-weighted spherical harmonics, defined as~\cite{Kidder:2007rt,Creighton:2011}
\begin{equation}
\begin{aligned}
\null_{-s} Y_{lm}(\theta, \varphi) =& (-1)^s \sqrt{\frac{2l+1}{4\pi}} \sqrt{\frac{(l+m)!(l-m)!}{(l+s)!(l-s)!}} \mathrm{e}^{\mathrm{i} m \varphi} \cos^{2l}{(\theta/2)}  \\
& \times \sum_{k_i}^{k_f} (-1)^k \binom{l-s}{k} \binom{l+s}{s-m+k} \tan^{s-m+2k}{(\theta/2)} \,,
\end{aligned}
\end{equation}
where $k_i = \max(0,m-s)$ and $k_f = \min(l+m,l-s)$. For $s=0$ they reduce to the scalar harmonics $Y_{lm}$. The spin-weighted harmonics satisfy the relations
\begin{equation}
\begin{gathered}
\null_{s}Y_{lm}(\theta, \varphi)=  (-1)^{s+m} \null_{-s}Y^*_{l,-m}(\theta, \varphi) \\[0.32cm]
\null_{s}Y_{lm}(\pi-\theta, \varphi+\pi)=  (-1)^{l} \null_{-s}Y_{lm}(\theta, \varphi) \,.
\end{gathered}
\end{equation}

\chapter{Supplementary material of the inverse stellar problem study}
\label{sec:appC}
\begin{figure}[]
\captionsetup[subfigure]{labelformat=empty}
\centering
{\includegraphics[width=0.32\textwidth]{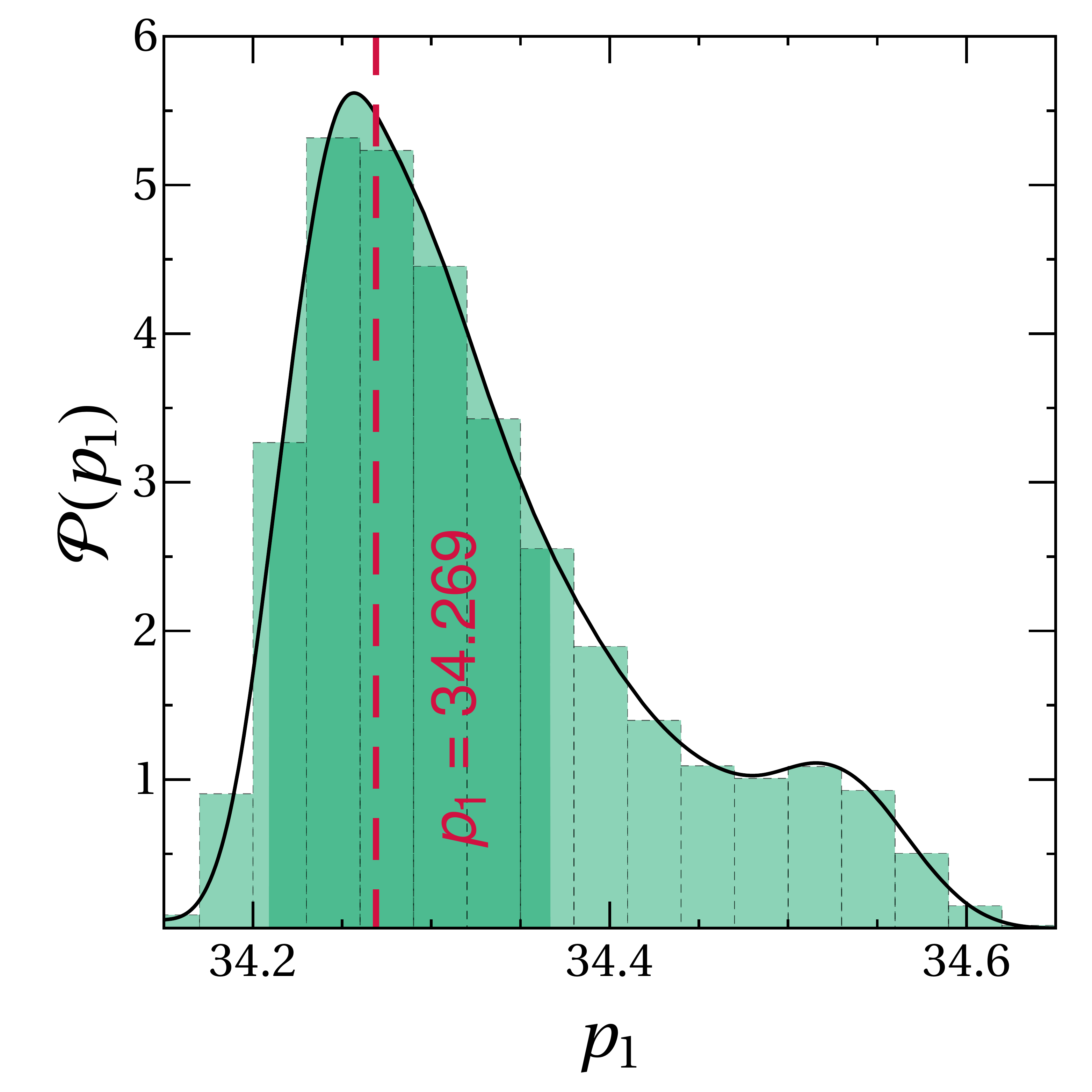}}  
{\includegraphics[width=0.32\textwidth]{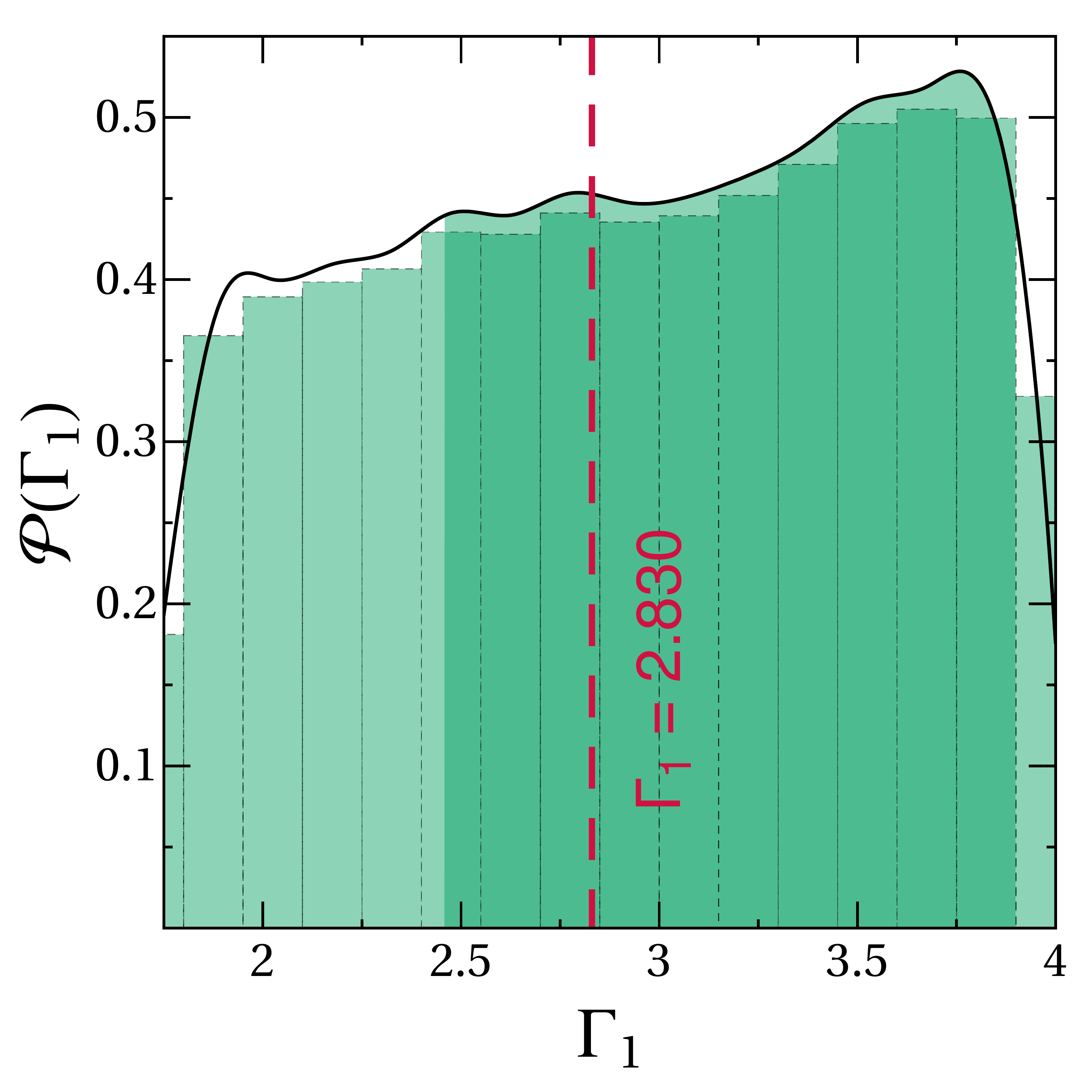}}
{\includegraphics[width=0.32\textwidth]{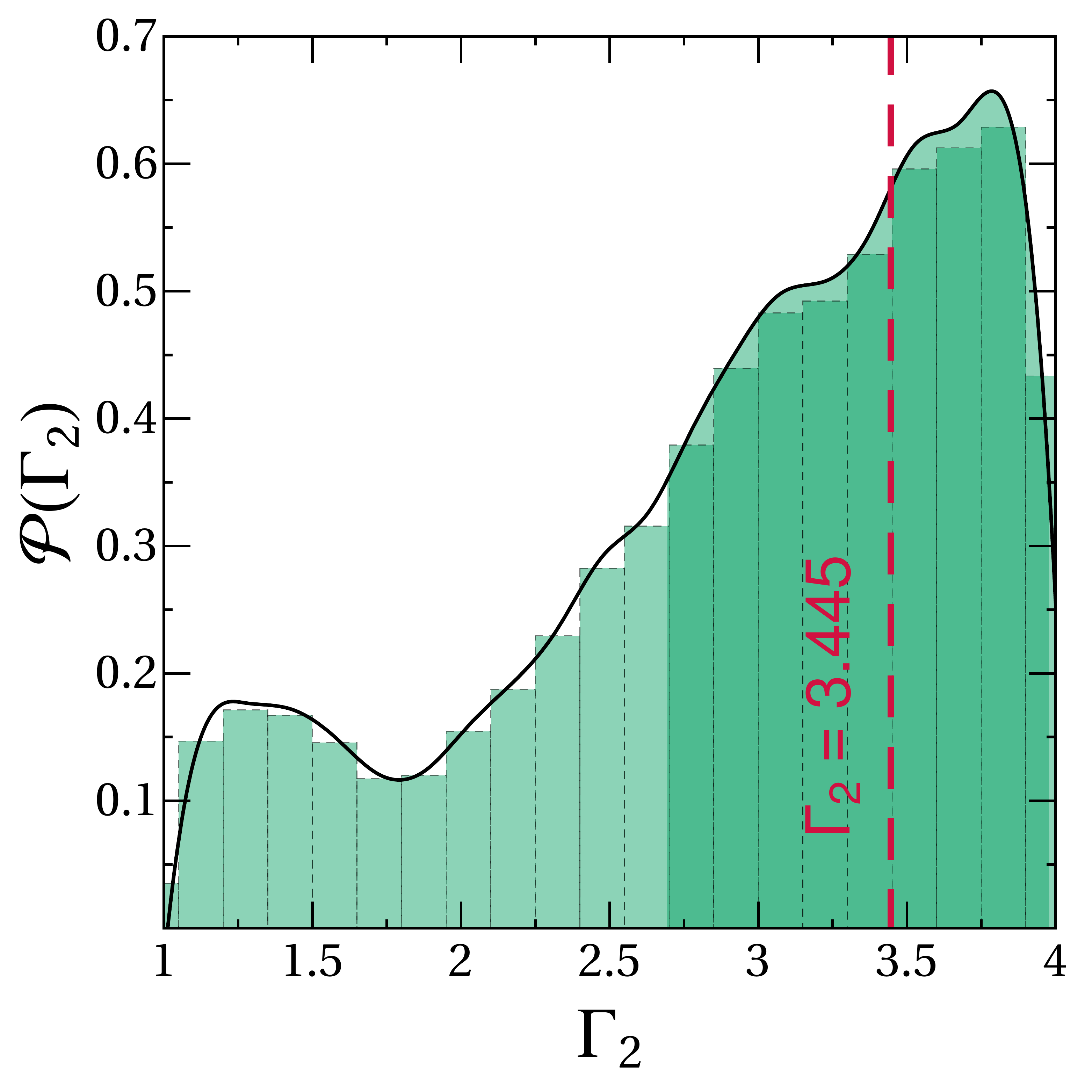}}
{\includegraphics[width=0.32\textwidth]{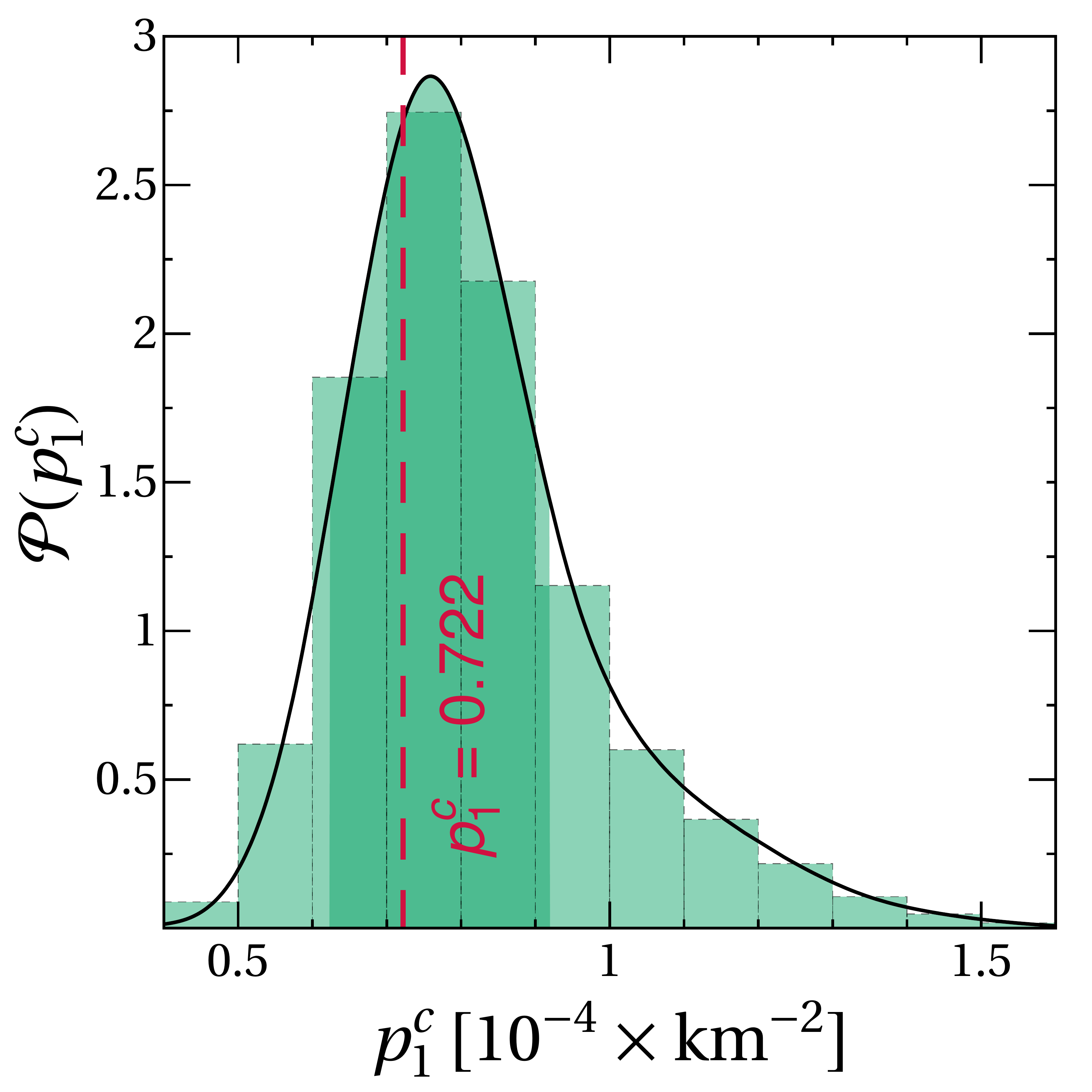}}  
{\includegraphics[width=0.32\textwidth]{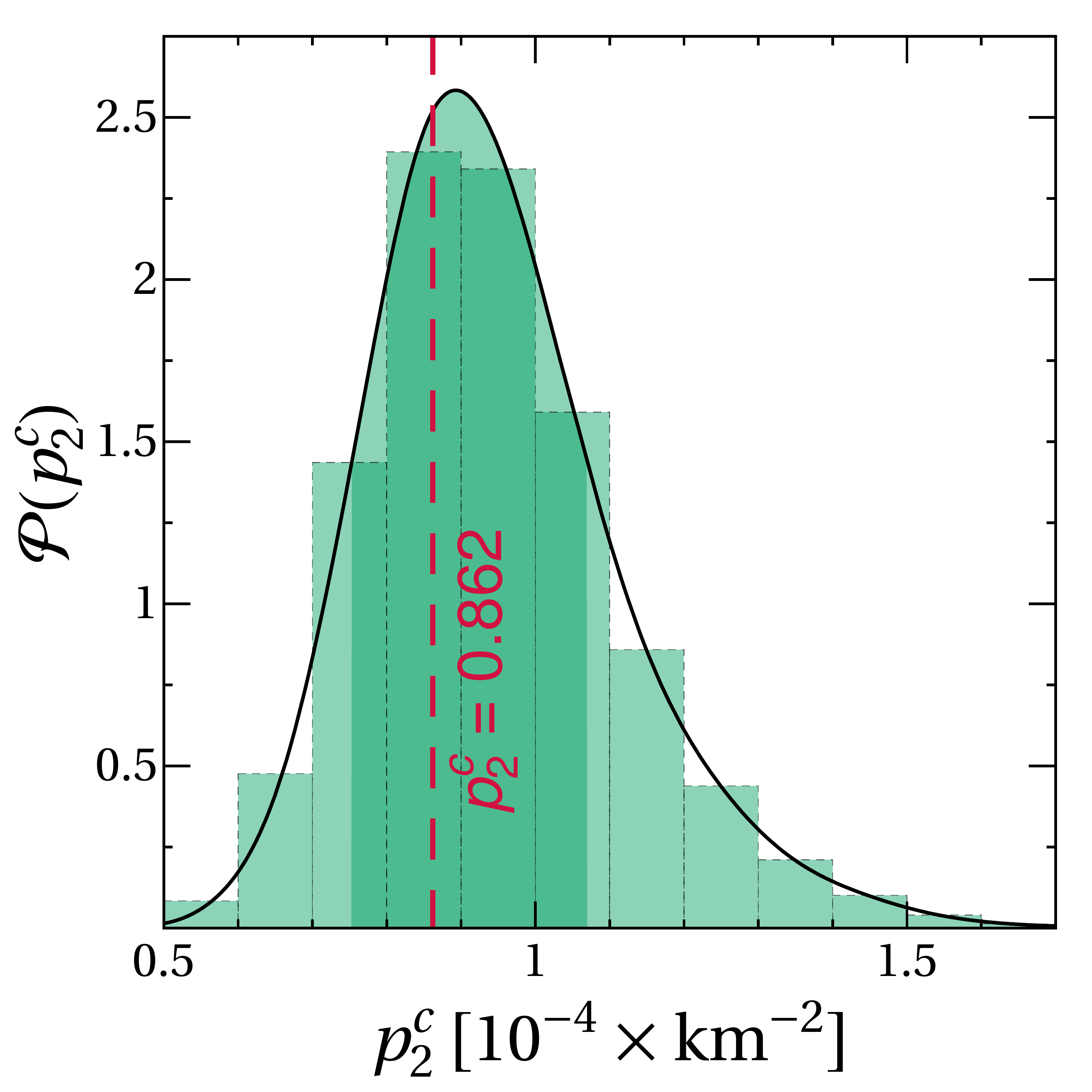}}
{\includegraphics[width=0.32\textwidth]{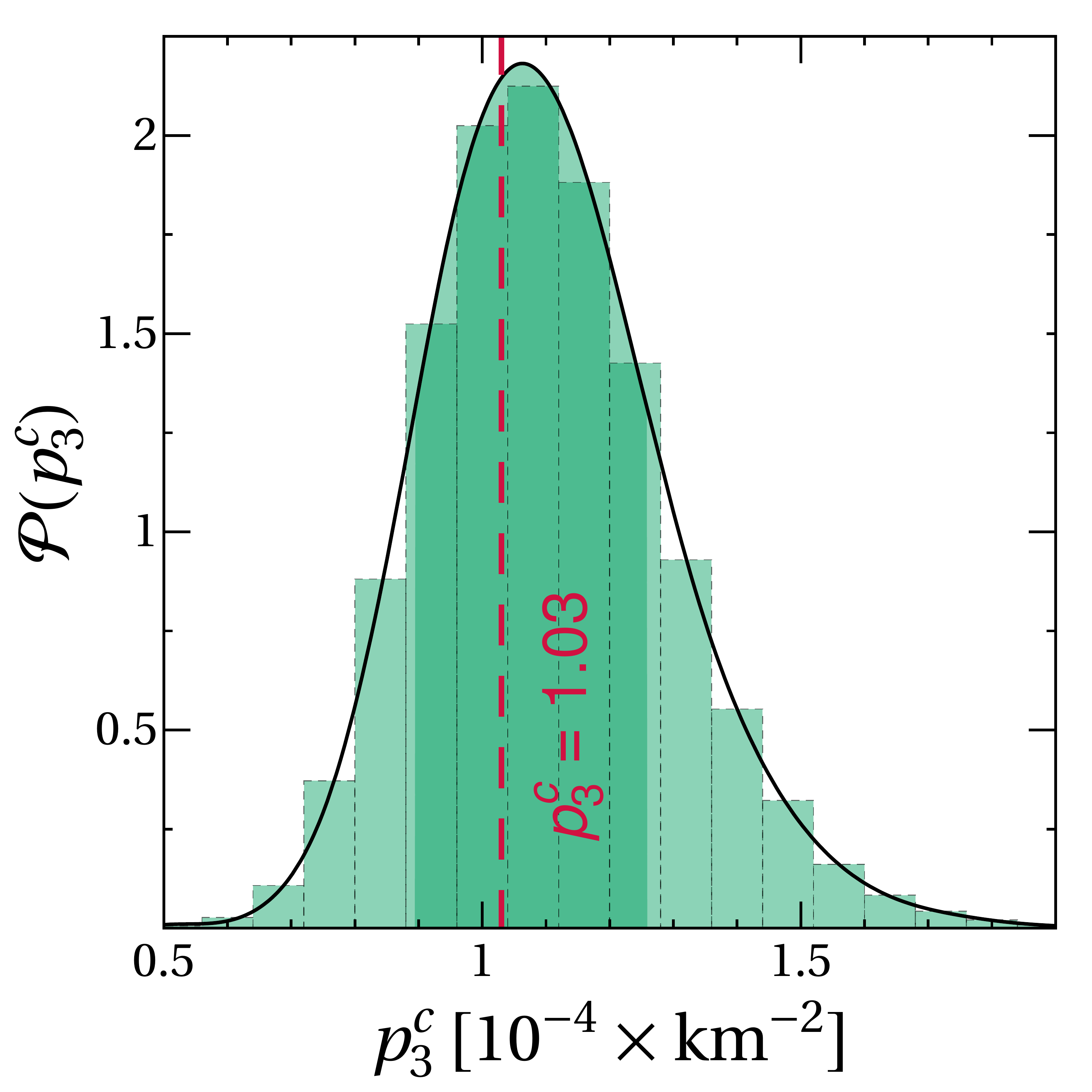}}
\caption{\textsl{Marginalized posterior PDF for the parameters of the \texttt{apr4} EOS, derived for the \texttt{m123} model with neutron stars masses $(1.1,1.2,1.3)M_\odot$. The histograms of the sampled points are shown below each function. The red, dashed vertical lines identify the injected true values, while the shaded bands correspond to the $1\sigma$ credible regions of each parameter.}}
\captionsetup{format=hang,labelfont={sf,bf}}
\label{fig:m123apr4}
\end{figure}

\begin{figure}[]
\captionsetup[subfigure]{labelformat=empty}
\centering
{\includegraphics[width=0.32\textwidth]{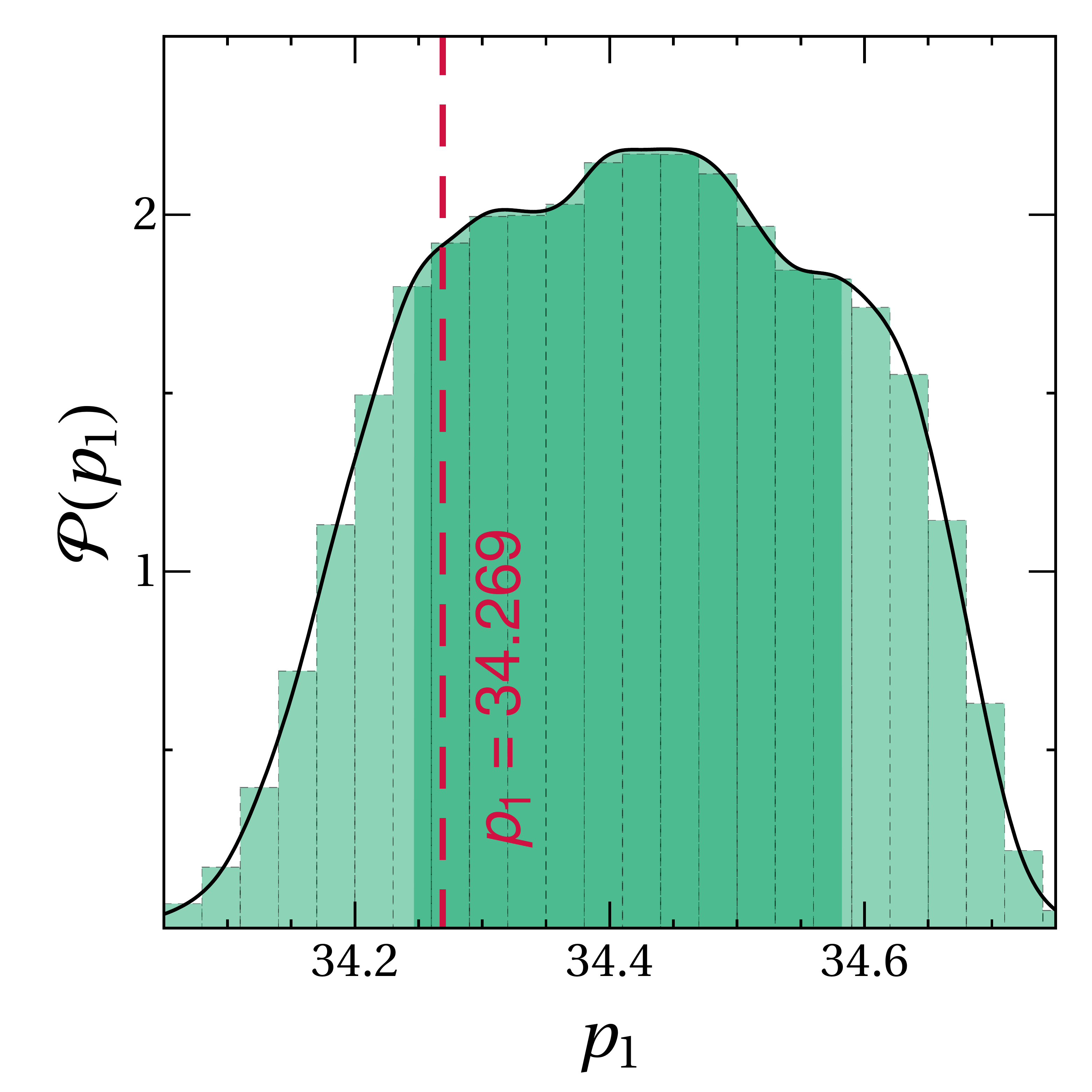}}  
{\includegraphics[width=0.32\textwidth]{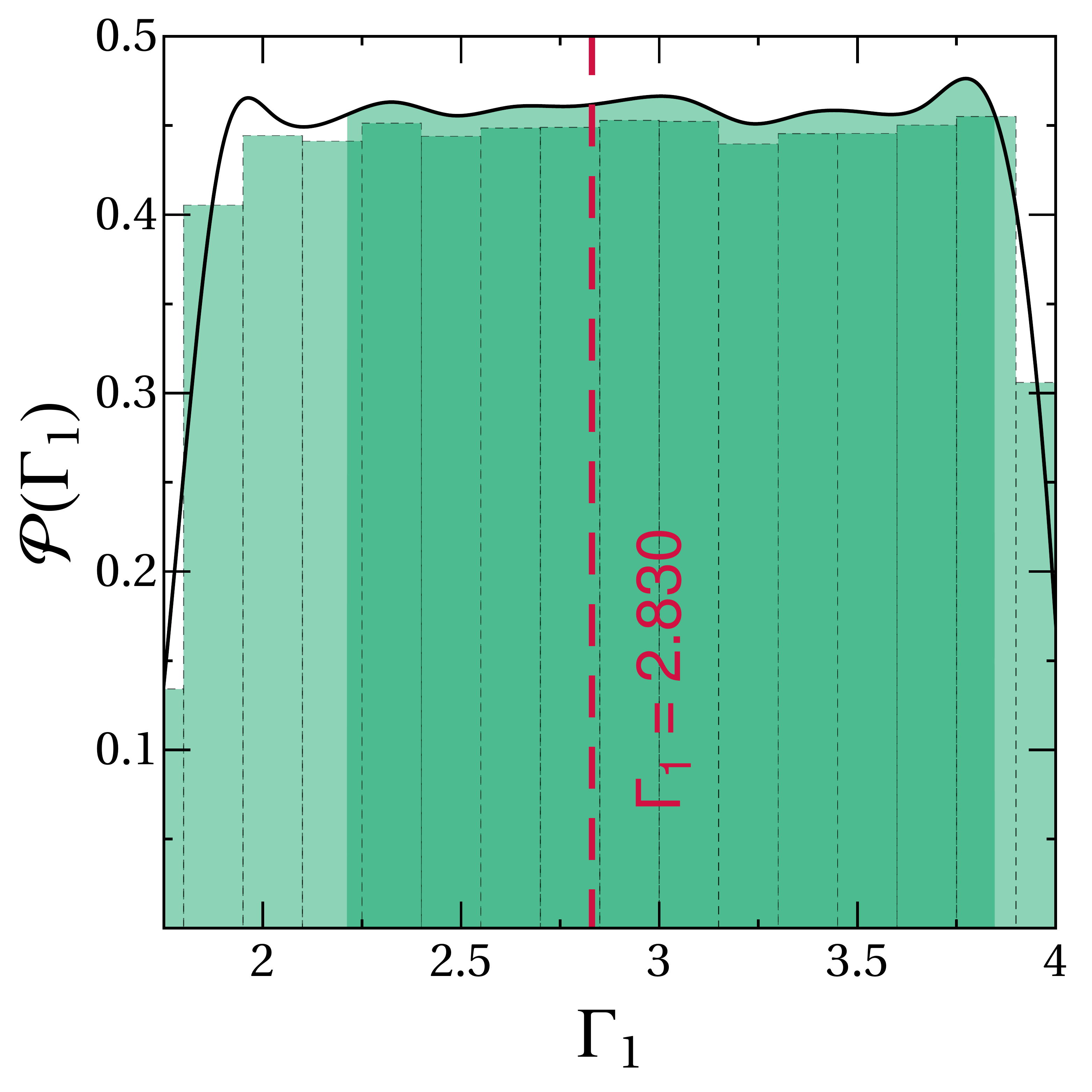}}
{\includegraphics[width=0.32\textwidth]{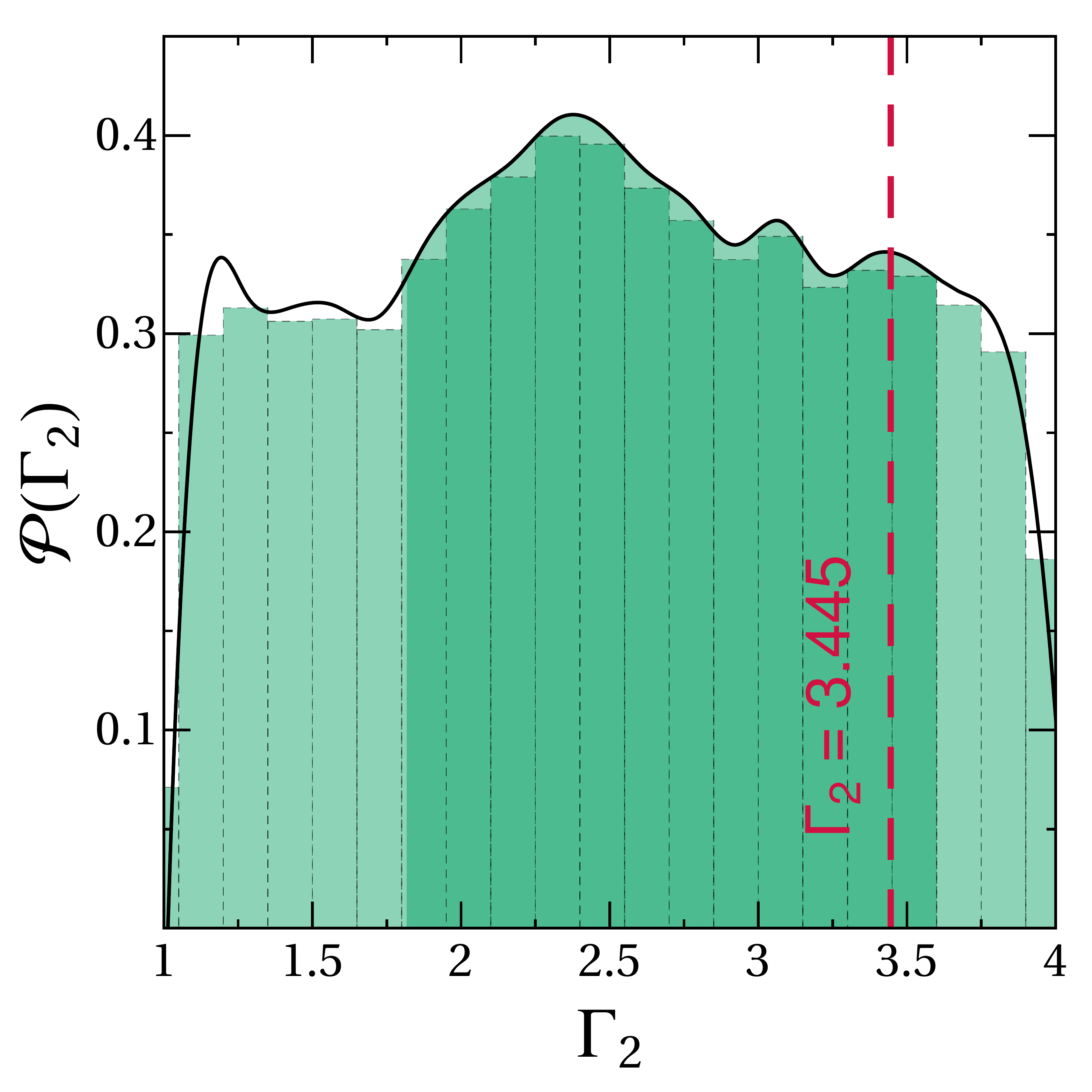}}
{\includegraphics[width=0.32\textwidth]{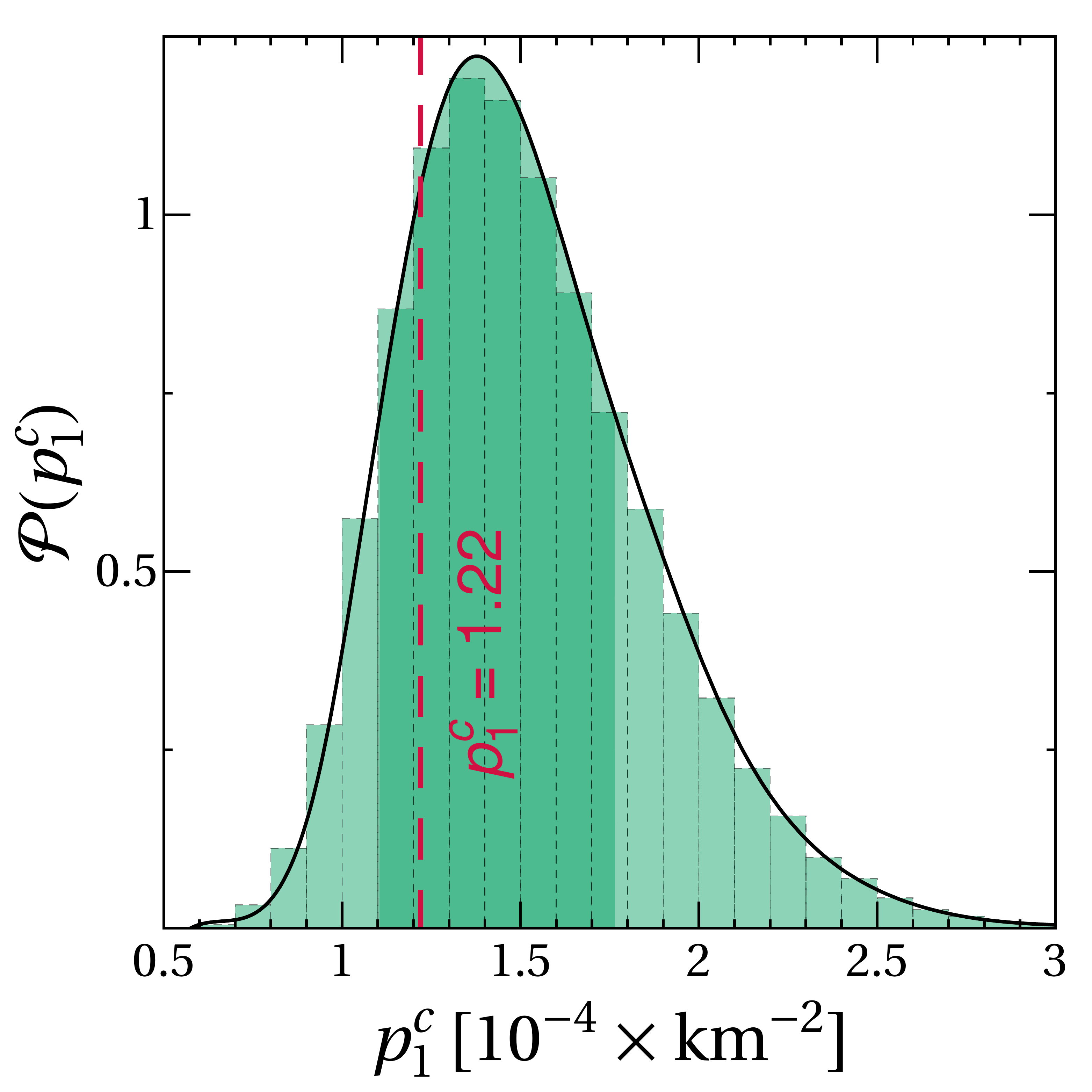}}  
{\includegraphics[width=0.32\textwidth]{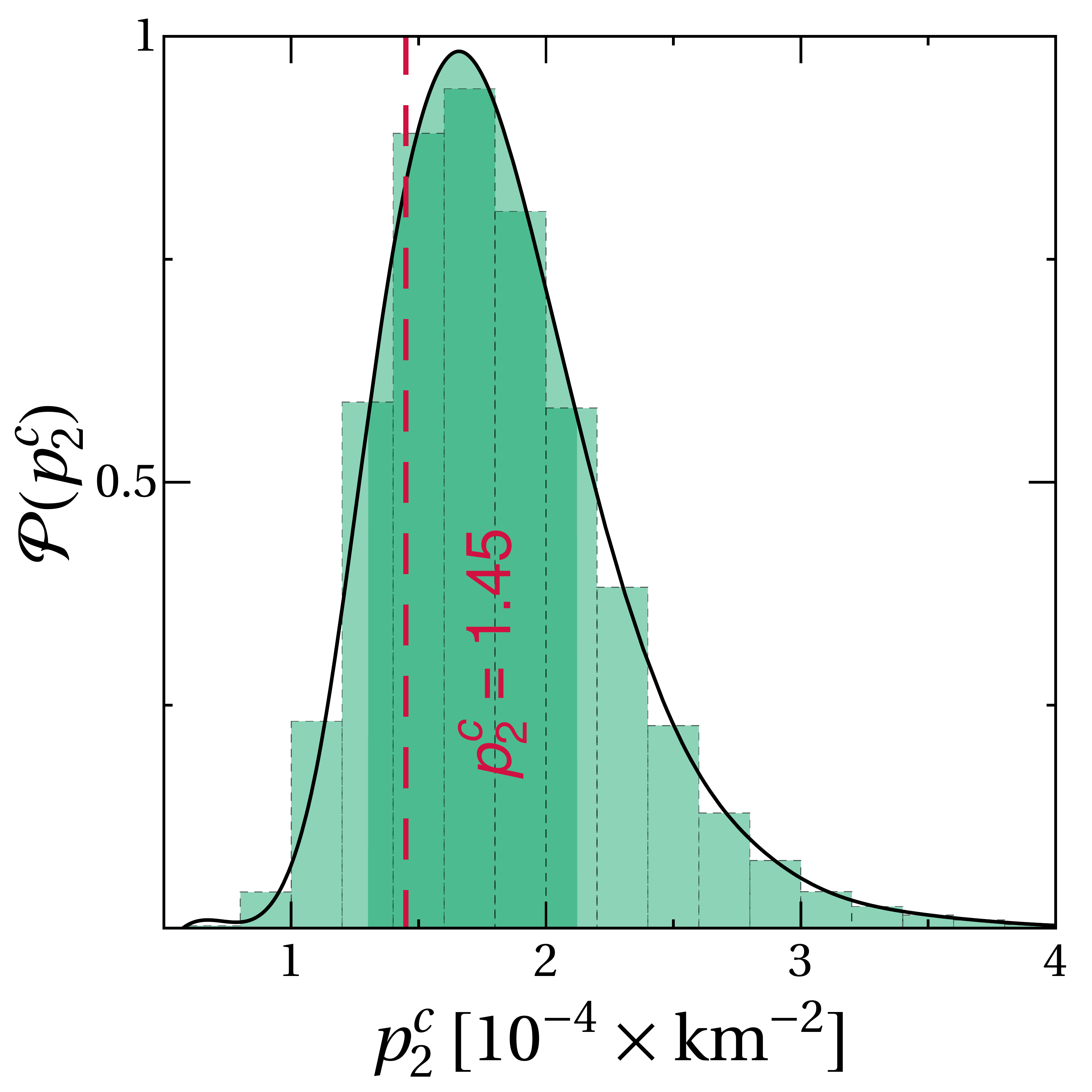}}
{\includegraphics[width=0.32\textwidth]{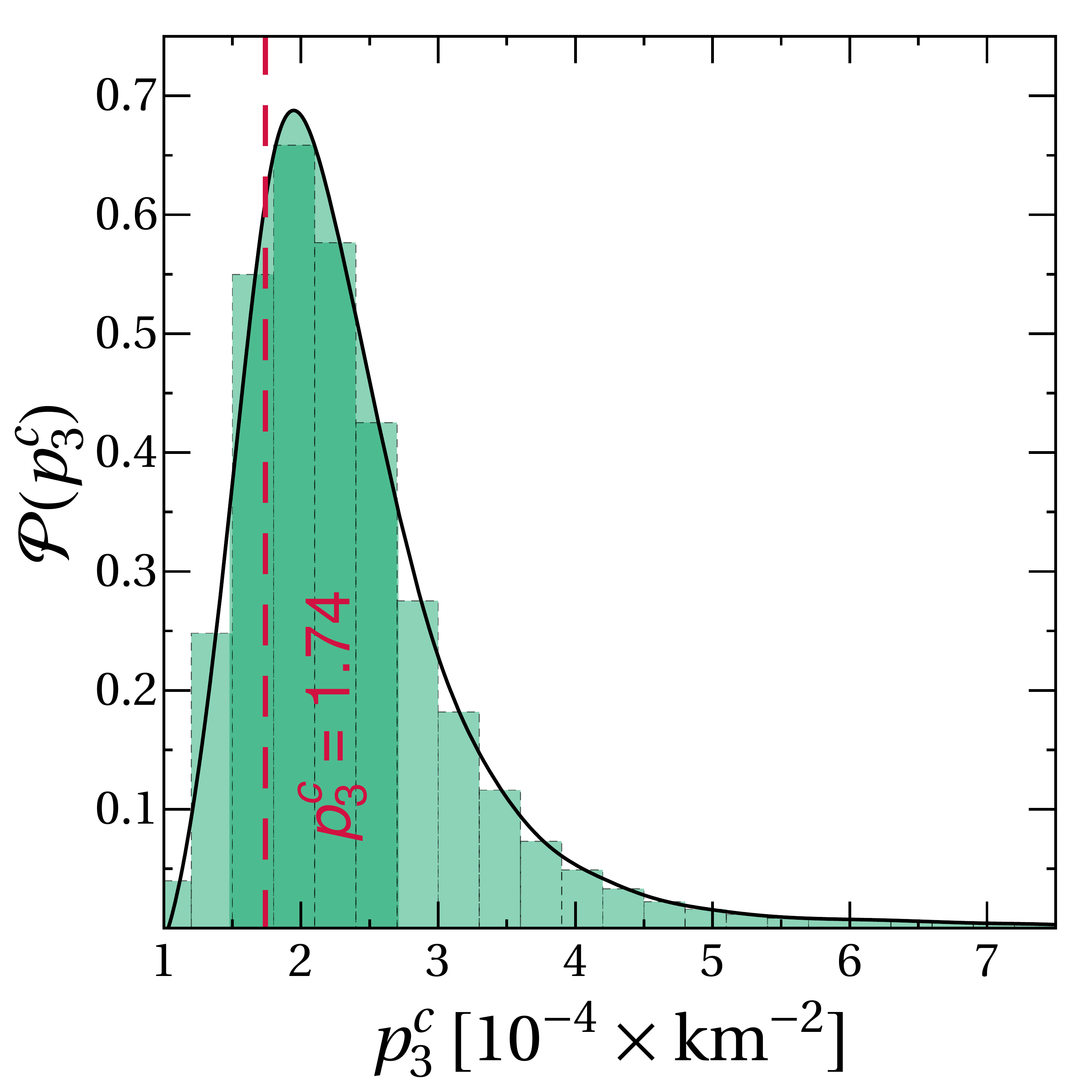}}
\caption{\textsl{Marginalized posterior PDF for the parameters of the \texttt{apr4} EOS, derived for the \texttt{m456} model with neutron stars masses $(1.4,1.5,1.6)M_\odot$. The histograms of the sampled points are shown below each function. The red, dashed vertical lines identify the injected true values, while the shaded bands correspond to the $1\sigma$ credible regions of each parameter.}}
\captionsetup{format=hang,labelfont={sf,bf}}
\label{fig:m456apr4}
\end{figure}

\begin{figure}[]
\captionsetup[subfigure]{labelformat=empty}
\centering
{\includegraphics[width=0.32\textwidth]{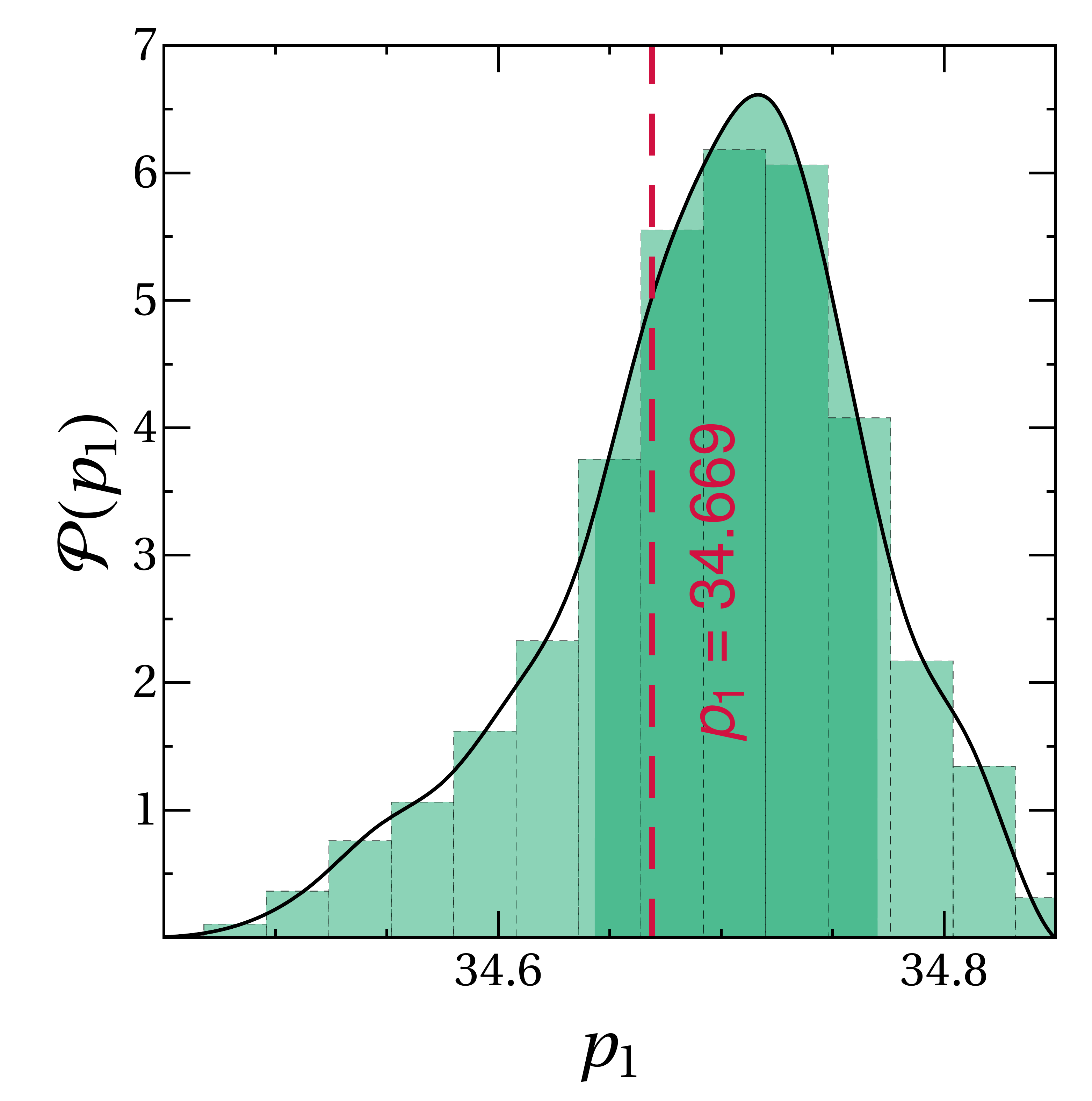}}  
{\includegraphics[width=0.32\textwidth]{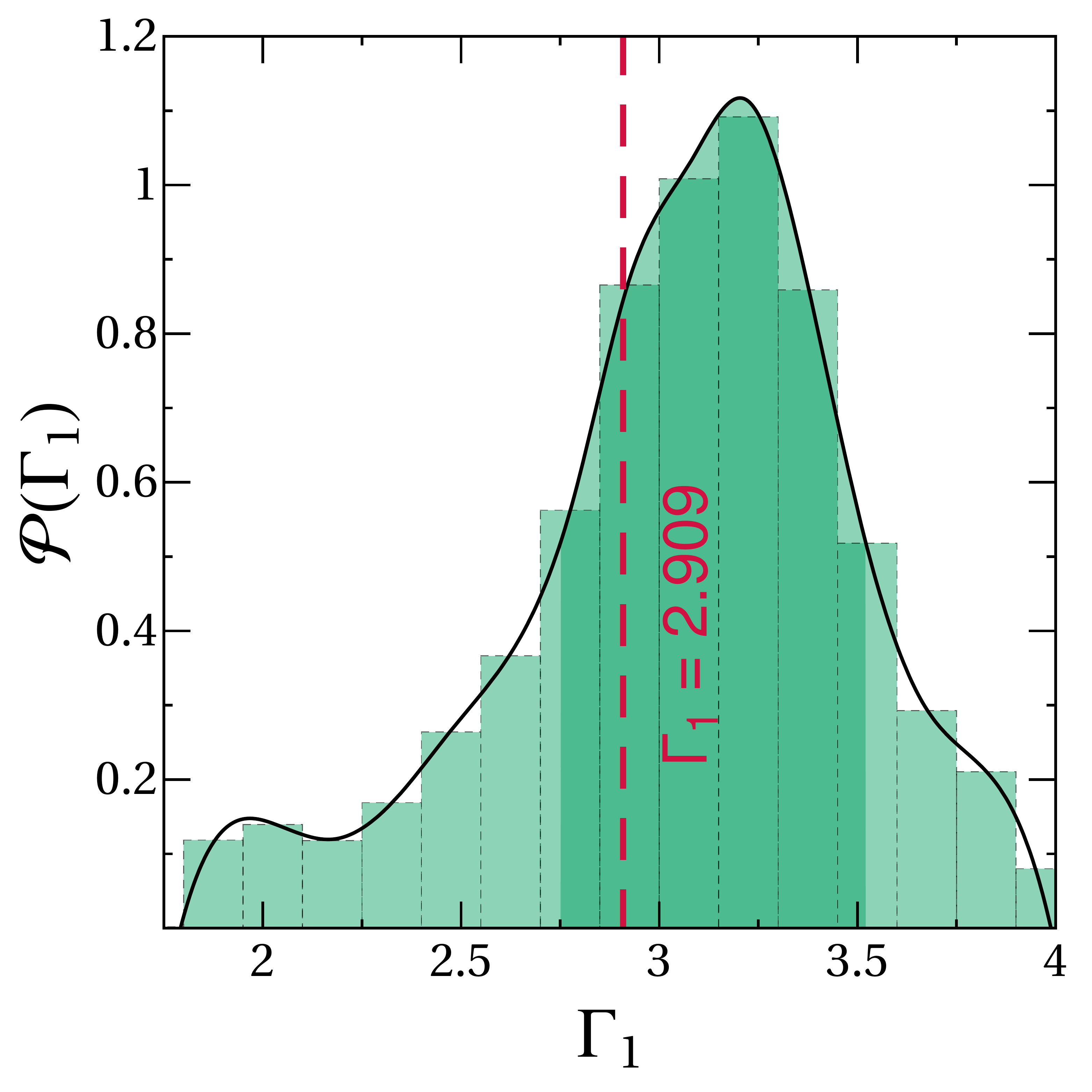}}
{\includegraphics[width=0.32\textwidth]{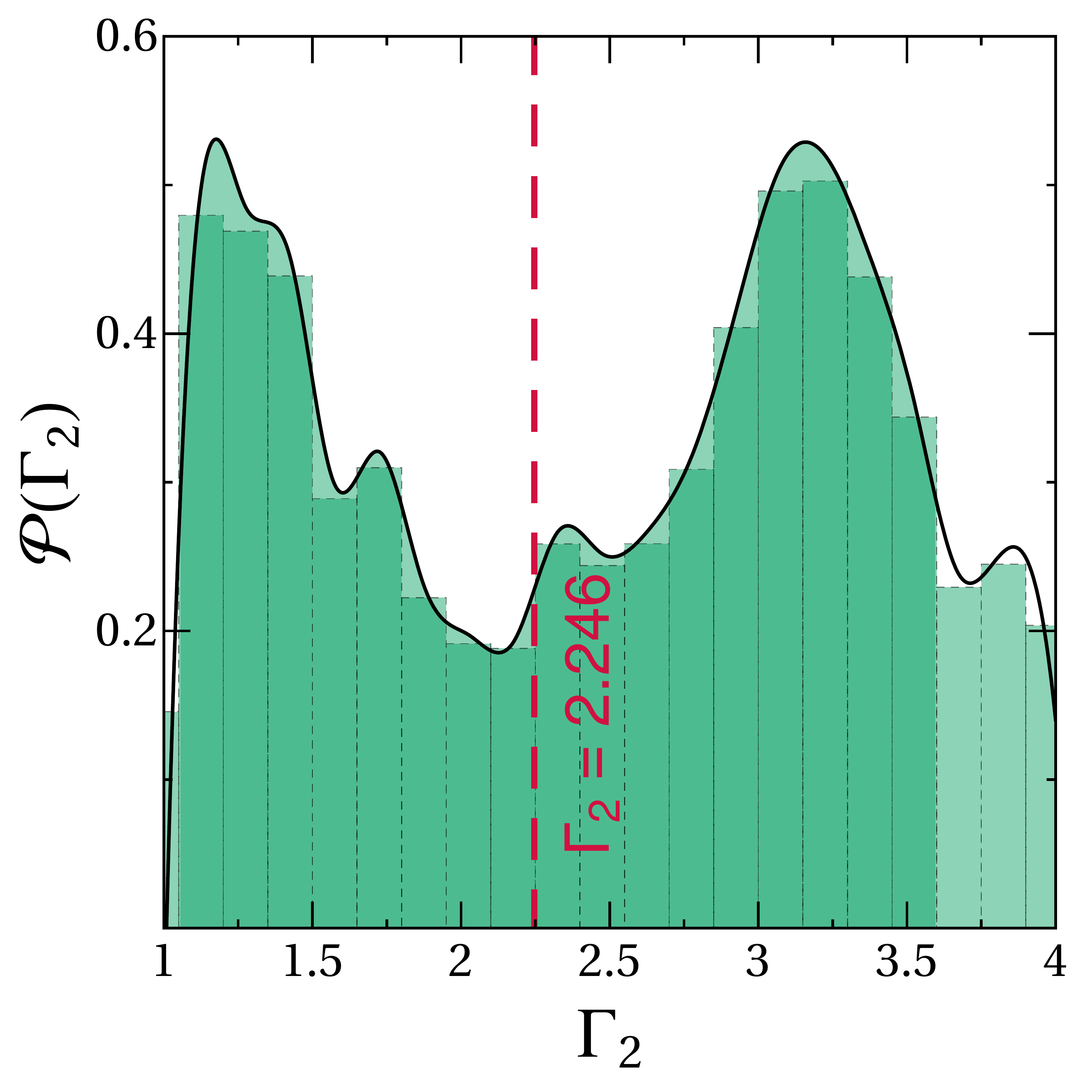}}
{\includegraphics[width=0.32\textwidth]{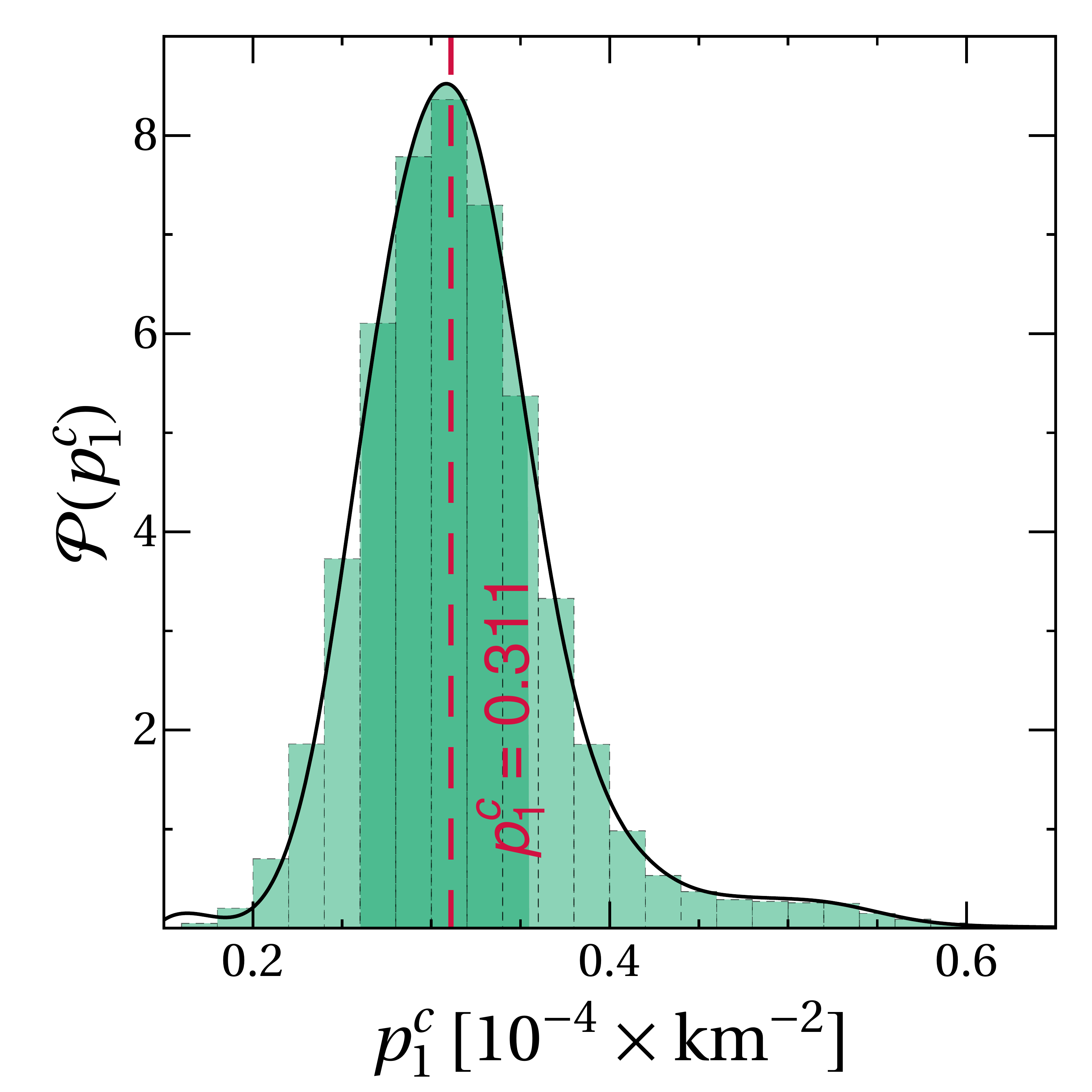}}  
{\includegraphics[width=0.32\textwidth]{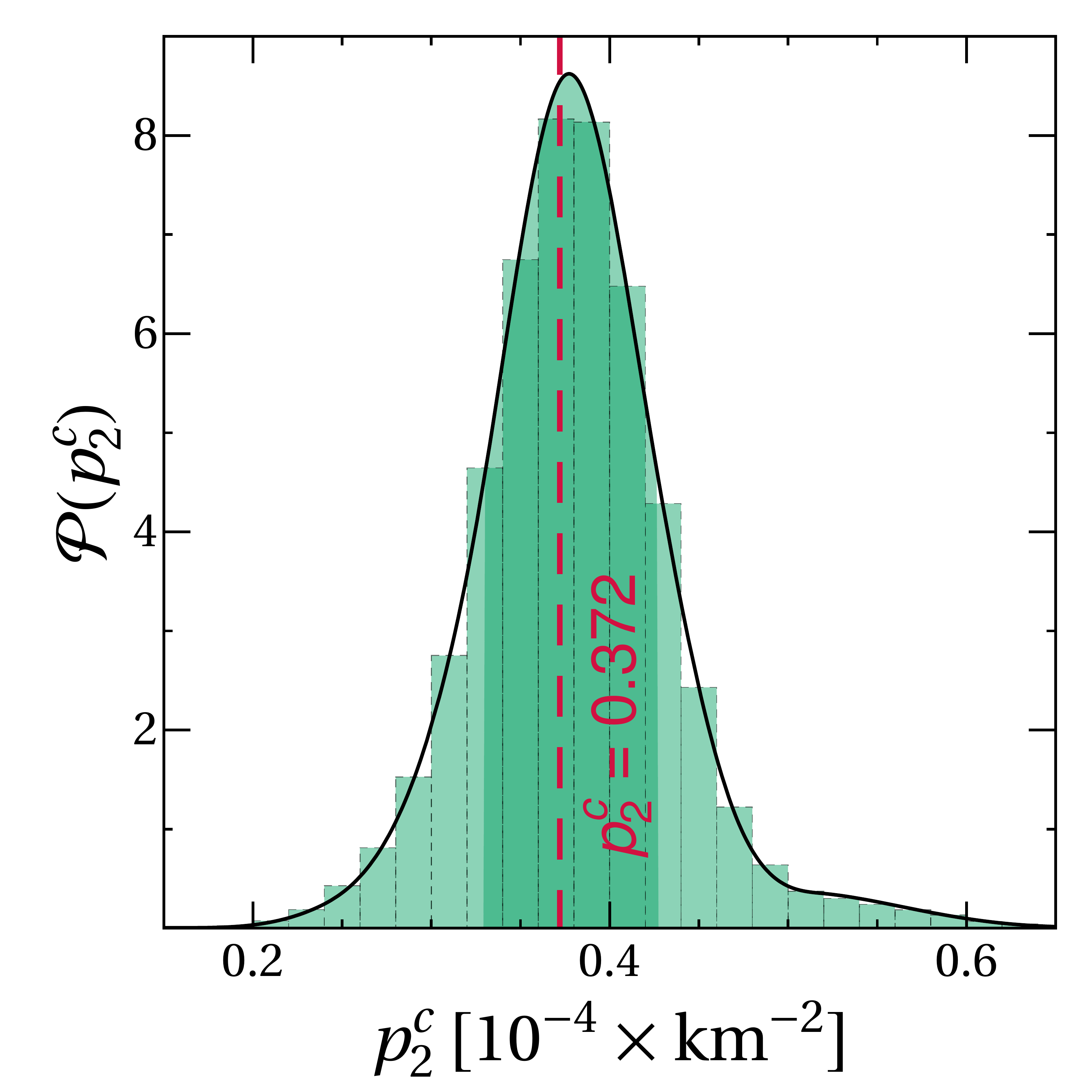}}
{\includegraphics[width=0.32\textwidth]{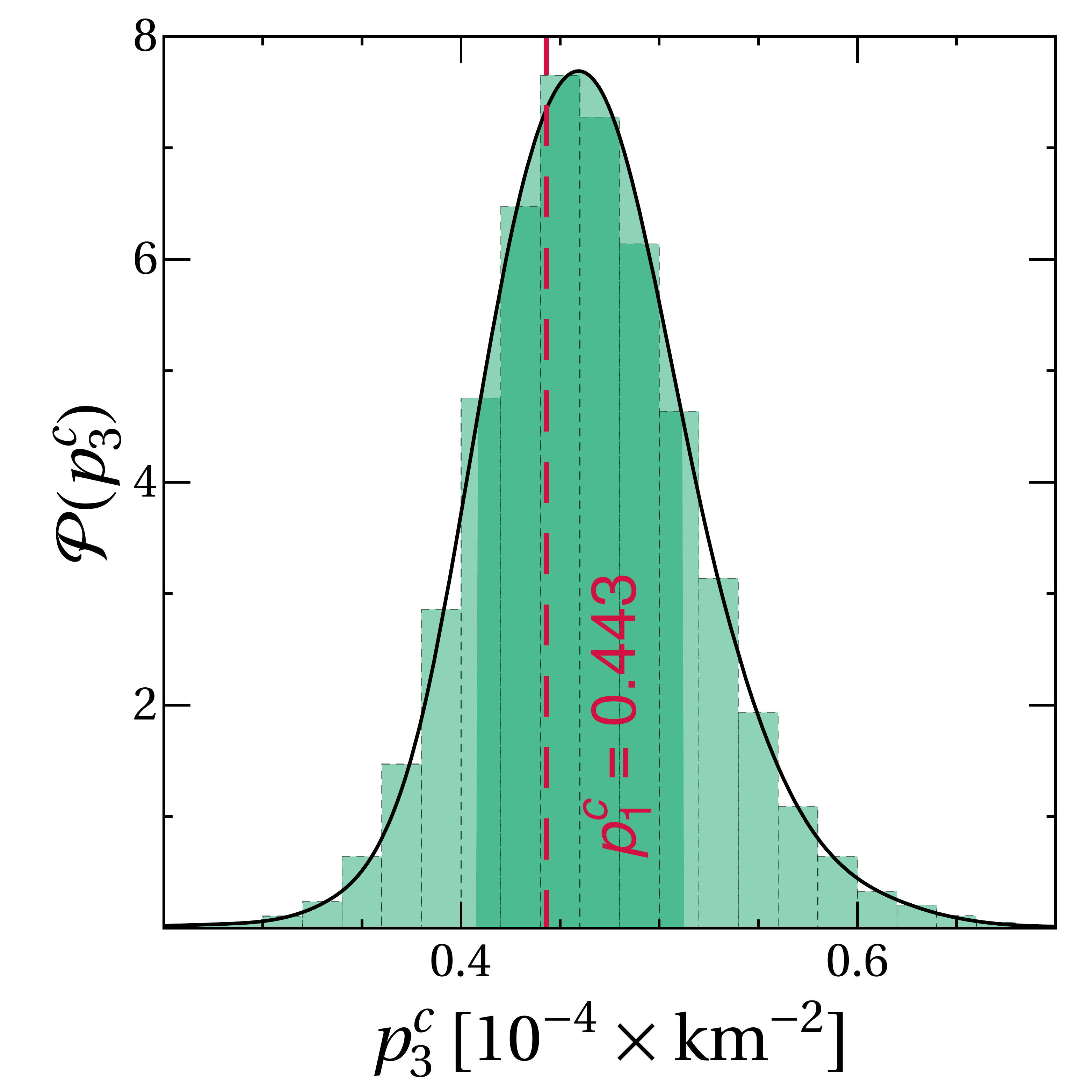}}
\caption{\textsl{Marginalized posterior PDF for the parameters of the \texttt{h4} EOS, derived for the \texttt{m123} model with neutron stars masses $(1.1,1.2,1.3)M_\odot$. The histograms of the sampled points are shown below each function. The red, dashed vertical lines identify the injected true values, while the shaded bands correspond to the $1\sigma$ credible regions of each parameter.}}
\captionsetup{format=hang,labelfont={sf,bf}}
\label{fig:m123h4}
\end{figure}

\begin{figure}[]
\captionsetup[subfigure]{labelformat=empty}
\centering
{\includegraphics[width=0.32\textwidth]{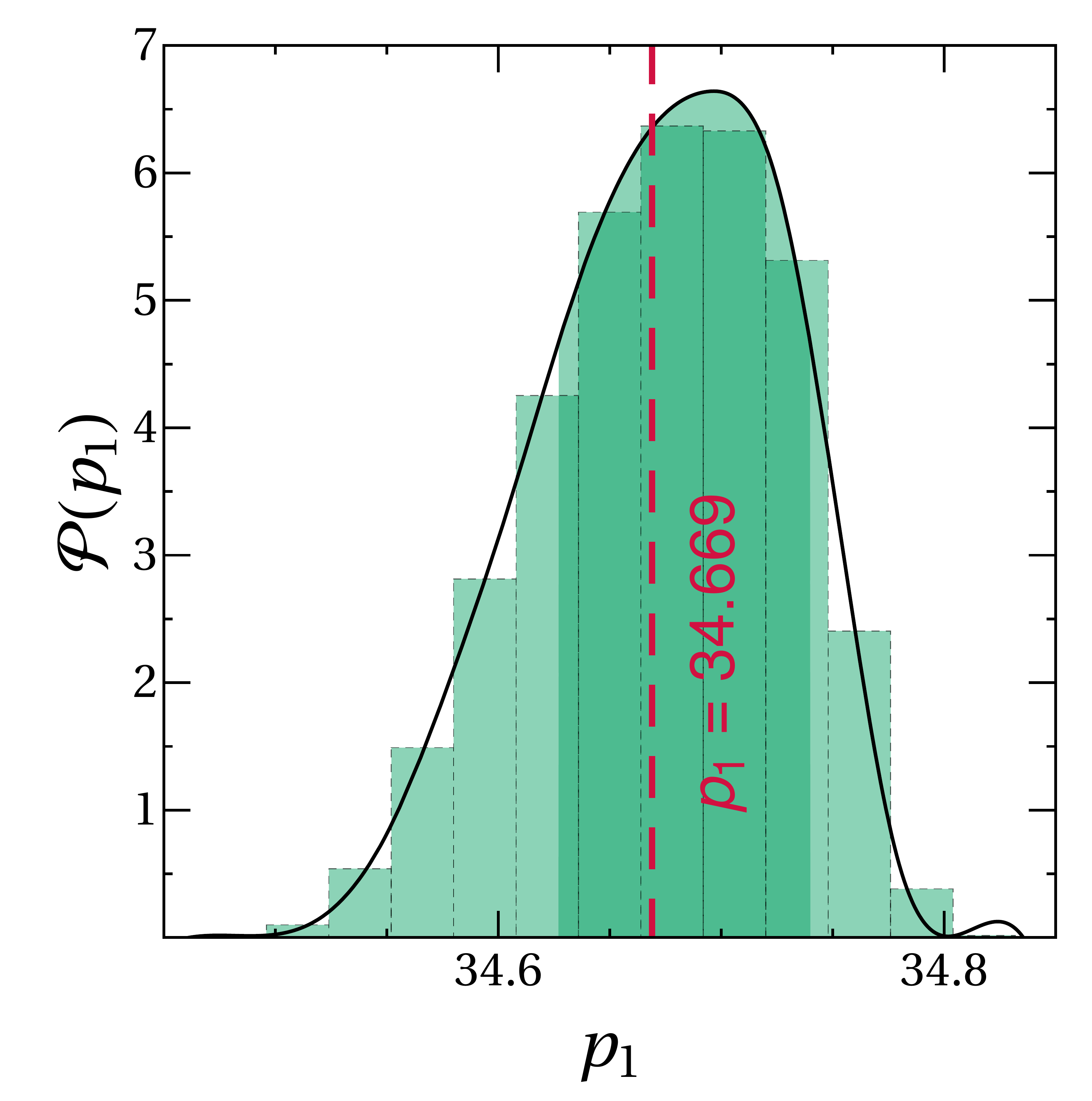}}  
{\includegraphics[width=0.32\textwidth]{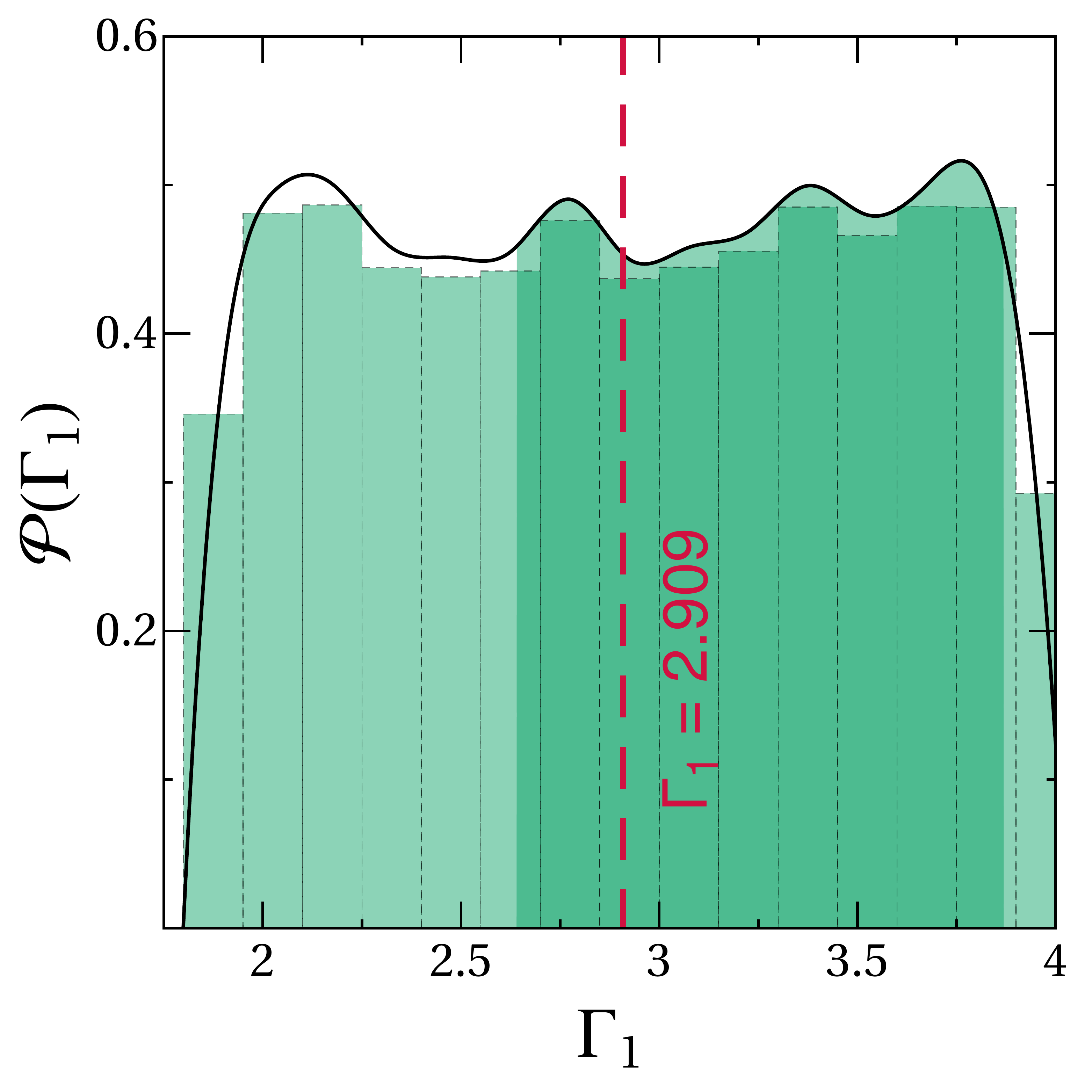}}
{\includegraphics[width=0.32\textwidth]{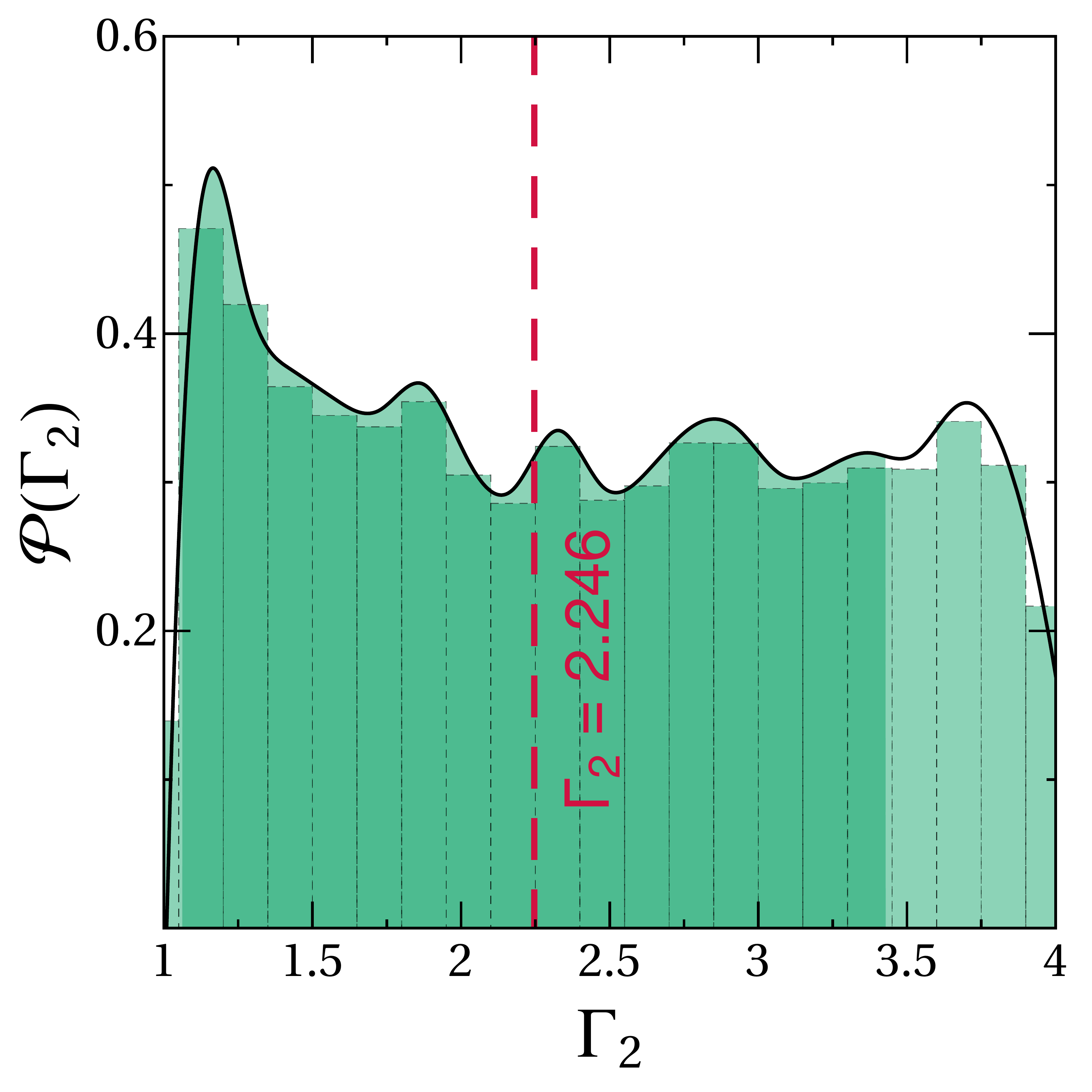}}
{\includegraphics[width=0.32\textwidth]{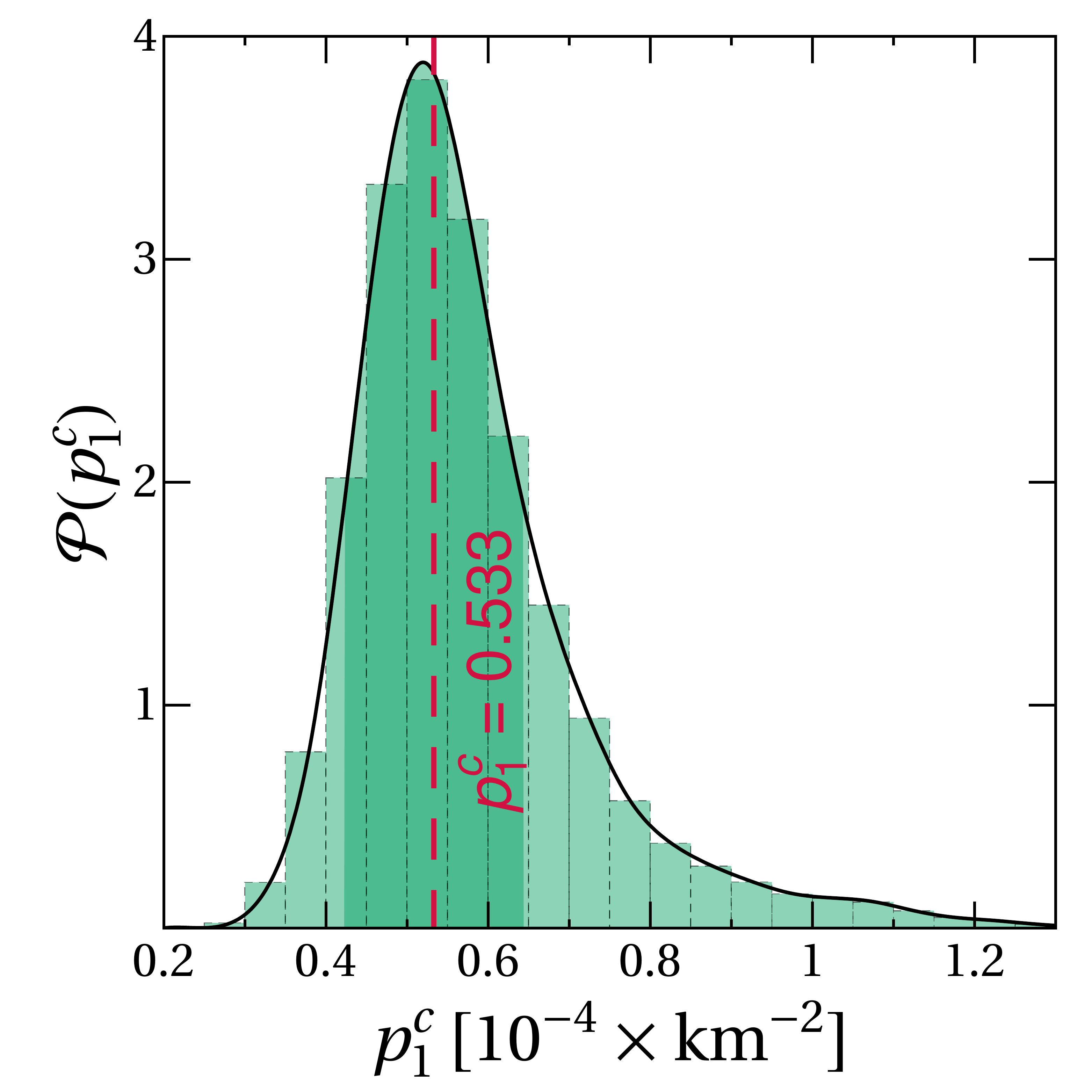}}  
{\includegraphics[width=0.32\textwidth]{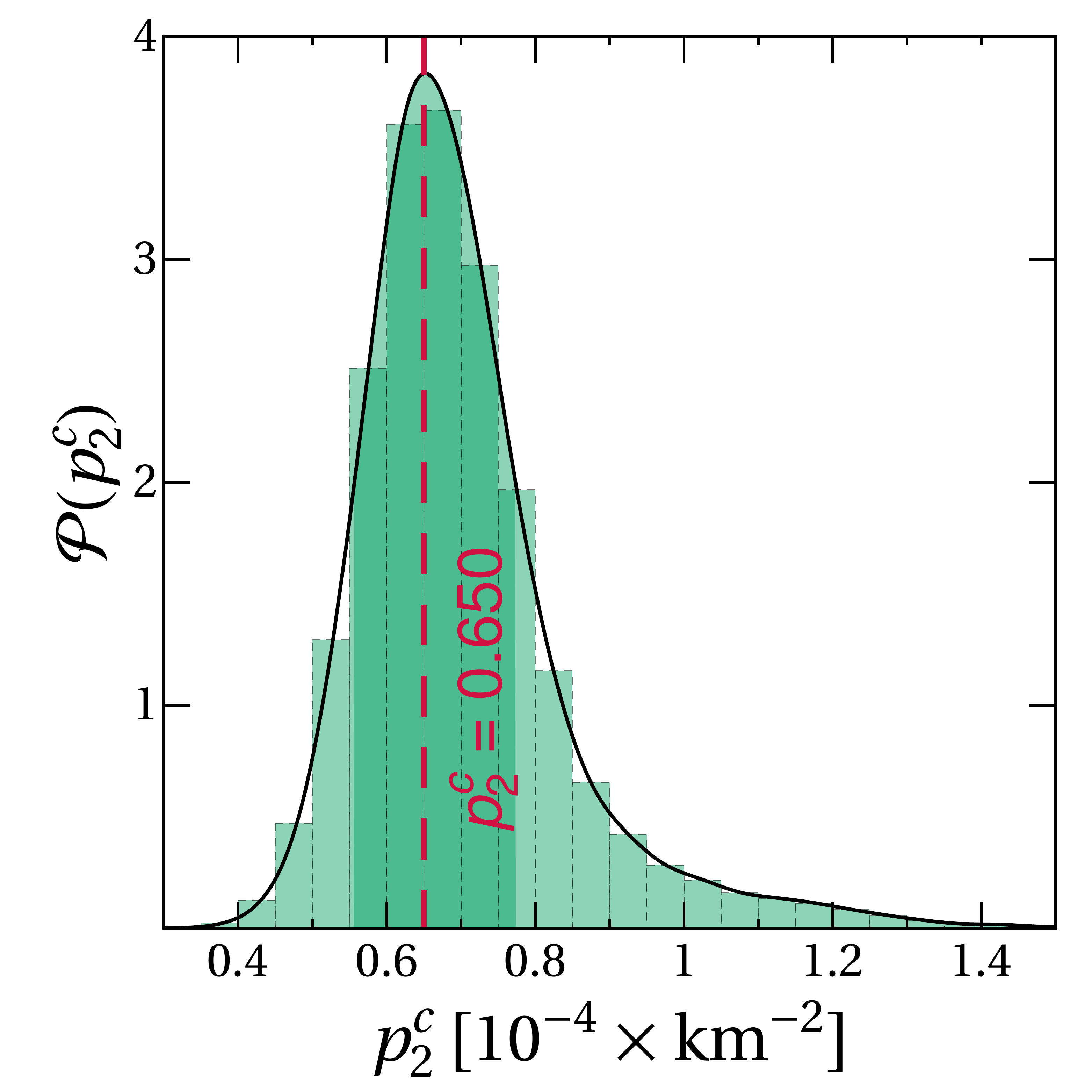}}
{\includegraphics[width=0.32\textwidth]{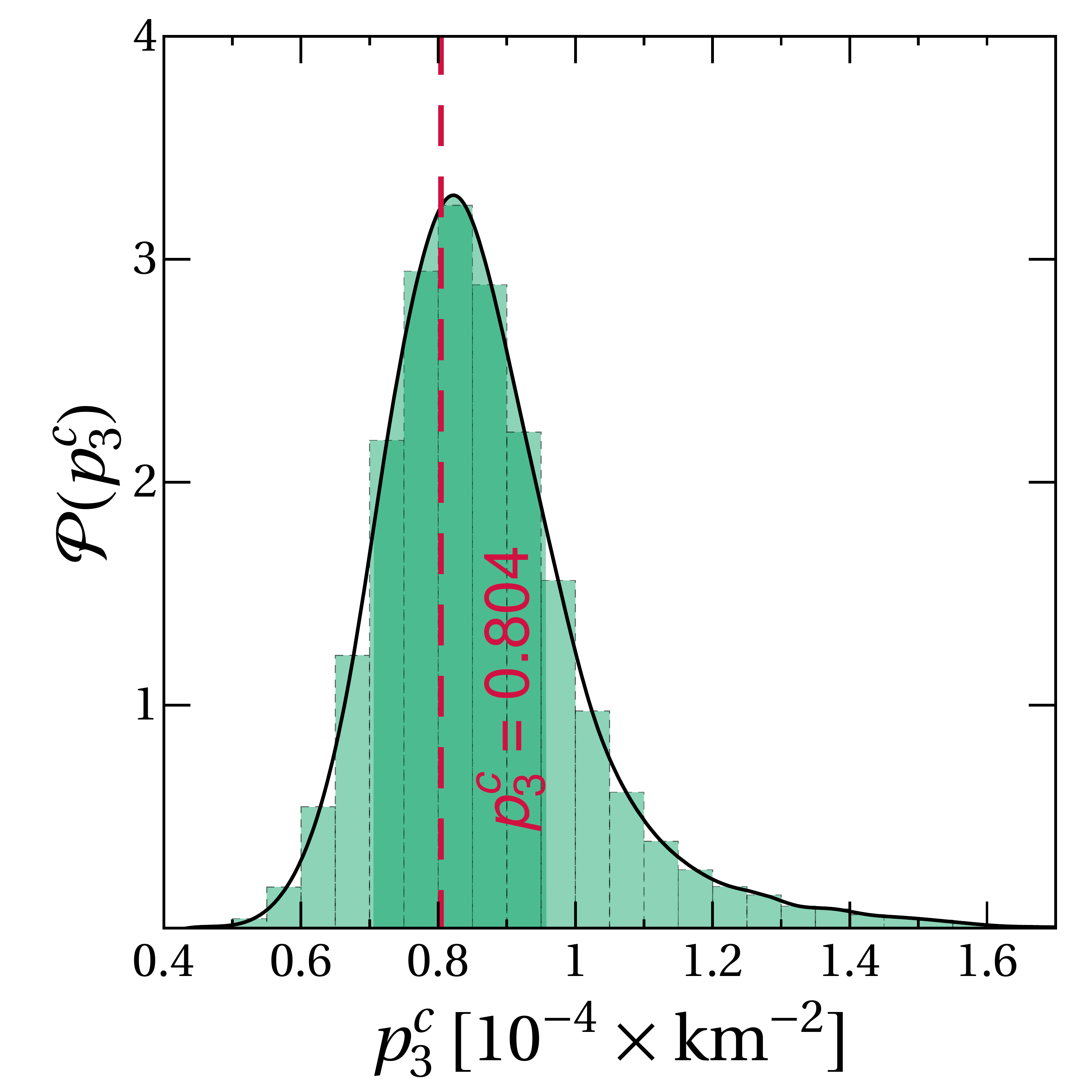}}
\caption{\textsl{Marginalized posterior PDF for the parameters of the \texttt{h4} EOS, derived for the \texttt{m456} model with neutron stars masses $(1.4,1.5,1.6)M_\odot$. The histograms of the sampled points are shown below each function. The red, dashed vertical lines identify the injected true values, while the shaded bands correspond to the $1\sigma$ credible regions of each parameter.}}
\captionsetup{format=hang,labelfont={sf,bf}}
\label{fig:m456h4}
\end{figure}

\begin{figure}[]
\centering
\includegraphics[width=0.99\textwidth]{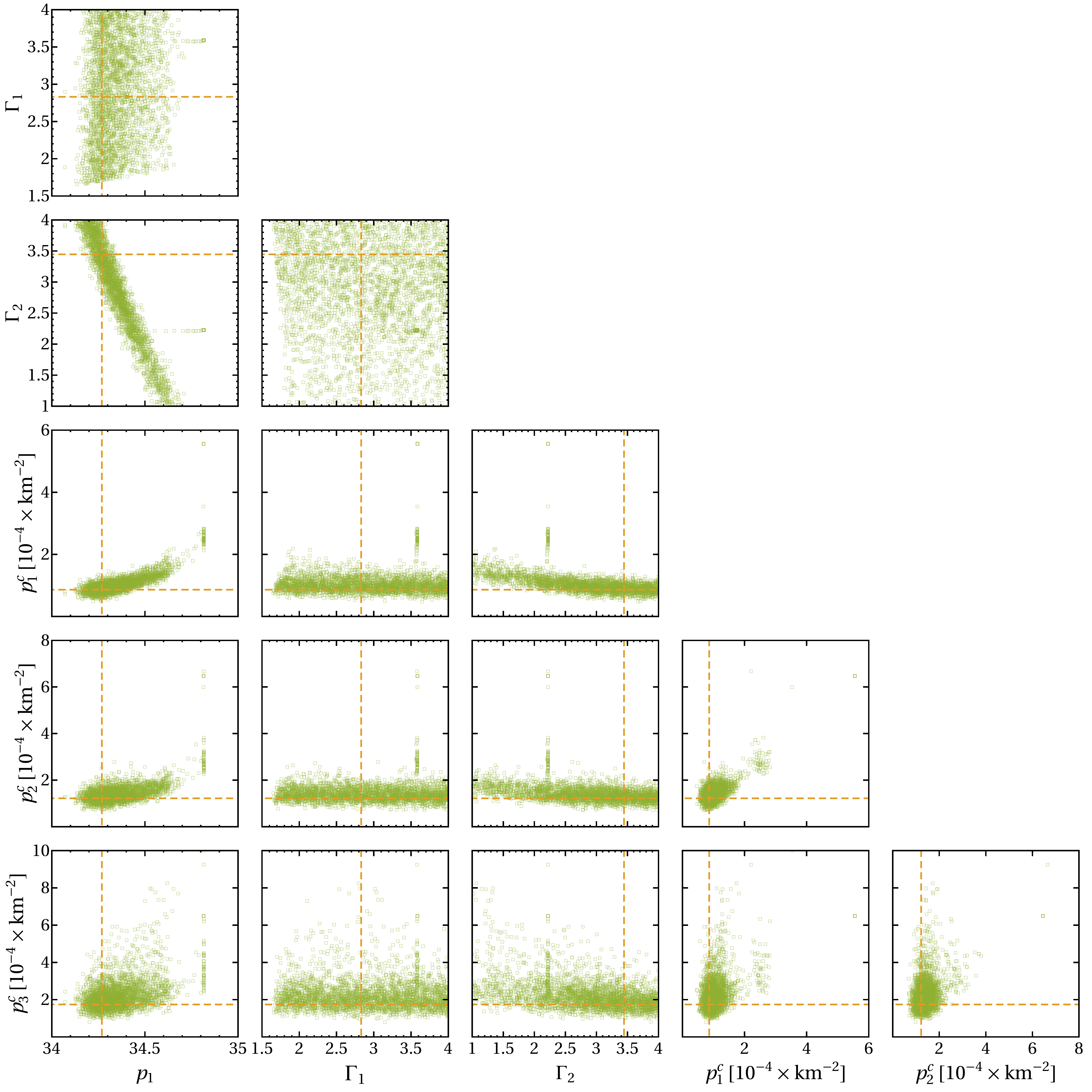}
\caption{\textsl{An example of the chains produced by the GaA algorithm for the model \texttt{m246} and the EOS \texttt{apr4}. The dashed lines denote the injected values.}}
\captionsetup{format=hang,labelfont={sf,bf}}
\label{fig:chain_apr4}
\end{figure}

\begin{figure}[]
\centering
\includegraphics[width=0.99\textwidth]{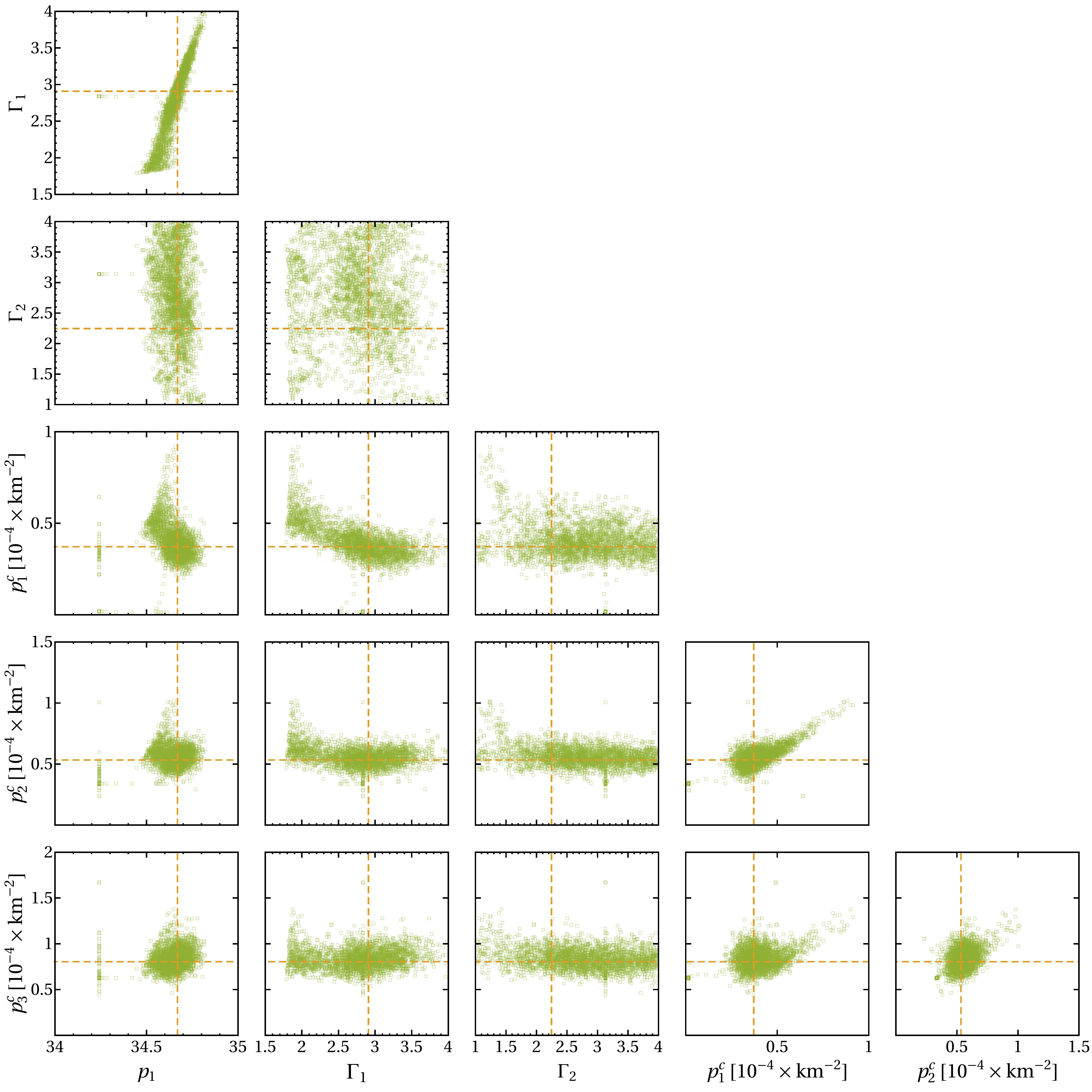}
\caption{\textsl{An example of the chains produced by the GaA algorithm for the model \texttt{m246} and the EOS \texttt{h4}. The dashed lines denote the injected values.}}
\captionsetup{format=hang,labelfont={sf,bf}}
\label{fig:chain_h4}
\end{figure}

In this appendix we report some additional plots of the results obtained through the MCMC simulations described in Chapter~\ref{sec:inverse}. In Figs.~\ref{fig:m123apr4}--\ref{fig:m456h4} we show the marginalized posterior distributions of the reconstructed parameters, as in Figs.~\ref{fig:m246apr4} and~\ref{fig:m246h4}, but for the configurations \texttt{m123} and \texttt{m456}. In Figs.~\ref{fig:chain_apr4} and~\ref{fig:chain_h4} we show two examples of the chains produced by the MCMC simulations using the GaA algorithm~\ref{mh}.

\cleardoublepage
\phantomsection
\addcontentsline{toc}{chapter}{\bibname}
\bibliography{biblio}
\bibliographystyle{h-physrev}

\end{document}